\documentclass[11pt,a4paper]{article}

\usepackage{authblk}
\usepackage{fancyhdr}

\usepackage[utf8]{inputenc}
\usepackage[T1]{fontenc}
\usepackage[margin=2cm]{geometry}
\usepackage{multicol}
\usepackage{parskip}

\usepackage{amsmath,amssymb}
\usepackage{empheq}
\usepackage{stmaryrd}

\usepackage{graphicx}
\usepackage{caption}
\usepackage{subcaption}
\usepackage{xcolor}
\definecolor{revisioncolor}{HTML}{A02030}

\DeclareRobustCommand{\revision}[1]{#1}%

\newcommand{\doubleboxeq}[1]{%
  \setlength{\fboxrule}{0.5pt}%
  \setlength{\fboxsep}{3pt}%
  \fbox{\setlength{\fboxsep}{2pt}\fbox{#1}}%
}%

\usepackage{booktabs}
\usepackage{tabularx}
\usepackage{array}

\newcolumntype{L}[1]{%
  >{\raggedright\arraybackslash\hsize=#1\hsize}X%
}

\usepackage{titlesec}

\usepackage{standalone}
\usepackage{tikz}
\usetikzlibrary{arrows.meta,calc,decorations.markings}

\usepackage{verbatim}

\newcommand{\showsectioncounts}[1]{%
  \immediate\write18{%
    texcount -merge -sub=section
    -sum=1,0,0,0,0,0,0 -q
    "#1.tex" > "#1-wcdetail.txt"
  }%
  \IfFileExists{#1-wcdetail.txt}{%
    \verbatiminput{#1-wcdetail.txt}%
  }{%
    \textbf{TeXcount report was not generated.
    Check the root-file name.}%
  }%
}

\usepackage[
  style=numeric-comp,
  sorting=none,
  backend=biber
]{biblatex}

\usepackage[
  colorlinks=true,
  linkcolor=blue,
  citecolor=blue,
  urlcolor=blue
]{hyperref}

\fancypagestyle{preprint}{
    \fancyhf{}
    
    \fancyhead[L]{%
        \color{gray}\small
        \textit{Bossard et al., arXiv:2608.00306}%
    }
    \fancyfoot[C]{\thepage}
}

\newsavebox{\EvidenceStatusHeader}

\titleformat{\section}
  {\large\bfseries}
  {\thesection.}
  {0.45em}
  {}

\titleformat{\subsection}
  {\normalsize\bfseries}
  {\thesubsection.}
  {0.45em}
  {}

\title{\vspace{-1.4cm}\Large\bfseries Mechanistic bridges from receptors to whole-brain dynamics: promise and limits of master-equation mean-field models}

\author[1,2,*]{Yannaël Bossard}
\author[3,4]{Lahna Bekri}
\author[2]{Alain Destexhe}

\affil[1]{\footnotesize Université Paris-Saclay, École Normale Supérieure Paris-Saclay, 91190 Gif-sur-Yvette, France.}

\affil[2]{\footnotesize Paris-Saclay Institute of Neuroscience (NeuroPSI), CNRS, Paris-Saclay University, Saclay, 91400, Île-de-France, France.}

\affil[3]{\footnotesize École Polytechnique, Institut Polytechnique de Paris, Palaiseau, 91120, Île-de-France, France.}

\affil[4]{\footnotesize Paris Brain Institute, Inserm, Sorbonne Université, CNRS, APHP, Hôpital de la Pitié-Salpêtrière, Paris, France.}

\affil[*]{\textit{Corresponding author:} \href{mailto:yannael.bossard@ens-paris-saclay.fr}{yannael.bossard@ens-paris-saclay.fr}}

\date{}

\begin{document}

\maketitle


\begin{center}
    \begin{minipage}{0.99\textwidth}
        \small
        \textbf{Abstract} -- Many pharmacological and pathological perturbations originate at molecular, synaptic, or cellular scales, yet their consequences are often observed through population and whole-brain signals. Bridging these scales requires reductions that remain tractable without severing the mechanisms relevant to the scientific question. This review asks which microscopic mechanisms remain explicit, interpretable, and testable after successive reductions, and what claims the resulting whole-brain models can support. We use the master-equation lineage culminating in a receptor-aware adaptive mean-field model as a worked case, tracing the passage from finite-size population statistics and semi-analytical transfer functions to conductance-based adaptive nodes and connectome-coupled dynamics \cite{el2009master,zerlaut2016heterogeneous,zerlaut2018modeling,di2019biologically,sacha2025computational}. \revision{Re-deriving its finite-size moment equations, we correct the original covariance equation, whose closure omitted a curvature term of the same order as the retained terms.} We critically compare this strategy with phenomenological neural masses, exact low-dimensional and population-density reductions, large-scale spiking models, and learned or hybrid surrogates, and examine computational work and memory traffic as additional benchmark dimensions. The analysis shows that \revision{effective excitatory and inhibitory synaptic kinetics}, conductance state, and spike-frequency adaptation can remain manipulable across scales, enabling mechanistically interpretable perturbations \revision{informed by pharmacology} and testable mesoscopic and macroscopic consequences. This continuity is conditional, however, on coarse-grained Markovianity, population homogeneity, quasi-stationary transfer functions, moment closure, regional uniformity, and measurement-specific observation models. First-order whole-brain implementations also discard endogenous covariance dynamics, and macroscopic agreement does not identify a unique molecular cause. Computationally, node-local biological detail mainly changes prefactors, whereas dense propagation of global covariances changes the scaling class. We therefore argue that cross-scale models should be judged by the intervention pathways and observables they preserve, their validity domain, identifiability, empirical adequacy, and computational burden. Receptor-aware mean fields are not a universal solution, and are framed not simply as a compromise between biological realism and computational cost, but as a transparent and tractable strategy for selected mechanistic questions when each link in the reduction chain is independently validated.
        \par\vspace{0.45em}
    \end{minipage}
\end{center}

\begin{center}
    \centering
    \includegraphics[width=0.95\linewidth]{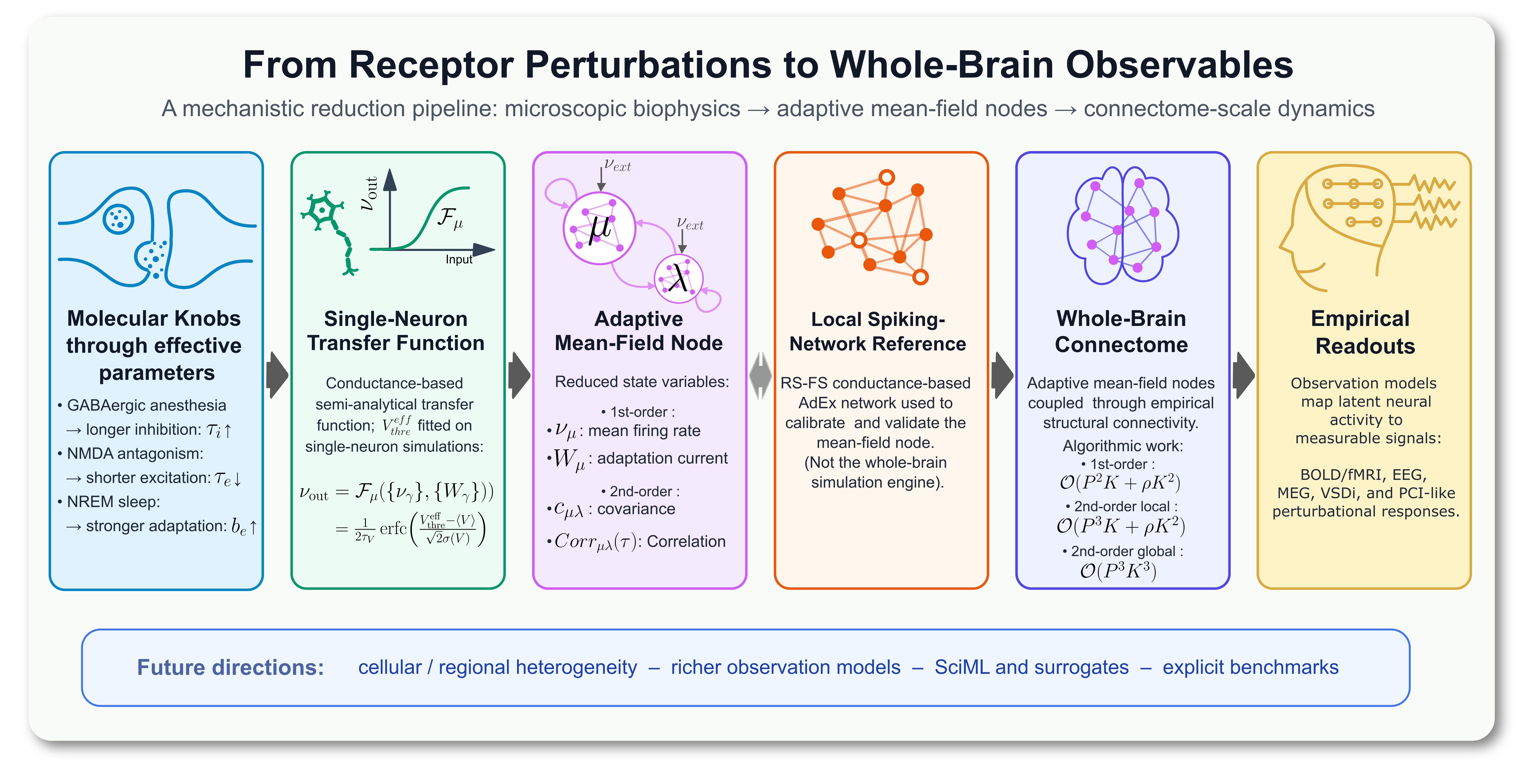}
    \vspace{0.01cm} 
\end{center}






\section{Introduction}

Understanding how microscopic neural mechanisms give rise to large-scale brain activity remains a central challenge in computational neuroscience. Detailed spiking-neuron and conductance-based circuit models can represent membrane dynamics, synaptic kinetics, adaptation, and cellular diversity, whereas EEG, MEG, fMRI, VSDi, and LFP report activity aggregated across populations, regions, or whole-brain networks. Cross-level computational integration therefore requires principled links among molecular and cellular mechanisms, collective network dynamics, and empirical observables. The challenge is not to maximize detail at every scale, but to preserve the mechanisms required by a given scientific question while enabling tractable analysis across scales.

This problem is especially consequential in pharmacology and brain-state transitions. Interventions of clinical or theoretical interest may alter excitatory or inhibitory receptor kinetics, spike-frequency adaptation, or neuromodulatory tone, while their consequences (including slow waves, altered responsiveness, perturbational complexity, and structure-function coupling) are measured at mesoscopic or macroscopic scales~\cite{sacha2025computational}. The central question of this review is therefore: which microscopic mechanisms remain interpretable and testable after successive reductions to population and whole-brain dynamics, and under which assumptions can the resulting models support mechanistic explanation or prediction?

Direct microscopic simulation is scientifically indispensable for many questions, but the cellular scale of the human brain is enormous~\cite{azevedo2009equal}; in practice, the dimensionality and computational cost of explicit simulations can make them poorly suited to extensive parameter exploration, inversion, or connectome-level embedding. Reduced population models instead replace explicit neuronal trajectories with a limited set of collective variables. This compression sacrifices microscopic detail but can make population dynamics analyzable and scalable. Because different reductions preserve different variables, fluctuations, and intervention pathways, however, they should be evaluated not only by their output but also by the biological information retained through the reduction.

We examine this issue through the receptor-aware whole-brain framework of Sacha et al.~\cite{sacha2025computational}, treated as a representative worked example rather than a complete or unique solution. This framework represents receptor-level interventions through effective changes in synaptic and cellular parameters, reduces a conductance-based spiking circuit to an adaptive mean-field node, and embeds these nodes in a connectome-driven model implemented in The Virtual Brain (TVB)~\cite{sanz2013virtual}. It thereby links GABAergic and NMDA-mediated synaptic kinetics and adaptation to whole-brain dynamics associated with wakefulness, anesthesia-like states, and NREM-like slow-wave activity~\cite{sacha2025computational}. Its scope is thus receptor-aware rather than molecularly exhaustive: detailed molecular dynamics and intracellular signaling are not modeled explicitly. These mappings are effective simplifications rather than one-to-one molecular descriptions, so their mechanistic interpretation remains conditional.

The adaptive mean-field node is the endpoint of a specific theoretical lineage. El Boustani and Destexhe~\cite{el2009master} formulated finite-size asynchronous irregular networks in terms of population mean activities and covariances. Zerlaut et al.~\cite{zerlaut2016heterogeneous} developed semi-analytical neuronal transfer functions from membrane-potential fluctuation statistics, then incorporated them into a conductance-based excitatory-inhibitory cortical mean-field model~\cite{zerlaut2018modeling}. Di Volo et al.~\cite{di2019biologically} subsequently introduced spike-frequency adaptation as an explicit mesoscopic dynamical variable. Sacha et al.~\cite{sacha2025computational} combined these elements in a whole-brain architecture. Following this lineage makes it possible to track, at each transition, the assumptions introduced, the quantities retained, and the information absorbed into effective functions or discarded.

This review has three aims. First, it reconstructs the logic and validity conditions of the master-equation reduction underlying this lineage. Second, it critically situates that lineage within the broader landscape of phenomenological neural masses, biophysically motivated and exact low-dimensional reductions, population-density and related finite-size approaches, and data-driven or hybrid surrogates~\cite{wilson1972excitatory,wong2006recurrent,brunel2000dynamics,montbrio2015macroscopic,schwalger2017towards,augustin2017low}. Third, it considers algorithmic work and memory traffic alongside simulator faithfulness, empirical adequacy, robustness, identifiability, and mechanistic traceability as distinct benchmark dimensions. The purpose is not to rank all mean-field approaches or advocate a universally optimal level of description, but to assess what this selected reduction strategy enables, where its interpretation becomes conditional, and when alternative models are preferable. \revision{Reconstructing this reduction also led us to correct its covariance equation. The original derivation~\cite{el2009master}, whose second-order equations were reused in the conductance-based and adaptive extensions~\cite{zerlaut2018modeling,di2019biologically}, evaluated the second jump moment of the master equation at the mean activity instead of averaging it over the activity distribution. This mean-value closure omits a curvature term of the same order as the retained terms. Section~\ref{subsec:second_order_closure} gives the corrected equation and summarizes its consequences; Supplementary~\ref{subsubsec:covariance_matrix} details its derivation, and Supplementary~\ref{subsubsec:covariance_closure_benchmarks} tests it against exact solutions of the master equation. The correction changes the predicted finite-size fluctuations, their stability condition and their transients. It leaves the form of the mean equation unchanged, and does not affect the first-order spatial and whole-brain models of the lineage~\cite{zerlaut2018modeling,sacha2025computational}.}

Section~\ref{sec:background_modeling_landscape} introduces the relevant modeling landscape and motivates the selected lineage; \revision{Section~\ref{subsec:under_hood_one_chain} gives a short roadmap of its reduction chain, which the subsequent sections trace from single-neuron transfer functions and finite-size mesoscopic dynamics to receptor-aware whole-brain embedding.} The final sections synthesize its limitations, complementary approaches, future extensions, and benchmarking requirements. Detailed derivations, model definitions, and computational-accounting assumptions are provided in three pedagogically structured Supplementary Notes, which make the reduction chain auditable and accessible across disciplinary backgrounds, including for readers without extensive mathematical or theoretical-physics training, while allowing the main text to remain a critical scientific review.

\section{Background: modeling scales and reduced-population approaches}
\label{sec:background_modeling_landscape}

Mechanistic brain models can be placed along a continuum of spatial and
dynamical resolution. At the microscopic end lie conductance-based and
morphology-aware spiking-neuron models, which represent membrane dynamics, synaptic kinetics, dendritic geometry, and cellular diversity with high biological realism. At the mesoscopic end lie neural-mass, neural-field, population-density, and mean-field models, which replace explicit spike-by-spike simulation by a small number of collective state variables. At the macroscopic end, such reduced nodes can be coupled through anatomical connectivity to form whole-brain dynamical systems. Here, \emph{whole-brain} does not necessarily refer to a model including every region, such as subcortical, cerebellar, or brainstem structures; rather it follows the terminology of the regional connectome employed, usually representing the cortex. Each level serves a distinct purpose: detailed spiking models are indispensable for questions of dendritic integration, synaptic microstructure, spike timing, or the synthesis of electrophysiological signals, whereas reduced population models are preferable for systematic parameter exploration, bifurcation analysis, model inversion, and large-scale connectome embedding. No class is universally
superior; each optimizes a different trade-off between realism,
interpretability, and tractability.

The framework of Sacha et al.~\cite{sacha2025computational} sits at the reduced end of this continuum: it embeds a biologically grounded, conductance-based mean-field node into a connectome-level whole-brain architecture. To situate it, this section does not attempt an exhaustive survey of reduced models, but is organized around three questions: why reductions remain necessary even as spiking simulations grow more powerful (Sections~\ref{subsec:background_snn_capabilities}-\ref{subsec:background_why_reduced_models}); what the main families of reduced models are (Section~\ref{subsec:background_reduced_model_families}); and why this review focuses specifically on the master-equation lineage (Section~\ref{subsec:background_why_master_equation}).

\subsection{What large-scale spiking network models already enable}
\label{subsec:background_snn_capabilities}

Detailed spiking models should not be dismissed as merely local toy systems. Driven by advances in high-performance computing, such as large CPU clusters, GPUs, or dedicated supercomputing resources, mechanistic simulations now reach organism-, microcircuit-, and even hemisphere-scale regimes. BAAIWorm couples a biophysically detailed model of a \textit{C.\ elegans} sensorimotor circuit to a closed-loop body-environment simulation \cite{zhao2024integrative,sarma2018openworm}; in \textit{Drosophila},
Shiu et al.~\cite{shiu2024drosophila} built a brainwide leaky integrate-and-fire model from the FlyWire connectome ($\sim10^5$ neurons, $5\times10^7$ synapses)~\cite{dorkenwald2024neuronal}; and human-hemisphere models now combine layered cortical columns, diffusion-imaging-based long-range connectivity, millions of neurons and billions of synapses~\cite{pronold2024multi}.

These simulations can generate synthetic data, test mechanistic hypotheses, validate analysis pipelines, and reveal how anatomical structure shapes dynamics. Yet they also expose the central practical limitation of explicit spiking approaches: even when feasible, they are computationally heavy, high-dimensional, difficult to invert, and poorly suited to repeated parameter sweeps or subject-specific fitting. The cost is concrete: the human-hemisphere model of Pronold et al.~\cite{pronold2024multi} requires on the order of $200$ core-hours on a $768$-core machine to produce only $10$\,s of biological activity. For $10$\,min of biological activity, the biophysically detailed model of rat primary somatosensory cortex of Laquitaine et al. \cite{laquitaine2024spike,reimann2026modeling,isbister2026modeling} requires approximately $6$\,days on large clusters containing 120 CPUs. The lesson for this review is therefore not that spiking models are inadequate, but that their growing power does not remove the need for reduced descriptions: the challenge is to retain enough biological meaning to study receptor-to-whole-brain mechanisms while shedding the full cost of spike-by-spike simulation. Further detail is given in
Supplementary~\ref{subsec:appendix_background_snn_capabilities}. These costs do not define an intrinsic ranking: small regional LIF networks can approach the algorithmic cost of elaborate reduced nodes, depending on scale, implementation, and target accuracy (Section ~\ref{subsec:algorithmic_simulation_cost}).

\subsection{Why reduced population models are especially relevant here}
\label{subsec:background_why_reduced_models}

Computational cost is only one reason to reduce. Neurons embedded in cortical networks receive fluctuating synaptic bombardment, and cortical responses show substantial trial-to-trial variability~\cite{faisal2008noise}. For many questions, reproducing each spike sequence is neither possible nor the relevant criterion of adequacy. A model may fail to predict microscopic trajectories yet reproduce the population distribution, variability, mean membrane potential, synaptic gating, or adaptation relevant to the phenomenon. Conversely, a detailed trajectory can be insufficient if the population statistic of interest is wrong. Adequacy therefore depends on the empirical quantity being explained, not simply on the amount of represented detail.

Reduced models formalize this change of target by replacing individual trajectories with variables such as firing rate, membrane-potential statistics, neuronal or refractory age, synaptic gating, adaptation, and finite-size covariances. These variables can expose collective mechanisms and are closer in scale to aggregate measurements, but are not directly observed. EEG, MEG, LFP, VSDi, and BOLD pool and biophysically filter neural activity; an observation model is therefore generally required to map latent population dynamics onto data. Scale proximity does not by itself guarantee empirical adequacy.

This intermediate resolution is relevant here because the question is how local synaptic kinetics, receptor timescales, and adaptation alter mesoscopic and whole-brain state transitions. Sacha et al.~\cite{sacha2025computational} provide one implementation, but the requirement is general: retain the intervention pathway, latent state variables, and observation mapping needed to relate microscopic perturbations to macroscopic consequences without simulating an explicit spiking connectome.

\subsection{Canonical families of reduced population models}
\label{subsec:background_reduced_model_families}

Reduced population models differ in their state variables, derivation assumptions, biological aims, and mathematical form. We distinguish five families for orientation, but they are neither mutually exclusive nor orthogonal. Some labels describe a modeling philosophy, others a mathematical formalism, and several approaches are formally related, particularly Fokker--Planck and master-equation descriptions. This pragmatic grouping is intended to prevent misleading equivalences while locating the lineage reviewed below. 

\subsubsection{Phenomenological neural-mass models: Wilson--Cowan}

Wilson and Cowan~\cite{wilson1972excitatory} modeled interacting excitatory and inhibitory populations using low-dimensional activity variables. Their equations describe each population's first-order relaxation, with characteristic time constants, toward nonlinear gain functions of weighted recurrent and external inputs [Eq.~\eqref{eq:wilson_cowan_ODEs}; Supplementary~\ref{subsec:wilson_cowan_model}]. The formulation is phenomenological: response functions and coupling parameters are specified top-down rather than derived from a particular spiking circuit. This makes the models analytically transparent, inexpensive, and readily scalable to spatial fields or connectome networks. They capture generic excitation-inhibition competition, multistability, oscillations, and propagation, but their parameters map only indirectly onto cellular biophysics; receptor- or conductance-level interpretations therefore require additional assumptions.

\subsubsection{Biophysically motivated neural masses: Wong--Wang type reductions}

Wong--Wang-type models~\cite{wong2006recurrent} retain a more specific link to an underlying recurrent spiking circuit through a separation of timescales. Because NMDA-receptor gating is slow relative to AMPA, GABA, and spike generation, fast variables are approximated by quasi-steady-state values and slow NMDA-mediated gating variables become the principal dynamical coordinates. Population input is converted through an effective current-to-rate relation, and the resulting rate drives the gating dynamics. The reduction, detailed in Supplementary~\ref{subsec:Wong_Wang_model}, thus yields ODEs for synaptic gating rather than for firing rates alone.

This compact reduction preserves slow recurrent integration and attractor dynamics and has been widely used for decision-making, working memory, and whole-brain or BOLD-scale dynamics. It is not a comprehensive conductance-level description: the adiabatic approximation privileges the target slow process, assumes that faster variables follow it sufficiently rapidly, and does not retain the microscopic fluctuation structure except through any phenomenological noise explicitly added.

\subsubsection{Exact low-dimensional reductions: Montbrió--Pazó--Roxin}

The Montbrió--Pazó--Roxin (MPR) formalism provides an exact low-dimensional reduction of globally coupled quadratic integrate-and-fire (QIF) neurons~\cite{montbrio2015macroscopic}. Under its defining assumptions (QIF dynamics, a Lorentzian distribution of neuronal excitabilities, global coupling, and the thermodynamic limit) the population is described exactly by two ODEs for firing rate and mean membrane potential; the derivation is summarized in Supplementary~\ref{subsec:MPR_model}. These variables jointly retain rate and voltage dynamics, including fast transients that rate-only descriptions may miss. Exactness means equality to that specified infinite-network model, not an assumption-free reduction of arbitrary spiking circuits.

This microscopic-macroscopic correspondence is powerful for bifurcation analysis and collective-state theory. Its limitation is model specificity: the standard closure relies on QIF structure and commonly used current-based, effectively instantaneous interactions, and does not automatically extend to conductance-based synapses, finite networks, or receptor-aware mechanisms. Exponential integrate-and-fire descriptions can also better capture the experimentally inferred voltage dependence of spike initiation than a quadratic form~\cite{badel2008dynamic}. More detailed neuron and synapse models therefore require extensions or other closures, and finite-size fluctuations must be introduced beyond the thermodynamic-limit system.

\subsubsection{Population-density and Fokker--Planck approaches}
\label{subsec:background_fokkerplanck}

Population-density approaches track a probability density over neuronal states rather than selecting a few moments. The state may be membrane potential $V$ alone or include conductances, adaptation, and refractory or neuronal age. For many weak, approximately independent inputs, a diffusion approximation replaces discrete synaptic events by stochastic drift and diffusion, yielding a Fokker--Planck equation for $P(V,t)$~\cite{longtin2010stochastic}. With threshold, reset, and refractory boundary conditions, firing rate is the probability flux through threshold. Self-consistency between input statistics and this flux can yield stationary rates and state diagrams, as in Brunel~\cite{brunel2000dynamics}; alternatively, the full density can be evolved to study transient redistribution through state space. See Supplementary~\ref{subsec:fokker_planck_reminder}.

These approaches retain rich distributional information that moment or rate models discard and remain close to stochastic single-neuron dynamics~\cite{ccetin2026deterministic}. They can resolve threshold occupancy, refractoriness, and multimodal or strongly non-Gaussian state distributions without assuming that two moments suffice. Their main cost is dimensionality: the PDEs require nontrivial boundary conditions, and the state space expands rapidly with adaptation, conductances, or other cellular variables. Low-dimensional reductions~\cite{augustin2017low} exchange some retained information for efficiency. Related refractory-density approaches organize neurons by time since their last spike and, in finite-size formulations, restore demographic fluctuations absent from infinite-population densities~\cite{schwalger2017towards}. Density methods are thus appropriate when the evolving neuronal-state distribution is itself informative.

\subsubsection{Master-equation-based stochastic population models}

Master equations describe probability evolution over stochastic neural states and form a broad tradition rather than a single model. They are connected to Fokker--Planck descriptions: under suitable conditions, second-order truncation of a Kramers--Moyal expansion yields a Fokker--Planck equation. Operationally, population-density methods emphasize continuous neuronal-state densities, often in a diffusion or infinite-size limit, whereas population master equations can use discrete finite-size activity, spike counts, or active-neuron fractions. The latter naturally retain demographic fluctuations and correlations.

Ohira and Cowan~\cite{ohira1993master} treated population activity as a stochastic process; later field-theoretic and system-size-expansion methods derived correlations and higher-order finite-size corrections around deterministic activity equations~\cite{buice2007field, buice2010systematic, bressloff2010stochastic}. Variables across this tradition include binary activity, active fractions, spike counts, and spatial neural fields, and closures range from stochastic differential equations to hierarchies of activity cumulants. Large-population limits commonly recover deterministic rate or field equations, whereas finite-size formulations retain covariance or higher-order fluctuations. The tradition is therefore defined by a probabilistic evolution law, not by the particular closure used below.

    The El Boustani--Destexhe approach is one biologically grounded branch of this tradition. Over a finite time bin, a representative neuron's transfer function specifies its firing probability conditional on the current population activity. Conditional independence and within-population homogeneity then give a finite-size transition kernel for population spike counts. The associated master equation is expanded and closed at second order, yielding equations for mean activities, covariances, and stationary lagged correlations while discarding higher moments. The transfer function maps synaptic-input statistics to stationary output rate and provides the microscopic closure (Supplementary~\ref{subsec:transfer_funtion_ME}). Because the final system propagates moments through ODEs rather than a full density, it lies close to deterministic ODE population models in \c{C}etin et al.~\cite{ccetin2026deterministic}, while retaining finite-size covariance dynamics. Its tractability thus comes from moment closure, not from solving the complete master equation. Its relevance here is the combination of finite-size population dynamics with semi-analytical transfer functions for conductance-based neurons: the master equation supplies mesoscopic stochastic structure, while the transfer function carries selected neuronal and synaptic biophysics. This pairing enabled the conductance-based, adaptation-aware, and whole-brain extensions of Zerlaut, Di Volo, and Sacha. We follow this branch without equating it with the entire master-equation tradition.

\subsection{Why this lineage, and the scope of the review}
\label{subsec:background_why_master_equation}

The El Boustani--Zerlaut--Di Volo--Sacha lineage exposes a traceable sequence from finite-size population statistics, through conductance-based transfer functions and explicit adaptation, to connectome-coupled dynamics. It is therefore relevant when synaptic kinetics, conductance state, or adaptation must remain interpretable after reduction. Its local mechanisms enter through explicit or fitted physiological parameters, its dynamics can be analyzed and validated against reference spiking neural networks (SNNs), and its nodes remain compact enough for whole-brain implementation. These properties provide biological informativeness without requiring every microscopic degree of freedom.

This choice does not imply universal superiority; the preceding families and SNNs remain preferable when their retained variables match the question. In particular, a first-order mesoscopic node cannot represent all finite-size correlations, spike-time structure, cellular heterogeneity, dendritic computation, or receptor and intracellular mechanisms. Other master-equation, stochastic neural-field, and field-theoretic formulations are outside the principal scope because the review follows the path leading specifically to the receptor-aware whole-brain node, rather than surveying nor ranking stochastic population theory exhaustively. Detailed molecular dynamics, morphology-resolved neurons, and predominantly data-driven surrogates are likewise not components of this lineage, although complementary approaches are considered later.

Two criteria delimit the analysis: the assumptions introduced by reduction must be identifiable, and enough microscopic structure must remain to formulate and test the intervention. Here the assumptions include coarse-grained Markovianity, quasi-stationary responses, conditional independence and within-population homogeneity, diffusion-like bombardment in the transfer function approximation, and second-order moment closure. Conductance and adaptation parameters can therefore influence population dynamics and support mechanistic perturbations, but remain effective descriptions rather than one-to-one molecular mappings. In fact, mesoscopic biological informativeness is not equivalent to maximal detail or empirical validation. \revision{Although the El Boustani--Destexhe derivation includes activities and covariances, the first-order spatial and whole-brain formulations discussed here omit explicit covariance dynamics for tractability \cite{zerlaut2018modeling,goldman2023comprehensive,sacha2025computational}.} The derivation remains relevant because it states the node's validity conditions and what first-order truncation discards. We evaluate this lineage as a conditional cross-scale strategy, not a complete account of neural dynamics.

\section{Overview of the modeling chain}
\label{subsec:under_hood_one_chain}

The whole-brain framework of Sacha et al.~\cite{sacha2025computational} connects microscopic neuronal biophysics to macroscopic brain activity through a sequence of reductions and embeddings. Its purpose is not to reproduce every neuron, but to retain enough cellular and synaptic detail for molecular perturbations to remain interpretable, while reducing the dynamics enough to simulate many interacting brain regions. Each step gains tractability at the cost of assumptions that delimit where the resulting model can be trusted. \revision{The chain is best justified for asynchronous irregular activity, with irregular firing, weak synchrony and rapidly decaying population correlations. Throughout, $\mu,\nu\in\llbracket 1,P\rrbracket$ index neuronal populations, each containing $N_\mu$ neurons.}

\revision{%
\begin{enumerate}
\item \textbf{Transfer function} (Section~\ref{sec:semi-analytical_TF}). When a neuron receives many weak, approximately independent synaptic inputs, its stationary firing rate can be approximated from a statistical description of these inputs, without tracking individual input spike times~\cite{brunel2000dynamics}. For specified input statistics and fixed neuronal and synaptic parameters, the transfer function $\mathcal{F}_\mu$ is this static input-output map: it gives the stationary output rate of a neuron of population $\mu$ as a function of the presynaptic population rates~\cite{zerlaut2018modeling}. It is the first reduction of the chain. The membrane, channel and synaptic dynamics of a neuron are summarized by one function, through which the selected cellular and synaptic parameters still act. For conductance-based neurons, it is obtained semi-analytically: the input rates determine the mean, amplitude and correlation time of the membrane-potential fluctuations, and a threshold-crossing formula, with an effective threshold fitted to single-neuron simulations or recordings, maps these statistics to an output rate.

\item \textbf{Mesoscopic state and master equation} (Section~\ref{sec:MFT_NN_formalism}). The state of population $\mu$ is its activity $m_\mu$: the fraction of its neurons that fire within a time bin $T$, divided by $T$. Given the current state, neurons are assumed to fire independently, each with probability $T\mathcal{F}_\mu$ during the next bin, so that the number of active neurons follows a binomial law. This law, the transition kernel, specifies how the population state jumps from one bin to the next. The master equation describes how the probability distribution of the population state evolves under these jumps.

\item \textbf{Moment closure} (Section~\ref{subsec:second_order_closure}). Rather than evolving the full distribution, the formalism follows its mean $\langle m_\mu\rangle$ and covariance $c_{\mu\nu}$. Their equations involve the expected change and the expected product of changes of the activity per unit time, i.e.\ the drift and diffusion of the jump process. Expanding these around the mean and neglecting higher-order moments closes the system. The resulting second-order closure retains finite-size fluctuations and is accurate when the activity distribution is narrow; neglecting covariance feedback gives the first-order rate model (e.g., for infinite populations near a stable stationary state in the regular weak-fluctuation regime).

\item \textbf{Adaptation variable} (Section~\ref{sec:DiVolo_Adaptation_MFT}). A stationary transfer function has no memory. It cannot reproduce effects that depend on past activity, such as the prolonged hyperpolarization that follows a strong stimulus, or the slow alternation between active and silent phases (Up-Down states). In cortical neurons, such effects involve slow spike-frequency adaptation currents. The population-averaged adaptation current $W_\mu$ is therefore added to the state. It changes little within one bin $T$, is driven by the population activity, and lowers excitability through the transfer function.

\item \textbf{Local model, embedding and observation} (Sections~\ref{sec:mean_field_macro_VSDi_model} and~\ref{sec:Sacha_Molecular_MFT}). The local excitatory-inhibitory model is compared with the spiking network it represents, and then used as a node of spatial or connectome-based networks. Molecular perturbations are represented by changes of parameters that remain explicit in the reduced model, such as synaptic decay times and adaptation strength. Separate observation models map the simulated activity to measurable signals, such as voltage-sensitive dye imaging, BOLD and perturbational complexity.
\end{enumerate}

The sections follow the biological scales, from single neurons to whole-brain signals, which is also the order in which each object is needed to define the next. This is not the historical order. Transfer functions of simple neuron models were used in population theories before this formalism~\cite{amit1997model,brunel2000dynamics}. El Boustani and Destexhe then formulated the master equation for any neuron model whose transfer function can be characterized, using analytical transfer functions for current-based neurons and a fitted one for conductance-based neurons~\cite{el2009master}. The semi-analytical transfer function of Section~\ref{sec:semi-analytical_TF} was introduced later~\cite{zerlaut2016heterogeneous}, and combined with the master equation~\cite{zerlaut2018modeling}, explicit adaptation~\cite{di2019biologically} and connectome embedding~\cite{goldman2023comprehensive,sacha2025computational} in subsequent work. Presenting this transfer function first does not tie the master equation to it. The formalism accepts any transfer function that describes a stationary response, depends only on the current population state (and on explicitly added slow variables such as $W_\mu$), is shared by the neurons of a population, is bounded by $1/T$, and is smooth enough for its first and second derivatives to be used (Section~\ref{subsec:finite_bin_transition_law}).}

The framework should therefore be viewed as a tractable and interpretable approximation chain, not as an exact derivation of whole-brain dynamics. Its principal strength is that cellular, synaptic, and receptor parameters remain connected to population and whole-brain predictions; its principal limitation is that this connection is conditional on a succession of statistical, dynamical, and anatomical approximations. \revision{The following sections unpack these steps in detail, beginning with the transfer function and the master-equation formalism, then introducing adaptation and whole-brain embedding, and finally examining the resulting limitations and possible extensions.}

\revision{Table~\ref{tab:evidence_status_chain} summarizes the construction, assumptions and evidential support of the modeling chain, from effective cellular parameters through local population dynamics to coupled regional activity and macroscopic observables. Its rows are ordered by biological scale. The pharmacological mapping comes first because it acts on cellular parameters, whereas Section~\ref{sec:Sacha_Molecular_MFT} presents it last because it requires the complete chain.}

\begin{table}[tbp]
\centering
\begingroup
\small
\setlength{\tabcolsep}{4pt}
\renewcommand{\arraystretch}{1.04}

\def\EvidenceStatusHead{%
  \textbf{Link or modeling step} &
  \textbf{Construction and assumptions} &
  \textbf{Calibration or evaluation} &
  \textbf{Interpretation and limitations}%
}

\sbox0{%
  \begin{tabularx}{\dimexpr\linewidth-8mm\relax}
    {@{}L{0.64}L{1.12}L{1.12}L{1.12}@{}}
    \toprule
    \EvidenceStatusHead \\
    \midrule
  \end{tabularx}%
}
\pgfmathsetlengthmacro{\EvidenceHeaderHeight}{\ht0+\dp0}

\revision{%
\begin{tikzpicture}
\node[
  inner sep=0pt,
  outer sep=0pt,
  anchor=north west
] (evidence) at (0,0) {%
\begin{tabularx}{\dimexpr\linewidth-8mm\relax}
  {@{}L{0.64}L{1.12}L{1.12}L{1.12}@{}}
\toprule
\EvidenceStatusHead \\
\midrule

Pharmacology
$\rightarrow$
effective parameters
&
Pharmacologically motivated choices of excitatory and inhibitory
decay times $\tau_e,\tau_i$ and excitatory spike-triggered
adaptation $b_e$.
&
Pharmacological rationale; spiking-network transition maps and
mean-field fixed-point comparisons~\cite{sacha2025computational}.
&
Effective intervention axes. Pooled excitatory kinetics and
composite parameter changes do not uniquely identify receptor
mechanisms.
\\
\addlinespace[4pt]

Input statistics
$\rightarrow$
voltage statistics
&
Poisson synaptic shot noise and an effective passive membrane
filter give mean voltage $\langle V\rangle$, standard deviation
$\sigma(V)$ and correlation time $\tau_V$.
&
Analytical relations in
Supplementary~\ref{subsec:calculs_subthreshold_membrane_statistics};
used in single-neuron calibration
(Sec.~\ref{sec:semi-analytical_TF}).
&
Conditional subthreshold statistics. Downstream firing-rate fits
do not independently test this link or remove its stationary-input
and filtering assumptions.
\\
\addlinespace[4pt]

Voltage statistics
$\rightarrow$
firing-rate transfer function $\mathcal{F}_\mu$
&
Gaussian fluctuation-based $\mathrm{erfc}$ template with a fitted
effective threshold; quasi-stationary neuronal response.
&
Fits to model-neuron responses and \textit{in vitro}
recordings~\cite{zerlaut2016heterogeneous}; adaptive exponential
integrate-and-fire (AdEx) neuron
calibration~\cite{zerlaut2018modeling}.
&
Biophysically constrained fit. Fit accuracy alone establishes
neither out-of-sample prediction nor recurrent-network accuracy.
\\
\addlinespace[4pt]

Transfer functions
$\rightarrow$
master equation
$\rightarrow$
population moments
&
Conditional independence and a Markov time bin $T$ define a
master equation for activity probabilities. Its second-order
closure evolves mean firing rates $\nu_\mu$ and population
covariances $c_{\mu\nu}$.
&
Published closure: spiking-network comparisons~\cite{el2009master,zerlaut2018modeling}.
Corrected closure: exact master-equation benchmarks
(Supplementary~\ref{subsubsec:covariance_closure_benchmarks}).
&
Retains finite-size contributions; neglects higher moments.
Requires narrow distributions away from instabilities.
$c_{\mu\nu}$ concerns local populations' binned activities,
not interregional covariances.
\\
\addlinespace[4pt]

Population moments
$\rightarrow$
adaptive local model
&
Extends $(\nu_\mu,c_{\mu\nu})$ with population-mean adaptation
current $W_\mu$, using averaged AdEx dynamics. Slow-adaptation assumption $\tau_w\gg T$.
&
Parent-network comparisons of rates, mean voltage, response
history and Up-Down dynamics~\cite{di2019biologically};
Eq.~\eqref{eq:diVolo2019_adaptive_second_order}.
&
Only excitatory cells adapt in the implementation described here.
Mean feedback omits adaptation fluctuations; external noise also
shapes Up-Down timing.
\\
\addlinespace[4pt]

Adaptive local model
$\rightarrow$
coupled regional nodes
&
One excitatory-inhibitory unit per region $k$, with first-order
neuronal state $(\nu_e(k),\nu_i(k),W(k))$, shared local parameters,
delayed excitatory coupling, local inhibition and external noise
(Sec.~\ref{sec:Sacha_Molecular_MFT}).
&
Empirical structural connectivity and tract lengths couple
68 regions~\cite{sacha2025computational}. Homogeneous regional
physiology is assumed; its separate validation is not established
here.
&
$W(k)$ retains excitatory mean adaptation; local covariance equations
are omitted. This does not imply absent interregional correlations,
which can be estimated from simulated regional time series.
\\
\addlinespace[4pt]

Whole-brain activity
$\rightarrow$
macroscopic observables
&
BOLD requires hemodynamic filtering of $\nu_e(k)$; functional
connectivity is computed from the resulting signals.
Perturbational complexity uses thresholded evoked activity patterns.
&
Reported qualitative structure-function and perturbational-complexity
comparisons; the magnitude of structure-function changes is
overestimated~\cite{sacha2025computational}.
&
Observation and analysis procedures add assumptions. Composite
agreement neither validates every upstream link nor uniquely
identifies receptor mechanisms.
\\

\bottomrule
\end{tabularx}%
};

\coordinate (evidence-top) at
  ([xshift=-2mm,
    yshift=-\EvidenceHeaderHeight]evidence.north west);
\coordinate (evidence-bottom) at
  ([xshift=-2mm,yshift=1pt]evidence.south west);

\draw[
  draw=.,
  line width=0.4pt,
  -{Latex[length=1.7mm,width=1mm]}
]
  (evidence-top) -- (evidence-bottom);

\path (evidence-top) -- (evidence-bottom)
  node[
    midway,
    xshift=-3mm,
    rotate=90,
    inner sep=0pt
  ] {Micro to macro};

\path[use as bounding box]
  ([xshift=-8mm]evidence.north west)
  rectangle (evidence.south east);
\end{tikzpicture}%
}
\endgroup

\caption{\revision{Construction and evidence across the modeling chain. A link can be derived, fitted, assumed or empirically evaluated, and these categories overlap. Calibration is distinguished from independent validation, and simulator agreement from biological evidence. The arrow indicates bottom-up organization; not every row introduces a new spatial scale or a rigorously derived coarse-graining.}}
\label{tab:evidence_status_chain}
\end{table}

\section{Single-neuron semi-analytical transfer functions}
\label{sec:semi-analytical_TF}

\revision{In this lineage, a transfer function is the stationary firing rate of a neuron expressed as a function of the statistics of its synaptic input, and ultimately of the presynaptic population rates, $\nu_{\mathrm{out}}=\mathcal{F}_\mu(\nu_e,\nu_i)$. It is a static nonlinear input-output map, related to the familiar $f$--$I$ curve but driven by fluctuating synaptic input rather than by a constant current. The term follows El Boustani and Destexhe~\cite{el2009master} and the subsequent models reviewed here. It should not be confused with the transfer function of signal processing, which represents a linear time-invariant system in the frequency domain. Temporal filtering is not part of $\mathcal{F}_\mu$; in the population equations of Section~\ref{sec:MFT_NN_formalism}, it enters separately, through the relaxation time $T$.

The transfer function is the first reduction of the modeling chain: the membrane, channel and synaptic dynamics of a neuron are summarized by one function of a few input variables, through which the selected cellular and synaptic parameters still act. Analytical expressions exist for simple neuron models driven by diffusive input~\cite{amit1997model,brunel2000dynamics}, but generally not for conductance-based neurons with active spike-generation mechanisms. For these neurons, the construction has two steps. The presynaptic rates are first converted into the statistics of the membrane potential, using shot-noise conductances and an effective membrane filter (Supplementary~\ref{subsec:calculs_subthreshold_membrane_statistics}). These statistics are then mapped to an output rate, as described below. Section~\ref{sec:MFT_NN_formalism} shows how any transfer function satisfying a few conditions defines the stochastic dynamics of a finite population; nothing in that construction is specific to the template described here.}

\revision{For conductance-based (COBA) neurons, El Boustani and Destexhe found that an analytical approximation of the transfer function gave poor quantitative predictions of network activity, and introduced} a phenomenological transfer function described by Eq.~\eqref{eq:phenomenological_TF_el_boustani_destexhe}, fitting a 2D surface $\mathcal{F}( \langle V \rangle, \sigma(V))$ with two free parameters to numerical data, the time constant $\tau$ and the corrective term $\Delta h$.
\begin{equation}
\begin{aligned}
\mathcal{F} = \frac{1}{2 \tau}(1+\mathrm{erf}(\frac{\langle V \rangle-V^{thresh}}{\sqrt{2}\sigma(V)} + \Delta h))
\end{aligned}
\label{eq:phenomenological_TF_el_boustani_destexhe}
\end{equation}

Zerlaut et al. (2016) \cite{zerlaut2016heterogeneous} made this strategy more systematic and experimentally grounded by introducing a fluctuation-dependent effective threshold. Their study does not yet build a full population mean-field model, but it provides the smooth, compact, and biophysically interpretable semi-analytical transfer function template required by later mesoscopic and whole-brain models. They validated this approach both on theoretical neuron models and on \textit{in vitro} recordings of layer-V pyramidal neurons. A central result was that neurons are heterogeneous not only in their baseline excitability, but also in their sensitivity to the statistical properties of membrane-potential fluctuations. Some neurons were strongly affected by the speed of fluctuations $\tau_V$, others were comparatively insensitive to it, while others responded more strongly to the amplitude of fluctuations $\sigma(V)$.

\subsection{The 3D Somatic State: Integrating Fluctuation Speed ($\tau_V$)}

The core idea of Zerlaut et al. \cite{zerlaut2016heterogeneous} is that, in the fluctuation-driven regime characteristic of awake-like asynchronous activity, the relevant state of a neuron is better described by the statistical properties of its somatic membrane-potential fluctuations than by the raw synaptic input itself. Using perforated-patch recordings combined with dynamic clamp\footnote{Patch clamp technique used for long, stable recordings, which allows injection of synthetic fluctuating current and conductance, such that they could artificially precisely control $(\langle V \rangle, \sigma(V),\tau_V)$. They recorded from 30 layer-V pyramidal neurons in juvenile mouse visual cortex. The original article uses the notation $(\mu_V, \sigma_V, \tau_V)$; here we keep $(\langle V \rangle, \sigma(V),\tau_V)$ for consistency with the rest of the review.}, they designed an experimental protocol allowing them to control three somatic variables: the mean membrane potential $\langle V \rangle$, the standard deviation of membrane fluctuations $\sigma(V)$, and their global autocorrelation time $\tau_V$\footnote{$
\tau_V = \frac{1}{2} \int_{\mathbb{R}}\frac{A(\tau)}{A(0)}d\tau = \frac{1}{2} \left ( \frac{\int_{\mathbb{R}} P_V(f) df}{P_V(0)}  \right)^{-1} = \tau_s + \tau_m^{eff}, \,\ P_V(f)= \sum_{s} K_s\nu_s |\widehat{PSP}_s(f)|^2
$: Calculating autocorrelation $A(\tau)$ in the time domain involves a convolution product, which is easy to evaluate numerically but not analytically, so they calculated $\tau_V$ via the power spectral density, which is the Fourier transform of the autocorrelation function. $\tau_s$ and $\tau_m^{eff}$ are the synaptic decay time and the effective membrane time constant.}. They also introduced the normalized quantity $\tau_V^N = \tau_V/\tau_m^0$, where $\tau_m^0=C_m/g_L$ is the resting membrane time constant, in order to compare cells with different passive membrane properties. 

This extends earlier approaches by adding the speed of fluctuations as an explicit determinant of firing, such that: $\nu_{out} = \mathcal{F}(\langle V \rangle, \sigma(V),\tau_V)$. Experimentally, this is important because two neurons with similar $\langle V\rangle$ and $\sigma(V)$ can have different firing responses if their membrane fluctuations occur on different fast or slow timescales.

Starting from a simple approximation analogous to El Boustani and Destexhe \cite{el2009master}, and Amit and Brunel \cite{amit1997model}, Zerlaut et al. assume that the time axis can be split into bins of size $\tau_V$, each bin being treated as approximately independent in the low-rate regime. If a spike occurs whenever the membrane potential crosses threshold, the firing rate is estimated as the probability of being above threshold divided by the fluctuation timescale (see Supplementary~\ref{subsec:fokker_planck_microscopic_level} for more details):
\begin{equation}
\begin{aligned}
\nu_{out} \approx \frac{P(V \geq V_{thresh})}{\tau_V}, \text{so}\footnotemark, \,\  \nu_{out}= \mathcal{F}(\nu_e,\nu_i)= \mathcal{F}(\langle V \rangle, \sigma(V),\tau_V) = \frac{1}{2\tau_V^N \tau_m^0}\mathrm{erfc}(\frac{V_{thresh} - \langle V\rangle}{\sqrt{2}\sigma(V)}) 
\end{aligned}
\label{eq:transfer_function_zerlaut2016}
\end{equation}

\footnotetext{This expression assumes approximately Gaussian membrane-potential fluctuations and treats successive bins of duration $\tau_V$ as approximately independent.}
This expression captures the expected qualitative behavior: firing increases when the mean depolarization rises, when fluctuations become larger, or when they become faster.

\subsection{Phenomenological Effective Threshold ($V_{\text{thresh}}^{\text{eff}}$)}

The Gaussian threshold-crossing approximation captures the correct qualitative dependencies of firing on \(\langle V\rangle\), \(\sigma(V)\), and \(\tau_V\), but it remains too rigid to account for the active nonlinearities of real neurons. Zerlaut et al. \cite{zerlaut2016heterogeneous} therefore introduced a phenomenological effective threshold: $V_{\mathrm{thresh}}^{\mathrm{eff}} = V_{\mathrm{thresh}}^{\mathrm{eff}}( \langle V \rangle,\sigma(V),\tau_V^N)$, which absorbs the effects of sodium-channel activation, sodium inactivation\footnote{For instance, if the voltage rises slowly, sodium channels inactivate, which effectively raises the threshold}, spike-frequency adaptation, and other active membrane properties. The idea is not to derive a new mathematically complex Fokker--Planck solution for every biological nonlinearity, which would be difficult because of the nonlinear threshold and boundary dynamics. Instead, the strategy is to preserve the convenient \(\mathrm{erfc}\)-based analytical form\footnote{See Supplementary~\ref{subsec:master_equation_CUBA} for the corresponding Gaussian-threshold approximation.} while allowing the threshold itself to depend on the fluctuation state of the neuron.  The effective threshold $V_{\mathrm{thresh}}^{\mathrm{eff}}$ does not relax the Gaussian assumption intrinsic to \textit{erfc} by itself; rather, it compensates and absorbs for non-ideal spike initiation and other active biophysical effects by nonlinearly warping the response, while the diffusion/Gaussian approximation for the membrane potential remains intrinsic to the framework.

Zerlaut et al. \cite{zerlaut2016heterogeneous} chose a linear phenomenological template rather than a formally derived biophysical law, which can be interpreted as a first-order local approximation around a reference fluctuation state $(\langle V^0 \rangle, \sigma(V^0), \tau_V^{N0})$. This gives:
\begin{equation}
\begin{aligned}
V_{\text{thresh}}^{\text{eff}} 
&= P_0 + P_\mu \delta\langle V \rangle + P_\sigma \delta\sigma(V) + P_\tau \delta\tau_V, \,\ \\
&= P_0 + P_\mu \frac{\langle V \rangle - \langle V^0 \rangle}{\delta\langle V^0 \rangle} + P_\sigma \frac{\sigma(V) - \sigma(V^0)}{\delta\sigma(V^0)} + P_\tau \frac{\tau_V^N - \tau_V^{N0}}{\delta\tau_V^{N0}}, \,\ \footnotemark
\end{aligned}
\end{equation}
\footnotetext{The quantities $(\langle V^0 \rangle,\delta\langle V^0 \rangle, \sigma(V^0), \delta\sigma(V^0), \tau_V^{N0}, \delta\tau_V^{N0})$ are reference values and rescaling factors used to normalize the $(\langle V \rangle,\sigma(V),\tau_V^N)$-space.}
In this formulation, the complex active biophysics of the neuron are compressed into four phenomenological parameters:
$(P_0,P_\mu,P_\sigma,P_\tau)$. Here, \(P_0\) represents the baseline effective threshold, while \(P_\mu\), \(P_\sigma\), and \(P_\tau\) quantify the sensitivities of the threshold to mean depolarization, fluctuation amplitude, and fluctuation speed, respectively.

\subsection{Fitting procedure and Validation}

A practical advantage of the approach is that the analytical expression can be inverted. Given the measured firing rate $\mathcal{F}(\langle V \rangle, \sigma(V),\tau_V)$ from a complex simulation or \textit{in vitro} data, one can compute the corresponding effective threshold:
$V_{\text{thresh}}^{\text{eff}} = \sqrt{2}\sigma(V) \cdot \text{erfc}^{-1}(2\tau_V \nu_{\text{out}}) + \langle V \rangle.$ This makes the fitting procedure straightforward: first, the effective threshold is reconstructed from data; then the coefficients $(P_0,P_\mu,P_\sigma,P_\tau)$ are estimated by linear regression and refined by nonlinear least squares optimization.

Zerlaut et al. \cite{zerlaut2016heterogeneous} validated this framework on several synthetic neuron models of the integrate-and-fire family. They found that a constant-threshold description achieved a mean goodness of fit of approximately \(84.6\%\), whereas the linear four-parameter effective threshold increased the fit to approximately \(99.0\%\). A second-order polynomial threshold with ten parameters further improved the fit to approximately \(99.6\%\). However, the four-parameter version was retained as a good compromise between accuracy, simplicity, and interpretability. The same template also provided a robust characterization of real layer-V pyramidal neurons recorded \textit{in vitro}. Importantly, the study revealed strong functional heterogeneity across cells: neurons differed not only in baseline excitability, but also in their sensitivities to \(\langle V \rangle \), \(\sigma(V)\), and \(\tau_V\). Zerlaut et al. \cite{zerlaut2016heterogeneous} therefore established a compact semi-analytical framework able to capture both theoretical and experimental firing responses in the fluctuation-driven regime.

\revision{The fitted effective threshold absorbs nonlinear spike initiation, reset and other active cellular effects that are absent from the subthreshold calculation. It does not remove the stationarity and diffusion approximations underlying that calculation. Accuracy therefore degrades when synaptic events are too rare, too strong or too correlated to resemble Gaussian background fluctuations, or when the inputs change faster than the stationary transfer function can track.}

\section{Master Equation formalism for finite-size mesoscopic dynamics}
\label{sec:MFT_NN_formalism}

\revision{A transfer function describes a single neuron in a stationary state. The master-equation formalism turns it into a stochastic description of finite, recurrent populations.} El Boustani and Destexhe~\cite{el2009master} introduced the theoretical starting point of the lineage examined here: a mesoscopic description of finite, recurrent neuronal populations in asynchronous irregular (AI) regimes (see Supplementary~\ref{subsubsec:AI_state}). The objective is not to reproduce individual spike trains, but to derive the dynamics of population activities and their finite-size covariances from a single-neuron response law. The construction thereby separates two components: a generic stochastic population formalism and a neuron-model-specific transfer function. This separation is the principal micro-to-mesoscopic bridge used in the following sections. \revision{The formalism applies to any transfer function compatible with assumptions (i)--(v) of Section~\ref{subsec:finite_bin_transition_law} and smooth enough for the second-order closure of Section~\ref{subsec:second_order_closure}, including the semi-analytical one of Section~\ref{sec:semi-analytical_TF}.} A detailed pedagogical derivation is provided in Supplementary~\ref{sec:appendix_master_equation_El_boustani_Destexhe}.

\subsection{Finite-bin transition law and micro-to-mesoscopic bridge}
\label{subsec:finite_bin_transition_law}

Consider $P$ homogeneous populations, with $N_\mu$ neurons in population $\mu$. During a finite bin of duration $T$, let $K_\mu$ be the number of neurons that fire and define the coarse-grained activity $m_\mu=K_\mu/(N_\mu T)$. The mesoscopic state is $\mathbf{m}=(m_1,\ldots,m_P)$. Conditioned on the preceding state $\mathbf{m}'$, the construction assumes: (i) effective Markovianity and time-homogeneity on the scale $T$; (ii) a quasi-stationary neuronal response; (iii) conditional independence of firing events; (iv) at most one spike per neuron and bin; and (v) statistical homogeneity within each population. If $\mathcal{F}_\mu(\mathbf{m}')$ is the stationary firing rate of a representative neuron, then

\begin{equation}
K_\mu\mid\mathbf{m}'\sim
\mathcal{B}\!\left(N_\mu,T\mathcal{F}_\mu(\mathbf{m}')\right),
\qquad
\begin{aligned}
\mathbb{E}[m_\mu\mid\mathbf{m}']&=\mathcal{F}_\mu(\mathbf{m}'),\\
\operatorname{Var}(m_\mu\mid\mathbf{m}')&=
\frac{\mathcal{F}_\mu(\mathbf{m}')\left[T^{-1}-\mathcal{F}_\mu(\mathbf{m}')\right]}{N_\mu}.
\end{aligned}
\label{eq:binomial_kernel_main}
\end{equation}

Thus, selected microscopic properties enter through $\mathcal{F}_\mu$, whereas finite population size determines the demographic sampling variance. For sufficiently large $N_\mu\mathcal{F}_\mu T$ and $N_\mu(1-\mathcal{F}_\mu T)$, the binomial kernel is approximated by a Gaussian with the same mean and variance. The joint kernel factorizes across populations under conditional independence. Writing the corresponding coarse-grained transition rate as $\mathcal{W}(\mathbf{m}\mid\mathbf{m}')=P_T(\mathbf{m}\mid\mathbf{m}')/T$ gives the gain--loss equation

\begin{equation}
\partial_t P_t(\mathbf{m})=
\int_{\Omega}\!\left[
P_t(\mathbf{m}')\mathcal{W}(\mathbf{m}\mid\mathbf{m}')-
P_t(\mathbf{m})\mathcal{W}(\mathbf{m}'\mid\mathbf{m})
\right]d\mathbf{m}',
\label{eq:master_equation_}
\end{equation}

where $\Omega=[0,T^{-1}]^P$ in the continuous approximation. This is already a reduced stochastic model: spike timing within the bin, membrane trajectories, detailed connectivity, common-input correlations not captured by the conditioned state, and within-population diversity have been integrated out or absorbed into $\mathcal{F}_\mu$.

\subsection{Second-order closure and finite-size interpretation}
\label{subsec:second_order_closure}

Solving Eq.~\eqref{eq:master_equation_} for the full probability density would largely defeat the reduction. The distribution is therefore projected onto the mean rates $\nu_\mu\equiv\langle m_\mu\rangle$ and covariances $c_{\mu\nu}=\langle(m_\mu-\nu_\mu)(m_\nu-\nu_\nu)\rangle$, and the moment hierarchy is truncated at second order. \revision{Greek indices label populations and run from $1$ to $P$. An index that is repeated within a term and absent from the left-hand side (here $\lambda$ and $\eta$) is summed over all populations (Einstein convention); for example, $\partial_\lambda\mathcal{F}_\mu\,\partial_\eta\mathcal{F}_\nu\,c_{\lambda\eta}=\sum_{\lambda,\eta=1}^{P}\partial_\lambda\mathcal{F}_\mu\,\partial_\eta\mathcal{F}_\nu\,c_{\lambda\eta}$. Here $\partial_\lambda\equiv\partial/\partial\nu_\lambda$, and all transfer functions and their derivatives are evaluated at $\boldsymbol{\nu}$.} The resulting equations are

\revision{%
\begin{empheq}[box=\fbox]{equation}
\begin{aligned}
T\partial_t\nu_\mu
&=\mathcal{F}_\mu-\nu_\mu
+\frac{1}{2}\,\partial_\lambda\partial_\eta\mathcal{F}_\mu\,c_{\lambda\eta},\\
T\partial_t c_{\mu\nu}
&=\delta_{\mu\nu}A^{-1}_{\mu\mu}
+\partial_\lambda\mathcal{F}_\mu\,\partial_\eta\mathcal{F}_\nu\,c_{\lambda\eta}
-c_{\mu\nu}
+T^2\,\partial_t\nu_\mu\,\partial_t\nu_\nu,\\
A^{-1}_{\mu\mu}
&=\frac{\mathcal{F}_\mu(T^{-1}-\mathcal{F}_\mu)}{N_\mu}.
\end{aligned}
\label{eq:final_ODEs_master_equation_}
\end{empheq}%
}

\revision{The mean equation is that of the original derivation~\cite{el2009master}: the mean relaxes toward the single-neuron transfer function prediction, with a covariance-dependent correction controlled by its curvature. The covariance equation is corrected. The published equation, also used by the second-order models of this lineage~\cite{zerlaut2018modeling,di2019biologically}, evaluated the second jump moment of the master equation at the mean state instead of averaging it over the activity distribution; this mean-value closure omits a curvature term of the same order as those retained (Supplementary~\ref{subsubsec:covariance_matrix}). With the gain matrix $J_{\mu\lambda}=\partial_\lambda\mathcal{F}_\mu$, $C=(c_{\mu\nu})$, $A^{-1}=\operatorname{diag}(A^{-1}_{\mu\mu})$, the displacement $\mathbf{g}=\boldsymbol{\mathcal{F}}-\boldsymbol{\nu}$ and the velocity of the mean $\mathbf{v}=T\partial_t\boldsymbol{\nu}$, the two forms read}
\revision{%
\begin{equation}
\begin{aligned}
\text{corrected:}\quad T\partial_t C&=A^{-1}+JCJ^{\mathsf T}-C+\mathbf{v}\mathbf{v}^{\mathsf T},\\
\text{published:}\quad T\partial_t C&=A^{-1}+JC+CJ^{\mathsf T}-2C+\mathbf{g}\mathbf{g}^{\mathsf T}.
\end{aligned}
\label{eq:covariance_corrected_vs_published_main}
\end{equation}%
}%
\revision{The corrected covariance equation retains the curvature contribution missing from the published mean-value closure. It follows from the complete second-order expansion after further omitting terms of orders $C/N$ and $C^2$, which are $\mathcal O(N^{-2})$ in a regular regime with $C=\mathcal O(N^{-1})$ (Supplementary~\ref{subsubsec:covariance_matrix}). In the corrected form, each transition injects the binomial sampling noise ($A^{-1}$), transmits the current fluctuations through the population gain ($JCJ^{\mathsf T}$) and erases the previous state ($-C$), while a moving mean adds the square of its velocity ($\mathbf{v}\mathbf{v}^{\mathsf T}$, zero at stationary states). At least three consequences follow. Stationary fluctuations are larger: for one population linearized about a stationary state with gain $J=\Delta$, the variance is $A^{-1}/(1-\Delta^2)$ instead of $A^{-1}/[2(1-\Delta)]$, so that inhibitory feedback amplifies fluctuations as much as excitatory feedback of the same strength. In this local weak-fluctuation approximation, stable stationary covariance dynamics require the eigenvalues $\zeta_a$ of $J$ to satisfy $|\zeta_a|<1$. The published covariance equation instead has the mean-dynamics stability condition $\operatorname{Re}\zeta_a<1$ and can therefore miss the loss of local fluctuation stability. The bounded nonlinear master equation may still have a stationary distribution outside this regime. During transients, the source term follows the actual velocity of the mean. Exact solutions of the master equation support the corrected form (Supplementary~\ref{subsubsec:covariance_closure_benchmarks}); the published structure is recovered only under specific conditions: when the gain matrix is close to the identity ($J\simeq I$, a marginally stable regime), or in a small-jump limit, in which $T$ is the time constant of an exponentially filtered activity rather than a bin in which each neuron fires at most once. This small-jump construction is, however, a different model: it describes exponentially filtered spike activity, with noise source $\mathcal F_\mu/(N_\mu T)$ and no squared-displacement source, whereas the published and corrected equations concern the same finite-bin master equation. Their resemblance therefore does not make the two closures equivalent or establish which observable a published calculation represents. Complete derivations of the transition kernel, moment closure, benchmark examples, and lagged-covariance equation are provided in Supplementary~\ref{sec:appendix_master_equation_El_boustani_Destexhe}.}

\revision{Near a stable stationary state in the regular weak-fluctuation regime, $A^{-1}_{\mu\mu}$ and $c_{\mu\nu}$ scale as $N_\mu^{-1}$. Neglecting covariance feedback then gives the first-order relation $T\partial_t\nu_\mu=\mathcal F_\mu(\boldsymbol\nu)-\nu_\mu$. Nevertheless, large population size alone does not eliminate the transient ensemble spread caused by random redraw times (Supplementary~\ref{subsubsec:covariance_matrix}).} This makes finite-size fluctuations an endogenous consequence of the finite population construction, not an arbitrary additive-noise term. Recurrent amplification can nevertheless make them important near a critical point\revision{, where an eigenvalue of $J$ approaches the unit circle}, precisely where the narrow-distribution and second-order assumptions may also become least reliable.

This correction restores consistency with the stated master equation at the retained order of the moment closure, and the benchmarks in Supplementary~\ref{subsubsec:covariance_closure_benchmarks} assess its accuracy against that stochastic description. It does not, by itself, guarantee improved agreement with the parent SNN, because the master-equation description also relies on approximations concerning population homogeneity, conditional independence, and neuronal responses. Whether the corrected equations better reproduce the SNN's population statistics therefore remains to be established through dedicated numerical comparisons, which is out of the scope of this review.

\subsection{Validity, retained mechanisms, and connection to subsequent models}
\label{subsec:ME_validity}

The closed equations are approximate at several distinct levels. The finite-bin kernel assumes that $T$ is long enough for unresolved microscopic memory and correlations to decay, yet short enough to resolve the dynamics, keep multiple spikes per neuron improbable, and justify a quasi-stationary response. \revision{The second-order truncation requires sufficiently populated bins and an activity distribution that is narrow compared with the range over which $\mathcal{F}_\mu$ bends, so that it is adequately characterized by its first two moments.  The local stationary covariance approximation requires the gain eigenvalues to lie inside the unit circle. Near this boundary, broadening and nonlinear saturation can invalidate the second-order Taylor closure, as illustrated by the sigmoid example in Supplementary~\ref{subsubsec:covariance_closure_benchmarks}.} Conditional independence is most plausible for sparse, weakly correlated AI activity; strong synchrony, multimodality, rapid forcing, or proximity to a transition can invalidate it. Homogeneity is also structural: an effective transfer function may approximate weak diversity at the level of the mean, but the binomial finite-size term is not generally preserved under strong or multimodal heterogeneity. Finally, accuracy remains bounded by the quality and calibration domain of $\mathcal{F}_\mu$. El Boustani and Destexhe obtained the clearest quantitative agreement for current-based networks and required an effective fitted transfer function for conductance-based networks~\cite{el2009master}; the validation details are retained in Supplementary~\ref{sec:appendix_master_equation_El_boustani_Destexhe}. \revision{These reported simulator comparisons used the published
covariance equation. They therefore do not independently evaluate
the corrected equation introduced here. Reassessment requires the
original implemented equations, activity observable and comparison
protocol; changing the interpretation of the observable alone (e.g., with the covariance of exponentially filtered activity) is
not a validation of either closure (Supplementary~\ref{subsubsec:covariance_closure_benchmarks}).}

Within these conditions, the formalism preserves the effects of neuronal and synaptic parameters that measurably alter $\mathcal{F}_\mu$, together with population size, mean activity, finite-size variance, covariance propagation, and stationary correlation times. It does not preserve individual spike timing, microscopic connectivity realizations, dendritic or intracellular states, or cellular diversity unless they are represented through additional populations, state variables, or transfer function parameters. \revision{The semi-analytical transfer functions of Section~\ref{sec:semi-analytical_TF} carry selected conductance-based biophysics;} Section~\ref{sec:mean_field_macro_VSDi_model} inserts them into Eq.~\eqref{eq:final_ODEs_master_equation_}\revision{, used there in their published form}; and Section~\ref{sec:DiVolo_Adaptation_MFT} promotes slow adaptation to an explicit mesoscopic state. The first-order spatial and whole-brain formulations discussed here omit population covariance dynamics, while the adaptive formulations retain mean adaptation. The second-order derivation remains scientifically informative because it identifies the endogenous fluctuation structure removed by that truncation and the regimes in which its omission is most consequential.

\section{Conductance-based mean-field model for VSDi-like cortical dynamics}
\label{sec:mean_field_macro_VSDi_model}

Zerlaut et al. (2018) \cite{zerlaut2018modeling} connected the two elements introduced above: the finite-size master-equation formalism of El Boustani and Destexhe \cite{el2009master} and the semi-analytical transfer function of Zerlaut et al. (2016) \cite{zerlaut2016heterogeneous}. Their aim was to obtain a model that remained interpretable in terms of cellular and synaptic properties, yet was inexpensive enough to describe the spatially extended cortical dynamics measured by voltage-sensitive dye imaging (VSDi). The resulting construction proceeds in two stages: a local conductance-based mean-field model is validated against its parent spiking network, then first-order local units are coupled spatially and linked to a population-voltage readout.

\subsection{Local conductance-based RS-FS circuit model}

The reference circuit contains \(10{,}000\) Adaptive Exponential Integrate-and-Fire (AdEx) neurons with \(5\%\) random connectivity: \(80\%\) are excitatory regular-spiking (RS) cells and \(20\%\) are inhibitory fast-spiking (FS) cells. The RS cells display subthreshold and spike-triggered adaptation, whereas the FS cells have sharper spike initiation and greater excitability. Adaptation is therefore present in the microscopic RS dynamics, but at this stage it affects the mean-field model only through a stationary transfer function rather than through an explicit history-dependent state variable.

For each population \(\mu\in\{e,i\}\), the transfer function \(\mathcal{F}_\mu(\nu_e,\nu_i)\) maps stationary presynaptic excitatory and inhibitory rates to the output rate of the corresponding AdEx neuron. As in Section~\ref{sec:semi-analytical_TF}, presynaptic rates determine conductance statistics, which determine the mean, amplitude, and autocorrelation time of membrane-potential fluctuations and, through the \(\mathrm{erfc}\)-based expression in Eq.~\eqref{eq:transfer_function_zerlaut2016}, the firing probability. The required subthreshold quantities \((\langle V\rangle,\sigma(V),\tau_V^N,\langle g\rangle)\) are given in Supplementary~\ref{subsec:calculs_subthreshold_membrane_statistics}.

The 2018 implementation used dense single-neuron simulations over the \((\nu_e,\nu_i)\) input plane to fit a richer effective threshold than the linear form used in 2016. Defining \(\mathcal{X}=\{\langle V\rangle,\sigma(V),\tau_V^N\}\), it takes the form
\begin{equation}
\begin{aligned}
V_{\mathrm{thresh}}^{\mathrm{eff}}
(\langle V\rangle,\sigma(V),\tau_V^N,\langle g\rangle)
&=
P_0
+\sum_{x\in\mathcal{X}}P_x\frac{x-x_0}{\delta x_0}
+P_{\langle g\rangle}\log\!\left(\frac{\langle g\rangle}{g_L}\right)
+\sum_{x,y\in\mathcal{X}}
P_{xy}
\frac{x-x_0}{\delta x_0}
\frac{y-y_0}{\delta y_0}.
\end{aligned}
\label{eq:effective_threshold_zerlaut2018}
\end{equation}
The quadratic terms improve flexibility outside the fluctuation-driven regime, particularly at low presynaptic rates where the diffusion approximation is less accurate, while the conductance term accounts for the influence of the total conductance state on effective spike initiation. The mapping of the semi-analytical transfer function is therefore partly mechanistic and partly calibrated: synaptic inputs are propagated analytically through conductance and voltage statistics, whereas active spike-generation nonlinearities are compressed into the fitted threshold coefficients.

\subsection{Master Equation closure and local validation}

Inserting \(\mathcal{F}_e\) and \(\mathcal{F}_i\) into the second-order master-equation closure gives the dynamics of the mean rates \((\nu_e,\nu_i)\), variances \((c_{ee},c_{ii})\), and covariance \(c_{ei}\)\revision{; with the corrected covariance equation, Eq.~\eqref{eq:final_ODEs_master_equation_}, they read}
\begin{equation}
\begin{aligned}
T\partial_t\nu_\mu
&=
\mathcal{F}_\mu-\nu_\mu
+\frac{1}{2}\partial_\lambda\partial_\eta
\mathcal{F}_\mu\,c_{\lambda\eta},
\\
T\partial_t c_{\mu\nu}
&=
\delta_{\mu\nu}\frac{\mathcal{F}_\mu(T^{-1}-\mathcal{F}_\mu)}{N_\mu}
\revision{+\partial_\lambda\mathcal{F}_\mu\,\partial_\eta\mathcal{F}_\nu\,c_{\lambda\eta}
-c_{\mu\nu}
+T^2\,\partial_t\nu_\mu\,\partial_t\nu_\nu}.
\end{aligned}
\label{eq:zerlaut2018_second_order}
\end{equation}
Repeated population indices are summed, and the transfer functions and their derivatives are evaluated at \((\nu_e,\nu_i)\). \revision{Zerlaut et al.~\cite{zerlaut2018modeling} integrated the published covariance equation, i.e.\ the second line of Eq.~\eqref{eq:covariance_corrected_vs_published_main}; the mean equation is the same.} The original implementation used \(T=5\,\mathrm{ms}\), chosen to resolve visually evoked transients while retaining the approximate Markov property.

Against direct simulations of the reference RS-FS network, the model reproduced the spontaneous excitatory and inhibitory activity distributions and the mean membrane-potential and conductance statistics, although it slightly overestimated mean population rates and underestimated the standard deviation of membrane-potential fluctuations. The latter discrepancy was attributed to residual synchrony not represented by the asynchronous closure. \revision{These activity-distribution comparisons used the published covariance equation and $T=5\,\mathrm{ms}$ activity bins~\cite{zerlaut2018modeling}. Validation of the corrected equation against the parent network therefore remains open and requires the same comparison protocol; out of the scope of this review. The magnitude of any change requires re-evaluation with the original comparison protocol. The mean equation keeps its form, but a change in covariance can alter its curvature feedback; that feedback is $\mathcal O(N^{-1})$ only in the regular weak-fluctuation regime.} Responses to low-frequency transient inputs were also captured sufficiently well for the intended visual-response application. However, the model missed the prolonged post-stimulus hyperpolarization generated by accumulated adaptation and did not reproduce all frequency-dependent effects, notably the approximately \(50\,\mathrm{Hz}\) resonance of the spiking network. These failures identify the consequence of compressing adaptation into a stationary transfer function: the Markov mean-field node can represent its steady-state effect on excitability, but not its evolving memory of previous activity.

\subsection{From local mean-field units to a spatial VSDi model}

Zerlaut et al. \cite{zerlaut2018modeling} next embedded the RS-FS units in a one-dimensional ring representing a translation-invariant cortical sheet. Excitatory and inhibitory activities are convolved with Gaussian kernels of lateral connectivity, $\mathcal{K}_\mu(x) = e^{-x^2/(2l_{\mu}^2)}/(\sqrt{2\pi}l_{\mu})$, and finite axonal speed \(v_c\) introduces distance-dependent delays. For this spatial model, only the first moments are retained:
\begin{equation}
\begin{aligned}
\nu_\mu^{\mathrm{input}}(x,t)
&=
\delta_{\mu e}\nu_e^{\mathrm{drive}} +
\int_\mathbb{R} \mathcal{K}_\mu(x-y)\nu_\mu(y,t-\Arrowvert y-x \Arrowvert/v_c)dy, \\
T\partial_t\nu_\mu(x,t)
&=
-\nu_\mu(x,t)
+\mathcal{F}_\mu\!\left(
\nu_e^{\mathrm{input}}(x,t)
+\delta_{\mu e}\nu_e^{\mathrm{aff}}(x,t),
\nu_i^{\mathrm{input}}(x,t)
\right).
\end{aligned}
\label{eq:zerlaut2018_spatial}
\end{equation}
Here, the afferent thalamic drive is Gaussian in space and has a double-Gaussian rise and decay in time.\footnote{A representative form is \(\nu_e^{\mathrm{aff}}(x,t)=A G_x(x)G_t(t)\), with \(G_x(x)=\exp[-(x-x_0)^2/(2l_{\mathrm{stim}}^2)]\). Before \(t_0\), \(G_t(t)=\exp[-(t-t_0)^2/(2\tau_1^2)]\); after \(t_0\), \(G_t(t)=\exp[-(t-t_0)^2/(2\tau_2^2)]\). Here, \(A\) is the amplitude, \((x_0,t_0)\) the stimulus center, \(l_{\mathrm{stim}}\) its spatial width, and \(\tau_1,\tau_2\) its rise and decay constants.}

The comparison with VSDi uses the normalized mean membrane-potential deflection $\delta V_N(x,t)=\frac{\langle V(x,t)\rangle-V_{\mathrm{rest}}}{V_{\mathrm{rest}}},$ rather than firing rate alone, where \(V_{\mathrm{rest}}\) denotes the mean membrane potential during spontaneous activity. This provides an observation proxy aligned with the population-averaged subthreshold voltage contribution to VSDi, although it is not a complete optical forward model. The ring reproduced qualitative features of awake-monkey V1 recordings, including localized afferent responses, propagating activity, and a broader spatial extent of subthreshold voltage than of multi-unit spiking activity \cite{zerlaut2018modeling}. In this sense, each RS-FS mean-field unit is not merely an abstract neural mass: it can be interpreted as a local VSDi-like pixel with an explicit biophysical correlate.

\subsubsection{Inferring physiological parameters from imaging data}

The spatial model was also fitted to recorded VSDi patterns by minimizing the least-squares discrepancy between simulated and experimental spatiotemporal responses. Fitted quantities included the propagation speed \(v_c\), excitatory and inhibitory spatial extents \(l_{\mathrm{exc}}\) and \(l_{\mathrm{inh}}\), stimulus width \(l_{\mathrm{stim}}\), and rise and decay constants \(\tau_1\) and \(\tau_2\). The resulting values were physiologically plausible, including broader excitatory than inhibitory coupling and a propagation speed compatible with experimental estimates. This demonstrates constrained inference of effective physiological parameters, but not their unique identification: estimates remain conditional on the ring geometry, Gaussian coupling kernels, calibrated transfer functions, and simplified VSDi readout.

This study therefore established a key step in the reduction chain: cell-type-specific conductance effects remain accessible through \(\mathcal{F}_\mu\), and the retained mean voltage provides a bridge to mesoscopic imaging, while individual spike timing, microscopic connectivity, and dynamic adaptation are averaged out. The first-order spatial truncation additionally removes explicit finite-size covariance dynamics; its implications are discussed in Section~\ref{subsec:first_order_truncation_limitations_section}.

\section{Adaptive mean-field model with an explicit mesoscopic adaptation variable}
\label{sec:DiVolo_Adaptation_MFT}

In the Zerlaut et al. (2018) model \cite{zerlaut2018modeling}, spike-frequency adaptation affected the stationary transfer function but was not propagated as a dynamical state. The model could therefore reproduce spontaneous activity and some transient responses, but not the slow response history generated by accumulated adaptation, notably the prolonged post-stimulus hyperpolarization of the parent spiking network. Di Volo et al. (2019) \cite{di2019biologically} addressed this limitation by adding the population-averaged adaptation current \(W\) to the master-equation state. This preserves a selected cellular memory mechanism at mesoscopic scale while retaining a low-dimensional population description.

\subsection{Mesoscopic equations with explicit spike-frequency adaptation}

The state is enlarged from the population activities $\mathbf{m}\equiv  \{m_\gamma\}$ to \(\{m_\gamma,W_\gamma\}\). To retain an approximately Markovian transition over the coarse-graining interval \(T\), adaptation is assumed to be slow, \(\tau_w\gg T\): \(W_\gamma\) is effectively constant during one transition step but evolves across successive steps. The second-order system then couples mean firing rates \(\nu_\mu\), finite-size covariances \(c_{\mu\nu}\), and population adaptation\revision{; with the corrected covariance equation, it reads}:
\begin{empheq}[box=\fbox]{equation}
\begin{aligned}
T\partial_t \nu_\mu
&= (\mathcal{F}_\mu - \nu_\mu) + \frac{1}{2}\partial_\lambda\partial_\eta \mathcal{F}_\mu\, c_{\lambda\eta} \\
T\partial_t c_{\mu\nu}
&= \delta_{\mu\nu} A_{\mu\mu}^{-1} \revision{+ \partial_\lambda \mathcal{F}_\mu\,\partial_\eta \mathcal{F}_\nu\, c_{\lambda\eta} - c_{\mu\nu} + T^2\,\partial_t\nu_\mu\,\partial_t\nu_\nu} \\
\tau_w\partial_t W_\mu
&= -W_\mu + b_{\mu}\tau_w\nu_\mu + a_{\mu}(\langle V(\nu_e, \nu_i, W_\mu)\rangle - E_{L,\mu})
\end{aligned}
\label{eq:diVolo2019_adaptive_second_order}
\end{empheq}
Here \(\mathcal{F}_\mu=\mathcal{F}_\mu(\nu_e,\nu_i,W_\mu)\), repeated population indices are summed\revision{, and derivatives are taken with respect to the rates at fixed \(W_\mu\)}. The first two equations retain the master-equation closure introduced above\revision{, here with the corrected covariance equation; Di Volo et al. used the published one, which also underlies the second-order covariance estimates reported with their simulations~\cite{di2019biologically}}. The third equation follows the population average of the subthreshold and spike-triggered AdEx adaptation terms, controlled by \(a_\mu\) and \(b_\mu\), respectively. In the RS-FS implementation, adaptation is assigned to excitatory RS cells, whereas \(a_i=b_i=0\) for inhibitory FS cells. The full derivation is given in Supplementary~\ref{subsec:adaptation_master_equation_formalism}.

The adaptation current enters the transfer function primarily by shifting the mean membrane potential. Writing \(\langle g\rangle=g_L+\langle G_e\rangle+\langle G_i\rangle\), the dynamic-state expression and its stationary form used to calibrate the transfer function are
\begin{equation}
\begin{aligned}
\langle V(\nu_e,\nu_i,w) \rangle = \frac{g_L E_L + \langle G_e \rangle E_e + \langle G_i \rangle E_i - w}{\langle g \rangle} = \frac{g_L E_L + \langle G_e \rangle E_e + \langle G_i \rangle E_i - \nu_{out}\tau_wb +aE_L }{\langle g \rangle + a}, \,\ \footnotemark
\end{aligned}
\label{eq:adaptation_mean_voltage}
\end{equation}
\footnotetext{See Supplementary~\ref{subsec:adaptation_master_equation_formalism}. The left expression corresponds to the dynamic-state formulation used during mean-field simulation, whereas the right expression is the steady-state algebraic solution used to calibrate the transfer function.}

Increasing activity therefore builds adaptation, lowers \(\langle V\rangle\), reduces \(\mathcal{F}_\mu\), and creates slow negative feedback. By contrast, \(\sigma(V)\) and \(\tau_V^N\) are kept as in the preceding conductance-based construction, on the assumption that slow adaptation and its fluctuations make a negligible contribution to fast voltage fluctuations. The transfer function also retains the \(\mathrm{erfc}\)-based form and quadratic effective threshold introduced previously, but removes the explicit logarithmic conductance term used by Zerlaut et al. (2018); \(W\) affects the threshold only indirectly through the voltage statistics. The closure is consequently mechanistic in how adaptation shifts the retained subthreshold state, but remains conditional on the calibrated threshold approximation.

\subsection{Dynamical consequences and validation against spiking networks}

Di Volo et al. used the adaptive model to describe alternations between low-activity Down states and depolarized Up states. The mechanism combines bistability, slow negative feedback, and noise. When \(W\) is low, a fluctuation can move the system from the Down state to the active branch; activity then accumulates adaptation until the Up state destabilizes and the system returns to the Down state.

For the irregular Up-Down regime studied in the article, stochastic external bombardment is represented by an Ornstein--Uhlenbeck (OU) modulation of the drive:
\begin{equation}
\begin{aligned}
\nu_{\mathrm{drive}}(t)
&= \nu_{\mathrm{drive}}+\sigma\xi(t),\\
d\xi(t)
&= -\frac{\xi(t)}{\tau_{\mathrm{OU}}}\,dt+dB_t,
\end{aligned}
\label{eq:OU_noise}
\end{equation}
where \(B_t\) is a Wiener process. Adaptation supplies the slow termination mechanism, whereas the external OU process contributes to transition initiation and irregular timing. The resulting mean field reproduces heterogeneous Up-state durations and internal rebound-like structure observed in the corresponding conductance-based spiking network. These transitions should therefore not be interpreted as arising from adaptation alone or from an endogenous finite-size stochastic closure.

The adaptive mean field was validated against the same family of conductance-based RS-FS AdEx networks used in the preceding section. In asynchronous irregular regimes, it reproduced excitatory and inhibitory firing rates and the mean membrane potential across changes in adaptation strength. Its advantage was clearest for time-dependent input: after transient excitation, the accumulated \(W\) generated a hyperpolarizing tail lasting hundreds of milliseconds, in agreement with the spiking network and unlike the stationary-adaptation approximation. The model also reproduced state-dependent responsiveness: the same perturbation evokes a weaker response from a highly active, strongly adapting conductance state than from a low-activity state. Thus, promoting \(W\) to a state variable retains a testable link between the AdEx adaptation parameters and mesoscopic response history and gain.

\subsection{Validity domain and limits of the reduction}

\revision{The extension assumes that adaptation changes little
within a bin ($\tau_w\gg T$) and separately neglects adaptation
variance and mixed rate-adaptation covariances. The latter
requires small fluctuation feedback and is not guaranteed by
$\tau_w/T$ alone.} The choice of \(T\) is consequently regime dependent: Di Volo et al. used \(T=\tau_m=20\,\mathrm{ms}\) by default and found that \(T=50\,\mathrm{ms}\) better reproduced Up-state durations, whereas the VSDi model used \(T=5\,\mathrm{ms}\) to resolve faster transients. However, inhibitory rates during Up-state can approach or exceed $20$ Hz, so the larger bins that improve Up-state durations simultaneously push the discrete-state construction toward, or beyond, its saturation limit. \revision{In the binned construction, \(T\mathcal{F}_\mu\) is a firing probability, and the transition variance \(A^{-1}_{\mu\mu}=\mathcal{F}_\mu(T^{-1}-\mathcal{F}_\mu)/N_\mu\) becomes negative once \(\mathcal{F}_\mu>1/T\). In the small-jump reading, whose covariance equation resembles the published one, \(m_\mu\) is instead a filtered rate without this ceiling, but the noise source is then \(\mathcal{F}_\mu/(N_\mu T)\) rather than \(A^{-1}_{\mu\mu}\) (Supplementary~\ref{subsubsec:covariance_closure_benchmarks}). In either reading, covariance equations with the binomial noise term should not be integrated above \(\mathcal{F}_\mu=1/T\).} A fixed \(T\) is therefore a modeling approximation rather than a universal physiological time constant, and agreement in different regimes is empirical. State-dependent response times, which vary with firing rate and transfer function gain \cite{ostojic2011spiking}, suggest a possible route toward a more adaptive closure.

Further limitations remain. The Up states extend the equations beyond the asynchronous regime in which the master-equation assumptions are best justified, so agreement with the spiking network is empirical rather than guaranteed by the derivation. The imposed OU drive reintroduces stochasticity but not the state-dependent covariance structure of the second-order closure. The RS and FS populations also remain internally homogeneous, and adaptation is represented only by its population mean; neuron-to-neuron variability, adaptation-current fluctuations, and correlations between adaptation and spiking are averaged out. Nevertheless, \(W\) survives the later first-order whole-brain truncation and provides the explicit intervention pathway through which changes in excitatory adaptation can be propagated to large-scale dynamics in Section~\ref{sec:Sacha_Molecular_MFT}.

\section{Molecular-to-Macro Mapping: A whole-brain framework}
\label{sec:Sacha_Molecular_MFT}

Sacha et al. (2025) \cite{sacha2025computational} provide the final integration step in the lineage examined here. Rather than introducing a new local mean-field theory, they combine receptor-aware conductance-based transfer functions, explicit mesoscopic adaptation, and connectome coupling within TVB. The resulting framework propagates effective changes in synaptic kinetics and cellular excitability from a reference cortical circuit to whole-brain dynamics associated with wakefulness, anesthesia, and NREM sleep. Its contribution is therefore a mechanistically traceable model chain, not a molecularly exhaustive or one-to-one description: microscopic interventions remain interpretable only through the parameters and state variables preserved by the successive reductions.

\subsection{From an adaptive cortical circuit to a whole-brain model}

\revision{The reference circuit contains $10{,}000$ conductance-based AdEx neurons: $80\%$ excitatory regular-spiking (RS) cells and $20\%$ inhibitory fast-spiking (FS) cells, with $5\%$ random connectivity. Synaptic input is represented by one pooled excitatory and one inhibitory conductance process, with quantal conductances $Q_e,Q_i$ and decay times $\tau_e,\tau_i$; AMPA and NMDA contributions are not separate dynamical variables. Single-neuron simulations calibrate the transfer functions, and recurrent-network simulations provide the reference for evaluating the adaptive mean-field node. The spiking circuit is not simulated within the whole-brain model.}

At the macroscopic scale, 68 cortical regions from the Desikan--Killiany atlas are coupled through empirical structural connectivity and tract-length-dependent delays. Each region $k\in \llbracket 1,68\rrbracket$ is represented by the same first-order adaptive RS-FS mean-field system: 
\begin{equation}
\begin{aligned}
T\frac{d\nu_e(k)}{dt} 
&= \mathcal{F}_e[\nu_e^{input}(k), \nu_i(k), W(k)] - \nu_e(k), \\
T\frac{d\nu_i(k)}{dt} 
&= \mathcal{F}_i[\nu_e^{input}(k), \nu_i(k), 0] - \nu_i(k), \\
\tau_w \frac{dW(k)}{dt} 
&= -W(k) + b\tau_w \nu_e(k) + a(\langle V(\nu_e(k), \nu_i(k), W(k))\rangle - E_L),
\end{aligned}
\label{eq:TF-based-adaptive-rate-first-order-model} 
\end{equation}
\revision{Here $W(k)$ is the excitatory adaptation current;
inhibitory adaptation is set to zero. Adaptation therefore
affects inhibitory activity indirectly through the coupled rates.}

\revision{The set of equations \ref{eq:TF-based-adaptive-rate-first-order-model} is an adaptive rate model obtained by first-order truncation of the master-equation moment system. The truncation implications are discussed in Section~\ref{subsec:first_order_truncation_limitations_section}.} Its derivation is detailed in Supplementary~\ref{subsec:adaptation_master_equation_formalism}. The excitatory input rate $\nu_e^{input}(k)$ includes the afferent noise described in Eq.~\eqref{eq:OU_noise}, together with delayed excitatory input from connected regions weighted by the structural connectome ($G\sum_jC_{kj}\nu_e(j,t-\ell_{kj}/v_c)$, with $\ell_{kj}$ the tract length between regions $j$ and $k$). Long-range interactions are therefore excitatory, whereas inhibition remains local to each node.

The whole-brain implementation thus preserves excitatory and inhibitory rates, excitatory adaptation, \revision{effective} synaptic kinetics within \(\mathcal{F}_\mu\), connectome weights, and propagation delays. It assigns the same local transfer functions and cellular parameters to every region and omits the second-order covariance dynamics; the consequences of these approximations are examined in the following section. The resulting neuronal states are not empirical signals. For comparison with fMRI, the TVB BOLD monitor convolves each region's excitatory rate with a hemodynamic response based on the first-order Volterra kernel of the Balloon--Windkessel model and downsamples the result to the acquisition timescale \cite{friston2000nonlinear}. This is a separate observation model whose assumptions condition the comparison with BOLD data.

\subsection{Effective mapping of molecular mechanisms}
\label{subsec:effective_mapping_molecular_mechanisms}

The framework represents molecular or neuromodulatory interventions through parameters that remain explicit after reduction. GABAergic anesthetics such as propofol are approximated by increasing the inhibitory synaptic decay time \(\tau_i\), representing prolonged GABA\(_A\)-mediated inhibition. NMDA antagonists such as ketamine and xenon are approximated by decreasing the excitatory decay time \(\tau_e\), representing shortened AMPA-NMDA excitatory kinetics. NREM sleep is modeled mainly by increasing the excitatory spike-triggered adaptation parameter \(b_e\), an effective representation of the increased adaptation associated with reduced cholinergic tone. \revision{We stress that both the spiking circuit and the reduced transfer functions distinguish one excitatory and one inhibitory conductance process; AMPA and NMDA contributions are not represented by separate conductance variables. The model therefore preserves selected pharmacologically motivated intervention parameters, rather than a receptor-resolved molecular description.}

\revision{Local parameter-interaction analyses show how synaptic kinetics and adaptation jointly affect the transition from sustained activity to Up-Down dynamics. Increasing $b_e$ strengthens slow negative feedback, whereas increasing $\tau_i$ or decreasing $\tau_e$ changes excitability and shifts the adaptation threshold for loss of the active fixed point. Spiking-network survival-time maps and first-order mean-field fixed-point analysis provide complementary descriptions of this transition; finite-size fluctuations make the spiking-network boundary more gradual. These analyses characterize interactions between model parameters. The whole-brain contrasts below instead compare composite parameter settings.}

\revision{The NMDA and AMPA pooling is already present in the reference spiking circuit and is inherited by the reduced transfer functions} (Supplementary~\ref{subsec:calculs_subthreshold_membrane_statistics}). The parameters $(\tau_e,\tau_i,b_e)$ therefore define effective intervention axes, not separate receptor mechanisms or a model of drug concentration and pharmacokinetics. The correspondences $(\tau_e,\tau_i,b_e)$ are effective and generally non-unique; they allow alternative mechanistic routes to be compared, but do not by themselves identify a molecular cause from a macroscopic signal.

\subsection{Whole-brain signatures and empirical comparison}

Sacha et al. tested whether the microscopic perturbations reproduce two macroscopic signatures of unconscious states. \revision{The drug-like conditions are composite parameter settings: relative to wakefulness, both drug-like conditions increased $b_e$ from 5 to 30 pA, together with either an increase in $\tau_i$ from 5 to 7 ms (for propofol-like) or a decrease in $\tau_e$ from 5 to 3.75 ms (for NMDA-blockade-like). Their macroscopic contrasts consequently cannot be assigned directly to the synaptic change alone}. First, they compared functional connectivity derived from empirical and simulated BOLD signals with anatomical structural connectivity. Structure-function correlation increased under propofol anesthesia and NREM sleep, but not under ketamine, in both the empirical datasets and the corresponding simulations. The model overestimated the magnitude of the increase, so this result supports qualitative state discrimination rather than quantitative validation.

Second, a localized perturbation was applied to one cortical node, and perturbational complexity was computed from the thresholded spatiotemporal pattern of evoked activity. Wake-like dynamics produced more widespread and sustained propagation and higher perturbational complexity index (PCI) values than the propofol-like, NMDA-blockade, and NREM-like conditions. The framework therefore links distinct effective microscopic interventions not only to spontaneous slow waves, but also to state-dependent large-scale responsiveness.

Together, these results establish the principal promise of the receptor-to-whole-brain construction: synaptic time constants and excitatory adaptation remain manipulable across the reduction and generate testable mesoscopic and macroscopic consequences. They do not establish a unique inverse mapping from BOLD or PCI to molecular mechanism. That interpretation remains conditional on the calibrated transfer functions, homogeneous regional nodes, first-order truncation, connectome, external noise, and observation or analysis models examined in Section~\ref{sec:Sacha_whole_brain_framework_limitation_extansion}.

\section{Discussion and perspectives}
\label{sec:Sacha_whole_brain_framework_limitation_extansion}

The lineage reviewed above provides a traceable, but conditional, bridge from cellular and synaptic mechanisms to whole-brain dynamics. Its main strength is that selected intervention variables (notably excitatory and inhibitory synaptic kinetics and excitatory adaptation) remain explicit after reduction and can therefore generate testable consequences across scales. Its limitations follow from the same reductions: finite-size covariances, cellular diversity, regional specialization, plasticity, and measurement biophysics are either simplified or omitted. This section examines these limitations and the complementary roles of mechanistic, data-driven, and hybrid extensions.

\subsection{First-order truncation and finite-size fluctuations}
\label{subsec:first_order_truncation_limitations_section}

The spatial and whole-brain implementations derived from the adaptive master equation discussed here retain only first-order population dynamics \cite{zerlaut2018modeling,di2019biologically,goldman2023comprehensive,sacha2025computational}. \revision{The truncation retains the calibrated conductance-based transfer functions and mean adaptation, and the neuron number still enters the transfer functions through the input connectivity (Supplementary~\ref{subsec:calculs_subthreshold_membrane_statistics}). It omits, however, the explicit evolution of the finite-size covariances and their curvature feedback on the mean activity $\boldsymbol{\nu}$.} \revision{As the linear model illustrates (Supplementary~\ref{subsec:master_equation_linear_model}), recurrent amplification can make these terms appreciable near an instability even when populations are large: with the corrected covariance equation, the stationary variances grow as $(1-\Delta^2)^{-1}$ with the net gain $\Delta$, for strong inhibitory ($\Delta\to-1$) as well as excitatory ($\Delta\to1$) feedback.} The second-order derivation therefore remains informative: it identifies the assumptions and discarded terms, and the regimes in which the first-order approximation is most vulnerable\revision{. It does not, by itself, give the first-order equations a unique dynamical or empirical advantage; their specificity rests mainly on the semi-analytical transfer functions.}

The motivation for truncation is primarily identifiability and computational, although the relevant distinction is between local and global covariances. As detailed in Section~\ref{subsec:algorithmic_simulation_cost}, retaining covariances only among the \(P\) populations within each of \(K\) nodes increases the node-local prefactor but, for fixed \(P\), preserves the whole-brain scaling \(\mathcal{O}(K+\rho K^2)\). Propagating covariances among all \(PK\) population variables instead requires \(\mathcal{O}(P^2K^2)\) storage and, under dense covariance algebra, up to \(\mathcal{O}(P^3K^3)\) work. A dense global second-order closure is therefore generally cost dominant and can become prohibitive for many regions $K$, whereas a node-local closure may remain tractable.

Externally imposed Ornstein--Uhlenbeck noise reintroduces variability into the first-order model \revision{, which can generate recurrently amplified activity and correlated activity}, but it is not equivalent to endogenous finite-size fluctuations: it does not reproduce state dependence or their population-size \(N_\mu^{-1}\) scaling. First-order dynamics are consequently appropriate when the target is scalable, qualitative mean behavior, but require caution for quantitative variance, transition statistics, or dynamics near critical points. A useful intermediate extension would retain second-order covariances within each region\revision{, using the corrected covariance equation,} while truncating inter-regional covariances.

\subsection{Heterogeneity across cells and regions}
\subsubsection{Microscopic heterogeneity: Intra-node diversity}

Each cortical node in Sacha et al. contains one homogeneous excitatory RS population and one homogeneous inhibitory FS population \cite{sacha2025computational}. This is a strong reduction because cortical neurons span diverse transcriptomic, morphological, and intrinsic electrophysiological phenotypes \cite{scala2021phenotypic}; such heterogeneity can alter population responsiveness, dynamical states, and recurrent information flow rather than merely add noise \cite{di2021optimal}. The heterogeneous firing responses reported by Zerlaut et al. \cite{zerlaut2016heterogeneous} suggest one practical extension: calibrated transfer functions or effective-threshold coefficients could be clustered into a small number of functional subpopulations instead of being averaged into a single excitatory or inhibitory response.

This extension is potentially consequential. Neuronal diversity can increase responsiveness and information propagation by positioning recurrent networks near dynamical transitions \cite{di2021optimal}. Similarly, Kim and Choi \cite{kim2026inhibitory} studied heterogeneous inhibitory populations, including PV, SST, and VIP interneurons, and showed analytically, with mean-field formalism, that inhibitory diversity can stabilize spatially structured cortical networks in ways that homogeneous E-I models cannot. Yet additional populations increase the number of states and poorly constrained parameters, and some cellular variability is averaged out at EEG, MEG, or fMRI scales. Heterogeneity should therefore be retained selectively: its inclusion is justified when it changes the node's dynamical regime, perturbation response, or target observable, and when the corresponding distributions can be constrained by data.

\subsubsection{Macroscopic heterogeneity: Regional Specialization}

The whole-brain model assigns the same local transfer functions and cellular parameters to all cortical regions; regions differ only through their connectome-derived inputs. Real regions also differ in cytoarchitecture, receptor expression, local circuitry, and intrinsic dynamics. Introducing a limited number of region- or tissue-specific node classes may therefore improve biological realism without resolving every local cell type. The thalamic mean-field model of Overwiening et al. \cite{overwiening2024multi}, whose responsiveness depends on physiological state, illustrates the value of a specialized node. Such extensions remain labor-intensive and create additional identifiability problems, but are a natural step from qualitative whole-brain reproduction toward quantitatively constrained, anatomically differentiated models.

\subsection{From latent dynamics to empirical signals}
\label{subsec:beyond_qualitative_comparison_for_empirical_signals}

Mean-field variables such as firing rate, adaptation, conductance statistics, and mean membrane potential are latent states rather than EEG, MEG, LFP, VSDi, or BOLD signals. The VSDi application in Section~\ref{sec:mean_field_macro_VSDi_model} and the BOLD monitor in Section~\ref{sec:Sacha_Molecular_MFT} already introduce observation mappings, but these mappings are themselves approximations and must be evaluated as part of the complete generative model.

For electrophysiology, Tesler et al. \cite{tesler2022mean} combined mean-field dynamics with unitary-LFP kernels, current-dipole models, and volume conduction, and validated the resulting LFP and MEG predictions against conductance-based AdEx spiking networks. For fMRI, Tesler et al. \cite{tesler2023modeling} proposed a more mechanistic pathway from excitatory activity through astrocytic calcium, vasomodulator release, vascular tone, cerebral blood flow, and Balloon-type hemodynamics. These approaches make direct multimodal tests possible, but add assumptions concerning tissue geometry, dipole orientation, neurovascular coupling, and hemodynamics. Observation models should therefore be treated as mechanistic layers with their own parameters and validity domains, not as neutral filters of neuronal activity.

\subsection{Plastic mean-field models beyond fixed local transfer functions}

In Sacha et al., synaptic efficacies and local transfer function parameters are fixed rather than modified by activity-dependent plasticity. Helson, Tanré, and Veltz \cite{helson2025mean} developed a McKean--Vlasov formulation for an all-to-all network of stochastic binary neurons with STDP. Rather than tracking $\mathcal{O}(N^2)$ individual synapses, their numerical mean-field approximation evolves representative neurons together with distributions of presynaptic states and incoming weights, yielding a reported $\mathcal{O}(N)$ scaling while retaining plasticity-induced synaptic heterogeneity. This provides a proof of principle for plastic mean-field modeling, but not yet a direct replacement for AdEx-based whole-brain nodes: the distribution-valued state remains high-dimensional, the neuronal dynamics are binary, and extensions to heterogeneous populations, receptor-specific mechanisms, and connectome-coupled regions remain open.

\subsection{Data-driven and hybrid extensions}
\label{subsec:data_driven_surrogates_MFT}

Because the transfer function is a central bottleneck of the mean-field framework, data-driven surrogates offer a natural extension. These strategies differ in how much of the mechanistic model they replace. One may replace only the single-neuron transfer function, learn the full local mesoscopic vector field, use scientific machine learning to train or augment the mechanistic equations, or bypass node-based modeling entirely with a whole-brain foundation model.

\subsubsection{Replacing only the transfer function}

The most conservative strategy replaces \(\mathcal{F}\) while preserving the master-equation scaffold. Spaeth et al. \cite{spaeth2024model} fitted a four-parameter \textit{Refractory SoftPlus} transfer function directly to simulated spike trains. The construction is model-agnostic and avoids requiring infinitesimal postsynaptic potentials, high presynaptic rates, or an analytically tractable diffusion approximation; it can therefore accommodate neuron models for which semi-analytical voltage statistics are unavailable, such as multi-compartment Hodgkin–Huxley neurons, and it is computationally lighter than the semi-analytical transfer function of Zerlaut et al. \cite{zerlaut2016heterogeneous}. The cost is a loss of mechanistic transparency: the explicit chain from synaptic conductances through membrane-potential moments to firing rate is compressed into a fitted input-output curve, which remains tied to the statistics of the input ensemble used during calibration. This weakens receptor-level interpretation and extrapolation to synchronized or out-of-distribution regimes.

\subsubsection{Learning the local mesoscopic vector field}

A broader strategy learns the complete local population dynamics from a microscopic simulator. Breyton et al. \cite{breyton2025data} trained a multilayer perceptron to reproduce the phase flow of mean firing rate and membrane potential generated by a spiking network, benchmarked it against the exact MPR reduction, and included connection probability as an explicit parameter. Because the learned system remains a low-dimensional vector field, it can be analyzed with bifurcation tools, embedded in a whole-brain model, and inverted against synthetic or empirical fMRI data.

The advantage is flexibility: any parameter that can be varied in the simulator can in principle be included in the learned macroscopic model, such as connection probabilities that may be difficult to include in standard analytical reductions which require strong assumptions and approximations. This flexibility transfers the validity problem to the training distribution. Learned dynamical-system reconstructions can fail to generalize to unsampled regions of phase space, particularly without
appropriate structural priors \cite{goring2024out}. Moreover, unconstrained neural ODEs can learn unnecessarily irregular vector fields, motivating Jacobian or kinetic regularization to improve their
regularity and numerical conditioning \cite{finlay2020train}. Nevertheless, encouraging extrapolation behavior was observed in practice, where Breyton et al. \cite{breyton2025data} recovered a chaotic regime comparable to the analytical solution, without explicit feeding chaotic data during training. The requirement for spiking-network simulations is partly offset by the fact that mechanistic reductions also require them for calibration and validation, although typically in far smaller quantity. This approach no longer answers the question "how can the mean-field equations be derived from the microscopic model?", but rather "how can a low-dimensional surrogate be trained to reproduce the microscopic simulator?" This is a legitimate and potentially powerful aim, where deep learning can excel, but this is a different objective from deriving a mechanistic population model presented in this present review.

\subsubsection{Scientific machine learning and hybrid equation learning}

Scientific machine learning (SciML) offers an intermediate route between fixed mechanistic models and unconstrained black-box surrogates. Rather than discarding the governing equations, it can make them differentiable or replace only selected, poorly characterized components with trainable functions. Neural ODEs \cite{chen2018neural} parameterize continuous-time dynamics and permit gradient-based training through a numerical solver, whereas physics-informed neural networks constrain learned solutions by penalizing violations of known equations. Universal differential equations (UDEs) \cite{rackauckas2020universal} more explicitly combine both strategies, $\dot{X} = f_{\mathrm{known}}(X,\theta)+f_{\mathrm{learned}}(X,\phi)$, retaining the mechanistic scaffold while learning a missing term. In receptor-aware mean-field models, the known component could describe adaptive conductance-based population dynamics and connectome coupling, while the learned component accounts for unresolved heterogeneity, state-dependent gain modulation, plasticity, or other systematic model discrepancies \cite{elgazzar2024universal}.

Related methods address complementary aspects of the problem. Neural SDEs \cite{kidger2021neural} can represent unresolved stochastic dynamics, Neural CDEs \cite{kidger2020neural} are well suited to irregularly sampled observations, and differentiable delay-equation solvers could retain inter-regional propagation delays during training. Neural operators instead learn mappings between functions, such as a map from parameters or initial conditions to complete spatiotemporal solutions. Fourier neural operators (FNOs) \cite{li2020fourier} may therefore provide fast surrogates for repeated simulation or parameter exploration. However, unlike UDEs, they do not necessarily preserve the decomposition of the original equations; moreover, standard FNOs are most natural on regular spatial domains, whereas graph-based operator variants may be better suited to irregular connectomes. These approaches complement recent proposals for functional whole-brain models that combine biologically grounded dynamics with gradient-based optimization \cite{senden2026functional}.

SciML may also facilitate parameter inference. Differentiable frameworks such as TVB-Optim \cite{pille2025fast} enable gradient-based fitting of whole-brain models to empirical observables while retaining their mechanistic structure. When the likelihood is unavailable but forward simulation is possible, simulation-based inference (SBI) can instead learn an approximate posterior from simulated parameter-observation pairs and then apply it to empirical data \cite{gonccalves2020training}. However, SciML approaches do not eliminate all the core difficulties of the problem. First, the resulting optimization can be numerically delicate, especially for stiff, chaotic, delayed, or weakly identifiable systems. Second, the learned correction term in a UDE or related hybrid model may absorb several distinct modeling errors at once, thereby reducing interpretability if it is left unconstrained. In all cases, however, identifiability remains a major limitation: many parameter combinations may generate similar macroscopic signals, failing to recover a unique mechanistic explanation. Therefore, optimization must be constrained by strong biological priors; nevertheless, while it allows to restrict the admissible parameter space, it cannot guarantee uniqueness when different parameter combinations produce indistinguishable macroscopic signals. Posterior correlations or likelihood ridges should then be treated as evidence of non-identifiability, motivating reparameterization around identifiable combinations, model reduction, or additional observables and perturbations \cite{wieland2021structural}. Learned correction terms likewise require structural constraints and out-of-sample validation, because they may conflate several sources of model error. SciML is thus best viewed as a promising way to fit, accelerate, and selectively extend mechanistic whole-brain models, not as a substitute for mechanistic assumptions or identifiability analysis.

\subsubsection{Beyond node-based mean-field: data-driven whole-brain foundation models}

Foundation models such as TRIBE v2~\cite{d2026foundation} or the visual-cortex model of Wang et al.~\cite{wang2025foundation} pursue a different objective. Rather than deriving local circuit equations and coupling them through a structural connectome, TRIBE v2 maps video, audio, and text representations to high-resolution fMRI responses across subjects and tasks. This supports prediction, subject generalization, in-silico stimulus exploration, protocol design, and possibly data-augmentation strategies and knowledge distillation from teacher models \cite{gou2021knowledge} which could be valuable in brain-computer interface.

Such models are not assumption-free: their predictions depend on pretrained feature extractors, architecture, training objectives, temporal windows, subject alignment, and large supervised datasets. They also do not directly explain how receptor kinetics, adaptation, excitation-inhibition balance, or anatomical coupling generate the predicted activity, and their fMRI target inherits BOLD's temporal and hemodynamic constraints. They should therefore be viewed as predictive complements and empirical benchmarks, not as direct replacements for receptor-aware mechanistic models. \revision{At the sensory-circuit scale, Wang et al.~\cite{wang2025foundation} developed a foundation model of mouse visual cortex whose shared core supported transfer to new mice with limited additional training and predicted neuronal responses to stimulus classes absent from training. This provides a complementary example of cross-animal and cross-stimulus generalization. Nevertheless, while relying on finer-grained scale of prediction, it still suffers from the same identifiability and causal interpretations.}

\subsubsection{Complementary objectives and model choice}
\label{subsubsec:mechanistic_vs_datadriven}

These approaches form a spectrum of retained mechanistic structure rather than a single performance ranking. Receptor-aware mean-field models preserve an explicit intervention path from synaptic kinetics, conductances, and adaptation through local dynamics and connectome coupling. Replacing only \(\mathcal{F}\) preserves most of that scaffold; learning the local vector field increases flexibility but weakens the derivational link; foundation models prioritize prediction and generalization while largely relinquishing the microscopic-to-macroscopic causal chain.

These approaches should not be compared through a simple "which is better?" criterion, because they optimize different objectives. A large data-driven model may define a strong predictive benchmark for the empirical observables included in its training objective, and may set a strong empirical baseline or approximate performance ceiling for the specified task and dataset, but this advantage comes with weaker parameter identifiability, reduced causal interpretability, high training cost, and limited ability to map receptor-level mechanisms onto model variables. The appropriate model therefore depends on the claim being tested. Mechanistic models are preferable for causal interpretation, pharmacological perturbation, and biologically grounded extrapolation when the retained mechanisms and validity domain are appropriate, but their parameters may still be non-identifiable and their empirical predictions inaccurate. Data-driven models may provide stronger predictive baselines without revealing a unique biological mechanism. Reverse-engineering their representations can expose useful computational constraints, but does not by itself recover biophysical causation. Hybrid approaches may combine mechanistic scaffolds with data-driven calibration or correction, provided that the learned component is constrained and separately validated \cite{senden2026functional}. These distinct objectives motivate the multi-axis benchmarks developed in Section~\ref{sec:benchmarks_surrogates}, which separate simulator faithfulness, empirical adequacy, generalization, identifiability, mechanistic traceability, and computational burden.

\section{Toward benchmarks for mesoscopic and whole-brain surrogates}
\label{sec:benchmarks_surrogates}

As the repertoire of mesoscopic and whole-brain models expands, model choice increasingly depends on the scientific objective. Predictive accuracy is necessary but not sufficient: model comparisons depend on the imaging modality~\cite{castaldo2023multi}, the statistic used to summarize activity~\cite{liu2025benchmarking}, the dynamical features selected for evaluation~\cite{bryant2024extracting}, and the fitting procedure~\cite{wischnewski2022towards}. Benchmarks should therefore be multi-axis and should report the scientific task, reference target, observable, fitting protocol, and computational workload. Table~\ref{tab:benchmark_axes} summarizes these dimensions without assigning model scores.

Two validation questions should be distinguished. \emph{Simulator-to-surrogate faithfulness} asks whether a reduced model preserves specified macroscopic properties of a microscopic simulator. \emph{Empirical adequacy} asks whether the complete generative pipeline (e.g., node dynamics, connectivity, delays, noise, and observation model) reproduces measured signals. Neither reference is complete biological ground truth: a simulator embodies modeling assumptions, whereas empirical data are partial and measurement-dependent. A surrogate can therefore be faithful to its spiking parent yet fail to reproduce BOLD or MEG, while a model fitted to empirical functional connectivity need not correspond to a controlled microscopic reduction.

\subsection{Evidence from targeted studies at both scales}
\label{subsec:benchmark_evidence}

\paragraph{Simulator-to-surrogate faithfulness.}
When a surrogate is intended as a reduction of a specified spiking network, the simulator provides the immediate reference. Empirical resources such as MICrONS~\cite{microns2025functional} add biological constraints, but do not observe every state variable, connection, or physiological parameter. Deschle et al.~\cite{deschle2021validity} compared Freeman-type neural masses with averaged noisy LIF networks across E/I composition and network degree. Agreement was restricted to particular regimes, notably near the onset of low-frequency synchronization, while amplitude, frequency, or synchrony differed elsewhere. Validity must therefore be tested across the regimes and observables relevant to the intended use. Baldy et al.~\cite{baldy2024inference} further show how dynamical noise, bistability, and partial observation affect parameter and vector-field inference and the difficulty to recover some parameters, though simulation-based inference and Neural ODE methods showed to respectively improve parameter recovery and reconstruction of macroscopic vector fields directly from microscopic simulations. \revision{The BRIDGE workflow~\cite{carannante2026bridge} integrates transfer-function calibration with spiking network comparisons. By evaluating transfer functions at measured network operating points alongside self-consistent mean-field predictions, it provides a practical diagnostic for distinguishing local prediction errors from discrepancies emerging in recurrent dynamics.} Analytical toolboxes such as NNMT~\cite{layer2022nnmt} can provide common baselines for such comparisons.

\paragraph{Empirical adequacy.}
At the whole-brain scale, the object of validation is the complete generative pipeline. Castaldo et al.~\cite{castaldo2023multi} found that comparisons between Stuart--Landau and Wilson--Cowan models depended on modality and evaluated feature. Liu et al.~\cite{liu2025benchmarking} showed that functional-connectivity statistics differ substantially across mapping objectives, while Bryant et al.~\cite{bryant2024extracting} demonstrated that alternative representations of local dynamics and inter-regional coupling capture complementary information. Calibration is itself part of the comparison: Wischnewski et al.~\cite{wischnewski2022towards} showed that optimization methods can approach dense grid-search solutions with markedly different computational costs and stability. Likewise, observation models must be evaluated rather than treated as neutral readouts. Shared platforms such as Neurolib~\cite{cakan2023neurolib} and Virtual Brain Inference~\cite{ziaeemehr2025virtual} facilitate comparisons under common simulation and inference pipelines.

These studies supply components of a benchmarking framework, but not a single protocol for all surrogate classes considered here. The next subsection examines one frequently underreported dimension: the cost of forward computation.

\begin{table*}[!t]
    \centering
    \small
    \renewcommand{\arraystretch}{0.95}
    \begin{tabularx}{\textwidth}{@{} L{0.72} L{1.35} L{1.35} L{0.53} @{}}
        \toprule
        \textbf{Benchmark dimension} &
        \textbf{What should be evaluated} &
        \textbf{Why it requires separate reporting} &
        \textbf{Support} \\
        \midrule
        Simulator faithfulness &
        Agreement with the specified microscopic simulator in rates, spectra, transients, fixed points, and bifurcations. &
        Agreement is feature- and regime-dependent; the simulator is a reference model, not biological ground truth. &
        \cite{deschle2021validity}; Sec.~\ref{subsec:under_hood_one_chain}; Sec.~\ref{subsec:benchmark_evidence} \\
        \addlinespace[1pt]
        Empirical adequacy &
        Agreement with modality-specific observables after applying the complete observation model. &
        Rankings depend on modality and summary statistic; similar FC fits can conceal differences in other measured features. &
        \cite{castaldo2023multi,liu2025benchmarking,bryant2024extracting} \\
        \addlinespace[1pt]
        Robustness and generalization &
        Performance across dynamical regimes, perturbations, parameter ranges, and out-of-fit conditions. &
        Accuracy within the calibration domain does not establish extrapolative validity, including for mechanistic models. &
        \cite{deschle2021validity}; Sec.~\ref{subsec:data_driven_surrogates_MFT} \\
        \addlinespace[1pt]
        Identifiability and fitting &
        Parameter recovery, uncertainty, degeneracy, sensitivity to noise and partial observation, and optimization cost. &
        Predictive fit does not imply unique parameter recovery; results can depend on the inference procedure. &
        \cite{baldy2024inference,wischnewski2022towards} \\
        \addlinespace[1pt]
        Mechanistic traceability &
        Whether variables and perturbations retain a testable mapping to biological quantities across scales. &
        Interpretability is task-dependent and cannot be placed on the same scale as predictive error. &
        Sec.~\ref{subsubsec:mechanistic_vs_datadriven} \\
        \addlinespace[1pt]
        Computational burden &
        Algorithmic work, memory traffic, runtime, and energy for forward simulation, fitting, and training. &
        Compute- and memory-limited regimes depend on workload, implementation, and hardware; training and inference are distinct costs. &
        Sec.~\ref{subsec:algorithmic_simulation_cost} \\
        \bottomrule
    \end{tabularx}
    \caption{Evaluation dimensions for mesoscopic and whole-brain surrogate models. The axes are reported separately because their relevance and measurement depend on the scientific task; the table is a checklist, not a model ranking.}
    \label{tab:benchmark_axes}
\end{table*}

\subsection{\revision{Algorithmic work and array accesses as benchmark dimensions}}
\label{subsec:algorithmic_simulation_cost}

Computational burden matters when models are used for parameter sweeps, inversion, or whole-brain embedding. However, comparisons are meaningful only for a specified workload and numerical accuracy. We therefore use an analytical proxy: the work required to process one biological second at a prescribed numerical resolution. The arithmetic cost per step \(W_{\Delta t}\) is counted in FLOP-equivalents, with special functions assigned explicit weights, and the streaming-memory term \(M_{\Delta t}\) counts bytes accessed under the numerical and storage assumptions stated in Supplementary~\ref{sec:algorithmic_cost_analysis_neural_simulation_appendix}. These are reproducible accounting conventions following the Roofline model~\cite{williams2009roofline}, but not hardware-independent runtime predictions. Unlike wall-clock runtime, these quantities do not directly depend on programming language, compiler, or hardware, and are therefore complementary to, rather than a substitute for, empirical runtime or energy measurements. In particular, estimated byte accesses need not equal traffic to main memory because caching, batching, precision, and data layout are implementation-dependent.

For a mesoscopic network with \(K\) regions and connectome density \(\rho\), the work separates into a node-local term and a long-range coupling term,
\begin{equation}
W_{T_{\mathrm{bio}}} \approx \frac{T_{\mathrm{bio}}}{\Delta t}\left[\, (W_{\mathrm{node}}^{\mathrm{RHS}}+W_{\mathrm{upd}}^{\mathrm{node}})\,K + c_{\mathrm{edge}}\,\rho K^2 \,\right].
\end{equation}
\revision{Here, $W_{\mathrm{node}}^{\mathrm{RHS}}$ counts one local right-hand-side evaluation, $W_{\mathrm{upd}}^{\mathrm{node}}$ the integration update, and $c_{\mathrm{edge}}$ one coupling operation. For fixed $P$, shared coupling conventions and fixed numerical resolution,} local biological detail changes the node-local prefactor but not its dependence on \(K\); \revision{first-order and node-local second-order models retain $\mathcal{O}(K+\rho K^2)$ work.} A dense global-covariance closure instead stores \(\mathcal{O}(P^2K^2)\) covariance entries and, under dense covariance algebra, requires \(\mathcal{O}(P^3K^3)\) work. This can rapidly dominate at whole-brain resolution, although exploitable sparsity or low-rank structure could alter the practical cost. 

\begin{figure}[htb]
    \centering
    \includegraphics[width=0.9\textwidth]{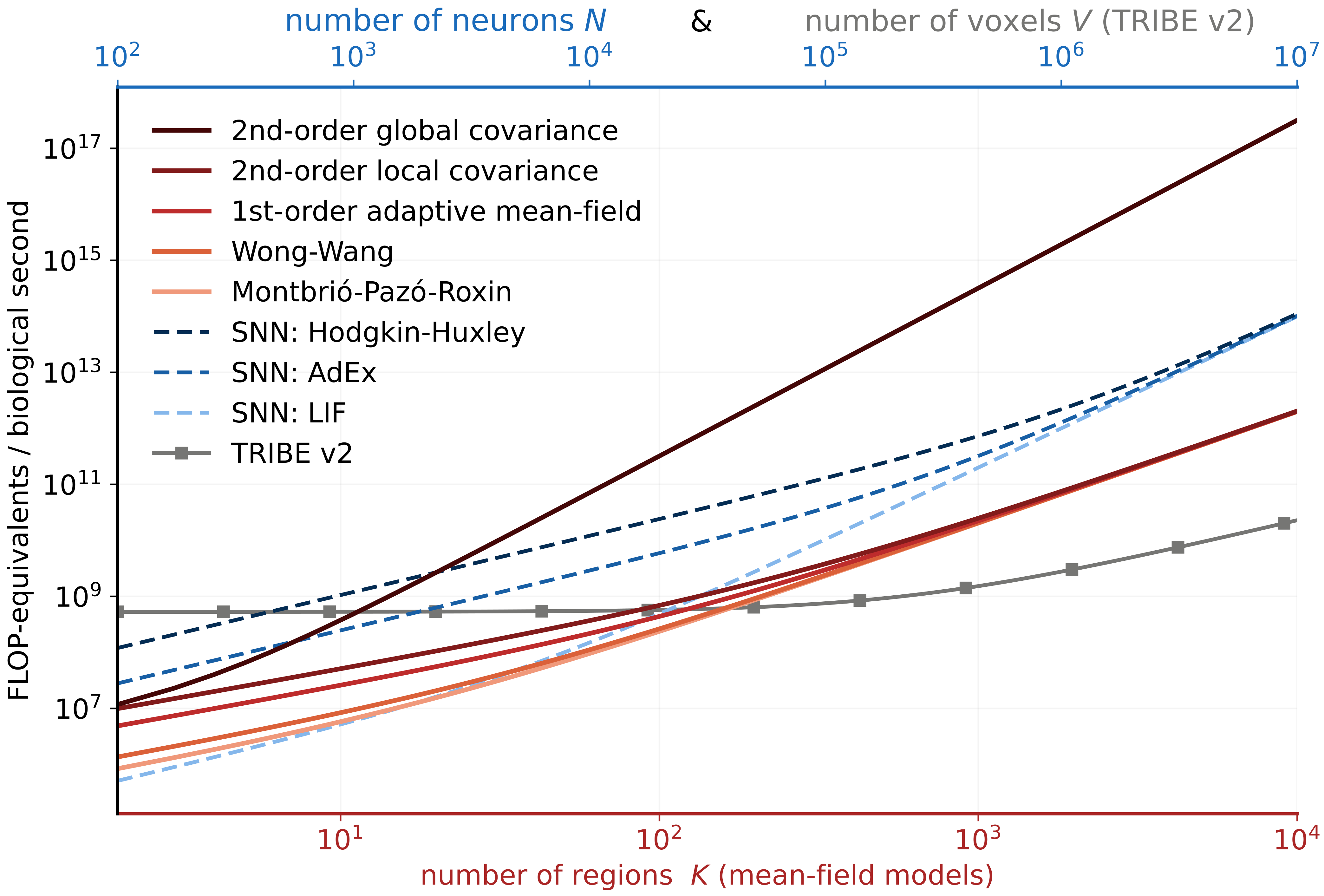}
    \caption{\revision{Illustrative FLOP-equivalent work for regional models (red, regions $K$), flat SNNs (blue dashed, neurons $N$), and cached-feature fMRI prediction (gray squares, targets $V$). The regional scenarios use $P=P_w=2$ and $\rho=1$; numerical resolutions and exclusions are specified in Supplementary~\ref{sec:algorithmic_cost_analysis_neural_simulation_appendix}. Work is per simulated biological second for the dynamical models and per second of input stimulus for TRIBE v2, excluding raw-feature extraction and training. The axis label "voxels" refers here to prediction targets, including cortical vertices. Each family uses its own logarithmic size axis; horizontal alignment, intersections and vertical separations do not establish biological equivalence or efficiency at matched accuracy or scientific performance. The global-covariance work curve is a schematic dense-algebra proxy.} We stress that the horizontal positions should therefore be compared only within the same colour family; the figure is intended to show scaling regimes and orders of magnitude, not a one-to-one equivalence between \(K\), \(N\), and \(V\).}
    \label{fig:algorithmic_cost_compute_P2}
\end{figure}

\revision{The plotted global-work curve is a schematic dense-algebra proxy, not a complete implementation estimate. Supplementary~\ref{sec:algorithmic_cost_analysis_neural_simulation_appendix} specifies these scenarios and their limitations, with work illustrations for $P=10$ (Fig.~\ref{fig:appendix_cost_compute_P10}) and array-access illustrations for $P=2$ and $P=10$ (Figs.~\ref{fig:appendix_memory_traffic_P2} and~\ref{fig:appendix_memory_traffic_P10}).} \revision{Figure~\ref{fig:algorithmic_cost_compute_P2} provides a compact view of how estimated work changes with each workload's own size parameter: regions $K$ for coupled population models, neurons $N$ for flat spiking networks, and output targets $V$ for cached-feature fMRI prediction. Costs are normalized per simulated biological second for the dynamical models and per second of input stimulus for TRIBE v2. We stress that the cross-family curves illustrate separately specified workloads, not comparative efficiency at equal scientific performance.}

A key point is that the per-node cost is independent of the number of microscopic neurons represented by each region, depending instead on the number of populations and the complexity of the local closure. Under the FLOP-equivalent convention of the Supplementary~\ref{sec:algorithmic_cost_analysis_neural_simulation_appendix}, the node-local RHS cost rises from about \(30\) operations for the exact Montbrió--Pazó--Roxin reduction to \(60\) for Wong--Wang, \(232\) for the first-order adaptive master equation (\(\mathcal{O}(P^2K+\rho K^2)\)), and \(472\) for its second-order local-covariance extension (\(\mathcal{O}(P^3K+\rho K^2)\)). This reflects the added cost of transfer functions, subthreshold moments, effective thresholds, and, in the second-order case, their derivatives. Because all these models share the same long-range coupling, their cost is increasingly dominated by the common \(\rho K^2\) term for large dense connectomes, and the associated memory traffic by the reading of connectivity weights and delayed states. The exception is the full global-covariance closure: propagating covariances between all \(PK\) population variables scales as \(\mathcal{O}(P^3K^3)\) in compute and \(\mathcal{O}(P^2K^2)\) in memory, which can become the dominant cost and may be prohibitive at high $P$, fine parcellation, or long duration at whole-brain resolution and motivates either omitting covariance dynamics through a first-order closure or retaining covariances only within individual nodes.

Spiking-network work contains a clock-driven term proportional to \(N/\Delta t\) and an event-driven term proportional to the number of synaptic deliveries, \(\langle\nu\rangle pN^2\). The illustrative integration steps adopted there also make detailed neuron models pay through both higher per-step work and more steps per biological second. At large event counts, synaptic delivery may dominate memory traffic, but the crossover remains implementation- and hardware-dependent. Nevertheless, as shown in the Supplementary~\ref{subsec:regional_snn_meanfield_benchmark}, the usual opposition between "spiking but local" and "mean-field but whole-brain" may be too simple. An intermediate architecture comprising $10\text{-}50$ regional microcircuits of $10^{3}\text{-}10^{4}$ point neurons each is no longer obviously out of computational reach, and a small LIF implementation can overlap with an elaborate mean-field closure in abstract work count. However, feasibility of one forward trajectory is not tractability for model fitting, nor is microscopic detail equivalent to biological information. The most informative validation is a matched comparison between an SNN and its own mean-field surrogate: if spike timing, finite-size fluctuations, heterogeneity, or plasticity do not alter the target observables or perturbation responses, the additional detail is primarily computational cost; if they do, the disagreement reveals precisely where the reduction ceases to be scientifically neutral.

TRIBE v2 is included only as a deliberately naive arithmetic proxy for a non-dynamical, fully data-driven whole-brain encoder. It maps stimulus windows to fMRI targets rather than integrating an autonomous neural system, so its cost is not task-equivalent to that of a spiking or mean-field simulation. The plotted workload covers prediction from cached multimodal features to fMRI and excludes the offline extraction of raw text, audio, and video features. It is neither a runtime benchmark nor a task-matched comparison: \(K\), \(N\), and $V$ represent different notions of model scale, and intersections between curves should not be interpreted as computational break-even points.

This benchmark is not a basis for ranking models in the absolute. A more expensive model may be preferable for its biological interpretability, dynamical fidelity, or access to mechanistic perturbations; a cheaper one for inversion, parameter sweeps, or clinical deployment. Throughout, we consider the inference regime only, the model is already parameterized and the task is to integrate it forward. Training and calibration costs can substantially exceed inference costs and should therefore be reported separately. The purpose of the metric is to make the trade-off between mechanistic detail, predictive performance, and computational burden explicit and to provide a systematic comparison between models.

\section{Conclusion}

The central conclusion is that mechanistic continuity across scales is possible without retaining microscopic trajectories, but only selectively and conditionally. In the lineage examined here, receptor-dependent synaptic kinetics, conductance-based input statistics, and spike-frequency adaptation are carried through calibrated neuronal transfer functions into mesoscopic and connectome-coupled whole-brain dynamics. The reduction thereby preserves intervention coordinates that can be perturbed and related to population states, large-scale propagation, and empirical observables. It supplies a mechanistic trace that an output fit alone cannot provide, but not an unbroken derivation from molecules to measurements.

The mapping between biological causes, effective model parameters, and macroscopic observables nevertheless remains non-unique, and quantitative reliability is regime-dependent. Molecular perturbations are represented by effective parameter changes; transfer functions and moment closures have bounded validity domains\revision{, and the original covariance closure had to be corrected}; first-order whole-brain nodes omit endogenous covariance dynamics; and cellular diversity, regional specialization, plasticity, connectivity, external noise, and observation models may produce partly overlapping macroscopic effects. Mechanistic models can therefore support causal analysis by predicting how interventions on defined variables propagate across scales. Agreement with functional connectivity or perturbational complexity establishes that a proposed mechanism is compatible with, and potentially sufficient to reproduce, the observations, but does not by itself demonstrate that this mechanism is the unique or actual biological cause. Computationally, node-local biological detail mainly changes prefactors, whereas dense global covariance propagation changes the scaling class.

Moving from causal predictions within the model to biologically supported causal explanations requires validation of each interface: molecular perturbations to cellular parameters, cellular dynamics to population closure, local nodes to large-scale coupling, and latent activity to measured signals. Matched comparisons with reference spiking networks, selective inclusion of heterogeneity and local fluctuations, multimodal perturbational data, uncertainty-aware inference, and out-of-regime testing are especially important. Learned surrogates and scientific machine learning can aid fitting and represent missing mechanisms, but must remain constrained when causal interpretation is claimed. The appropriate model is thus the least elaborate one that preserves the variables, fluctuations, intervention pathways, and observables required by the question, with its failures and computational costs reported as explicitly as its successes.

\section*{Author contributions}

YB: Conceptualization, Formal analysis, Investigation, Methodology, Visualization, Writing -- original draft, Writing -- review \& editing.
LB: Writing -- review \& editing.
AD: Conceptualization, Supervision, Writing -- review \& editing. All authors contributed to the article and approved the submitted version.

\section*{Data availability statement}

The code reproducing the covariance-closure benchmarks and the computational-cost figures is openly available at \href{https://github.com/YannaelB/mechanistic-bridges-mean-field-review}{YannaelB/mechanistic-bridges-mean-field-review}.

\section*{Funding}

\revision{Research supported by the CNRS and Paris-Saclay University, and by grants from the European Community (Virtual Brain Twin project 101137289, EBRAINS 2.0 project 101147319).}

\section*{Conflict of interest}

The authors declared that this work was conducted in the absence of any commercial or financial relationships that could be construed as a potential conflict of interest.

\printbibliography

@article{el2009master,
  title={A master equation formalism for macroscopic modeling of asynchronous irregular activity states},
  author={El Boustani, Sami and Destexhe, Alain},
  journal={Neural computation},
  volume={21},
  number={1},
  pages={46--100},
  year={2009},
  doi={10.1162/neco.2008.02-08-710}
}

@article{di2019biologically,
  title={Biologically realistic mean-field models of conductance-based networks of spiking neurons with adaptation},
  author={Di Volo, Matteo and Romagnoni, Alberto and Capone, Cristiano and Destexhe, Alain},
  journal={Neural computation},
  volume={31},
  number={4},
  pages={653--680},
  year={2019},
  doi={10.1162/neco_a_01173}
}

@article{sacha2025computational,
  title={A computational approach to evaluate how molecular mechanisms impact large-scale brain activity},
  author={Sacha, Maria and Tesler, Federico and Cofre, Rodrigo and Destexhe, Alain},
  journal={Nature Computational Science},
  volume={5},
  number={5},
  pages={405--417},
  year={2025},
  doi={10.1038/s43588-025-00796-8}
}

@article{brunel2000dynamics,
  title={Dynamics of sparsely connected networks of excitatory and inhibitory spiking neurons},
  author={Brunel, Nicolas},
  journal={Journal of computational neuroscience},
  volume={8},
  number={3},
  pages={183--208},
  year={2000},
  doi={10.1023/A:1008925309027}
}

@article{d2026foundation,
  title         = {A foundation model of vision, audition, and language for in-silico neuroscience},
  author        = {d'Ascoli, St{\'e}phane and Rapin, J{\'e}r{\'e}my and Benchetrit, Yohann and Brookes, Teon and Begany, Katelyn and Raugel, Jos{\'e}phine and Banville, Hubert and King, Jean-R{\'e}mi},
  journal       = {arXiv [Preprint]},
  year          = {2026},
  eprint        = {2605.04326},
  archiveprefix = {arXiv},
  primaryclass  = {q-bio.NC},
  doi           = {10.48550/arXiv.2605.04326},
  url           = {https://arxiv.org/abs/2605.04326}
}

@article{breyton2025data,
  title         = {Data-driven mean-field within whole-brain models},
  author        = {Breyton, Martin and Sip, Viktor and Woodman, Marmaduke and Hashemi, Meysam and Petkoski, Spase and Jirsa, Viktor},
  journal       = {arXiv [Preprint]},
  year          = {2025},
  eprint        = {2509.02799},
  archiveprefix = {arXiv},
  primaryclass  = {q-bio.NC},
  doi           = {10.48550/arXiv.2509.02799},
  url           = {https://arxiv.org/abs/2509.02799}
}

@article{kuhn2004neuronal,
  title={Neuronal integration of synaptic input in the fluctuation-driven regime},
  author={Kuhn, Alexandre and Aertsen, Ad and Rotter, Stefan},
  journal={Journal of Neuroscience},
  volume={24},
  number={10},
  pages={2345--2356},
  year={2004},
  doi={10.1523/JNEUROSCI.3349-03.2004}
}

@article{zerlaut2016heterogeneous,
  title={Heterogeneous firing rate response of mouse layer V pyramidal neurons in the fluctuation-driven regime},
  author={Zerlaut, Yann and Tele{\'n}czuk, Bartosz and Deleuze, Charlotte and Bal, T and Ouanounou, G and Destexhe, Alain},
  journal={The Journal of physiology},
  volume={594},
  number={13},
  pages={3791--3808},
  year={2016},
  doi={10.1113/JP272317}
}

@article{zerlaut2018modeling,
  title={Modeling mesoscopic cortical dynamics using a mean-field model of conductance-based networks of adaptive exponential integrate-and-fire neurons},
  author={Zerlaut, Yann and Chemla, Sandrine and Chavane, Frederic and Destexhe, Alain},
  journal={Journal of computational neuroscience},
  volume={44},
  number={1},
  pages={45--61},
  year={2018},
  doi={10.1007/s10827-017-0668-2}
}

@article{amit1997model,
  title={Model of global spontaneous activity and local structured activity during delay periods in the cerebral cortex.},
  author={Amit, Daniel J and Brunel, Nicolas},
  journal={Cerebral cortex (New York, NY: 1991)},
  volume={7},
  number={3},
  pages={237--252},
  year={1997},
  doi={10.1093/cercor/7.3.237}
}

@article{ostojic2011spiking,
  title={From spiking neuron models to linear-nonlinear models},
  author={Ostojic, Srdjan and Brunel, Nicolas},
  journal={PLoS computational biology},
  volume={7},
  number={1},
  pages={e1001056},
  year={2011},
  doi={10.1371/journal.pcbi.1001056}
}

@article{overwiening2024multi,
  title={A multi-scale study of thalamic state-dependent responsiveness},
  author={Overwiening, Jorin and Tesler, Federico and Guarino, Domenico and Destexhe, Alain},
  journal={PLOS Computational Biology},
  volume={20},
  number={12},
  pages={e1012262},
  year={2024},
  doi={10.1371/journal.pcbi.1012262}
}

@article{friston2000nonlinear,
  title={Nonlinear responses in fMRI: the Balloon model, Volterra kernels, and other hemodynamics},
  author={Friston, Karl J and Mechelli, Andrea and Turner, Robert and Price, Cathy J},
  journal={NeuroImage},
  volume={12},
  number={4},
  pages={466--477},
  year={2000},
  doi={10.1006/nimg.2000.0630}
}

@article{kim2026inhibitory,
  title   = {Inhibitory cell type heterogeneity in a spatially structured mean-field model of {V1}},
  author  = {Kim, Soon Ho and Choi, Hannah},
  journal = {Physical Review E},
  volume  = {113},
  number  = {5},
  pages   = {054406},
  year    = {2026},
  doi     = {10.1103/hmyq-c1j2}
}

@article{di2021optimal,
  title={Optimal responsiveness and information flow in networks of heterogeneous neurons},
  author={Di Volo, Matteo and Destexhe, Alain},
  journal={Scientific reports},
  volume={11},
  number={1},
  pages={17611},
  year={2021},
  doi={10.1038/s41598-021-96745-2}
}

@article{helson2025mean,
  title         = {Mean-field analysis of a neural network with stochastic {STDP}},
  author        = {Helson, Pascal and Tanr{\'e}, Etienne and Veltz, Romain},
  journal       = {arXiv [Preprint]},
  year          = {2025},
  eprint        = {2510.02545},
  archiveprefix = {arXiv},
  primaryclass  = {q-bio.NC},
  doi           = {10.48550/arXiv.2510.02545},
  url           = {https://arxiv.org/abs/2510.02545}
}

@article{tesler2022mean,
  title={Mean-field based framework for forward modeling of LFP and MEG signals},
  author={Tesler, Federico and Tort-Colet, N{\'u}ria and Depannemaecker, Damien and Carlu, Mallory and Destexhe, Alain},
  journal={Frontiers in computational neuroscience},
  volume={16},
  pages={968278},
  year={2022},
  doi={10.3389/fncom.2022.968278}
}

@article{tesler2023modeling,
  title={Modeling the relationship between neuronal activity and the BOLD signal: contributions from astrocyte calcium dynamics},
  author={Tesler, Federico and Linne, Marja-Leena and Destexhe, Alain},
  journal={Scientific Reports},
  volume={13},
  number={1},
  pages={6451},
  year={2023},
  doi={10.1038/s41598-023-32618-0}
}

@article{spaeth2024model,
  title={Model-agnostic neural mean field with a data-driven transfer function},
  author={Spaeth, Alex and Haussler, David and Teodorescu, Mircea},
  journal={Neuromorphic Computing and Engineering},
  volume={4},
  number={3},
  pages={034013},
  year={2024},
  doi={10.1088/2634-4386/ad787f}
}

@article{gou2021knowledge,
  title={Knowledge distillation: A survey},
  author={Gou, Jianping and Yu, Baosheng and Maybank, Stephen J and Tao, Dacheng},
  journal={International journal of computer vision},
  volume={129},
  number={6},
  pages={1789--1819},
  year={2021},
  doi={10.1007/s11263-021-01453-z}
}

@article{wilson1972excitatory,
  title={Excitatory and inhibitory interactions in localized populations of model neurons},
  author={Wilson, Hugh R and Cowan, Jack D},
  journal={Biophysical journal},
  volume={12},
  number={1},
  pages={1--24},
  year={1972},
  doi={10.1016/S0006-3495(72)86068-5}
}

@article{deschle2021validity,
  title   = {On the validity of neural mass models},
  author  = {Deschle, Nicol{\'a}s and Gossn, Juan Ignacio and Tewarie, Prejaas and Schelter, Bj{\"o}rn and Daffertshofer, Andreas},
  journal = {Frontiers in Computational Neuroscience},
  volume  = {14},
  pages   = {581040},
  year    = {2021},
  doi     = {10.3389/fncom.2020.581040}
}

@article{baldy2024inference,
  title={Inference on the macroscopic dynamics of spiking neurons},
  author={Baldy, Nina and Breyton, Martin and Woodman, Marmaduke M and Jirsa, Viktor K and Hashemi, Meysam},
  journal={Neural Computation},
  volume={36},
  number={10},
  pages={2030--2072},
  year={2024},
  doi={10.1162/neco_a_01701}
}

@article{layer2022nnmt,
  title={NNMT: mean-field based analysis tools for neuronal network models},
  author={Layer, Moritz and Senk, Johanna and Essink, Simon and van Meegen, Alexander and Bos, Hannah and Helias, Moritz},
  journal={Frontiers in neuroinformatics},
  volume={16},
  pages={835657},
  year={2022},
  doi={10.3389/fninf.2022.835657}
}

@article{bryant2024extracting,
  title={Extracting interpretable signatures of whole-brain dynamics through systematic comparison},
  author={Bryant, Annie G and Aquino, Kevin and Parkes, Linden and Fornito, Alex and Fulcher, Ben D},
  journal={PLoS computational biology},
  volume={20},
  number={12},
  pages={e1012692},
  year={2024},
  doi={10.1371/journal.pcbi.1012692}
}

@article{castaldo2023multi,
  title={Multi-modal and multi-model interrogation of large-scale functional brain networks},
  author={Castaldo, Francesca and Dos Santos, Francisco P{\'a}scoa and Timms, Ryan C and Cabral, Joana and Vohryzek, Jakub and Deco, Gustavo and Woolrich, Mark and Friston, Karl and Verschure, Paul and Litvak, Vladimir},
  journal={NeuroImage},
  volume={277},
  pages={120236},
  year={2023},
  doi={10.1016/j.neuroimage.2023.120236}
}

@article{wischnewski2022towards,
  title={Towards an efficient validation of dynamical whole-brain models},
  author={Wischnewski, Kevin J and Eickhoff, Simon B and Jirsa, Viktor K and Popovych, Oleksandr V},
  journal={Scientific reports},
  volume={12},
  number={1},
  pages={4331},
  year={2022},
  doi={10.1038/s41598-022-07860-7}
}

@article{ziaeemehr2025virtual,
  title   = {Virtual Brain Inference ({VBI}), a flexible and integrative toolkit for efficient probabilistic inference on whole-brain models},
  author  = {Ziaeemehr, Abolfazl and Woodman, Marmaduke and Domide, Lia and Petkoski, Spase and Jirsa, Viktor and Hashemi, Meysam},
  journal = {eLife},
  volume  = {14},
  pages   = {RP106194},
  year    = {2025},
  doi     = {10.7554/eLife.106194.4}
}

@article{liu2025benchmarking,
  title={Benchmarking methods for mapping functional connectivity in the brain},
  author={Liu, Zhen-Qi and Luppi, Andrea I and Hansen, Justine Y and Tian, Ye Ella and Zalesky, Andrew and Yeo, BT Thomas and Fulcher, Ben D and Misic, Bratislav},
  journal={Nature Methods},
  volume={22},
  number={7},
  pages={1593--1602},
  year={2025},
  doi={10.1038/s41592-025-02704-4}
}

@article{cakan2023neurolib,
  title={neurolib: A simulation framework for whole-brain neural mass modeling},
  author={Cakan, Caglar and Jajcay, Nikola and Obermayer, Klaus},
  journal={Cognitive Computation},
  volume={15},
  number={4},
  pages={1132--1152},
  year={2023},
  doi={10.1007/s12559-021-09931-9}
}

@article{senden2026functional,
  title         = {Functional whole-brain models: A new framework for unifying brain structure and cognitive function},
  author        = {Senden, Mario and Dalla Porta, Leonardo and Fousek, Jan and Mejias, Jorge F. and Zamora-L{\'o}pez, Gorka},
  journal       = {arXiv [Preprint]},
  year          = {2026},
  eprint        = {2605.18118},
  archiveprefix = {arXiv},
  primaryclass  = {q-bio.NC},
  doi           = {10.48550/arXiv.2605.18118},
  url           = {https://arxiv.org/abs/2605.18118}
}

@article{rackauckas2020universal,
  title         = {Universal differential equations for scientific machine learning},
  author        = {Rackauckas, Christopher and Ma, Yingbo and Martensen, Julius and Warner, Collin and Zubov, Kirill and Supekar, Rohit and Skinner, Dominic and Ramadhan, Ali and Edelman, Alan},
  journal       = {arXiv [Preprint]},
  year          = {2020},
  eprint        = {2001.04385},
  archiveprefix = {arXiv},
  primaryclass  = {cs.LG},
  doi           = {10.48550/arXiv.2001.04385},
  url           = {https://arxiv.org/abs/2001.04385}
}

@article{elgazzar2024universal,
  title   = {Universal differential equations as a unifying modeling language for neuroscience},
  author  = {El-Gazzar, Ahmed and van Gerven, Marcel},
  journal = {Frontiers in Computational Neuroscience},
  volume  = {19},
  pages   = {1677930},
  year    = {2025},
  doi     = {10.3389/fncom.2025.1677930}
}

@inproceedings{chen2018neural,
  title     = {Neural ordinary differential equations},
  author    = {Chen, Ricky T. Q. and Rubanova, Yulia and Bettencourt, Jesse and Duvenaud, David K.},
  booktitle = {Advances in Neural Information Processing Systems},
  volume    = {31},
  pages     = {6572--6583},
  year      = {2018},
  doi       = {10.5555/3327757.3327764}
}

@inproceedings{kidger2021neural,
  title     = {Neural {SDEs} as infinite-dimensional {GANs}},
  author    = {Kidger, Patrick and Foster, James and Li, Xuechen and Lyons, Terry J.},
  booktitle = {Proceedings of the 38th International Conference on Machine Learning},
  series    = {Proceedings of Machine Learning Research},
  volume    = {139},
  pages     = {5453--5463},
  year      = {2021},
  url       = {https://proceedings.mlr.press/v139/kidger21b.html}
}

@inproceedings{kidger2020neural,
  title     = {Neural controlled differential equations for irregular time series},
  author    = {Kidger, Patrick and Morrill, James and Foster, James and Lyons, Terry},
  booktitle = {Advances in Neural Information Processing Systems},
  volume    = {33},
  pages     = {6696--6707},
  year      = {2020},
  doi       = {10.5555/3495724.3496286}
}

@article{ccetin2026deterministic,
  title={Deterministic, stochastic, and mean-field PDE models in neuroscience},
  author={{\c{C}}etin, Co{\c{s}}kun and Piqueira, Jose Roberto Castilho and {\.I}zgi, Burhaneddin and Peker-Dobie, Ayse and Ahmetolan, Semra and {\"O}zkaya, Murat},
  journal={Frontiers in computational neuroscience},
  volume={20},
  pages={1762692},
  year={2026},
  doi={10.3389/fncom.2026.1762692}
}

@article{sarma2018openworm,
  title   = {{OpenWorm}: overview and recent advances in integrative biological simulation of {Caenorhabditis elegans}},
  author  = {Sarma, Gopal P. and Lee, Chee Wai and Portegys, Tom and Ghayoomie, Vahid and Jacobs, Travis and Alicea, Bradly and Cantarelli, Matteo and Currie, Michael and Gerkin, Richard C. and Gingell, Shane and others},
  journal = {Philosophical Transactions of the Royal Society B: Biological Sciences},
  volume  = {373},
  number  = {1758},
  pages   = {20170382},
  year    = {2018},
  doi     = {10.1098/rstb.2017.0382}
}

@article{zhao2024integrative,
  title={An integrative data-driven model simulating C. elegans brain, body and environment interactions},
  author={Zhao, Mengdi and Wang, Ning and Jiang, Xinrui and Ma, Xiaoyang and Ma, Haixin and He, Gan and Du, Kai and Ma, Lei and Huang, Tiejun},
  journal={Nature Computational Science},
  volume={4},
  number={12},
  pages={978--990},
  year={2024},
  doi={10.1038/s43588-024-00738-w}
}

@article{shiu2024drosophila,
  title={A Drosophila computational brain model reveals sensorimotor processing},
  author={Shiu, Philip K and Sterne, Gabriella R and Spiller, Nico and Franconville, Romain and Sandoval, Andrea and Zhou, Joie and Simha, Neha and Kang, Chan Hyuk and Yu, Seongbong and Kim, Jinseop S and others},
  journal={Nature},
  volume={634},
  number={8032},
  pages={210--219},
  year={2024},
  doi={10.1038/s41586-024-07763-9}
}

@article{dorkenwald2024neuronal,
  title={Neuronal wiring diagram of an adult brain},
  author={Dorkenwald, Sven and Matsliah, Arie and Sterling, Amy R and Schlegel, Philipp and Yu, Szi-Chieh and McKellar, Claire E and Lin, Albert and Costa, Marta and Eichler, Katharina and Yin, Yijie and others},
  journal={Nature},
  volume={634},
  number={8032},
  pages={124--138},
  year={2024},
  doi={10.1038/s41586-024-07558-y}
}

@article{reimann2026modeling,
  title   = {Modeling and simulation of neocortical micro- and mesocircuitry (Part I, anatomy)},
  author  = {Reimann, Michael W. and Bola{\~n}os-Puchet, Sirio and Courcol, Jean-Denis and Egas Santander, Daniela and Arnaudon, Alexis and Coste, Beno{\^\i}t and Delalondre, Fabien and Delemontex, Thomas and Devresse, Adrien and Dictus, Hugo and others},
  journal = {eLife},
  volume  = {13},
  pages   = {RP99688},
  year    = {2026},
  doi     = {10.7554/eLife.99688.3}
}

@article{isbister2026modeling,
  title   = {Modeling and simulation of neocortical micro- and mesocircuitry (Part II, physiology and experimentation)},
  author  = {Isbister, James B. and Ecker, Andr{\'a}s and Pokorny, Christoph and Bola{\~n}os-Puchet, Sirio and Egas Santander, Daniela and Arnaudon, Alexis and Awile, Omar and Barros-Zulaica, Natali and Blanco Alonso, Jorge and Boci, Elvis and others},
  journal = {eLife},
  volume  = {13},
  pages   = {RP99693},
  year    = {2026},
  doi     = {10.7554/eLife.99693.3}
}

@article{laquitaine2024spike,
  title   = {Spike sorting biases and information loss in a detailed cortical model},
  author  = {Laquitaine, Steeve and Imbeni, Milo and Tharayil, Joseph and Isbister, James B. and Reimann, Michael W.},
  journal = {bioRxiv [Preprint]},
  year    = {2024},
  doi     = {10.1101/2024.12.04.626805},
  url     = {https://www.biorxiv.org/content/10.1101/2024.12.04.626805}
}

@article{pronold2024multi,
  title={Multi-scale spiking network model of human cerebral cortex},
  author={Pronold, Jari and van Meegen, Alexander and Shimoura, Renan O and Vollenbr{\"o}ker, Hannah and Senden, Mario and Hilgetag, Claus C and Bakker, Rembrandt and van Albada, Sacha J},
  journal={Cerebral Cortex},
  volume={34},
  number={10},
  pages={bhae409},
  year={2024},
  doi={10.1093/cercor/bhae409}
}

@article{pille2025fast,
  title   = {Fast and easy whole-brain network model parameter estimation with automatic differentiation},
  author  = {Pille, Marius and Martin, Leon and Richter, Emilius and Perdikis, Dionysios and Schirner, Michael and Ritter, Petra},
  journal = {bioRxiv [Preprint]},
  year    = {2025},
  doi     = {10.1101/2025.11.18.689003},
  url     = {https://www.biorxiv.org/content/10.1101/2025.11.18.689003}
}

@article{wilson1973mathematical,
  title={A mathematical theory of the functional dynamics of cortical and thalamic nervous tissue},
  author={Wilson, Hugh R and Cowan, Jack D},
  journal={Kybernetik},
  volume={13},
  number={2},
  pages={55--80},
  year={1973},
  doi={10.1007/BF00288786}
}

@article{pinto1996quantitative,
  title={A quantitative population model of whisker barrels: re-examining the Wilson-Cowan equations},
  author={Pinto, David J and Brumberg, Joshua C and Simons, Daniel J and Ermentrout, G Bard and Traub, Roger},
  journal={Journal of computational neuroscience},
  volume={3},
  number={3},
  pages={247--264},
  year={1996},
  doi={10.1007/BF00161134}
}

@article{wang2002probabilistic,
  title={Probabilistic decision making by slow reverberation in cortical circuits},
  author={Wang, Xiao-Jing},
  journal={Neuron},
  volume={36},
  number={5},
  pages={955--968},
  year={2002},
  doi={10.1016/S0896-6273(02)01092-9}
}

@article{wong2006recurrent,
  title={A recurrent network mechanism of time integration in perceptual decisions},
  author={Wong, Kong-Fatt and Wang, Xiao-Jing},
  journal={Journal of Neuroscience},
  volume={26},
  number={4},
  pages={1314--1328},
  year={2006},
  doi={10.1523/JNEUROSCI.3733-05.2006}
}

@article{montbrio2015macroscopic,
  title={Macroscopic description for networks of spiking neurons},
  author={Montbri{\'o}, Ernest and Paz{\'o}, Diego and Roxin, Alex},
  journal={Physical Review X},
  volume={5},
  number={2},
  pages={021028},
  year={2015},
  doi={10.1103/PhysRevX.5.021028}
}

@article{deco2013resting,
  title={Resting-state functional connectivity emerges from structurally and dynamically shaped slow linear fluctuations},
  author={Deco, Gustavo and Ponce-Alvarez, Adri{\'a}n and Mantini, Dante and Romani, Gian Luca and Hagmann, Patric and Corbetta, Maurizio},
  journal={Journal of Neuroscience},
  volume={33},
  number={27},
  pages={11239--11252},
  year={2013},
  doi={10.1523/JNEUROSCI.1091-13.2013}
}

@article{coombes2023next,
  title={Next generation neural population models},
  author={Coombes, Stephen},
  journal={Frontiers in Applied Mathematics and Statistics},
  volume={9},
  pages={1128224},
  year={2023},
  doi={10.3389/fams.2023.1128224}
}

@article{schwalger2017towards,
  title={Towards a theory of cortical columns: From spiking neurons to interacting neural populations of finite size},
  author={Schwalger, Tilo and Deger, Moritz and Gerstner, Wulfram},
  journal={PLoS computational biology},
  volume={13},
  number={4},
  pages={e1005507},
  year={2017},
  doi={10.1371/journal.pcbi.1005507}
}

@article{augustin2017low,
  title={Low-dimensional spike rate models derived from networks of adaptive integrate-and-fire neurons: comparison and implementation},
  author={Augustin, Moritz and Ladenbauer, Josef and Baumann, Fabian and Obermayer, Klaus},
  journal={PLoS computational biology},
  volume={13},
  number={6},
  pages={e1005545},
  year={2017},
  doi={10.1371/journal.pcbi.1005545}
}

@article{gerstner2000population,
  title={Population dynamics of spiking neurons: fast transients, asynchronous states, and locking},
  author={Gerstner, Wulfram},
  journal={Neural computation},
  volume={12},
  number={1},
  pages={43--89},
  year={2000},
  doi={10.1162/089976600300015899}
}

@article{badel2008dynamic,
  title={Dynamic IV curves are reliable predictors of naturalistic pyramidal-neuron voltage traces},
  author={Badel, Laurent and Lefort, Sandrine and Brette, Romain and Petersen, Carl CH and Gerstner, Wulfram and Richardson, Magnus JE},
  journal={Journal of Neurophysiology},
  volume={99},
  number={2},
  pages={656--666},
  year={2008},
  doi={10.1152/jn.01107.2007}
}

@article{ohira1993master,
  title={Master-equation approach to stochastic neurodynamics},
  author={Ohira, Toru and Cowan, Jack D},
  journal={Physical Review E},
  volume={48},
  number={3},
  pages={2259},
  year={1993},
  doi={10.1103/PhysRevE.48.2259}
}

@article{buice2007field,
  title={Field-theoretic approach to fluctuation effects in neural networks},
  author={Buice, Michael A and Cowan, Jack D},
  journal={Physical Review E—Statistical, Nonlinear, and Soft Matter Physics},
  volume={75},
  number={5},
  pages={051919},
  year={2007},
  doi={10.1103/PhysRevE.75.051919}
}

@article{buice2010systematic,
  title={Systematic fluctuation expansion for neural network activity equations},
  author={Buice, Michael A and Cowan, Jack D and Chow, Carson C},
  journal={Neural computation},
  volume={22},
  number={2},
  pages={377--426},
  year={2010},
  doi={10.1162/neco.2009.02-09-960}
}

@article{bressloff2010stochastic,
  title={Stochastic neural field theory and the system-size expansion},
  author={Bressloff, Paul C},
  journal={SIAM Journal on Applied Mathematics},
  volume={70},
  number={5},
  pages={1488--1521},
  year={2010},
  doi={10.1137/090756971}
}

@article{microns2025functional,
  title   = {Functional connectomics spanning multiple areas of mouse visual cortex},
  author  = {{The MICrONS Consortium}},
  journal = {Nature},
  volume  = {640},
  number  = {8058},
  pages   = {435--447},
  year    = {2025},
  doi     = {10.1038/s41586-025-08790-w}
}

@article{williams2009roofline,
  title={Roofline: an insightful visual performance model for multicore architectures},
  author={Williams, Samuel and Waterman, Andrew and Patterson, David},
  journal={Communications of the ACM},
  volume={52},
  number={4},
  pages={65--76},
  year={2009},
  doi={10.1145/1498765.1498785}
}

@article{izhikevich2004model,
  title={Which model to use for cortical spiking neurons?},
  author={Izhikevich, Eugene M},
  journal={IEEE transactions on neural networks},
  volume={15},
  number={5},
  pages={1063--1070},
  year={2004},
  doi={10.1109/TNN.2004.832719}
}

@article{brette2007simulation,
  title={Simulation of networks of spiking neurons: a review of tools and strategies},
  author={Brette, Romain and Rudolph, Michelle and Carnevale, Ted and Hines, Michael and Beeman, David and Bower, James M and Diesmann, Markus and Morrison, Abigail and Goodman, Philip H and Harris Jr, Frederick C and others},
  journal={Journal of computational neuroscience},
  volume={23},
  number={3},
  pages={349--398},
  year={2007},
  doi={10.1007/s10827-007-0038-6}
}

@article{kaplan2020scaling,
  title         = {Scaling laws for neural language models},
  author        = {Kaplan, Jared and McCandlish, Sam and Henighan, Tom and Brown, Tom B. and Chess, Benjamin and Child, Rewon and Gray, Scott and Radford, Alec and Wu, Jeffrey and Amodei, Dario},
  journal       = {arXiv [Preprint]},
  year          = {2020},
  eprint        = {2001.08361},
  archiveprefix = {arXiv},
  primaryclass  = {cs.LG},
  doi           = {10.48550/arXiv.2001.08361},
  url           = {https://arxiv.org/abs/2001.08361}
}

@article{brette2005adaptive,
  title={Adaptive exponential integrate-and-fire model as an effective description of neuronal activity},
  author={Brette, Romain and Gerstner, Wulfram},
  journal={Journal of neurophysiology},
  volume={94},
  number={5},
  pages={3637--3642},
  year={2005},
  doi={10.1152/jn.00686.2005}
}

@article{azevedo2009equal,
  title={Equal numbers of neuronal and nonneuronal cells make the human brain an isometrically scaled-up primate brain},
  author={Azevedo, Frederico AC and Carvalho, Ludmila RB and Grinberg, Lea T and Farfel, Jos{\'e} Marcelo and Ferretti, Renata EL and Leite, Renata EP and Filho, Wilson Jacob and Lent, Roberto and Herculano-Houzel, Suzana},
  journal={Journal of comparative neurology},
  volume={513},
  number={5},
  pages={532--541},
  year={2009},
  doi={10.1002/cne.21974}
}

@article{sanz2013virtual,
  title={The Virtual Brain: a simulator of primate brain network dynamics},
  author={Sanz Leon, Paula and Knock, Stuart A and Woodman, M Marmaduke and Domide, Lia and Mersmann, Jochen and McIntosh, Anthony R and Jirsa, Viktor},
  journal={Frontiers in neuroinformatics},
  volume={7},
  pages={10},
  year={2013},
  doi={10.3389/fninf.2013.00010}
}

@article{faisal2008noise,
  title={Noise in the nervous system},
  author={Faisal, A Aldo and Selen, Luc PJ and Wolpert, Daniel M},
  journal={Nature reviews neuroscience},
  volume={9},
  number={4},
  pages={292--303},
  year={2008},
  doi={10.1038/nrn2258}
}

@article{goldman2023comprehensive,
  title={A comprehensive neural simulation of slow-wave sleep and highly responsive wakefulness dynamics},
  author={Goldman, Jennifer S and Kusch, Lionel and Aquilue, David and Yal{\c{c}}{\i}nkaya, Bahar Hazal and Depannemaecker, Damien and Ancourt, Kevin and Nghiem, Trang-Anh E and Jirsa, Viktor and Destexhe, Alain},
  journal={Frontiers in Computational Neuroscience},
  volume={16},
  pages={1058957},
  year={2023},
  doi={10.3389/fncom.2022.1058957}
}

@article{nykamp2000population,
  title={A population density approach that facilitates large-scale modeling of neural networks: Analysis and an application to orientation tuning},
  author={Nykamp, Duane Q and Tranchina, Daniel},
  journal={Journal of computational neuroscience},
  volume={8},
  number={1},
  pages={19--50},
  year={2000},
  doi={10.1023/A:1008912914816}
}

@article{omurtag2000simulation,
  title={On the simulation of large populations of neurons},
  author={Omurtag, Ahmet and Knight, Bruce W. and Sirovich, Lawrence},
  journal={Journal of computational neuroscience},
  volume={8},
  number={1},
  pages={51--63},
  year={2000},
  doi={10.1023/A:1008964915724}
}

@article{longtin2010stochastic,
  title={Stochastic dynamical systems},
  author={Longtin, Andre},
  journal={Scholarpedia},
  volume={5},
  number={4},
  pages={1619},
  year={2010},
  doi={10.4249/scholarpedia.1619}
}

@book{gerstner2014neuronal,
  title     = {Neuronal dynamics: From single neurons to networks and models of cognition},
  author    = {Gerstner, Wulfram and Kistler, Werner M. and Naud, Richard and Paninski, Liam},
  location  = {Cambridge},
  publisher = {Cambridge University Press},
  year      = {2014},
  doi       = {10.1017/CBO9781107447615}
}

@inproceedings{li2020fourier,
  title     = {Fourier neural operator for parametric partial differential equations},
  author    = {Li, Zongyi and Kovachki, Nikola and Azizzadenesheli, Kamyar and Liu, Burigede and Bhattacharya, Kaushik and Stuart, Andrew and Anandkumar, Anima},
  booktitle = {International Conference on Learning Representations},
  year      = {2021},
  url       = {https://openreview.net/forum?id=c8P9NQVtmnO}
}

@article{gonccalves2020training,
  title={Training deep neural density estimators to identify mechanistic models of neural dynamics},
  author={Gon{\c{c}}alves, Pedro J and Lueckmann, Jan-Matthis and Deistler, Michael and Nonnenmacher, Marcel and {\"O}cal, Kaan and Bassetto, Giacomo and Chintaluri, Chaitanya and Podlaski, William F and Haddad, Sara A and Vogels, Tim P and others},
  journal={elife},
  volume={9},
  pages={e56261},
  year={2020},
  doi={10.7554/eLife.56261}
}

@article{wieland2021structural,
  title={On structural and practical identifiability},
  author={Wieland, Franz-Georg and Hauber, Adrian L and Rosenblatt, Marcus and T{\"o}nsing, Christian and Timmer, Jens},
  journal={Current opinion in systems biology},
  volume={25},
  pages={60--69},
  year={2021},
  doi={10.1016/j.coisb.2021.03.005}
}

@article{scala2021phenotypic,
  title={Phenotypic variation of transcriptomic cell types in mouse motor cortex},
  author={Scala, Federico and Kobak, Dmitry and Bernabucci, Matteo and Bernaerts, Yves and Cadwell, Cathryn Ren{\'e} and Castro, Jesus Ramon and Hartmanis, Leonard and Jiang, Xiaolong and Laturnus, Sophie and Miranda, Elanine and others},
  journal={Nature},
  volume={598},
  number={7879},
  pages={144--150},
  year={2021},
  doi={10.1038/s41586-020-2907-3}
}

@inproceedings{goring2024out,
  title     = {Out-of-domain generalization in dynamical systems reconstruction},
  author    = {G{\"o}ring, Niclas Alexander and Hess, Florian and Brenner, Manuel and Monfared, Zahra and Durstewitz, Daniel},
  booktitle = {Proceedings of the 41st International Conference on Machine Learning},
  series    = {Proceedings of Machine Learning Research},
  volume    = {235},
  pages     = {16071--16114},
  year      = {2024},
  doi       = {10.5555/3692070.3692711},
  url       = {https://proceedings.mlr.press/v235/goring24a.html}
}

@inproceedings{finlay2020train,
  title     = {How to train your neural {ODE}: the world of {Jacobian} and kinetic regularization},
  author    = {Finlay, Chris and Jacobsen, J{\"o}rn-Henrik and Nurbekyan, Levon and Oberman, Adam},
  booktitle = {Proceedings of the 37th International Conference on Machine Learning},
  series    = {Proceedings of Machine Learning Research},
  volume    = {119},
  pages     = {3154--3164},
  year      = {2020},
  doi       = {10.5555/3524938.3525234},
  url       = {https://proceedings.mlr.press/v119/finlay20a.html}
}

@book{van1992stochastic,
  author    = {van Kampen, N. G.},
  title     = {Stochastic Processes in Physics and Chemistry},
  year      = {1992},
  publisher = {North-Holland}
}

@article{peretto1984collective,
  title={Collective properties of neural networks: a statistical physics approach},
  author={Peretto, Pierre},
  journal={Biological cybernetics},
  volume={50},
  number={1},
  pages={51--62},
  year={1984},
  publisher={Springer}
}

@article{richardson2005synaptic,
  title={Synaptic shot noise and conductance fluctuations affect the membrane voltage with equal significance},
  author={Richardson, Magnus JE and Gerstner, Wulfram},
  journal={Neural computation},
  volume={17},
  number={4},
  pages={923--947},
  year={2005},
  publisher={MIT Press}
}

@article{sanchezvives2000longlasting,
  author  = {Sanchez-Vives, Maria V. and Nowak, Lionel G. and McCormick, David A.},
  title   = {Cellular Mechanisms of Long-Lasting Adaptation in Visual Cortical Neurons In Vitro},
  journal = {Journal of Neuroscience},
  year    = {2000},
  volume  = {20},
  number  = {11},
  pages   = {4286--4299},
  doi     = {10.1523/JNEUROSCI.20-11-04286.2000}
}

@article{pozzorini2013temporal,
  author  = {Pozzorini, Christian and Naud, Richard and Mensi, Skander and Gerstner, Wulfram},
  title   = {Temporal whitening by power-law adaptation in neocortical neurons},
  journal = {Nature Neuroscience},
  volume  = {16},
  number  = {7},
  pages   = {942--948},
  year    = {2013},
  doi     = {10.1038/nn.3431}
}

@article{carannante2026bridge,
  title={BRIDGE: A Computational Workflow from Single Neurons to Network of Mean-Field Models},
  author={Carannante, Ilaria and Depannemaecker, Damien and Woodman, Marmaduke and Purohit, Pratik and Destexhe, Alain},
  journal={bioRxiv},
  year={2026},
  doi={10.64898/2026.07.31.742067},
  publisher={Cold Spring Harbor Laboratory},
  note={Preprint, version 1}
}

@article{wang2025foundation,
  title={Foundation model of neural activity predicts response to new stimulus types},
  author={Wang, Eric Y and Fahey, Paul G and Ding, Zhuokun and Papadopoulos, Stelios and Ponder, Kayla and Weis, Marissa A and Chang, Andersen and Muhammad, Taliah and Patel, Saumil and Ding, Zhiwei and others},
  journal={Nature},
  volume={640},
  number={8058},
  pages={470--477},
  year={2025},
  doi={10.1038/s41586-025-08829-y},
  publisher={Nature Publishing Group UK, London}
}




\clearpage
\pagestyle{plain}
\setcounter{section}{0}
\setcounter{equation}{0}
\setcounter{figure}{0}
\setcounter{table}{0}

\renewcommand{\thesection}{\Alph{section}}  
\renewcommand{\theequation}{S\arabic{equation}}
\renewcommand{\thefigure}{S\arabic{figure}}
\renewcommand{\thetable}{S\arabic{table}}

\renewcommand{\theHsection}{supp.\Alph{section}}
\renewcommand{\theHequation}{supp.\arabic{equation}}
\renewcommand{\theHfigure}{supp.\arabic{figure}}
\renewcommand{\theHtable}{supp.\arabic{table}}

\titleformat{\section}
  {\large\bfseries}
  {Supplementary Note \thesection.}
  {0.5em}
  {}
\titleformat{\subsection}
  {\normalfont\large\bfseries}
  {\thesubsection}
  {1em}
  {}

\phantomsection
\pdfbookmark[1]{Supplementary Material}{supplementary-material}
\begin{center}
    \Large\bfseries Supplementary Material
\end{center}

\section{Detailed derivation of the finite-size Master Equation formalism and its adaptive extension}
\label{sec:appendix_master_equation_El_boustani_Destexhe}

This note gives the detailed pedagogical derivational support for Section~\ref{sec:MFT_NN_formalism}. The derivation follows the formalism of El Boustani and Destexhe~\cite{el2009master}, using $\mathcal{F}_\gamma$ for the single-neuron transfer function and $P$ for the number of neuronal populations, consistently with the main text. Here, we keep their singleton notation $\{\cdots\}$ that indicates the set of all populations (for example, $\{m_\gamma\}$ denotes the joint state of $m_E$ and $m_I$ in a two-population model). In a vector notation, one may write:
$$\{m_\gamma\} \equiv \mathbf{m}=(m_1,\ldots,m_P),  \quad \nu_\gamma(t)\equiv\langle m_\gamma\rangle_t, \quad \langle\mathbf{m}\rangle \equiv \boldsymbol{\nu}.$$

They provided the theoretical framework to describe asynchronous irregular cortical activity at a mesoscopic scale, between microscopic spiking simulations and macroscopic neural-mass models. The key idea is to model the population activity of finite-size networks as a stochastic process governed by a master equation. In contrast with first-order rate models, this formalism explicitly keeps track not only of mean population activity, but also of finite-size fluctuations, covariances, and temporal correlations. We stress that we sometimes intentionally employ the notation $\{m_\gamma\}$ to maintain a direct parallel with the original paper \cite{el2009master}, though the vector notation $\mathbf{m}$ lightens the equations.

\revision{\paragraph{Index notation and summation convention.}
Greek indices ($\mu,\nu,\lambda,\eta,\alpha,\gamma$) label populations and run from $1$ to $P$. In compact equations, we use the Einstein summation convention in the following form. An index that appears twice within a single term and does not appear on the left-hand side of the equation is a \emph{dummy} index: it is summed from $1$ to $P$. The other indices are \emph{free}: the equation holds for each of their values, and they are never summed. For example, in
\[
\partial_t c_{\mu\nu}=\dots+\partial_\lambda a_\mu\,c_{\nu\lambda}+\dots
\qquad\text{i.e.}\qquad
\partial_t c_{\mu\nu}=\dots+\sum_{\lambda=1}^{P}\frac{\partial a_\mu}{\partial m_\lambda}\,c_{\nu\lambda}+\dots,
\]
$\mu$ and $\nu$ are free, and $\lambda$ is summed. Likewise, $\frac{1}{2}\partial_\lambda\partial_\eta a_\mu\,c_{\lambda\eta}$ is a double sum over $\lambda$ and $\eta$, whereas in $\delta_{\mu\nu}A^{-1}_{\mu\mu}$ the free index $\mu$ is not summed: $A^{-1}_{\mu\mu}$ is the $\mu$-th diagonal entry of $A^{-1}$. We write $\partial_\lambda=\partial/\partial m_\lambda$ for the derivative with respect to the activity of population $\lambda$; unless stated otherwise, derivatives are evaluated at the mean state $\boldsymbol\nu$. For two populations $E$ and $I$, for instance,
\[
\frac{1}{2}\,\partial_\lambda\partial_\eta\mathcal{F}_E\,c_{\lambda\eta}
=\frac{1}{2}\Big(\partial_E^2\mathcal{F}_E\,c_{EE}+2\,\partial_E\partial_I\mathcal{F}_E\,c_{EI}+\partial_I^2\mathcal{F}_E\,c_{II}\Big),
\]
using $c_{IE}=c_{EI}$. In matrix form, with the gain matrix $J_{\mu\lambda}=\partial_\lambda\mathcal{F}_\mu$ and $C=(c_{\mu\nu})$, the contractions used below read $\partial_\lambda\mathcal{F}_\mu\,c_{\lambda\nu}=(JC)_{\mu\nu}$ and $\partial_\lambda\mathcal{F}_\mu\,\partial_\eta\mathcal{F}_\nu\,c_{\lambda\eta}=(JCJ^{\mathsf T})_{\mu\nu}$.}

\subsection{Asynchronous irregular states and the need for second-order statistics}
\label{subsubsec:AI_state}

In awake, behaving animals, cortical activity is characterized by considerable subthreshold fluctuations of the membrane potential and highly irregular, asynchronous spiking \cite{el2009master}. It is widely believed that it is within this dynamic, noisy regime that the brain performs its primary computational tasks. In computational neuroscience, this behavior corresponds to the Asynchronous Irregular (AI) state, a regime typically exhibited in networks where dense recurrent connectivity creates a dynamic balance between excitatory and inhibitory synaptic inputs \cite{brunel2000dynamics}. The AI state is of particular theoretical interest because its macroscopic characteristics closely replicate the spontaneous activity observed in the awake cortex.

Dynamically, the AI state is defined by several core properties: very low average firing rates, highly irregular spike trains, and remarkably weak synchrony between individual neurons. Crucially for macroscopic modeling, the activity autocorrelation exhibits an exponential decay, and when sampled over suitable macroscopic time bins, the population activity distributions in the AI regime approximate a Gaussian profile. This exponential-like decay of correlations and approximately Gaussian distribution of coarse-grained activity motivate, but do not establish yet, a Markovian second-order closure. Its adequacy remained to be tested against the underlying network in the relevant regime.

\revision{Classical first-order population equations propagate mean activity
without explicitly evolving finite-population covariances. Nevertheless, fluctuation-driven single-neuron firing can enter their transfer functions through membrane-voltage statistics.} El Boustani and Destexhe~\cite{el2009master} additionally proposed a second-order mesoscopic formalism in which the state of the network is described not only by the mean activity of each population but also by their population-rate variances and covariance $c_{\mu\nu}$ through a second-order moment closure. \revision{We correct its covariance equation in Supplementary~\ref{subsubsec:covariance_matrix}.} In this framework, fluctuations are not treated as negligible background noise; finite-size fluctuations and their feedback on mean activity can be evaluated under the closure assumptions as part of the dynamical mechanism. \revision{AI states, in which activity is strongly fluctuation-driven and where population variance and correlation structure carry important dynamical information \cite{el2009master}, nevertheless should not be interpreted as describable only by population covariance dynamics.}

\subsection{Coarse-grained Markovian dynamics over a finite time bin}
\label{subsec:memoryless_assumption}

Biological neural systems are not intrinsically memoryless. Refractoriness, membrane filtering, synaptic time constants, adaptation, and plasticity all introduce history dependence at their own characteristic timescales. However, a process can often be treated as approximately Markovian after an appropriate coarse-graining: if the observation window $T$ is long enough compared with the relevant microscopic correlations, the future state can be approximated as depending mainly on the present coarse-grained state rather than on the full preceding history.

In the master-equation formalism, this Markovian\footnotemark assumption is therefore not an absolute claim about neurons or networks, but a modeling approximation applied to the mesoscopic activity variables over a finite time bin $T$. In AI states, population activity autocorrelations decay exponentially, making such a coarse-grained description plausible. The choice of $T$ is crucial and has to be done on the order of the relevant network time constants: if $T$ is too small, residual spike-history effects and correlations violate the approximation; if $T$ is too large, relevant fluctuations are averaged out. The formalism is thus valid only on an intermediate timescale where microscopic memory has sufficiently decayed but mesoscopic dynamics are still resolved.

\footnotetext{A Markovian process satisfies: $P(X_{t+1} \mid X_t, X_{t-1}, X_{t-2}, \dots) = P(X_{t+1} \mid X_t)$. Here this identity is not assumed exactly at all scales. Rather, it is used as an effective approximation after coarse-graining over a finite time bin $T$, so that the activity at $t+T$ depends primarily on the activity at $t$. Accounting for the entire history of a neural network to predict its next step would lead to unsolvable systems; Markov property makes the math tractable.}

\subsection{Finite-bin transition kernel and transfer-function closure}

In this formalism, both the membrane potential of the neurons and the network activity are described as stochastic processes. For a population $\gamma$ of size $N_\gamma$, with $K_\gamma(t)$ the number of spikes emitted during the interval $(t-\Delta t,t]$, under the assumption that each neuron fires at most once during the bin. The formal instantaneous population activity is defined as:
\begin{equation}
\begin{aligned}
m_\gamma(t) 
= \lim_{\Delta t \to 0} \frac{K_\gamma(t -\Delta t, t)}{N_\gamma\,\Delta t}  
\end{aligned}
\label{eq:instantaneous_network_activity}
\end{equation}

Let's consider $P$ homogeneous populations of neurons, indexed by $\gamma = 1,\cdots, P$. Over a finite time bin $T$, the Markovian description is expressed through the conditional probability $P(\{m_\gamma(t)\} \mid \{m'_\gamma(t-T)\})$, where $\{m_\gamma\} \equiv \mathbf{m}$ denotes the joint activity state of all populations and $\{m'_\gamma\}$ denotes the previous state. Assuming time-invariant dynamics, this transition probability depends only on the duration $T$, not on the absolute time $t$, and is written $P_T(\{m_\gamma\} | \{m'_\gamma\})$. This assumption is valid as long as the microscopic parameters and transfer functions are not themselves changing over time.\footnote{Here, time-invariant means that the transition rules do not depend explicitly on clock time. The same state reached at two different moments is assumed to evolve according to the same transition law.}

Brunel \cite{brunel2000dynamics} showed that for networks where connectivity is sparse and balanced, the shared input between neurons becomes vanishingly small as the network gets larger, which can support AI regimes with weak pairwise correlations. Thus, their membrane-potential fluctuations need not synchronize, and pairwise correlations are negligible at the microscopic level. Therefore, if the neurons are assumed conditionally independent on timescale $T$ (and by extension the populations too), the complex joint probability of the whole network can be chopped up into a simple product of individual marginal probabilities:
$$P_T(\{m_\gamma\} | \{m'_\gamma\}) = P_T(m_1 | \{m'_\gamma\})... P_T(m_P | \{m'_\gamma\}) = \prod_{\alpha}^{P} P_T(m_\alpha | \{m'_\gamma\})  , \,\ \footnotemark $$\footnotetext{We stress that this conditional independence does not mean that population activities have zero covariance over time. If the previous state increases the drive to both excitatory and inhibitory populations, both may increase their activity. The assumption is only that, conditioned on the previous global state and over the bin $T$, their instantaneous firing events are sampled independently.}

As the network is assumed to be memoryless beyond the time interval $T$ (see Supplementary~\ref{subsec:memoryless_assumption}), we can define a Markovian transition operator $\mathcal{W}(\{m_\mu\}|\{m'_\gamma\})$ and write down the continuous master equation for population activities:

\begin{equation}
\begin{aligned}
\partial_t P_t(\{m_\gamma\}) &= \int^{\frac{1}{T}}_{0}  [P_t(\{m'_\gamma \})\mathcal{W}(\{m_\gamma\} | \{m'_\gamma\}) - P_t(\{m_\gamma \})\mathcal{W}(\{m'_\gamma\} | \{m_\gamma\})]d\{m'_\gamma\} \\
&= \prod_{\alpha = 1,...,P} \int^{\frac{1}{T}}_{0}  [P_t(\{m'_\gamma \})\mathcal{W}(\{m_\gamma\} | \{m'_\gamma\}) - P_t(\{m_\gamma \})\mathcal{W}(\{m'_\gamma\} | \{m_\gamma\})]dm'_\alpha
\label{eq:master_equation}
\end{aligned}\footnotemark
\end{equation}
\footnotetext{This equation governs the temporal evolution of the probability density of the network being in state ${m_\gamma}$. It is a \textit{gain-minus-loss} equation: the first term reflects the probability of flowing into the current state, and the second term reflects the probability of flowing out. Thus, the change in probability of finding the network in state $A$ is equal to the sum of all probabilities moving from state $B$ to $A$, minus all probabilities moving from state $A$ to $B$ (where $B$ is $\Omega \setminus A$). See Supplementary~\ref{subsec:master_equation_reminder} for deeper insights.}

A key point of the El Boustani--Destexhe formalism is that the time bin $T$ is not sent to zero. In many continuous-time Markov descriptions, transition rates are obtained from an infinitesimal-time limit\footnotemark. Here, however, the transition operator is defined over a finite mesoscopic time window. This finite bin is required because the population activity $m_\gamma$ is a rate variable: if $T$ were taken arbitrarily small, isolated spike events would be divided by a near-zero time interval, producing discontinuous activity jumps and artificially large fluctuations. Conversely, if $T$ were too large, relevant mesoscopic fluctuations would be averaged out. Thus, $T$ must be chosen as an intermediate coarse-graining scale, long enough for microscopic correlations to decay but short enough to preserve the population dynamics of interest. \footnotetext{In standard statistical physics, infinitesimal time steps can often be used because macroscopic particle numbers are extremely large. In finite neural populations, however, the tracked variable $m_\gamma$ is a rate rather than a simple particle count, which is inversely proportional to the time bin $dt=T$. Dividing isolated spikes by an arbitrarily small time bin would artificially inflate the variance. The finite bin $T$ therefore acts as a biologically meaningful smoothing window, comparable to the timescale over which neurons and synapses integrate inputs, while the master equation remains defined for all $T$.}

Additionally, the finite-bin description keeps the population-level formalism compatible with the single-neuron diffusion approximation. Fokker--Planck-based transfer functions are expressed in terms of smooth firing rates, mean membrane potentials, and voltage fluctuations (see Section~\ref{subsec:background_fokkerplanck}). Treating $m_\gamma$ as a continuous activity variable over a finite bin $T$ therefore allows the moment equations to be written with standard derivatives and closed as ODEs for the mean activity and covariance. Finally, if $T$ is too small, residual spike-history effects, refractoriness, synaptic delays, and microscopic correlations can no longer be neglected, weakening the coarse-grained Markovian approximation. The transition operator is therefore defined over this finite time bin $T$:
\begin{equation}
\begin{aligned}
\mathcal{W}(\{m_\gamma\} | \{m'_\gamma\})
&=  \frac{\prod_{\alpha = 1,...,P} P_T(m_\alpha | \{m'_\gamma\})}{T}
\end{aligned}
\label{eq:transition_operator_continuous_definition}
\footnotemark
\end{equation}
\footnotetext{The transition operator $\mathcal{W}(\{m_\gamma\} \mid \{m'_\gamma\})$ provides the rate of transition from the previous state $\{m'_\gamma\}$ to the current state $\{m_\gamma\}$.}

At this point, the model is not yet closed: one still needs an expression for the conditional distribution $P_T(m_\alpha | \{m'_\gamma\})$. This is where the single-neuron transfer function enters.

\subsubsection{Transfer-function closure of the transition kernel}
\label{subsec:transfer_funtion_ME}

Over the bin $T$, neurons in population $\gamma$ are assumed to fire at most once, so that $0 \le m_\gamma \le \frac{1}{T}$. The activity $m_\gamma$ can therefore be interpreted as the fraction of neurons that fired during the previous time bin, divided by $T$, different from instantaneous activity described by Eq.~\eqref{eq:instantaneous_network_activity}.

This one-spike-per-bin assumption imposes an upper activity bound $T^{-1}$ Hz, so $T$ cannot be chosen too large. This creates a ceiling issue, sacrificing the ability to mathematically track multiple spikes per neuron per time window, in exchange for a simplified set of equations. Conversely, if $T$ is too small, the activity resolution $\Delta m=\frac{1}{N_\gamma T}$ becomes too coarse and residual microscopic correlations may no longer be negligible. Thus, as discussed above, $T$ must be chosen as an intermediate coarse-graining scale: large enough to smooth in invidual spike events into a population activity, but small enough to preserve the relevant mesoscopic fluctuations. Conveniently, AI states are naturally behaving at low firing rates, often below ceiling. Increasing $N_\gamma$ can partly compensate for a small $T$, which would otherwise induce finite-size temporal correlations \cite{brunel2000dynamics}, by keeping $\Delta m$ small as required for a continuous description; it does not, however, restore the Markov approximation. However, very large sparsely connected homogeneous population are not biologically relevant and second-order description collapses toward first-order one as $N$ goes to infinity near a stable stationary state in the regular
weak-fluctuation regime

To close the transition kernel, one needs an expression for the conditional distribution $P_T(m_\alpha \mid \{m'_\gamma\})$ of the activity of each population $\alpha$ during the next bin. This is where the single-neuron transfer function enters.
Let $\mathcal{F}_\alpha (\{ m'_\gamma \})$ denote the stationary output firing rate of a neuron in population $\alpha$, given the previous  population state $\{ m'_\gamma \}$. Under the quasi-stationary approximation, neurons are assumed to respond during the bin $T$ as if they were in the stationary regime determined by the previous state (adiabatic approximation\footnotemark). The probability that a neuron in population $\alpha$ fires during the next interval is therefore approximated by:
\footnotetext{Physics vocabulary to say that the neurons are assumed to react smoothly/quickly enough such that over one coarse-grained bin, they are almost at their stationary input-output response. However, this assumption breaks if the network is stimulated with signal that changes faster than $T^{-1}$, where the system won't be able to settle into an equilibrium, and the master equations will not predict valid outputs.}
$$ p_\alpha(\{ m'_\gamma \}) \approx \mathcal{F}_\alpha(\{ m'_\gamma \})T, \,\  p_\alpha \leq 1. \,\ \footnotemark$$
\footnotetext{El Boustani and Destexhe use the notation $\nu_\alpha$ for the transfer function. Here we write $\mathcal{F}_\alpha$ for consistency with the rest of the review.}

Since population $\alpha$ contains $N_\alpha$ neurons, each assumed conditionally independent over the bin $T$ with one-spike-per-bin assumption, the number of neurons that fire during this bin follows a binomial law\footnote{Equivalently, this is a sampling of $N_\alpha$ Bernoulli trials, each with success probability $p_\alpha$.}. Because the number of spikes in the bin is $m_{\alpha}N_{\alpha}T$, one obtains: $P_T(m_\alpha | \{m'_\gamma\}) \sim \mathcal{B}\left(N_{\alpha}, \mathcal{F}_\alpha(\{ m'_\gamma \})T\right)$

More explicitly, the probability that $m_\alpha N_\alpha T$ neurons fire follows:
\begin{equation}
\begin{aligned}
P_{T}(m_{\alpha}|\{m_{\gamma}'\}) = \Pr\!\left[K_\alpha=N_\alpha Tm_\alpha\mid\{m_{\gamma}'\}\right]
&= \binom{N_{\alpha}}{m_{\alpha}N_{\alpha}T} p_\alpha( \{ m'_\gamma \})^{m_{\alpha}N_{\alpha}T} (1 - p_\alpha( \{ m'_\gamma \}))^{N_{\alpha}(1 - m_{\alpha}T)} \\
&= \binom{N_{\alpha}}{m_{\alpha}N_{\alpha}T} (\mathcal{F}_{\alpha} T)^{m_{\alpha}N_{\alpha}T} (1 - \mathcal{F}_{\alpha} T)^{N_{\alpha}(1 - m_{\alpha}T)}. \,\ \footnotemark
\end{aligned}
\label{eq:binomial_conditional_probability}
\end{equation}
\footnotetext{The quantity $m_\alpha N_\alpha T$ is assumed to be an integer, or is treated as such in the continuum approximation. For readability: $\mathcal{F}_{\alpha}$ denotes $\mathcal{F}_{\alpha}(\{m'_\gamma\})$.}

The conditional moments follow directly from the binomial distribution:
\begin{equation}
\mathbb{E}[m_\mu\mid\{m_{\gamma}'\}]=\mathcal{F}_\mu(\{m_{\gamma}'\}),
\qquad
\operatorname{Var}(m_\mu\mid \{m_{\gamma}'\})=
\frac{\mathcal{F}_\mu(\{m_{\gamma}'\})\left[T^{-1}-\mathcal{F}_\mu(\{m_{\gamma}'\})\right]}{N_\mu}.
\label{eq:supp_conditional_moments}
\end{equation}

This step is the core \textbf{micro-to-meso bridge} of the formalism. This single-neuron transfer function determines the firing probability of each cell during the bin, and the binomial sampling of $N_\alpha$ such cells determines the population-level transition law. This is conceptually powerful: instead of massive simulation of a network to get the population transfer function, one can simulate a single spiking neuron and extract its input-output firing rate law \footnote{The main assumptions behind this construction are: (1) coarse-grained Markovianity over the bin $T$; (2) quasi-stationarity, so that neurons respond according to their stationary transfer function; (3) conditional independence of neurons given the previous population state; (4) binary firing within one bin; and (5) homogeneity within each population, so that all neurons in population $\alpha$ share the same transfer function $\mathcal{F}_\alpha$. In practice, El Boustani and Destexhe showed that moderate violations of these assumptions do not necessarily invalidate the model, but they reduce its quantitative accuracy.}. The transfer function carries the selected single-neuron and synaptic mechanisms, whereas sampling $N_\mu$ neurons produces an explicit finite-size variance. The relation is nevertheless conditional on the preceding assumptions: it does not retain the membrane trajectories, spike times within the bin, detailed adjacency matrix, or correlations that are not mediated by the conditioned state.

For large enough $N_\alpha p_\alpha\gg1$ and $N_\alpha(1-p_\alpha)\gg1$, Stirling's approximation\footnote{$n! \sim \sqrt{2\pi n}(n/e)^{n} $} allows the binomial distribution to be approximated by a Gaussian distribution:

$$P_{T}(m_{\alpha}|\{m_{\gamma}'\}) \sim \mathcal{N} \left(\mathcal{F}_\alpha(\{ m'_\gamma \}), \frac{\mathcal{F}_\alpha(\{ m'_\gamma \})(1/T - \mathcal{F}_\alpha(\{ m'_\gamma \}))}{N_\alpha} \right) $$
Equivalently,
\begin{equation}
\begin{aligned}
P_{T}(m_{\alpha}|\{m_{\gamma}'\}) \simeq \sqrt{\frac{N_{\alpha}}{2\pi \mathcal{F}_{\alpha}(\{ m'_\gamma \})(1/T - \mathcal{F}_{\alpha}(\{ m'_\gamma \}))}} \exp \left[ -N_{\alpha}\frac{(m_{\alpha} - \mathcal{F}_{\alpha}(\{ m'_\gamma \}))^{2}}{2\mathcal{F}_{\alpha}(\{ m'_\gamma \})(1/T - \mathcal{F}_{\alpha}(\{ m'_\gamma \}))} \right]. \,\ \footnotemark
\end{aligned}
\label{eq:gaussian_conditional_probability}
\end{equation}
\footnotetext{The derivation is given in Supplementary~\ref{subsec:from_Binomial_to_Gaussian_distrbution}.
In the binomial notation, $Npq = N_\alpha (\mathcal{F}_\alpha T) (1 - \mathcal{F}_\alpha T)$ and $\delta = k - Np = m_\alpha N_\alpha T - N_\alpha \mathcal{F}_\alpha T = N_\alpha T(m_\alpha - \mathcal{F}_\alpha)$}

The variance $\frac{\mathcal{F}_{\alpha}(1/T - \mathcal{F}_{\alpha})}{N_{\alpha}}$ is inversely proportional to the population size. Thus, in the limit $N_\alpha \to \infty$, the conditional distribution collapses to a Dirac mass centered on $\mathcal{F}_{\alpha}$, and the model recovers the first-order mean-field limit. The new activity is then deterministic conditional on the previous state, while the continuous-time redraw times remain random. Keeping $N_{\alpha}$ finite is therefore precisely what allows the formalism to retain finite-size fluctuations.

One may think that the homogeneous-population assumption is biologically unrealistic. In fact, this assumption can be relaxed qualitatively. If neuronal properties vary weakly and continuously within a population, by using the Central Limit Theorem, the activity distribution may still be approximated by an effective Gaussian law, with its mean captured by an effective population transfer function. However, strong heterogeneity would require either multiple subpopulations or a more explicit distribution of transfer functions.

Finally, we can recover the joint transition rate density $\mathcal{W}$ for the entire network by dividing by $T$:

\begin{equation}
\begin{aligned}
\mathcal{W}(\{m_\gamma\} \mid \{m'_\gamma\}) = \frac{1}{T} \sqrt{\frac{det(A)}{(2\pi)^P}} \exp \left[ -\frac{1}{2} (m_\mu - \mathcal{F}_\mu(\{ m'_\gamma \})) A_{\mu\nu} (m_\nu - \mathcal{F}_\nu(\{ m'_\gamma \})) \right]
\end{aligned}
\label{eq:joint_transition_probability}
\end{equation}

$A$ is the precision matrix defined as $A_{\mu\nu} = \delta_{\mu\nu} \frac{N_\mu}{\mathcal{F}_\mu(\{ m'_\gamma \})(1/T - \mathcal{F}_\mu(\{ m'_\gamma \}))}$ \footnote{$\delta_{\mu\nu}$ is the Kronecker delta. $A$, the inverse of the covariance matrix of the transition, is a $P \times P$ diagonal matrix because the populations are assumed conditionally independent over the bin $T$. It is invertible as long as ($N_\alpha >0)$ and ($0<\mathcal{F}_\alpha<1/T$).}. The integral of $\mathcal{W}$ over the transition state space is $1/T$, as expected for a transition rate. This normalization also emphasizes again that the choice of $T$ is part of the model: too small $T$ will capture irrelevant fluctuations, too large $T$ will remove the finite-size fluctuations that the second-order formalism is designed to capture.

\subsection{Second-order moment equations and covariance dynamics}

The master equation defines the full probability density $P_t(\{m_\gamma\})$ over all possible population-activity. Solving this full density directly is generally intractable. Instead, the dynamics can be projected onto its low-order statistical moments: the mean activity $\langle m_\mu \rangle$, the covariance matrix $c_{\mu\nu}=\langle(m_\mu-\langle m_\mu \rangle)(m_\nu-\langle m_\nu \rangle)\rangle$, and, at stationarity, the time-lagged correlation matrix $Corr_{\mu\nu}(\tau)$. In this way, the full probability distribution is approximated through its center and its second-order shape \footnote{The probability density $P_t(\{m_\gamma\})$ may be viewed as a cloud of possible network activity states. The first moment $\langle m_\mu\rangle$ describes the center of this cloud, while the covariance matrix $c_{\mu\nu}$ describes its width and orientation. Moment equations usually form a hierarchy: the mean can depend on covariance, covariance can depend on higher-order moments, and so on. A closure approximation is therefore required. Under a Gaussian approximation, higher-order cumulants are neglected, yielding a second-order closure.}.

In balanced networks operating in the AI regime, the stationary distribution of population activity is close to Gaussian for suitable choices of $N$ and $T$ \cite{el2009master}. This motivates a second-order closure of the master equation\revision{: cumulants of order three and higher are neglected, and the jump moments are expanded around the mean state to second order in the fluctuations (Supplementary~\ref{sub:generic_second_order_master_eq_moments}). Gaussianity does not, however, justify replacing the average of the second jump moment by its value at the mean state: its curvature term is of the same order as the retained terms}. This leads to:

\begin{equation}
\begin{aligned}
\partial_t\langle m_\mu \rangle
&= a_\mu(\{\langle m_\gamma \rangle \}) +
\frac{1}{2}
\partial_{\lambda}\partial_{\eta} a_\mu(\{\langle m_\gamma \rangle \})c_{\lambda \eta}\\
\partial_t c_{\mu\nu} 
&=  a_{\mu\nu}(\{ \left\langle m_\gamma \right\rangle \}) \revision{+ \frac{1}{2}\partial_{\lambda}\partial_{\eta} a_{\mu\nu}(\{\langle m_\gamma \rangle \})c_{\lambda \eta}}\\
&\quad + \partial_{\lambda} a_\mu(\{\langle m_\gamma \rangle \})c_{\nu \lambda} + \partial_{\lambda} a_\nu(\{\langle m_\gamma \rangle \})c_{\mu \lambda}
\end{aligned}
\label{eq:second_order_statistical_moments}
\footnotemark
\end{equation}
\footnotetext{\revision{Einstein summation convention, as defined at the beginning of Supplementary~\ref{sec:appendix_master_equation_El_boustani_Destexhe}: $\lambda$ and $\eta$, which are repeated and absent from the left-hand side, are summed from $1$ to $P$, whereas $\mu$ and $\nu$ are free. For instance, $\partial_\lambda a_\mu\,c_{\nu\lambda}=\sum_{\lambda=1}^{P}\partial_\lambda a_\mu\,c_{\nu\lambda}$ and $\frac{1}{2}\partial_\lambda\partial_\eta a_{\mu\nu}\,c_{\lambda\eta}=\frac{1}{2}\sum_{\lambda=1}^{P}\sum_{\eta=1}^{P}\partial_\lambda\partial_\eta a_{\mu\nu}\,c_{\lambda\eta}$.}}

where $\langle m_\mu \rangle$ is the mean population activity and $c_{\mu\nu}$ is the activity covariance matrix. The functions $a_\mu$ and $a_{\mu\nu}$ are the first and second jump moments of the transition kernel. They play the role of drift and diffusion terms for the population activity process. The derivation of these generic moment equations is given in Supplementary~\ref{sub:generic_second_order_master_eq_moments}. \revision{The curvature term $\frac{1}{2}\partial_\lambda\partial_\eta a_{\mu\nu}\,c_{\lambda\eta}$ of the second jump moment is of the same order as the other terms of the covariance equation. The original derivation~\cite{el2009master} omitted it by evaluating the second jump moment at the mean state (mean-value closure); Supplementary~\ref{subsubsec:covariance_matrix} computes it and derives the corrected covariance equation.}

Substituting the finite-bin transition operator defined in Eq.~\eqref{eq:transition_operator_continuous_definition}, and evaluating the moments at the mean state $\{\langle m_\gamma\rangle\}$, gives:

\begin{equation}
\begin{aligned}
a_\mu(\{ \langle m_\gamma \rangle \})
&= \prod_{\alpha = 1}^{P} \int^{\frac{1}{T}}_{0}  (m'_\mu-\langle m_\mu \rangle)\mathcal{W}(\{m'_\gamma\} \mid \{ \langle m_\gamma \rangle \}) dm'_\alpha\\
&= \frac{1}{T} \prod_{\alpha = 1}^{P} \int^{\frac{1}{T}}_{0} dm'_\alpha (m'_\mu-\langle m_\mu \rangle) P_T(m'_\alpha \mid \{ \langle m_\gamma \rangle \}), \,\ \footnotemark \\
&= \frac{1}{T} \int dm'_\mu (m'_\mu - \langle m_\mu \rangle) P_T(m'_\mu \mid \{ \langle m_\gamma \rangle \}) \times \left[ \prod_{\alpha \neq \mu} \int dm'_\alpha P_T(m'_\alpha \mid \{ \langle m_\gamma \rangle \}) \right]\\
&= \frac{1}{T} \int_0^{\frac{1}{T}} dm'_\mu (m'_\mu - \langle m_\mu \rangle) P_T(m'_\mu \mid \{ \langle m_\gamma \rangle \})\\
a_{\mu\nu}(\{ \langle m_\gamma \rangle \}) 
&= \prod_{\alpha = 1}^{P} \int^{\frac{1}{T}}_{0}  (m'_\mu-\langle m_\mu \rangle)(m'_\nu-\langle m_\nu \rangle)\mathcal{W}(\{m'_\gamma\} \mid \{ \langle m_\gamma \rangle \}) dm'_\alpha\\
&= \frac{1}{T} \prod_{\alpha = 1}^{P} \int^{\frac{1}{T}}_{0} dm'_\alpha  (m'_\mu-\langle m_\mu \rangle)(m'_\nu-\langle m_\nu \rangle) P_T(\{m'_\gamma\} \mid \{ \langle m_\gamma \rangle \})\\
a_{\mu\nu}(\{ \langle m_\gamma \rangle \}) 
&= \frac{1}{T} \int^{\frac{1}{T}}_{0} dm'_\mu \int^{\frac{1}{T}}_{0} dm'_\nu  (m'_\mu-\langle m_\mu \rangle)(m'_\nu-\langle m_\nu \rangle) P_T(m'_\mu \mid \{ \langle m_\gamma \rangle \}) P_T(m'_\nu \mid \{ \langle m_\gamma \rangle \}) \quad (\mu\neq\nu),\\ 
a_{\mu\mu}(\{ \langle m_\gamma \rangle \})
&= \frac{1}{T} \int^{\frac{1}{T}}_{0} dm'_\mu\, (m'_\mu-\langle m_\mu \rangle)^2\, P_T(m'_\mu \mid \{ \langle m_\gamma \rangle \}) \quad (\mu=\nu).
\end{aligned}
\label{eq:step_drift_diffusion_moment_functions}
\end{equation}
\footnotetext{\revision{This notation can look tricky. Write $q_\alpha(x)=P_T(x\mid\boldsymbol\nu)$ for the
next-state marginal of population $\alpha$ given the entire
present state $\boldsymbol\nu$. Conditional independence gives
$P_T(\mathbf m'\mid\boldsymbol\nu)=\prod_\alpha q_\alpha(m'_\alpha)$. Here $\prod_{\alpha=1}^P \int dm'_\alpha$ denotes the multiple integral $\int \int \dots \int dm'_1 dm'_2 \dots dm'_P$, not a product of separate integrals. A transparent notation would be $a_\mu = \left( \prod_{\alpha=1}^P \int dm'_\alpha \right)(m'_\mu - \langle m_\mu \rangle) \frac{1}{T} \left( \prod_{\beta=1}^P P_T(m'_\beta \mid \dots) \right)$.}}
\revision{The integrals denote integration over the next-state support (sums for the exact discrete kernel). The last two lines of the second jump moment distinguish two cases. For $\mu\neq\nu$, the integrand depends on $m'_\mu$ and $m'_\nu$ only. Because the populations are conditionally independent over one bin, the joint transition density factorizes, the $P-2$ other integrals equal one, and the two remaining integrals give the product of the two mean displacements, $(\mathcal{F}_\mu-\langle m_\mu\rangle)(\mathcal{F}_\nu-\langle m_\nu\rangle)$. For $\mu=\nu$, the integrand depends on $m'_\mu$ only, a single marginal density remains, and the integral gives the conditional variance plus the squared displacement, $A^{-1}_{\mu\mu}+(\mathcal{F}_\mu-\langle m_\mu\rangle)^2$. Equation~\eqref{eq:final_diffusion_moments} below evaluates both cases and combines them with the Kronecker delta $\delta_{\mu\nu}$.}

Equation~\eqref{eq:second_order_statistical_moments} gave the generic \revision{second-order moment equations} of the master equation. One can complete the model with the correlation matrix of the network $Corr_{\mu\nu}(t,t+\tau)$, developed in Eq.~\eqref{eq:diff_equation_correlation_populations}, \revision{restricted to the stationary state}. The correlation function then depends only on the time lag $\tau$. Under this stationary approximation, its dynamics are governed by:
\begin{equation}
\begin{aligned}
\partial_{\tau}Corr_{\mu\nu}(\tau) 
&= \partial_{\lambda}a_{\nu}(\{\langle m_{\gamma}(t+\tau)\rangle\}) Corr_{\mu\lambda}(\tau) \\
&= \partial_{\lambda}a_{\nu}(\{\langle m_{\gamma}^{stat}\rangle\}) Corr_{\mu\lambda}(\tau)
\end{aligned}
\label{eq:time_lagged_correlation}
\end{equation}

This last equation assumes that the AI state is globally stable and that the system remains close to a stationary baseline, without strong transient stimulation. Together, Eqs.~\eqref{eq:second_order_statistical_moments} and \eqref{eq:time_lagged_correlation} reduce the full master equation to a tractable mesoscopic description in terms of mean activity, covariance, and lagged correlations.

\subsection{Final Mesoscopic Moment Equations}

The joint transition rate in Eq.~\ref{eq:joint_transition_probability} fully specifies the finite-bin stochastic process. Because the Gaussian distribution is sharply peaked with a variance that vanishes away from the boundaries ($0$ and $1/T$), the integrals can be extended over $\mathbb{R}$ without diverging. The boundary correction terms are of order $\mathcal{O}(e^{-N})$ and can be safely ignored for large enough populations. We can now evaluate the jump moments, and substitute them into the generic second-order moment expansion derived above:

\begin{equation}
\begin{aligned}
a_\mu({ \{ \langle m_\gamma \rangle \} })
&= \frac{1}{T} \int_{-\infty}^{\infty} (m'_\mu - \langle m_\mu \rangle) P_T(m'_\mu \mid \{ \langle m_\gamma \rangle \}) dm'_\mu  \\
&= \frac{1}{T} [ \int_{-\infty}^{\infty} m'_\mu P_T(m'_\mu \mid \{\langle m_\gamma \rangle\}) dm'_\mu - \langle m_\mu \rangle \int_{-\infty}^{\infty}  P_T(m'_\mu \mid \{\langle m_\gamma \rangle\}) dm'_\mu ]\\
&= \frac{1}{T} (\mathcal{F}_\mu - \langle m_\mu \rangle). \,\ \footnotemark \\
\end{aligned}
\label{eq:final_drift_moments}
\end{equation}
\footnotetext{The conditional density $P_T(m'_\mu \mid \{\langle m_\gamma \rangle\})$ is a Gaussian centered on $\mathcal{F}_\mu$.}
$\mathcal{F}_\mu = \mathcal{F}_\mu(\{\langle m_\gamma \rangle\})$ is the transfer function evaluated at the mean population state. Thus, the first moment $a_\mu$ measures the expected relaxation of the current mean activity toward the transfer function value.

\begin{equation}
\begin{aligned}
a_{\mu\nu}(\{ \langle m_\gamma \rangle \})
&=\frac1T\,
\mathbb E\!\left[(m'_\mu-\nu_\mu)(m'_\nu-\nu_\nu)
\mid\mathbf m=\boldsymbol\nu\right]\\
&= \int_{-\infty}^{\infty} dm'_\mu \int_{-\infty}^{\infty} dm'_\nu (m'_\mu - \langle m_\mu \rangle)(m'_\nu - \langle m_\nu \rangle) \mathcal{W}(\{m'_\gamma\} \mid \{\langle m_\gamma \rangle\}) \\
&= \frac{\delta_{\mu\nu}}{T} \int_{-\infty}^{\infty} dm'_\mu (m'^2_\mu + \mathcal{F}^2_\mu - \mathcal{F}^2_\mu - 2m'_\mu \langle m_\mu \rangle + \langle m_\mu \rangle^2) P_T(m'_\mu \mid \{\langle m_\gamma \rangle\}) \\
&\quad + \frac{(1 - \delta_{\mu\nu})}{T} (\mathcal{F}_\mu - \langle m_\mu \rangle)(\mathcal{F}_\nu - \langle m_\nu \rangle). \,\ \footnotemark \\
&= \frac{1}{T} \left[ \delta_{\mu\nu} \frac{\mathcal{F}_\mu(1/T - \mathcal{F}_\mu)}{N_\mu} + \delta_{\mu\nu}(\mathcal{F}_\mu - \langle m_\mu \rangle)^2 + (1 - \delta_{\mu\nu})(\mathcal{F}_\mu - \langle m_\mu \rangle)(\mathcal{F}_\nu - \langle m_\nu \rangle) \right] \\
&= \frac{1}{T} \left[ \delta_{\mu\nu} \frac{\mathcal{F}_\mu(1/T - \mathcal{F}_\mu)}{N_\mu} + (\mathcal{F}_\mu - \langle m_\mu \rangle)(\mathcal{F}_\nu - \langle m_\nu \rangle) \right] \\
&= \frac{1}{T} \left[ \delta_{\mu\nu} A_{\mu\mu}^{-1} + T^2 a_\mu a_\nu \right]
\end{aligned}
\label{eq:final_diffusion_moments}
\end{equation}
\footnotetext{This step can look tricky. It's actually because the calculation is different whether we look at the variance or covariance, hence the use of Kronecker delta $\delta_{\mu\nu}$. For covariance, because populations are independent we just split the two integrals and perform same calculus as in Eq.~\eqref{eq:final_drift_moments}. For variance, we have a single integral with squared integrand variable. We add $\mathcal{F}^2_\mu$ to get the identity $E[(X - c)^2] = \text{Var}(X) + (E[X] - c)^2$, then comes the gaussian variance and $\int (\mathcal{F}^2_\mu - 2m'_\mu \langle m_\mu \rangle + \langle m_\mu \rangle^2) P_T dm'_\mu = (\mathcal{F}_\mu - \langle m_\mu \rangle)^2$.}

\revision{This second jump moment is a raw, non-centred moment. It contains the transition variance $\delta_{\mu\nu}A^{-1}_{\mu\mu}$ and the product of the mean displacements $(\mathcal{F}_\mu-\langle m_\mu\rangle)(\mathcal{F}_\nu-\langle m_\nu\rangle)$. The latter is quadratic in the state, which is why the average of $a_{\mu\nu}$ over the activity distribution differs from its value at the mean state (Supplementary~\ref{subsubsec:covariance_matrix}).}

To complete the second-order mean-field expansion, we require the first and second partial derivatives of the first jump moment function $a_\mu$. Taking the derivatives with respect to the macroscopic activities gives:
\begin{equation*}
\begin{aligned}
\partial_\lambda a_\mu(\{ \langle m_\gamma \rangle \}) 
&= \frac{1}{T} \left( \frac{\partial \mathcal{F}_\mu}{\partial \langle m_\lambda \rangle} - \frac{\partial \langle m_\mu \rangle}{\partial \langle m_\lambda \rangle} \right)
=\frac{1}{T}(\partial_\lambda \mathcal{F}_\mu - \delta_{\mu\lambda})\\
\partial_\lambda \partial_\eta a_\mu(\{ \langle m_\gamma \rangle \}) 
&= \frac{1}{T} \partial_\lambda \partial_\eta \mathcal{F}_\mu
\end{aligned}
\end{equation*}

\revision{The second derivatives of $a_{\mu\nu}$, which enter the curvature term of the covariance equation, are computed in Supplementary~\ref{subsubsec:covariance_matrix}, Eq.~\eqref{eq:curvature_term_explicit}.} Finally, we substitute these derived moments and their derivatives back into the generic moment expansion defined earlier in Eq.~\eqref{eq:second_order_statistical_moments}\revision{, and neglect the contributions of order $N^{-2}$ (Supplementary~\ref{subsubsec:covariance_matrix})}:

\begin{empheq}[box=\fbox]{equation}
\begin{aligned}
T\partial_t \langle m_\mu \rangle
&= (\mathcal{F}_\mu - \langle m_\mu \rangle) + \frac{1}{2}\partial_\lambda\partial_\eta \mathcal{F}_\mu c_{\lambda\eta} \\
T\partial_t c_{\mu\nu}
&= \delta_{\mu\nu} A_{\mu\mu}^{-1} \revision{+ \partial_\lambda \mathcal{F}_\mu\,\partial_\eta \mathcal{F}_\nu\, c_{\lambda\eta} - c_{\mu\nu} + T^2\,\partial_t\langle m_\mu\rangle\,\partial_t\langle m_\nu\rangle} \\
T\partial_\tau Corr_{\mu\nu}(\tau)
&= (\partial_\lambda \mathcal{F}_\nu(\{\langle m_{\gamma}^{stat}\rangle\}) - \delta_{\lambda\nu}) Corr_{\mu\lambda}(\tau)
\end{aligned}
\label{eq:final_ODEs_master_equation}
\end{empheq}

These equations define the mesoscopic closure of the master-equation formalism. The first equation describes the relaxation of the mean population activity toward the transfer function prediction, corrected by the curvature of the transfer function and by population covariances. \revision{The second equation is the corrected covariance equation~\eqref{eq:covariance_consistent_closure}. At each transition, the diagonal term $\delta_{\mu\nu}A^{-1}_{\mu\mu}$ injects the finite-size sampling noise, the gain term $\partial_\lambda\mathcal{F}_\mu\,\partial_\eta\mathcal{F}_\nu\,c_{\lambda\eta}$ transmits the current fluctuations through the network and shapes them into covariances between populations, and $-c_{\mu\nu}$ erases the previous state. The last term is the product of the velocities of the mean activities, given by the first equation: it makes the covariance grow transiently while the mean moves, and vanishes exactly at stationary states. The published covariance equation~\cite{el2009master}, Eq.~\eqref{eq:covariance_published_closure}, differs in both the gain and the source terms (Supplementary~\ref{subsubsec:covariance_matrix}).} The third equation then describes how these correlations decay or persist over time around the stationary AI state\revision{; it only involves the first jump moment and is unchanged, but its initial value at $\tau=0$ is the equal-time covariance $c_{\mu\nu}$}.

The micro-to-meso bridge is therefore explicit: microscopic neuronal properties enter through the transfer function $\mathcal{F}_\mu$, while finite-size mesoscopic fluctuations are governed by its first and second derivatives and by the transition covariance $A_{\mu\mu}^{-1}$.

\subsection{Limitations and Validation of the Framework: From Toy Models to spiking networks}
\label{sec:validating_MFT_framework}

The derivation above provides a formal second-order mean-field description, but the framework also needs to be tested on explicit examples. El Boustani and Destexhe \cite{el2009master} therefore validated the approach in two complementary settings: a linear toy model, which clarifies the role of finite-size fluctuations analytically, and spiking-neuron networks, which test whether transfer functions extracted from microscopic dynamics can reproduce population-level behavior.

\paragraph{Linear model:}
The linear model is not intended to be biologically realistic; it is a pedagogical benchmark for the master-equation formalism. 
Because the transfer function is linear, the mean dynamics decouple from the covariance, allowing the fixed point and its stability to be solved exactly. \revision{The mean dynamics are stable for a net gain $\Delta<1$, exactly as in the first-order analysis, and correlation times diverge at $\Delta=1$. With the corrected covariance equation, however, stationary fluctuations exist only for $|\Delta|<1$: their variance grows as $(1-\Delta^2)^{-1}$ and diverges for strong inhibitory ($\Delta\to-1$) as well as excitatory ($\Delta\to1$) feedback. For $\Delta<-1$, the mean is stable but the fluctuations are not, a threshold that neither the first-order analysis nor the published covariance equation detects.} The model also shows that these fluctuations vanish as $N\to \infty$, thereby recovering the first-order mean-field limit. More details are given in Supplementary~\ref{subsec:master_equation_linear_model}.

\paragraph{Spiking Neural Network:}
El Boustani and Destexhe \cite{el2009master} then tested the formalism on explicit sparse excitatory/inhibitory spiking networks. The main question was whether a transfer function derived from microscopic neuron dynamics is sufficient to reproduce the mesoscopic behavior of recurrent populations. They considered two standard cases: current-based (CUBA) and conductance-based (COBA) integrate-and-fire networks.

In the CUBA case, synaptic inputs are treated as Poisson shot noise, and under a diffusion approximation, the membrane potential dynamics can be summarized by their mean and variance. This leads to a tractable Fokker--Planck-based approximation of the transfer function, which can then be inserted into the mesoscopic ODEs. The resulting mean-field predictions reproduce the activity statistics of simulated current-based networks over a broad region of the asynchronous irregular regime. This case therefore provides the cleanest demonstration of the micro-to-meso bridge: microscopic neuronal and synaptic properties determine a transfer function, and this transfer function determines the mesoscopic population dynamics.

In the COBA case, synaptic inputs act through conductance changes rather than additive currents. This is more biologically realistic, but mathematically harder, because no exact analytical transfer function is generally available. The authors therefore introduced an effective approximation and then improved it using a phenomenological transfer function fitted to simulations. This demonstrates both the flexibility and the limitation of the formalism: the mesoscopic equations can in principle accommodate more realistic neuron models, but their quantitative accuracy depends on how well the transfer function captures the underlying microscopic dynamics.

Taken together, these examples show that the master-equation framework is not tied to a single neuron model. The same mesoscopic ODE structure is preserved, while the neuron model enters through the transfer function and its derivatives. Current-based networks provide a relatively clean validation case, whereas conductance-based networks reveal the central bottleneck of the approach: once a reliable transfer function is available, the framework can reproduce finite-size population dynamics at much lower computational cost than direct spiking simulation; when the transfer function is poor, the mesoscopic predictions degrade accordingly. More details are given in Supplementary~\ref{subsec:master_equation_SNN_case}.

\paragraph{Limitations:} The main limitation of the master-equation formalism is that its quantitative accuracy depends on the transfer function used to close the model. While this function can be derived or approximated relatively cleanly for current-based integrate-and-fire networks, more realistic networks (e.g., conductance-based ones) require stronger approximations or phenomenological fitting. Structurally, the formalism is relatively flexible as long as its assumptions remain satisfied. It does not require one specific connectivity matrix, but it does require sparse enough interactions for conditional independence and weak correlations to remain plausible. El Boustani and Destexhe \cite{el2009master} also showed that local connectivity and heterogeneous transmission delays can be incorporated. In particular, random delays on the order of the coarse-graining time $T$ can reduce global oscillations and enlarge the stable asynchronous irregular region. In addition, the formalism is best justified in asynchronous irregular regimes, where coarse-grained Markovianity, weak correlations, conditional independence, and approximate Gaussian activity distributions are plausible. By contrast, asynchronous regular, synchronous regular, or strongly synchronous irregular states introduce periodicity, synchrony, or residual correlations that can violate the assumptions of the model. The original formalism should therefore be understood primarily as a theory of finite-size asynchronous irregular activity, rather than as a universal description of all brain states. Extensions to sleep, anesthesia, or strongly oscillatory regimes require additional mechanisms and specific validation.

\subsection{Master Equation Reminder}
\label{subsec:master_equation_reminder}

Master equations are used to describe the time evolution of a system that can be modeled as being in a probabilistic combination of states at any given time, and the switching between states is determined by a transition rate matrix $\mathcal{W}$. The equations are a set of differential equations, over time, of the probabilities that the system occupies each of the different states.

The master equation describes how the probability of the system being in state x changes over time:

$$
\frac{dP(x,t)}{dt} = \sum_{x'} \left[ \mathcal{W}(x|x')P(x',t) - \mathcal{W}(x'|x)P(x,t) \right], \,\ \footnotemark
$$
\footnotetext{$P(x,t)$ is the probability of being in state x at time t. $\mathcal{W}(x|x')$ is the transition rate (rule) for jumping from state $x'$ to state $x$. The right-hand side is therefore a \textit{gain-minus-loss} balance: the rate of change of $P(x,t)$ is \textit{everything transitioning to $x$} (ingoing probability) minus \textit{everything transitioning out of state $x$} (outgoing probability).}


Thus, we have the change of probability being the probability of flowing IN minus the probability of flowing OUT.
\\

\textbf{Derivation:}\\

Let $P_k$ be the probability of the system being in state $k$ and $A$ be the transition rate matrix. For each state $k$, the change in occupation probability depends on the contribution from all other states to $k$, and is given by:
$$\frac{dP_k}{dt} = \sum_\ell A_{k\ell} P_\ell =  \sum_{\ell \neq k} A_{k\ell} P_\ell + A_{kk}P_k$$

In a closed system, the system must be in one of the possible states. Thus, if we add up the probability of the system being in one of all the states, the sum must be equal to one.

$$
\sum_{\ell} P_\ell = 1
\quad \text{and} \quad
\frac{d}{dt} \sum_{\ell} P_\ell = 0
$$
Moreover, 
$$\frac{d}{dt} \sum_\ell P_\ell = \sum_\ell \left( \frac{dP_\ell}{dt} \right) = \sum_{\ell} (\sum_{k} A_{\ell k} P_k ) = \sum_{k} (\sum_{\ell} A_{\ell k})P_k = 0$$

Because the last statement holds for any probability distribution, we can test the specific case where $P_k = \delta_{kj}$ (i.e., $P_j = 1$ and $P_k = 0$ for all $k \neq j$). Plugging this into the sum collapses it to a single term, implying that $\sum_{\ell} A_{\ell j} = 0$ for any arbitrary state $j$.

So, $A_{kk} = - \sum_{\ell \neq k} A_{\ell k}$. This means that the rate of staying in state $k$ is exactly equal to the negative sum of the rates of leaving state $k$.

Thus, combining previous results, we get:

$$ \frac{dP_k}{dt} = \sum_{\ell \neq k} (A_{k\ell} P_\ell - A_{\ell k} P_k)$$
Therefore, the change in state $k$ over time is simply the sum of everything flowing in, minus everything flowing out.

\subsection{Generic Second-Order moments of the Master Equation formalism}
\label{sub:generic_second_order_master_eq_moments}

\revision{
\subsubsection{First and Second jump moments}

Let $\{m_\gamma\} = (m_1,\ldots,m_P) = \mathbf{m}$ denote the current activity of the $P$ populations, and let $\{m'_\gamma\} = \mathbf{m'}$ denote a possible destination state. The transition-rate density $\mathcal W(\mathbf \{m'_\gamma\}\mid \{m_\gamma\})$ specifies how frequently transitions from $\{m_\gamma\}$ to $\{m'_\gamma\}$ occur. Let $ \{ \langle m_\gamma \rangle \} = \boldsymbol\nu$ denote the current mean of the activity. Let $\Omega=[0,T^{-1}]^P$ denote the integration space.

A jump $\Delta\mathbf m=\mathbf m'-\mathbf m$ is a change between two states. It should be distinguished from the fluctuation $\delta\mathbf m=\mathbf m-\boldsymbol\nu$ around the current mean.

The \emph{first jump moment} $a_\mu(\{ m_\gamma \})$, or \emph{drift}, is:
\begin{equation}
a_\mu(\mathbf m)
=\int_\Omega (m'_\mu-m_\mu)
\mathcal W(\mathbf m'\mid\mathbf m)\,\mathrm d\mathbf m'.
\label{eq:first_jump_moment}
\end{equation}

It represents the expected signed change in population activity per unit time, conditional on the current state $\mathbf{m}$. A positive (resp. negative) value indicates an average tendency to increase (resp. decrease).

We approximate the expectation by expanding $a_\mu$ around $\boldsymbol\nu$, retaining terms through second degree in $\delta\mathbf m$:
\begin{equation}
\begin{aligned}
a_\mu(\{ m_\gamma \})
&=
a_\mu(\{\langle m_\gamma \rangle \})
+ \sum_{\lambda=1}^{P}
\frac{\partial a_{\mu}(\boldsymbol \nu)}{\partial m_{\lambda}}
\left(m_{\lambda}-\langle m_{\lambda}\rangle\right) \\
&\quad
+ \frac{1}{2}
\sum_{\lambda=1}^{P}\sum_{\eta=1}^{P}
\frac{\partial^{2} a_{\mu}(\boldsymbol \nu)}
{\partial m_{\lambda}\partial m_{\eta}}
\left(m_{\lambda}-\langle m_{\lambda}\rangle\right)
\left(m_{\eta}-\langle m_{\eta}\rangle\right)
+ \mathcal{O}(\delta \mathbf{m}^3) \\
\\
a_\mu(\mathbf m)
&= a_\mu(\boldsymbol\nu) +\sum_{\lambda=1}^P \partial_\lambda a_\mu(\boldsymbol\nu)\,\delta m_\lambda
+\frac12\sum_{\lambda=1}^P\sum_{\eta=1}^P
\partial_\lambda\partial_\eta a_\mu(\boldsymbol\nu)
\,\delta m_\lambda\delta m_\eta
+\mathcal O(\|\delta\mathbf m\|^3). \, \footnotemark
\end{aligned}
\label{eq:Taylor_expansion_drift_term}
\end{equation}
\footnotetext{Here $\partial_\lambda$ means differentiation with respect to $m_\lambda$; all displayed derivatives are evaluated at $\mathbf m=\boldsymbol\nu$.}

Because $\langle m_\mu - \langle m_\mu \rangle \rangle  = \langle \delta m_\mu \rangle = 0$ and $ c_{\lambda \eta} = \langle( m_\lambda -  \langle m_\lambda \rangle)(m_\eta - \langle m_\eta \rangle) \rangle $, after averaging, we obtain:

\begin{equation}
\begin{aligned}
\langle a_\mu(\mathbf m)\rangle
\simeq a_\mu(\boldsymbol\nu)
+\frac12\sum_{\lambda=1}^P\sum_{\eta=1}^P
\partial_\lambda\partial_\eta a_\mu(\boldsymbol\nu)
\,c_{\lambda\eta}.
\end{aligned}
\label{eq:averaged_Taylor_expansion_drift_term}
\end{equation}

The curvature term accounts for the fact that the average drift over a distribution need not equal the drift evaluated at its mean. The expansion does not make the distribution narrow: it is accurate only if $P_t(\mathbf m)$ is already concentrated in a region where $a_\mu$ is well approximated by a quadratic function, that is, if the width of the activity distribution is small compared with the range of activities over which the transfer function bends. We work under this narrow-distribution assumption throughout. For large populations in a stable asynchronous state, $c_{\lambda\eta}$ is assumed $\sim \mathcal{O}(N^{-1})$, and the neglected terms involve third- and fourth-order central moments, of order $N^{-2}$, against $N^{-1}$ for the retained curvature term: the truncation is then controlled by the population size. The assumption fails near instabilities, where the distribution broadens, and in multistable regimes, where it becomes multimodal (Supplementary~\ref{subsubsec:covariance_closure_benchmarks}). Under this assumption, the expectation of $a_\mu$ only requires the mean and the covariance of the activity.

The \emph{second jump moment} $a_{\mu\nu}(\{ m_\gamma \})$, or \emph{diffusion}, is:

\begin{equation}
a_{\mu\nu}(\mathbf m)
=\int_\Omega (m'_\mu-m_\mu)(m'_\nu-m_\nu)
\mathcal W(\mathbf m'\mid\mathbf m)\,\mathrm d\mathbf m'.
\label{eq:second_jump_moment}
\end{equation}

It represents the expected product of activity jumps per unit time, conditional on the current state. For $\mu=\nu$, it measures squared jump sizes and is nonnegative, so upward and downward jumps both contribute positively. Consequently, the first jump moment can vanish because opposite jumps cancel on average, while the second remains positive. For $\mu\ne\nu$, opposite-direction changes contribute negatively. If activity is measured in Hz and time in seconds, $a_\mu$ has units of Hz/s and $a_{\mu\nu}$ has units of Hz$^2$/s. These are raw jump moments: the jump increments are not centred on their conditional mean. The second jump moment is therefore distinct from the conditional covariance of a finite update or from $c_{\mu\nu}$, which describes the spread of the current activity distribution.

Applying the same Taylor expansion to the second jump moment, and because $\langle m_{\lambda}-\langle m_{\lambda}\rangle \rangle = 0$, we get:
\begin{equation}
\begin{aligned}
a_{\mu\nu}(\{ m_\gamma \}) 
&= a_{\mu\nu}(\{\langle m_\gamma \rangle\}) + \sum_{\lambda=1}^{P} \frac{\partial a_{\mu\nu}(\boldsymbol\nu)}{\partial m_{\lambda}} \left(m_{\lambda}-\langle m_{\lambda}\rangle\right) + \\
&\frac{1}{2} \sum_{\lambda,\eta} \frac{\partial^{2} a_{\mu\nu}(\boldsymbol\nu)}{\partial m_{\lambda}\partial m_{\eta}} \left(m_{\lambda}-\langle m_{\lambda}\rangle\right)\left(m_{\eta}-\langle m_{\eta}\rangle\right) +
\mathcal{O}\!\left(
\|\delta\mathbf{m}\|^{3}
\right)\\
\langle a_{\mu\nu}(\mathbf m)\rangle
&\simeq a_{\mu\nu}(\boldsymbol\nu)
+\frac12\sum_{\lambda=1}^P\sum_{\eta=1}^P
\partial_\lambda\partial_\eta a_{\mu\nu}(\boldsymbol\nu)
\,c_{\lambda\eta}.
\end{aligned}
\label{eq:average_second_jump_moment}
\end{equation}

Similarly, the curvature term accounts for the fact that the average diffusion over a distribution need not equal the diffusion evaluated at its mean. The same narrow-distribution assumption is required, and under it the expectation of $a_{\mu\nu}$ only requires the mean and the covariance of the activity.

}

\subsubsection{Mean Activity}

The neurons are assumed to spike at most once within a time bin $T$, so that $0 \le m_\mu \le \frac{1}{T}$.
We denote the mean activity $\langle m_\mu \rangle$ as the average firing rate of population $\mu$ across all possible network states:

\begin{align*}
\nu_\mu = \langle m_\mu\rangle
&= \int^{\frac{1}{T}}_{0} m_\mu P_t(\{m_\gamma \})d\{m_\gamma\}, \,\ \footnotemark\\
&= \prod_{\alpha = 1,...,P} \int^{\frac{1}{T}}_{0} m_\mu P_t(\{m_\gamma \})dm_\alpha \\
&=  \int_{\Omega} m_\mu P_t(\mathbf{m})d\mathbf{m}
\end{align*}
\footnotetext{Each state $\{m_{\gamma}\} =(m_1,..,m_P)$ has probability density $P_t(m_1,...,m_P)$ and contributes proportionally to the mean. This is a joint probability distribution over all populations of the network. $d\{m_{\gamma}\}=dm_1\,dm_2\cdots dm_P$}

where $P_t(\{m_\gamma\})$ is the probability of the network being in this state at time $t$.

Since the integration bounds are constant and $m_\gamma$ does not depend on time, the time derivative acts only on the probability distribution. Then, by substituting the master equation \eqref{eq:master_equation}, we get:

\revision{
\begin{equation}
\begin{aligned}
\partial_t\langle m_\mu \rangle
&= \int_\Omega m_\mu\,\partial_t P_t(\mathbf{m})\,d\mathbf{m},
\qquad \int_\Omega d\mathbf{m}\equiv\prod_{\alpha=1}^{P}\int_0^{1/T}dm_\alpha,\\
&= \int_\Omega d\mathbf{m}\; m_\mu \int_\Omega d\mathbf{m}'\,\Big[P_t(\mathbf{m}')\,\mathcal{W}(\mathbf{m}\mid\mathbf{m}') - P_t(\mathbf{m})\,\mathcal{W}(\mathbf{m}'\mid\mathbf{m})\Big]\\
&= \int_\Omega d\mathbf{m}\int_\Omega d\mathbf{m}'\,(m'_\mu-m_\mu)\,\mathcal{W}(\mathbf{m}'\mid\mathbf{m})\,P_t(\mathbf{m}) \,\ \footnotemark \\
&= \int_\Omega a_\mu(\mathbf{m})\,P_t(\mathbf{m})\,d\mathbf{m}\\
\partial_t\langle m_\mu \rangle &= \langle a_\mu(\mathbf{m}) \rangle
\end{aligned}
\label{eq:mean_activity_drift}
\end{equation}
\footnotetext{We switch in the first term $m'$ and $m$. This is possible because these integrals run over all possible states, such that $\int_{\Omega} f(x,y)dxdy=\int_{\Omega} f(y,x)dydx$. Therefore, the time derivative of the mean is the mean drift, being the mean of the change of activity times the transition rate.}
}

Finally, using equation \eqref{eq:averaged_Taylor_expansion_drift_term}, we have:
\begin{equation}
\begin{aligned}
\partial_t\langle m_\mu \rangle
&\simeq a_\mu(\{\langle m_\gamma \rangle \}) +
\frac12\sum_{\lambda,\eta=1}^{P}
\partial_{\lambda}\partial_{\eta} a_\mu(\{\langle m_\gamma \rangle \})c_{\lambda \eta}\\ 
\text{also written as}, \ \ \qquad
\partial_t\nu_\mu
&\simeq a_\mu(\boldsymbol{\nu}) +
\frac12\sum_{\lambda,\eta=1}^{P}
\partial_{\lambda}\partial_{\eta} a_\mu(\boldsymbol{\nu})c_{\lambda \eta}
\end{aligned}
\label{eq:mean_equation_mean_field}
\end{equation}

\subsubsection{Covariance Matrix}
\label{subsubsec:covariance_matrix}

\revision{We now derive the evolution equation of the covariance matrix $c_{\mu\nu}=\langle (m_\mu-\langle m_\mu\rangle)(m_\nu-\langle m_\nu\rangle)\rangle$. The derivation has two steps. We first obtain an exact identity, in which no approximation is made. We then expand this identity to second order in the fluctuations $\delta m_\lambda = m_\lambda-\langle m_\lambda\rangle$. The second step contains a subtlety that is not handled correctly in the original derivation~\cite{el2009master}, so we treat it in detail.}

\revision{By definition of the covariance,
\begin{equation}
\begin{aligned}
\partial_t c_{\mu\nu}
&= \partial_t \left\langle
(m_\mu - \langle m_\mu \rangle)
(m_\nu - \langle m_\nu \rangle)
\right\rangle \\
&= \partial_t \langle m_\mu m_\nu \rangle
- \langle m_\nu \rangle \partial_t \langle m_\mu \rangle
- \langle m_\mu \rangle \partial_t \langle m_\nu \rangle .
\end{aligned}
\label{eq:covariance_time_derivative}
\end{equation}

Similarly to the mean activity, we substitute the master equation~\eqref{eq:master_equation} and exchange $\mathbf m$ and $\mathbf m'$ in the gain term, as in Eq.~\eqref{eq:mean_activity_drift}:
\begin{equation*}
\begin{aligned}
\partial_t\langle m_\mu m_\nu \rangle
&= \prod_{\alpha = 1,...,P} \int_0^{1/T} dm_\alpha m_\mu m_\nu \partial_t P_t(\{ m_\gamma \})\\
&= \int_\Omega m_\mu m_\nu\,\partial_t P_t(\mathbf{m})\,d\mathbf{m}\\
&= \int_\Omega d\mathbf{m}\; m_\mu m_\nu \int_\Omega d\mathbf{m}'\,\Big[P_t(\mathbf{m}')\,\mathcal{W}(\mathbf{m}\mid\mathbf{m}') - P_t(\mathbf{m})\,\mathcal{W}(\mathbf{m}'\mid\mathbf{m})\Big]\\
&= \int_\Omega d\mathbf{m}\int_\Omega d\mathbf{m}'\,(m'_\mu m'_\nu-m_\mu m_\nu)\,\mathcal{W}(\mathbf{m}'\mid\mathbf{m})\,P_t(\mathbf{m}).
\end{aligned}
\end{equation*}
 
Using the definition \eqref{eq:second_jump_moment} and the relations
$$
m'_\mu m'_\nu - m_\mu m_\nu = (m'_\mu - m_\mu)(m'_\nu - m_\nu) + m_\mu(m'_\nu-m_\nu) + m_\nu(m'_\mu-m_\mu)
$$
we obtain:
\begin{equation}
\begin{aligned}
\partial_t\langle m_\mu m_\nu \rangle = \langle a_{\mu\nu}(\mathbf{m})\rangle + \langle m_\mu a_\nu(\mathbf{m})\rangle + \langle m_\nu a_\mu(\mathbf{m})\rangle .
\end{aligned}
\label{eq:time_derivative_co_product}
\end{equation}

Therefore, using equations \eqref{eq:mean_activity_drift}, \eqref{eq:covariance_time_derivative}, and \eqref{eq:time_derivative_co_product}, we get
\begin{equation}
\begin{aligned}
\partial_t c_{\mu\nu}
&=
\left\langle a_{\mu\nu}(\mathbf{m}) \right\rangle+
\left\langle m_\nu a_\mu(\mathbf{m}) \right\rangle+
\left\langle m_\mu a_\nu(\mathbf{m}) \right\rangle -
\langle m_\nu \rangle \partial_t \langle m_\mu \rangle-
\langle m_\mu \rangle \partial_t \langle m_\nu \rangle \\
&=
\left\langle a_{\mu\nu}(\mathbf{m}) \right\rangle+
\left\langle a_\mu(\mathbf{m}) \cdot (m_\nu-\langle m_\nu\rangle) \right\rangle +
\left\langle a_\nu(\mathbf{m}) \cdot (m_\mu-\langle m_\mu\rangle) \right\rangle .
\end{aligned}
\label{eq:covariance_exact_identity}
\end{equation}
This identity is exact: it only uses the master equation~\eqref{eq:master_equation} and the definitions of the jump moments.

By injecting the Taylor expansion of Eq.~\eqref{eq:Taylor_expansion_drift_term} into $\left\langle a_\mu(\mathbf{m}) \cdot (m_\nu-\langle m_\nu\rangle) \right\rangle$, we get:
\begin{equation*}
\begin{aligned}
\left\langle a_\mu(\mathbf{m}) \cdot (m_\nu-\langle m_\nu\rangle) \right\rangle
&= \left\langle [a_\mu(\boldsymbol \nu) + \partial_{\lambda} a_\mu(\boldsymbol \nu)
\left(m_{\lambda}-\langle m_{\lambda}\rangle\right)]\cdot (m_\nu-\langle m_\nu\rangle) \right\rangle \\
&\quad + \mathcal{O}\!\left(
\left\langle\|\delta\mathbf{m}\|^{3}\right\rangle
\right) \\
&= \partial_{\lambda} a_\mu(\boldsymbol \nu)c_{\nu \lambda} + \mathcal{O}\!\left(
\left\langle\|\delta\mathbf{m}\|^{3}\right\rangle
\right). \,\ \footnotemark
\end{aligned}
\end{equation*}
\footnotetext{The constant term vanishes because $\langle\delta m_\nu\rangle=0$. The quadratic drift term contributes third central moments, which are omitted in this second-order Taylor closure. Their smallness is an assumption about the activity distribution. They vanish for a Gaussian distribution and can be $\mathcal O(N^{-2})$ in a regular system-size expansion; this order does not follow from finite population size alone.}
 
Finally, we have:
\begin{equation}
\begin{aligned}
\partial_t c_{\mu\nu}
&= \left\langle a_{\mu\nu}(\{m_\gamma\}) \right\rangle + \partial_{\lambda} a_\mu(\{\langle m_\gamma \rangle \})c_{\nu \lambda} + \partial_{\lambda} a_\nu(\{\langle m_\gamma \rangle \})c_{\mu \lambda}
+ \mathcal{O}\!\left(\left\langle\|\delta\mathbf{m}\|^{3}\right\rangle\right)
\end{aligned}
\label{eq:covariance_intermediate_expression}
\end{equation}

One may notice that Eq.~\eqref{eq:covariance_intermediate_expression} is not yet closed. It contains the second jump moment averaged over all network states, $\langle a_{\mu\nu}(\{m_\gamma\})\rangle$, not the second jump moment at the mean state, $a_{\mu\nu}(\{\langle m_\gamma\rangle\})$. The original derivation~\cite{el2009master} (their Appendix~A, Eq.~A.5) replaces the former by the latter, invoking Van Kampen's approximation \cite{van1992stochastic}. However, the average of a function is not the function of the average, unless specific functions (e.g., linear ones). The second jump moment is not linear in the state: we will see that it is quadratic in the deviation $\mathcal{F}_\mu - m_\mu$. The fact that $a_{\mu\nu}$ is called a second jump moment does not, by itself, make its Hessian contribution of fourth order. A Gaussian conditional transition law does not by itself make the time-dependent marginal distribution $P_t(\mathbf{m})$ Gaussian, and even a Gaussian $P_t(\mathbf{m})$ would not make this difference vanish.

Writing the Taylor expansion \eqref{eq:average_second_jump_moment} as

\begin{equation*}
\begin{aligned}
\langle a_{\mu\nu}(\mathbf m)\rangle
&= a_{\mu\nu}(\boldsymbol\nu)
+ \mathcal{K}_{\mu\nu} +\mathcal{O}\!\left(
\left\langle\|\delta\mathbf{m}\|^{3}\right\rangle
\right),
\qquad
\mathcal{K}_{\mu\nu}
\equiv
\frac{1}{2}
\sum_{\lambda,\eta=1}^{P}
\left.
\frac{\partial^{2}a_{\mu\nu}}
{\partial m_{\lambda}\partial m_{\eta}}
\right|_{\{m_\gamma\}=\{\langle m_\gamma\rangle\}}
c_{\lambda\eta}.
\end{aligned}
\end{equation*}

$\mathcal{K}_{\mu\nu}$ is the curvature correction to the expected second jump moment. It depends on the covariance itself, and is of the same degree in the fluctuations as the two drift-covariance terms. Dropping it is an additional approximation with specific conditions, distinct from the neglect of third-order moments. Whether this approximation is acceptable depends on the size of $\mathcal{K}_{\mu\nu}$ compared with the terms that are kept. We now compute it explicitly for the transition operator of the master equation.

\paragraph{Curvature of the second jump moment at an arbitrary state.}
Replacing $\{\langle m_\gamma \rangle\}$ by an arbitrary state $\{m_\gamma\}$ in Eq.~\eqref{eq:final_diffusion_moments}, the same steps give, for any state,
\begin{equation}
\begin{aligned}
a_\mu(\{m_\gamma\}) &= \frac{1}{T}\left(\mathcal{F}_\mu(\{m_\gamma\})-m_\mu\right),\\
a_{\mu\nu}(\{m_\gamma\}) &= \frac{1}{T}\Big[\delta_{\mu\nu}A^{-1}_{\mu\mu}(\{m_\gamma\})
+\left(\mathcal{F}_\mu(\{m_\gamma\})-m_\mu\right)\left(\mathcal{F}_\nu(\{m_\gamma\})-m_\nu\right)\Big],\\
A^{-1}_{\mu\mu}(\{m_\gamma\})&=\frac{\mathcal{F}_\mu(\{m_\gamma\})\left[1/T-\mathcal{F}_\mu(\{m_\gamma\})\right]}{N_\mu}. \,\ \footnotemark
\end{aligned}
\label{eq:jump_moments_any_state}
\end{equation}
\footnotetext{The second jump moment has two contributions. The first, $\delta_{\mu\nu}A^{-1}_{\mu\mu}$, is the variance of the binomial draw (the transition noise). The second is the square of the deterministic displacement: starting from $\{m_\gamma\}$, the network does not jump to a state close to $\{m_\gamma\}$, but to a state close to $\mathcal{F}(\{m_\gamma\})$. It is this second contribution that makes $a_{\mu\nu}$ curved.}

In what follows, $\partial_\lambda=\partial/\partial m_\lambda$, we write $\mathcal{F}_\mu$ for $\mathcal{F}_\mu(\{m_\gamma\})$ and $\partial_\lambda m_\mu=\delta_{\mu\lambda}$. Let us differentiate the terms in Eq.~\eqref{eq:jump_moments_any_state} with respect to $m_\lambda$ to find an expression of $\mathcal{K}_{\mu\nu}$:

\begin{equation*}
\partial_\lambda\Big[(\mathcal{F}_\mu-m_\mu)(\mathcal{F}_\nu-m_\nu)\Big]
=(\partial_\lambda\mathcal{F}_\mu-\delta_{\mu\lambda})(\mathcal{F}_\nu-m_\nu)
+(\mathcal{F}_\mu-m_\mu)(\partial_\lambda\mathcal{F}_\nu-\delta_{\nu\lambda}).
\end{equation*}
Differentiating once more with respect to $m_\eta$, and using $\partial_\eta\delta_{\mu\lambda}=0$:
\begin{equation*}
\begin{aligned}
(a) \quad \partial_\eta\partial_\lambda\Big[(\mathcal{F}_\mu-m_\mu)(\mathcal{F}_\nu-m_\nu)\Big]
={}&\partial_\eta\partial_\lambda\mathcal{F}_\mu\,(\mathcal{F}_\nu-m_\nu)
+(\partial_\lambda\mathcal{F}_\mu-\delta_{\mu\lambda})(\partial_\eta\mathcal{F}_\nu-\delta_{\nu\eta})\\
&+(\partial_\eta\mathcal{F}_\mu-\delta_{\mu\eta})(\partial_\lambda\mathcal{F}_\nu-\delta_{\nu\lambda})
+(\mathcal{F}_\mu-m_\mu)\,\partial_\eta\partial_\lambda\mathcal{F}_\nu .
\end{aligned}
\end{equation*}

For the transition variance,
\begin{equation*}
(b) \quad \partial_\eta\partial_\lambda A^{-1}_{\mu\mu}
=\frac{1}{N_\mu}\Big[\left(\tfrac{1}{T}-2\mathcal{F}_\mu\right)\partial_\eta\partial_\lambda\mathcal{F}_\mu
-2\,\partial_\lambda\mathcal{F}_\mu\,\partial_\eta\mathcal{F}_\mu\Big].
\end{equation*}

Combining $(a)$ and $(b)$, we now evaluate all functions and these derivatives at the mean state $\{\langle m_\gamma\rangle\}$, multiply by $\frac{1}{2}c_{\lambda\eta}$ and sum over $\lambda$ and $\eta$. Hence:
\begin{equation}
\begin{aligned}
T\,\mathcal{K}_{\mu\nu}
={}&\underbrace{\frac{\delta_{\mu\nu}}{2}\,\partial_\lambda\partial_\eta A^{-1}_{\mu\mu}\,c_{\lambda\eta}}_{\text{(i)}}
+\underbrace{(\partial_\lambda\mathcal{F}_\mu-\delta_{\mu\lambda})\,c_{\lambda\eta}\,(\partial_\eta\mathcal{F}_\nu-\delta_{\nu\eta})}_{\text{(ii)}}\\
&+\underbrace{\frac{1}{2}\Big[(\mathcal{F}_\nu-\langle m_\nu\rangle)\,\partial_\lambda\partial_\eta\mathcal{F}_\mu
+(\mathcal{F}_\mu-\langle m_\mu\rangle)\,\partial_\lambda\partial_\eta\mathcal{F}_\nu\Big]c_{\lambda\eta}}_{\text{(iii)}} ,
\end{aligned}
\label{eq:curvature_term_explicit}
\end{equation}

In what follows, we use the two notations
\begin{equation}
g_\mu\equiv\mathcal{F}_\mu-\langle m_\mu\rangle,
\qquad
h_\mu\equiv\partial_\lambda\partial_\eta\mathcal{F}_\mu\,c_{\lambda\eta},
\label{eq:shorthand_g_h}
\end{equation}
With these notations, term (iii) reads $\frac{1}{2}\left(g_\nu h_\mu+g_\mu h_\nu\right)$.

\paragraph{Complete second-order covariance equation.}
Inserting Eqs.~\eqref{eq:average_second_jump_moment}-\eqref{eq:curvature_term_explicit} into Eq.~\eqref{eq:covariance_intermediate_expression}, multiplying by $T$, and using $T\partial_\lambda a_\mu=\partial_\lambda\mathcal{F}_\mu-\delta_{\mu\lambda}$ (from Eq.~\eqref{eq:final_drift_moments}), we obtain
\begin{equation*}
\begin{aligned}
T\partial_t c_{\mu\nu}
={}&\delta_{\mu\nu}A^{-1}_{\mu\mu}+g_\mu g_\nu
+(\partial_\lambda\mathcal{F}_\mu-\delta_{\mu\lambda})c_{\nu\lambda}
+(\partial_\lambda\mathcal{F}_\nu-\delta_{\nu\lambda})c_{\mu\lambda}\\
&+\text{(i)}+\text{(ii)}+\text{(iii)}
+\mathcal{O}\!\left(\left\langle\|\delta\mathbf{m}\|^{3}\right\rangle\right).
\end{aligned}
\end{equation*}
The terms that involve the gain $\partial_\lambda\mathcal{F}_\mu$ simplify. Expanding term (ii) term by term,
\begin{equation*}
(\partial_\lambda\mathcal{F}_\mu-\delta_{\mu\lambda})\,c_{\lambda\eta}\,(\partial_\eta\mathcal{F}_\nu-\delta_{\nu\eta})
=\partial_\lambda\mathcal{F}_\mu\,\partial_\eta\mathcal{F}_\nu\,c_{\lambda\eta}
-\partial_\lambda\mathcal{F}_\mu\,c_{\lambda\nu}
-\partial_\eta\mathcal{F}_\nu\,c_{\mu\eta}
+c_{\mu\nu},
\end{equation*}
and expanding the two drift terms,
\begin{equation*}
(\partial_\lambda\mathcal{F}_\mu-\delta_{\mu\lambda})c_{\nu\lambda}
+(\partial_\lambda\mathcal{F}_\nu-\delta_{\nu\lambda})c_{\mu\lambda}
=\partial_\lambda\mathcal{F}_\mu\,c_{\nu\lambda}-c_{\nu\mu}
+\partial_\lambda\mathcal{F}_\nu\,c_{\mu\lambda}-c_{\mu\nu}.
\end{equation*}
When the two lines are added, the terms linear in the gain cancel pairwise\footnote{$\partial_\lambda\mathcal{F}_\mu c_{\lambda\nu}=\partial_\lambda\mathcal{F}_\mu c_{\nu\lambda}$ by symmetry of $c$, and $\partial_\eta\mathcal{F}_\nu c_{\mu\eta}=\partial_\lambda\mathcal{F}_\nu c_{\mu\lambda}$ after renaming the dummy index $\eta\to\lambda$. The remaining terms give $c_{\mu\nu}-c_{\nu\mu}-c_{\mu\nu}=-c_{\mu\nu}$.}, and only $\partial_\lambda\mathcal{F}_\mu\,\partial_\eta\mathcal{F}_\nu\,c_{\lambda\eta}-c_{\mu\nu}$ remains. The complete second-order covariance equation is therefore
\begin{empheq}[box=\fbox]{equation}
\begin{aligned}
T\partial_t c_{\mu\nu}
={}&\delta_{\mu\nu}\Big[A^{-1}_{\mu\mu}+\tfrac{1}{2}\,\partial_\lambda\partial_\eta A^{-1}_{\mu\mu}\,c_{\lambda\eta}\Big]
+\partial_\lambda\mathcal{F}_\mu\,\partial_\eta\mathcal{F}_\nu\,c_{\lambda\eta}
-c_{\mu\nu}\\
&+g_\mu g_\nu+\tfrac{1}{2}\left(g_\mu h_\nu+h_\mu g_\nu\right).
\end{aligned}
\label{eq:covariance_complete}
\end{empheq}

It keeps the three curvature terms (i), (ii) and (iii). It neglects only the central moments of order three and higher, together with the corresponding Taylor remainders, and it makes no assumption on the size of the displacement $g_\mu$. It is exact, at all times, for constant and affine transfer functions, whose jump moments are polynomials of degree one and two, as long as $T\mathcal{F}_\mu$ remains in $[0,1]$ (Supplementary~\ref{subsubsec:covariance_closure_benchmarks}).
}

\revision{

\paragraph{Order of magnitude of each contribution:}
Let $N_\mu$ grow proportionally to $N$, with $T$ fixed and bounded transfer-function derivatives. Finite-size fluctuations scale with the population size, and in a regular weak-fluctuation regime, $c_{\lambda\eta}=\mathcal{O}(N^{-1})$, and so does the transition variance, $A^{-1}_{\mu\mu}=\mathcal{O}(N^{-1})$. The transfer function and its derivatives are $\mathcal{O}(1)$, so that $h_\mu=\mathcal{O}(N^{-1})$. Therefore:
\begin{itemize}
    \item term (i) is $C\times\mathcal{O}(N^{-1})=\mathcal{O}(N^{-2})$, because $\partial^2A^{-1}_{\mu\mu}$ carries a factor $N_\mu^{-1}$;
    \item term (ii) is $\mathcal{O}(1)\times\mathcal{O}(N^{-1})=\mathcal{O}(N^{-1})$, like the retained terms $\delta_{\mu\nu}A^{-1}_{\mu\mu}$ and $(\partial_\lambda\mathcal{F}_\mu-\delta_{\mu\lambda})c_{\nu\lambda}$. It does not become negligible for large populations: the factor $N^{-1}$ carried by $c_{\lambda\eta}$ is shared by all the terms of the equation\footnote{Both the omitted term and the retained terms vanish as $N\to\infty$, but their \emph{ratio} does not. The correct comparison is between $\mathcal{K}_{\mu\nu}$ and the other terms of the covariance equation, not between $\mathcal{K}_{\mu\nu}$ and the $\mathcal{O}(1)$ prefactor of $a_{\mu\nu}$.};
    \item the source term $g_\mu g_\nu$ and term (iii), $g\times\mathcal{O}(N^{-1})$, depend on the displacement $g_\mu$, which is not set by the population size but by the distance to a stationary state.
\end{itemize}

These estimates are conditional. Under abrupt forcing, the random transition times of the master-equation process (each transition redraws the whole state) can generate an ensemble spread of order one: some realizations have already changed state while others have not. Population size then does not control the moment truncation. The Taylor remainders must instead be assessed against the width of the distribution and the variation of the transfer function over that width.\\
\emph{At a stationary state}, the mean equation \eqref{eq:mean_equation_mean_field} gives $g_\mu=-\frac{1}{2}h_\mu=\mathcal{O}(N^{-1})$, so that $g_\mu g_\nu$ and term (iii) are both $\mathcal{O}(N^{-2})$. \\
\emph{In slow transients}, where the mean is displaced from the fixed point by an amount comparable with the fluctuations (relaxation of a spontaneous fluctuation, slowly varying input), $g_\mu=\mathcal{O}(N^{-1/2})$: the source $g_\mu g_\nu$ is $\mathcal{O}(N^{-1})$, like the other retained terms, and term (iii) is $\mathcal{O}(N^{-3/2})$, the leading correction. \\
\emph{In fast transients}, after an abrupt change of the input, $g_\mu=\mathcal{O}(1)$ for a duration of order $T$. The source term is then $\mathcal{O}(1)$, and the covariance grows far beyond its $\mathcal{O}(N^{-1})$ stationary value. In the master equation as stated, each realization keeps its state until its next transition, which occurs at a random time. Shortly after the change of input, some realizations have already jumped towards the new fixed point while others have not, and the spread of the ensemble is of the order of the displacement itself. The expansion is then no longer controlled by $N$, but by the width of the distribution compared with the range over which $\mathcal{F}$ bends, and term (iii) is of the same order as term (ii).

\paragraph{Corrected covariance equation:}
Expanding the product
\begin{equation*}
\left(g_\mu+\tfrac{1}{2}h_\mu\right)\left(g_\nu+\tfrac{1}{2}h_\nu\right)
=g_\mu g_\nu+\tfrac{1}{2}g_\mu h_\nu+\tfrac{1}{2}h_\mu g_\nu+\tfrac{1}{4}h_\mu h_\nu ,
\end{equation*}
we can write
\begin{equation*}
g_\mu g_\nu+\tfrac{1}{2}\left(g_\mu h_\nu+h_\mu g_\nu\right)
=\left(g_\mu+\tfrac{1}{2}h_\mu\right)\left(g_\nu+\tfrac{1}{2}h_\nu\right)-\tfrac{1}{4}h_\mu h_\nu .
\end{equation*}
The bracket is the right-hand side of the mean equation~\eqref{eq:mean_equation_mean_field},
\begin{equation*}
g_\mu+\tfrac{1}{2}h_\mu=\mathcal{F}_\mu-\langle m_\mu\rangle+\tfrac{1}{2}\partial_\lambda\partial_\eta\mathcal{F}_\mu\,c_{\lambda\eta}=T\partial_t\langle m_\mu\rangle .
\end{equation*}
The last term, $\frac{1}{4}h_\mu h_\nu$, is of order $C^2$ quantities: like term (i), it is $\mathcal{O}(N^{-2})$ under the scaling above. Neglecting these two terms in Eq.~\eqref{eq:covariance_complete} gives the \emph{corrected covariance equation}:

\begin{empheq}[box=\doubleboxeq]{equation}
\begin{aligned}
T\partial_t c_{\mu\nu}
&=\delta_{\mu\nu}A^{-1}_{\mu\mu}
+\partial_\lambda\mathcal{F}_\mu\,\partial_\eta\mathcal{F}_\nu\,c_{\lambda\eta}
-c_{\mu\nu}
+T^2\,\partial_t\langle m_\mu\rangle\,\partial_t\langle m_\nu\rangle ,
\end{aligned}
\label{eq:covariance_consistent_closure}
\end{empheq}
where $T\partial_t\langle m_\mu\rangle$ is given by the mean equation \eqref{eq:mean_equation_mean_field}, which is unchanged. Written with the same quantities as the published equation, the source term reads
\begin{equation}
T^2\,\partial_t\langle m_\mu\rangle\,\partial_t\langle m_\nu\rangle
=\Big(\mathcal{F}_\mu-\langle m_\mu\rangle+\tfrac{1}{2}\partial_\lambda\partial_\eta\mathcal{F}_\mu\,c_{\lambda\eta}\Big)
\Big(\mathcal{F}_\nu-\langle m_\nu\rangle+\tfrac{1}{2}\partial_\lambda\partial_\eta\mathcal{F}_\nu\,c_{\lambda\eta}\Big).
\label{eq:consistent_source_term}
\end{equation}
Term (iii) thus has a simple meaning: it replaces the squared displacement $(\mathcal{F}_\mu-\langle m_\mu\rangle)(\mathcal{F}_\nu-\langle m_\nu\rangle)$ by the product of the actual velocities of the mean activities, which include the curvature correction of the mean equation.

\paragraph{Stationary covariance equation:}
Near a stationary state, the source term is $\mathcal{O}(N^{-2})$, and it vanishes at the stationary state itself. To leading order in $N^{-1}$, Eqs.~\eqref{eq:covariance_complete} and~\eqref{eq:covariance_consistent_closure} then reduce to
\begin{equation}
T\partial_t c_{\mu\nu}
=\delta_{\mu\nu}A^{-1}_{\mu\mu}
+\partial_\lambda\mathcal{F}_\mu\,\partial_\eta\mathcal{F}_\nu\,c_{\lambda\eta}
-c_{\mu\nu}.
\label{eq:covariance_stationary}
\end{equation}
This is a reduced form in a specific regime. It gives the stationary covariance and the linear stability of the fluctuations, but it misses the growth of the covariance caused by the motion of the mean during transients. Using this equation during a transient additionally requires the squared mean velocity to be negligible at the accuracy sought.

\paragraph{The published equation (mean-value closure):}
The mean-value closure, used in El Boustani and Destexhe~\cite{el2009master}, gives instead
\begin{equation}
T\partial_t c_{\mu\nu}
=\delta_{\mu\nu}A^{-1}_{\mu\mu}+(\mathcal{F}_\mu-\langle m_\mu\rangle)(\mathcal{F}_\nu-\langle m_\nu\rangle)
+\partial_\lambda\mathcal{F}_\mu\,c_{\nu\lambda}+\partial_\lambda\mathcal{F}_\nu\,c_{\mu\lambda}
-2c_{\mu\nu}.
\label{eq:covariance_published_closure}
\end{equation}
It contains none of the three curvature terms. The omission is not justified by the large $N$ alone. It differs from Eq.~\eqref{eq:covariance_consistent_closure} by term (ii) which is generally of the same order as the retained terms, and by term (iii) during transients. Since term (ii) is of the same order as the retained terms, Eq.~\eqref{eq:covariance_published_closure} is not the leading order of the master equation~\eqref{eq:master_equation}, except when the gain matrix is close to the identity ($J\simeq I$, a marginally stable regime). The mean equation \eqref{eq:mean_equation_mean_field} is not affected: Eq.~\eqref{eq:mean_activity_drift} only involves the first jump moment $a_\mu$, and its expansion already keeps the curvature term $\frac{1}{2}\partial_\lambda\partial_\eta a_\mu\,c_{\lambda\eta}$.
 
\begin{table}[ht]
\centering
\small
\begin{tabularx}{\textwidth}{@{}>{\raggedright\arraybackslash}p{3.2cm} >{\raggedright\arraybackslash}p{2.1cm} >{\raggedright\arraybackslash}X >{\raggedright\arraybackslash}X@{}}
\toprule
Covariance equation & Curvature terms kept & Neglected & Use \\
\midrule
Exact identity, Eq.~\eqref{eq:covariance_exact_redraw} & -- & Nothing, but not closed: it involves the whole distribution & Reference for benchmarks \\
Complete, Eq.~\eqref{eq:covariance_complete} & (i), (ii), (iii) & Central moments of order $\geq3$ & Reference second-order equation; exact for affine $\mathcal{F}$ \\
Corrected, Eq.~\eqref{eq:covariance_consistent_closure} & (ii), (iii) & In addition, (i) and $\frac{1}{4}h_\mu h_\nu$, both $\mathcal{O}(N^{-2})$ & Recommended form, for stationary states and transients \\
Stationary, Eq.~\eqref{eq:covariance_stationary} & (ii) & In addition, the source term $T^2\partial_t\langle m_\mu\rangle\partial_t\langle m_\nu\rangle$ & Stationary covariance, stability of the fluctuations \\
\midrule
Published, Eq.~\eqref{eq:covariance_published_closure} & None & (i), (ii), (iii) & Requires additional restrictions, resembles to small jumps or $J\simeq I$ \\
\bottomrule
\end{tabularx}
\caption{\revision{Covariance equations derived from the master equation~\eqref{eq:master_equation}. The first four lines are successive truncations of the exact identity; the last line is the published mean-value closure \cite{el2009master}. All second-order equations require fluctuations that are narrow compared with the range over which $\mathcal{F}$ bends.}}
\label{tab:covariance_hierarchy}
\end{table}

\paragraph{Which form should be used? A hierarchy of truncations:}
Table~\ref{tab:covariance_hierarchy} summarizes the four equations. We stress that none of them is an exact reduction of the master equation, except the complete equation~\eqref{eq:covariance_complete} for an affine transfer function. They are truncations, valid as long as the distribution of the activity remains narrow compared with the range over which $\mathcal{F}$ bends. Within this domain, we recommend the corrected equation~\eqref{eq:covariance_consistent_closure} as the working form, for four reasons:

\begin{enumerate}
    \item \emph{It is as accurate as the complete equation.} Equations~\eqref{eq:covariance_complete} and~\eqref{eq:covariance_consistent_closure} differ only by term (i) and by $\frac{1}{4}h_\mu h_\nu$, which are smaller than the retained terms by a factor of order $N^{-1}$ under the assumed scaling. Keeping term (i) would not make the equation more accurate in general, because the second-order truncation also neglects other contributions of the same order $N^{-2}$: the third-order central moments in $\langle a_\mu\,\delta m_\nu\rangle$, and the third- and fourth-order Taylor terms of $\langle a_{\mu\nu}\rangle$, which involve $\langle\delta m^3\rangle$ and $\langle\delta m^4\rangle\simeq3c^2$. Keeping term (i) alone would single out one of several neglected terms of the same order. The only exception is an affine transfer function, for which all these other terms vanish and the complete equation is exact. Even then, the corrected equation only overestimates the stationary variance by the factor $1+\Delta^2/[N(1-\Delta^2)]$, i.e.\ by $1.2\%$ at $|\Delta|=0.99$ for $N=4000$ (Table~\ref{tab:covariance_closure_affine}).
    \item \emph{It has the structure of the exact identity.} Equation~\eqref{eq:covariance_consistent_closure} has, term by term, the structure of the exact identity~\eqref{eq:covariance_exact_redraw} derived below: injected transition noise, fluctuations transmitted through the transfer function, erasure of the current state, and squared velocity of the mean. Its source term is exact for a quadratic transfer function, and it vanishes identically at stationary states.
    \item \emph{It remains valid during transients}, unlike the stationary equation~\eqref{eq:covariance_stationary}, which has no source term.
    \item \emph{It costs no more than the published equation}, since $T\partial_t\langle m_\mu\rangle$ is computed anyway for the mean equation.
\end{enumerate}

A direct test confirms these statements\footnote{\label{fn:transient_check}We used the sigmoid of Eq.~\eqref{eq:sigmoid_F} with $m_0=1/(2T)=100\,\mathrm{Hz}$, $\Delta=-0.5$, $N=4000$ and $T=5\,\mathrm{ms}$, with an additional input $\theta$ inside the hyperbolic tangent. A step of $\theta$ at $t=0$ moves the fixed point from $60$ to $100\,\mathrm{Hz}$. The exact variance of the master equation is obtained by propagating the probability distribution itself, $d\boldsymbol{\pi}/dt=(P^{\mathsf T}-I)\boldsymbol{\pi}/T$, with no sampling.}. After a step of the input that moves the fixed point of a sigmoid from $60$ to $100\,\mathrm{Hz}$, the exact variance rises from $2.6$ to $752\,\mathrm{Hz}^2$ within $0.6\,T$, before relaxing towards its new stationary value. The complete and corrected equations follow it within $2.2\,\mathrm{Hz}^2$ ($0.3\%$ of the peak), and differ from each other by less than $0.05\,\mathrm{Hz}^2$. Without term (iii) the error is $5.4\,\mathrm{Hz}^2$. The stationary equation misses the transient entirely (its variance stays close to $3\,\mathrm{Hz}^2$), and the published equation deviates by up to $492\,\mathrm{Hz}^2$ ($65\%$ of the peak).

\paragraph{Stationary covariance differences:}
In what follows, we use the matrix form notations. Let us introduce the gain matrix $J_{\mu\lambda}=\partial_\lambda\mathcal{F}_\mu$, the covariance matrix $C=(c_{\mu\nu})$, the diagonal matrix of transition variances $A^{-1}=\operatorname{diag}(A^{-1}_{\mu\mu})$, the displacement vector $\mathbf{g}=\boldsymbol{\mathcal{F}}-\langle\mathbf{m}\rangle$, the velocity of the mean $\mathbf{v}=T\partial_t\langle\mathbf{m}\rangle$, and the identity $I$. The corrected, stationary and published equations read
\begin{equation}
\begin{aligned}
\text{corrected:}\quad T\partial_t C&=A^{-1}+JCJ^{\mathsf T}-C+\mathbf{v}\mathbf{v}^{\mathsf T},\\
\text{stationary:}\quad T\partial_t C&=A^{-1}+JCJ^{\mathsf T}-C,\\
\text{published\footnotemark :}\quad T\partial_t C&=A^{-1}+JC+CJ^{\mathsf T}-2C+\mathbf{g}\mathbf{g}^{\mathsf T}.
\end{aligned}
\label{eq:covariance_matrix_form}
\end{equation}
\footnotetext{'Published' refers to the equation obtained by EL Boustani and Destexhe \cite{el2009master}.}
The difference between the corrected and published equations is $(J-I)C(J-I)^{\mathsf T}+\big(\mathbf{v}\mathbf{v}^{\mathsf T}-\mathbf{g}\mathbf{g}^{\mathsf T}\big)$, namely term (ii)\footnote{Indeed, $(J-I)C+C(J-I)^{\mathsf T}+(J-I)C(J-I)^{\mathsf T}=JCJ^{\mathsf T}-C$.}, plus term (iii) up to the $\mathcal{O}(N^{-2})$ term $\frac{1}{4}h_\mu h_\nu$. At a stationary state, $\mathbf{v}=0$ and $\mathbf{g}\mathbf{g}^{\mathsf T}=\mathcal{O}(N^{-2})$, and the equations predict different stationary covariances:
\begin{equation}
\begin{aligned}
\text{corrected and stationary:}\quad C&=A^{-1}+JCJ^{\mathsf T},\\
\text{published:}\quad 0&=A^{-1}+(J-I)C+C(J-I)^{\mathsf T}.
\end{aligned}
\label{eq:stationary_covariances}
\end{equation}

\paragraph{A derivation without expanding $a_{\mu\nu}$:}
As a cross-check, we show that we can obtain the corrected equation, including term (iii), without any Taylor expansion of $a_{\mu\nu}$. Because $\mathcal{W}=P_T/T$ integrates to $1/T$ for every state (see the normalization below Eq.~\eqref{eq:joint_transition_probability}), the master equation describes a simple process. At random times, exponentially distributed with mean $T$, the whole state $\{m_\gamma\}$ is replaced by a new state drawn from $P_T(\cdot\mid\{m_\gamma\})$ (a redraw process). For any function $\varphi$ of the state, the master equation then gives\footnote{From Eq.~\eqref{eq:master_equation}, exchanging $m$ and $m'$ in the gain term as in Eq.~\eqref{eq:mean_activity_drift}: $\partial_t\langle \varphi\rangle=\iint[\varphi(\mathbf{m}')-\varphi(\mathbf{m})]\,\mathcal{W}(\mathbf{m}'\mid\mathbf{m})P_t(\mathbf{m})\,d\mathbf{m}'d\mathbf{m}$. Using $\mathcal{W}=P_T/T$ and $\int P_T(\mathbf{m}'\mid\mathbf{m})d\mathbf{m}'=1$ yields the result.}
$$
T\partial_t\langle \varphi\rangle=\Big\langle \mathbb{E}\big[\varphi(\mathbf{m}')\mid\mathbf{m}\big]\Big\rangle-\langle \varphi\rangle .
$$
With $\varphi=m_\mu$ and $\varphi=m_\mu m_\nu$, and the conditional moments of Eq.~\eqref{eq:supp_conditional_moments}, $\mathbb{E}[m'_\mu\mid\mathbf{m}]=\mathcal{F}_\mu(\mathbf{m})$ and $\mathbb{E}[m'_\mu m'_\nu\mid\mathbf{m}]=\delta_{\mu\nu}A^{-1}_{\mu\mu}(\mathbf{m})+\mathcal{F}_\mu(\mathbf{m})\mathcal{F}_\nu(\mathbf{m})$:
\begin{equation*}
T\partial_t\langle m_\mu\rangle=\langle\mathcal{F}_\mu\rangle-\langle m_\mu\rangle,
\qquad
T\partial_t\langle m_\mu m_\nu\rangle=\delta_{\mu\nu}\langle A^{-1}_{\mu\mu}\rangle+\langle\mathcal{F}_\mu\mathcal{F}_\nu\rangle-\langle m_\mu m_\nu\rangle .
\end{equation*}
Both relations are exact. Writing $f_\mu=\langle\mathcal{F}_\mu(\mathbf{m})\rangle$, the first one reads $T\partial_t\langle m_\mu\rangle=f_\mu-\langle m_\mu\rangle$. Inserting both into Eq.~\eqref{eq:covariance_time_derivative},
\begin{equation*}
\begin{aligned}
T\partial_t c_{\mu\nu}
&=\delta_{\mu\nu}\langle A^{-1}_{\mu\mu}\rangle+\langle\mathcal{F}_\mu\mathcal{F}_\nu\rangle-\langle m_\mu m_\nu\rangle
-\langle m_\nu\rangle\big(f_\mu-\langle m_\mu\rangle\big)-\langle m_\mu\rangle\big(f_\nu-\langle m_\nu\rangle\big)\\
&=\delta_{\mu\nu}\langle A^{-1}_{\mu\mu}\rangle
+\underbrace{\big(\langle\mathcal{F}_\mu\mathcal{F}_\nu\rangle-f_\mu f_\nu\big)}_{\operatorname{Cov}(\mathcal{F}_\mu,\mathcal{F}_\nu)}
-\underbrace{\big(\langle m_\mu m_\nu\rangle-\langle m_\mu\rangle\langle m_\nu\rangle\big)}_{c_{\mu\nu}}
+\big(f_\mu-\langle m_\mu\rangle\big)\big(f_\nu-\langle m_\nu\rangle\big),
\end{aligned}
\end{equation*}
and, using the exact mean equation in the last term,
\begin{equation}
T\partial_t c_{\mu\nu}
=\delta_{\mu\nu}\langle A^{-1}_{\mu\mu}\rangle
+\operatorname{Cov}(\mathcal{F}_\mu,\mathcal{F}_\nu)
-c_{\mu\nu}
+T^2\,\partial_t\langle m_\mu\rangle\,\partial_t\langle m_\nu\rangle .
\label{eq:covariance_exact_redraw}
\end{equation}
Equation~\eqref{eq:covariance_exact_redraw} is exact. Its structure is transparent. Each transition (a)~injects the transition noise $A^{-1}$; (b)~transmits the current fluctuations through the transfer function, $\operatorname{Cov}(\mathcal{F}_\mu,\mathcal{F}_\nu)$; (c)~erases the current state entirely, $-c_{\mu\nu}$; and (d)~while the mean moves, separates the realizations that have already jumped from those that have not, which adds the square of the velocity of the mean. The corrected equation~\eqref{eq:covariance_consistent_closure} follows from three approximations, each consistent with a second-order truncation:
\begin{itemize}
    \item $\langle A^{-1}_{\mu\mu}(\mathbf{m})\rangle\simeq A^{-1}_{\mu\mu}(\{\langle m_\gamma\rangle\})$, which neglects term (i);
    \item $\operatorname{Cov}(\mathcal{F}_\mu,\mathcal{F}_\nu)\simeq\partial_\lambda\mathcal{F}_\mu\,\partial_\eta\mathcal{F}_\nu\,c_{\lambda\eta}$ \footnote{$\mathcal{F}_\mu - f_\mu=\partial_\lambda\mathcal{F}_\mu \delta m_\lambda + \frac{1}{2}\partial_\lambda\partial_\eta \mathcal{F}_\mu (\delta m_\lambda \delta m_\eta - c_{\lambda\eta}) + \mathcal{O}(\delta m^3)$ }, which linearizes the transfer function and neglects third- and fourth-order central moments;
    \item $T\partial_t\langle m_\mu\rangle\simeq\mathcal{F}_\mu-\langle m_\mu\rangle+\frac{1}{2}\partial_\lambda\partial_\eta\mathcal{F}_\mu\,c_{\lambda\eta}$, which is the second-order mean equation.
\end{itemize}
The complete equation~\eqref{eq:covariance_complete} differs only in the treatment of $\mathcal{O}(N^{-2})$ pieces: it keeps $\langle A^{-1}_{\mu\mu}\rangle$ to second order, i.e.\ term (i), and drops the $\frac{1}{4}h_\mu h_\nu$ part of the squared velocity. For a quadratic transfer function, $f_\mu=\mathcal{F}_\mu(\{\langle m_\gamma\rangle\})+\frac{1}{2}\partial_\lambda\partial_\eta\mathcal{F}_\mu c_{\lambda\eta}$ exactly, so that the source term of Eq.~\eqref{eq:covariance_consistent_closure} is exact at all times. The mean-value closure of El Boustani and Destexhe \cite{el2009master} would instead require $\operatorname{Cov}(\mathcal{F}_\mu,\mathcal{F}_\nu)-c_{\mu\nu}$ to equal $\partial_\lambda\mathcal{F}_\mu c_{\nu\lambda}+\partial_\lambda\mathcal{F}_\nu c_{\mu\lambda}-2c_{\mu\nu}$, which holds only if $(J-I)C(J-I)^{\mathsf T}=0$, and the velocity of the mean to equal $\mathcal{F}_\mu-\langle m_\mu\rangle$, which ignores its curvature correction.

We stress that Eq.~\eqref{eq:covariance_exact_redraw} being exact does not mean that its implementation would capture exactly the covariance of a spiking neural network. The covariance equation is exact regarding the master equation formalism and all its hypotheses, not regarding a neural network. Moreover, it is an identity equation, not a closed equation for $c$. Its right-hand side contains three averages over the whole distribution $P_t(\boldsymbol{m})$. For nonlinear $\mathcal{F}$, these averages are not functions of the mean and covariance alone. Here it is used to see what each closure approximates and interpret each term, not to employ it in the framework.

The second line of Eq.~\eqref{eq:stationary_covariances} is instead the stationary covariance of an Ornstein--Uhlenbeck process, $T\,d\delta\mathbf{m}=(J-I)\,\delta\mathbf{m}\,dt+\sqrt{T}\,(A^{-1})^{1/2}\,d\mathbf{W}_t$. It describes a population activity that diffuses continuously around the fixed point, with a state-independent noise, whereas the master equation~\eqref{eq:master_equation} makes it jump. Replacing the jump process by this linear diffusion is the linear-noise approximation of van Kampen's system-size expansion (see, e.g.,~\cite{bressloff2010stochastic,van1992stochastic}). It is accurate only when individual jumps are small compared with the width of the activity distribution\footnote{Keeping the full state dependence of the first two jump moments, i.e.\ truncating the Kramers--Moyal expansion at second order to obtain a Fokker--Planck equation, would not be enough. Such an equation has exactly the same first- and second-moment identities as the master equation, and therefore leads to Eq.~\eqref{eq:covariance_complete}. The mean-value closure additionally freezes the second jump moment at the mean state.}. The two processes also have different stability criteria. The stationary solutions of the complete, corrected and stationary equations exist only if all eigenvalues $\zeta_a$ of the gain matrix satisfy $|\zeta_a|<1$ (up to corrections of order $N^{-1}$), whereas the published equation only requires $\operatorname{Re}\zeta_a<1$. The next subsection shows on explicit examples what these differences imply.

}

\revision{ 
\subsubsection{Benchmarks and validity domain of the covariance closures}
\label{subsubsec:covariance_closure_benchmarks}

We now compare the covariance equations with exact results of the master equation~\eqref{eq:master_equation}, for one population ($P=1$) at a stationary state. There, $\partial_t\langle m\rangle=0$: the source term vanishes, the corrected and stationary equations coincide, and the comparison isolates term (ii). At a fixed point $\nu$ with gain $J=\mathcal{F}'(\nu)$, the stationary solutions of the corrected (or stationary) and published equations are
\begin{equation}
c_{\text{corrected}}=\frac{A^{-1}}{1-J^2}\quad(\text{exists only if }|J|<1),
\qquad
c_{\text{published}}=\frac{A^{-1}}{2(1-J)}\quad(\text{exists only if }J<1),
\label{eq:one_population_closures}
\end{equation}
with $A^{-1}=\mathcal{F}(1/T-\mathcal{F})/N$ evaluated at $\nu$. When no stationary solution exists, the variance predicted by the corresponding equation grows without bound\footnote{For one population at a fixed point, the two equations read $T\partial_tc=A^{-1}+(J^2-1)c$ and $T\partial_tc=A^{-1}+2(J-1)c$. They relax towards Eq.~\eqref{eq:one_population_closures} when the coefficient of $c$ is negative, and diverge exponentially when it is positive.}. The complete equation~\eqref{eq:covariance_complete} adds term (i) to the corrected one. For the two transfer functions below, $\mathcal{F}''(m_0)=0$ and its stationary solution is $A^{-1}/(1-J^2+J^2/N)$.
 
Throughout, we use $N=4000$, $T=5\,\mathrm{ms}$ and a fixed point at the middle of the admissible range, $m_0=1/(2T)=100\,\mathrm{Hz}$, so that the firing probability at the fixed point is $Tm_0=1/2$ and $A^{-1}=m_0(1/T-m_0)/N=2.5\,\mathrm{Hz}^2$. This choice is not biologically motivated. It keeps the affine transfer function within the admissible range for every $|\Delta|<1$, and it makes the stationary distribution of the sigmoid explicit, so that every value in Tables~\ref{tab:covariance_closure_affine} and~\ref{tab:covariance_closure_sigmoid} is a closed form or an explicit finite sum.

}

\revision{

\paragraph{An exactly solvable counter-example: constant transfer function.}
The simplest way to show that the omitted term is not a small correction is to consider a case in which the master equation can be solved exactly, without any moment closure. Consider a single population whose transfer function does not depend on the network state:
$$\mathcal{F}(m)=r,\qquad 0<r<1/T .$$
This corresponds to a population driven by a constant input, without recurrent coupling ($J=0$). It is biologically trivial, but it is a legitimate instance of the formalism, and it isolates the closure step from any property of the transfer function.

\textit{1: exact solution of the master equation.} Since $\mathcal{F}$ is constant, the binomial kernel of Eq.~\eqref{eq:binomial_conditional_probability} no longer depends on the current state:
$$
P_T(m'\mid m)=q(m'),\qquad q:\ \text{binomial law of mean } r \text{ and variance } A^{-1}=\frac{r(1/T-r)}{N}.
$$
The master equation~\eqref{eq:master_equation} then becomes
\begin{equation*}
\begin{aligned}
\partial_t P_t(m)
&=\int_0^{\frac{1}{T}}\left[P_t(m')\,\frac{q(m)}{T}-P_t(m)\,\frac{q(m')}{T}\right]dm'\\
&=\frac{1}{T}\left[q(m)\int_0^{\frac{1}{T}}P_t(m')\,dm'-P_t(m)\int_0^{\frac{1}{T}}q(m')\,dm'\right]\\
&=\frac{1}{T}\big[q(m)-P_t(m)\big], \,\ \footnotemark
\end{aligned}
\end{equation*}
\footnotetext{Both integrals equal $1$, because $P_t$ and $q$ are normalized probability densities.}%
This is an ordinary linear ODE whose solution is $P_t(m)=q(m)+\big[P_0(m)-q(m)\big]e^{-t/T}$. Whatever the initial distribution, the stationary distribution is $q$ itself. Hence, each transition draws a completely new state, independently of the previous one. The exact stationary moments are therefore
\begin{equation}
\langle m\rangle_{\infty}=r,\qquad c_\infty=A^{-1}=\frac{r(1/T-r)}{N}.
\label{eq:constant_F_exact}
\end{equation}

\textit{2: exact moment equations.} With $\mathcal{F}=r$, Eq.~\eqref{eq:jump_moments_any_state} gives
$$
a_1(m)=\frac{r-m}{T},\qquad a_{11}(m)=\frac{A^{-1}+(r-m)^2}{T}.
$$
The first jump moment is linear in $m$ ($\partial_m a_1=-1/T$) and the second is quadratic ($\partial_m^2 a_{11}=2/T$). All higher derivatives vanish, so every Taylor expansion used above is exact here, with no remainder. The counter-example does not rely on neglecting any higher-order moment. Inserting these moments into the exact identity~\eqref{eq:covariance_exact_identity}:
\begin{equation*}
\begin{aligned}
T\partial_t c
&=T\langle a_{11}(m)\rangle+2T\big\langle a_1(m)\,(m-\langle m\rangle)\big\rangle\\
&=A^{-1}+\big\langle(r-m)^2\big\rangle-2\big\langle (m-r)(m-\langle m\rangle)\big\rangle\\
&=A^{-1}+\big[c+(r-\langle m\rangle)^2\big]-2c\\
&=A^{-1}+(r-\langle m\rangle)^2-c, \,\ \footnotemark
\end{aligned}
\end{equation*}
\footnotetext{We used $\langle(r-m)^2\rangle=\operatorname{Var}(m)+(\langle m\rangle-r)^2=c+(r-\langle m\rangle)^2$ (same identity as in the footnote of Eq.~\eqref{eq:final_diffusion_moments}) and $\langle(m-r)(m-\langle m\rangle)\rangle=\langle(m-\langle m\rangle)^2\rangle=c$, since $\langle m-\langle m\rangle\rangle=0$.}%
whose stationary solution is $c=A^{-1}$, in agreement with Eq.~\eqref{eq:constant_F_exact}. Since $\mathcal{F}$ is constant, terms (i) and (iii) vanish, and this exact equation coincides with the complete and corrected equations ($J=0$, $T\partial_t\langle m\rangle=r-\langle m\rangle$).
 
\textit{3: mean-value closure.} Replacing $\langle a_{11}(m)\rangle$ by $a_{11}(\langle m\rangle)=[A^{-1}+(r-\langle m\rangle)^2]/T$, as in Eq.~\eqref{eq:covariance_published_closure}, gives instead
$$
T\partial_t c=A^{-1}+(r-\langle m\rangle)^2-2c
\quad\Longrightarrow\quad
c_\infty^{\text{published}}=\frac{A^{-1}}{2}.
$$
The mean-value closure underestimates the stationary variance by a factor of two. The missing half is exactly the curvature term: $T\mathcal{K}_{11}=\frac{1}{2}\,T\,\partial_m^2a_{11}\;c=c$, i.e.\ term (ii) of Eq.~\eqref{eq:curvature_term_explicit} with $J=0$.
 
\textit{Why the closure fails.} A transition from $m$ to $m'$ has mean-square size $\langle(m'-m)^2\mid m\rangle=A^{-1}+(r-m)^2$ \footnote{$\mathbb{E}[X^2] =\text{Var}(X) + (\mathbb{E}[X])^2$}. It depends on how far the current state is from $r$. Averaged over the stationary distribution, the current deviation contributes $c$. Evaluating the jump moment at the mean state sets the current deviation to zero and forgets this contribution. The relative error is independent of $N$, although its absolute
size is $A^{-1}/2=\mathcal O(N^{-1})$. It is not caused by the Gaussian approximation (the binomial law was never approximated), and it is not a truncation of higher moments (all expansions are exact).

\paragraph{Affine transfer function:}
Let us now add recurrent coupling with an affine transfer function, as in the linear model of Supplementary~\ref{subsec:master_equation_linear_model}. We parametrize it by its fixed point $m_0$ and its slope $\Delta$:
\begin{equation}
\mathcal{F}(m)=m_0+\Delta\,(m-m_0),
\qquad\text{i.e.}\qquad
\mathcal{F}(m)=\nu_0+\Delta\,m\ \ \text{with}\ \ \nu_0=m_0(1-\Delta).
\label{eq:affine_F}
\end{equation}
$\Delta>0$ describes a net excitatory feedback, $\Delta<0$ a net inhibitory feedback, and $\Delta=0$ is the constant transfer function. With $m_0=1/(2T)$ and $|\Delta|\le1$, $\mathcal{F}$ maps the admissible range $[0,1/T]$ into itself.

For an affine $\mathcal{F}$, the first jump moment $a_1(m)=(\mathcal{F}(m)-m)/T$ is linear in $m$, and $T a_{11}(m)=A^{-1}(m)+(\mathcal{F}(m)-m)^2$ is quadratic. Every Taylor expansion of Supplementary~\ref{subsubsec:covariance_matrix} therefore terminates, no third-order moment ever appears, and the complete equation~\eqref{eq:covariance_complete} holds without any approximation:
\begin{equation}
T\partial_t c
=A^{-1}\big(\langle m\rangle\big)
\underbrace{-\frac{\Delta^2}{N}\,c}_{\text{(i)}}
+\big(\mathcal{F}(\langle m\rangle)-\langle m\rangle\big)^2
+\underbrace{2(\Delta-1)\,c+(\Delta-1)^2\,c}_{=(\Delta^2-1)\,c},
\label{eq:affine_exact_moment_equation}
\end{equation}
where $A^{-1}(\langle m\rangle)=\mathcal{F}(\langle m\rangle)[1/T-\mathcal{F}(\langle m\rangle)]/N$. Term (i) is $\frac{1}{2}\partial_m^2A^{-1}\,c$ with $\partial_m^2A^{-1}=-2\Delta^2/N$; term (ii) is $(\Delta-1)^2c$; term (iii) vanishes, since $\mathcal{F}''=0$. Equation~\eqref{eq:affine_exact_moment_equation} holds at all times, including transients.
 
\textit{Results.} Together with Eq.~\eqref{eq:one_population_closures} at $J=\Delta$, we obtain
\begin{equation}
c_{\text{exact}}=\frac{A^{-1}}{1-\Delta^2+\Delta^2/N},
\qquad
c_{\text{corrected}}=\frac{A^{-1}}{1-\Delta^2},
\qquad
c_{\text{published}}=\frac{A^{-1}}{2(1-\Delta)},
\label{eq:linear_one_population_variances}
\end{equation}
\begin{equation*}
\frac{c_{\text{corrected}}}{c_{\text{exact}}}=1+\frac{\Delta^2}{N(1-\Delta^2)},
\qquad
\frac{c_{\text{published}}}{c_{\text{exact}}}\simeq\frac{1+\Delta}{2}.
\end{equation*}
These values are reported in Table~\ref{tab:covariance_closure_affine}. The exact and corrected variances depend on $\Delta^2$ only: an excitatory and an inhibitory feedback of the same strength produce the same fluctuations. The published variance depends on the sign of $\Delta$. It is $25\%$ too small at $\Delta=0.5$, two times too small at $\Delta=0$ (the counter-example), and $4$, $20$ and $200$ times too small at $\Delta=-0.5$, $-0.9$ and $-0.99$. For $\Delta<-1$, each transition would overshoot the fixed point by more than the previous deviation, alternately above and below it, so that the fixed point cannot be a stationary state. The corrected equation reproduces this (its variance grows at the rate $(\Delta^2-1)/T$), whereas the published one predicts a finite variance, $0.5\,\mathrm{Hz}^2$ at $\Delta=-1.5$. However, this globally affine binomial model is admissible only for $|\Delta|\leq1$ at the chosen $m_0=1/(2T)$: outside this range, some firing probabilities lie outside $[0,1]$, so there is no corresponding stochastic kernel to benchmark. At $\Delta=-1$, the endpoints form a closed alternating class, with stationary weights $1/2,1/2$ and finite variance $m_0^2$. At $\Delta=1$, both endpoints are absorbing and stationary mixtures exist; the centered mixture also has variance $m_0^2$. These boundary cases lie outside the weak-fluctuation approximation used for $c_{\mathrm{corrected}}$.
 
\begin{table}[ht]
\centering
\small
\begin{tabular}{@{}r ccc cc@{}}
\toprule
& \multicolumn{3}{c}{stationary variance $c$ (Hz$^2$)} & \multicolumn{2}{c}{ratio to exact} \\
\cmidrule(lr){2-4}\cmidrule(l){5-6}
$\Delta$ & exact & corrected & published & corrected & published \\
\midrule
$+1.50$ & -- & $\infty$ & $\infty$ & -- & -- \\
$+0.99$ & $124.100$ & $125.628$ & $125.000$ & $1.012$ & $1.007$ \\
$+0.90$ & $13.144$ & $13.158$ & $12.500$ & $1.001$ & $0.951$ \\
$+0.50$ & $3.333$ & $3.333$ & $2.500$ & $1.000$ & $0.750$ \\
$0$ & $2.500$ & $2.500$ & $1.250$ & $1.000$ & $0.500$ \\
$-0.50$ & $3.333$ & $3.333$ & $0.833$ & $1.000$ & $0.250$ \\
$-0.90$ & $13.144$ & $13.158$ & $0.658$ & $1.001$ & $0.050$ \\
$-0.99$ & $124.100$ & $125.628$ & $0.628$ & $1.012$ & $5.1\times10^{-3}$ \\
$-1.50$ & -- & $\infty$ & $0.500$ & -- & -- \\
\bottomrule
\end{tabular}
\caption{\revision{
Stationary variance of one population with the affine transfer function~\eqref{eq:affine_F} ($N=4000$, $T=5\,\mathrm{ms}$, $m_0=1/(2T)=100\,\mathrm{Hz}$, $A^{-1}=2.5\,\mathrm{Hz}^2$). Exact: closed form~\eqref{eq:linear_one_population_variances}, identical to the stationary solution of the complete equation~\eqref{eq:covariance_complete}. Corrected and published: stationary solutions~\eqref{eq:one_population_closures} of Eqs.~\eqref{eq:covariance_consistent_closure} and~\eqref{eq:covariance_published_closure}. The
row $\Delta=0$ is the constant-transfer counterexample. Slopes
outside $[-1,1]$ do not define this binomial kernel: their exact column is left empty, and only the closure predictions are shown (the variance grows without bound).}}
\label{tab:covariance_closure_affine}
\end{table}

}

\revision{

\paragraph{Sigmoid transfer function:}
An affine transfer function has no curvature. It cannot show what happens when fluctuations become large enough to feel the nonlinearity of $\mathcal{F}$, or when the fixed point loses its stability. We therefore consider a sigmoid centred on the fixed point,
\begin{equation}
\mathcal{F}(m)=m_0\left[1+\tanh\!\left(\frac{\Delta\,(m-m_0)}{m_0}\right)\right],
\label{eq:sigmoid_F}
\end{equation}
We note four useful properties of this transfer function. (a)~$\mathcal{F}(m_0)=m_0$ for every $\Delta$, so $m_0$ is always a fixed point. (b)~With $m_0=1/(2T)$, $\mathcal{F}$ spans the whole admissible range $(0,1/T)$. (c)~The slope at the fixed point is $\mathcal{F}'(m_0)=\Delta$, as for the affine function~\eqref{eq:affine_F}: $\Delta>0$ gives an increasing sigmoid (net excitation), $\Delta<0$ a decreasing one (net inhibition). (d)~$m_0$ is the inflection point.
 
Let $\xi=(m-m_0)/m_0$ be the relative deviation and $t=\tanh(\Delta\xi)$, so that $\mathcal{F}=m_0(1+t)$. Using $dt/dm=(\Delta/m_0)(1-t^2)$,
\begin{equation}
\mathcal{F}'(m)=m_0\,\frac{\Delta}{m_0}\,(1-t^2)=\Delta\,(1-t^2),
\qquad
\mathcal{F}''(m)=\Delta\,(-2t)\,\frac{\Delta}{m_0}(1-t^2)=-\frac{2\Delta^2}{m_0}\,t\,(1-t^2).
\label{eq:sigmoid_derivatives}
\end{equation}
At the fixed point ($t=0$), $\mathcal{F}'=\Delta$ and $\mathcal{F}''=0$. The second-order mean equation is therefore satisfied by $\langle m\rangle=m_0$ whatever $c$, term (iii) vanishes, and the closures give exactly the same values as for the affine function with the same slope, Eq.~\eqref{eq:linear_one_population_variances}. Any difference between the affine and sigmoid exact values is due to the nonlinearity of $\mathcal{F}$, i.e.\ to moments of order higher than two, which this second-order Taylor expansion about $m_0$ does not capture.

\textit{Exact stationary law:} At the fixed point, the sigmoid and the affine function have the same slope, so the closures predict the same variances for both. To know whether these predictions are right, we need the exact stationary variance of the master equation for the sigmoid.\\
With $m_0=1/(2T)$, the stationary distribution of the master equation is explicit. In each bin, the population emits $K$ spikes, with $K=0,1,\dots,N$, so the activity can only take the $N+1$ values $m_K=K/(NT)$. We write $\xi_K=(m_K-m_0)/m_0=2K/N-1$. Let $\pi_K(t)$ be the probability that the network is in state $K$ at time $t$, and $P_{KK'}$ the probability that a transition starting from state $K$ leads to state $K'$. This is the binomial probability of $K'$ spikes, with the firing probability $p_K=T\mathcal{F}(m_K)$ of the starting state:
$$
p_K=T\mathcal{F}(m_K)=\tfrac{1}{2}\big[1+\tanh(\Delta\xi_K)\big]=\frac{e^{\Delta\xi_K}}{2\cosh(\Delta\xi_K)},
\qquad
1-p_K=\frac{e^{-\Delta\xi_K}}{2\cosh(\Delta\xi_K)},
$$
and, since $K'-(N-K')=N\xi_{K'}$, the binomial kernel takes the form
\begin{equation}
\begin{aligned}
P_{KK'}=\binom{N}{K'}p_K^{K'}(1-p_K)^{N-K'}
=\binom{N}{K'}\frac{e^{N\Delta\,\xi_K\xi_{K'}}}{2^N\cosh^N(\Delta\xi_K)} .  
\end{aligned}
\label{eq:kernel_symmetric_form}
\end{equation}

Transitions occur at rate $1/T$, and on these states the master equation~\eqref{eq:master_equation} becomes
\begin{equation}
\frac{d\pi_K}{dt}
=\frac{1}{T}\sum_{K'=0}^{N}\Big[\pi_{K'}P_{K'K}-\pi_KP_{KK'}\Big]
=\frac{1}{T}\Big[\sum_{K'=0}^{N}\pi_{K'}P_{K'K}-\pi_K\Big].
\label{eq:discrete_master_equation}
\end{equation}
In the stationary state, the left-hand side vanishes, and the stationary distribution $\pi_K$ satisfies
\begin{equation}
\sum_{K'=0}^{N}\pi_{K'}P_{K'K}=\pi_K\qquad\text{for every state }K .
\label{eq:stationarity_condition}
\end{equation}

Let us define the weights $w_K=\binom{N}{K}\cosh^N(\Delta\xi_K)$. Then
$$
w_K\,P_{KK'}=2^{-N}\binom{N}{K}\binom{N}{K'}\,e^{N\Delta\,\xi_K\xi_{K'}},
$$
which is symmetric under 
\begin{equation}
\begin{aligned}
K\leftrightarrow K': w_KP_{KK'}=w_{K'}P_{K'K}
\end{aligned}
\label{eq:detailed_balance}
\end{equation}
for every pair of states $(K, K')$; so the probability that flows from $K$ to $K'$ is equal to the probability that flows back. It is stronger than Eq.~\eqref{eq:stationarity_condition}: instead of balancing the total flow into each state, it balances the exchanges between every pair of states. Summing over $K$ gives $\sum_Kw_KP_{KK'}=w_{K'}\sum_KP_{K'K}=w_{K'}$. The weights $w_K$ satisfy detailed balance~\eqref{eq:detailed_balance}, hence the stationarity condition~\eqref{eq:stationarity_condition}. Because equations \eqref{eq:detailed_balance}-\eqref{eq:stationarity_condition} remain true when all $\pi_K$ are multiplied by the same number, we can normalize the weights $w_K$ by their sum $Z$ to make them probabilities. The normalized weights become therefore the stationary distribution of the sequence of visited states and, since the jump rate $1/T$ does not depend on the state, of the master equation:
\begin{equation}
\pi_K=\frac{w_K}{Z}=\frac{1}{Z}\binom{N}{K}\cosh^N\!\left(\Delta\,\frac{m_K-m_0}{m_0}\right),
\qquad
Z=\sum_{K=0}^{N}\binom{N}{K}\cosh^N\!\left(\Delta\,\frac{m_K-m_0}{m_0}\right).
\label{eq:logistic_law}
\end{equation}

Its mean and variance are finite sums, $\langle m\rangle=\sum_Km_K\pi_K$ and $c=\sum_K(m_K-\langle m\rangle)^2\pi_K$, which give the exact values of Table~\ref{tab:covariance_closure_sigmoid}. The two factors of $\pi_K$ have a simple meaning\footnote{Readers familiar with statistical physics will recognize in Eq.~\eqref{eq:logistic_law} a Gibbs (Boltzmann) distribution, $\pi_K\propto e^{-H_K}$, with the "energy" $H_K=-\ln\binom{N}{K}-N\ln\cosh(\Delta\xi_K)$. Our master equation is the simultaneous (parallel) dynamics of a network of binary neurons with symmetric couplings, for which Peretto derived this type of stationary distribution~\cite{peretto1984collective}.}. The binomial coefficient $\binom{N}{K}$ counts the ways of choosing which $K$ of the $N$ neurons fire. Alone, it gives the binomial distribution of independent neurons. For $\Delta=0$, Eq.~\eqref{eq:logistic_law} indeed reduces to the distribution $q$ of the constant transfer function. The factor $\cosh^N(\Delta\xi_K)$ is the effect of the recurrent coupling. It depends on $\Delta$ only through $\cosh(\Delta\xi)$, which is even. An increasing and a decreasing sigmoid with opposite slopes therefore have the same stationary distribution, and the same exact variance, as already found for the affine function, whose exact variance depends on $\Delta^2$ only.

The construction relied on one property: in Eq.~\eqref{eq:kernel_symmetric_form}, the departure and arrival states interact only through the symmetric product $\xi_K\xi_{K'}$. This happens because the logarithm of the odds of firing, $\ln[p_K/(1-p_K)]=2\Delta\xi_K$, is an affine function of the activity. Among sigmoids, only the logistic function that spans the whole admissible range $(0,1/T)$ has this property\footnote{Write $\ell_K=\ln[p_K/(1-p_K)]$, so that $P_{KK'}=\binom{N}{K'}(1-p_K)^Ne^{\ell_KK'}$: the two states interact through the product $\ell_KK'$. If $\ell_K=a+bK$, this product is $aK'+bKK'$, the sum of a term that depends on $K'$ only, which the weights absorb, and of a symmetric term. If $\ell_K$ contained, say, a term $dK^2$, the product would contain $dK^2K'$, which is neither symmetric nor separable. Formally, look for weights of the form $w_K=\binom{N}{K}\varphi_K/(1-p_K)^N$. Detailed balance requires $\varphi_Ke^{\ell_KK'}=\varphi_{K'}e^{\ell_{K'}K}$ for all $K,K'$. With $K'=0$, this gives $\ln\varphi_K=\ln\varphi_0+\ell_0K$, and then $(\ell_K-\ell_0)K'=(\ell_{K'}-\ell_0)K$, i.e.\ $\ell_K$ is affine in $K$.}. Equation~\eqref{eq:sigmoid_F} belongs to this family only when its maximum, $2m_0$, equals $1/T$, which is why we chose $m_0=1/(2T)$. For a sigmoid of smaller amplitude, or for an $\operatorname{erfc}$-type transfer function, detailed balance fails. A unique stationary distribution still exists, but it must be computed numerically, by solving the linear system~\eqref{eq:stationarity_condition}.
 
\textit{Exact distribution:} Equation~\eqref{eq:logistic_law} can be summed numerically, but its shape is not obvious at first sight. It becomes transparent for a large population. Both factors of $\pi_K$ are powers of order $N$: $\cosh^N(\Delta\xi)$ explicitly, and $\binom{N}{K}$, which Stirling's formula\footnote{Stirling's formula, $\ln n!\simeq n\ln n-n$, gives $\ln\binom{N}{K}\simeq N\big[\ln2-\tfrac12(1+\xi)\ln(1+\xi)-\tfrac12(1-\xi)\ln(1-\xi)\big]$, with $\xi=2K/N-1$, up to corrections of order $\ln N$.} turns into the exponential of $N$ times a smooth function of $\xi$. Taking the logarithm turns the product of the two factors into a sum, and gives

\begin{equation}
\begin{aligned}
V(\xi)&=\tfrac12(1+\xi)\ln(1+\xi)
+\tfrac12(1-\xi)\ln(1-\xi)-\ln\cosh(\Delta\xi),\\
\pi_K&=\frac{B_N}{\sqrt{1-\xi_K^2}}
\exp[-NV(\xi_K)]\,[1+\mathcal O(N^{-1})].
\end{aligned}
\label{eq:logistic_potential}
\end{equation}
Here $B_N$ is independent of $K$; the expansion applies away from
the endpoints. Thus $\pi(\xi)\propto e^{-N\,V(\xi)}$, where potential $V$, independent of $N$, gives the leading exponential dependence of the exact discrete law. It is a probability landscape (Fig.~\ref{fig:sigmoid_landscape}). The activity behaves like a ball in this landscape, shaken by a noise whose strength decreases as $1/N$. For $N=4000$, the factor $e^{-NV}$ is so steep that the distribution is concentrated near the lowest points of $V$, and its width is set by the shape of $V$ around them. Near $\xi=0$, $V$ can be expanded\footnote{$(1+\xi)\ln(1+\xi)+(1-\xi)\ln(1-\xi)=\xi^2+\xi^4/6+\mathcal{O}(\xi^6)$ and $\ln\cosh(\Delta\xi)=\Delta^2\xi^2/2-\Delta^4\xi^4/12+\mathcal{O}(\xi^6)$.}:
\begin{equation}
V(\xi)=\frac{1-\Delta^2}{2}\,\xi^2+\frac{1+\Delta^4}{12}\,\xi^4+\mathcal{O}(\xi^6).
\label{eq:logistic_laplace}
\end{equation}
The terms $\xi^2/2$ and $\xi^4/12$ come from the binomial factor: the transition noise alone keeps the activity close to $m_0$. The term $-\Delta^2\xi^2/2$ comes from the feedback, which amplifies deviations and flattens the landscape. The term $+\Delta^4\xi^4/12$ comes from the saturation of the sigmoid, which limits large deviations. $V$ is symmetric, so $\xi=0$ is always a flat point, but whether it is a valley bottom or a hilltop depends on the curvature there, $V''(0)=1-\Delta^2$. Hence, the sign of the quadratic coefficient, $1-\Delta^2$, decides the shape of the landscape. Three cases follow, illustrated in Fig.~\ref{fig:sigmoid_landscape}.
\begin{itemize}

    \item \emph{$|\Delta|<1$: one narrow valley.} At fixed $|\Delta|<1$ and large $N$, the leading Gaussian variance is $c=m_0^2/[N(1-\Delta^2)]$, as predicted by the corrected stationary equation. Including both the quartic potential and the Stirling prefactor gives
    \[
    c=\frac{m_0^2}{N(1-\Delta^2)}
    \left[1-\frac{\Delta^2(1+\Delta^2)}
    {N(1-\Delta^2)^2}+\mathcal O(N^{-2})\right].
    \]
    The remainder estimate here is for fixed $\Delta$ away from the critical boundary. A sufficient condition for the local quadratic approximation is
    \begin{equation}
    N(1-\Delta^2)^2\gg1+\Delta^4.
    \label{eq:sigmoid_validity}
    \end{equation}
    It is satisfied well at $|\Delta|=0.9$ ($144\gg1.7$ for $N=4000$: the corrected equation is within $1\%$), but not at $|\Delta|=0.99$ ($1.6$ against $2.0$: it is $63\%$ too high).

    \item \emph{$|\Delta|=1$: a flat-bottomed valley.} The quadratic coefficient vanishes: nothing pulls small deviations back, and only the quartic term confines the activity. The distribution, $\pi(\xi)\propto e^{-N\xi^4/6}$, is flat-topped and much wider than a Gaussian. Its variance is\footnote{With $s^4=6/N$ and the substitution $u=\xi^4/s^4$: $\int_0^\infty\xi^2e^{-\xi^4/s^4}d\xi=\frac{s^3}{4}\Gamma\big(\tfrac{3}{4}\big)$ and $\int_0^\infty e^{-\xi^4/s^4}d\xi=\frac{s}{4}\Gamma\big(\tfrac{1}{4}\big)$.}
    \begin{equation}
    c_{\text{crit}}=m_0^2\,\frac{\Gamma(3/4)}{\Gamma(1/4)}\sqrt{\frac{6}{N}}=130.9\,\mathrm{Hz}^2 ,
    \label{eq:critical_variance}
    \end{equation}
    against $131.7\,\mathrm{Hz}^2$ for the exact finite sum. 
    This variance is finite and scales as $N^{-1/2}$ rather than $N^{-1}$. The regular Gaussian, local linear-noise approximation therefore fails at this critical point. For this sigmoid at $|\Delta|=1$, the local second-order Taylor closure about $m_0$ misses the nonlinear confinement and does not give the stationary variance. This does not rule out alternative closures that evaluate nonlinear averages under an assumed distribution.

    \item \emph{$|\Delta|>1$: two valleys.} The quadratic coefficient is negative, and $m_0$ becomes the top of a hill between two valleys. The bottoms of the valleys, where $dV/d\xi=0$, satisfy $\operatorname{artanh}\xi=\Delta\tanh(\Delta\xi)$, and they are located at $\xi=\pm u$, where $u$ is the positive root of\footnote{If $u=\tanh(|\Delta|u)$, then $\operatorname{artanh}u=|\Delta|u$ and $\Delta\tanh(\Delta u)=|\Delta|\tanh(|\Delta|u)=|\Delta|u$.}
    \begin{equation}
    u=\tanh\big(|\Delta|\,u\big).
    \label{eq:two_state_amplitude}
    \end{equation}
    The distribution has two narrow peaks, at $m_0\pm a$ with $a=m_0u$. These are exactly the two states that the network occupies: the two fixed points of $\mathcal{F}$ when $\Delta>1$, and the two states of its period-two cycle when $\Delta<-1$. The variance is close to $a^2$: $a=65.86\,\mathrm{Hz}$ and $a^2=4337\,\mathrm{Hz}^2$ for $|\Delta|=1.2$, against $4334\,\mathrm{Hz}^2$ for the exact finite sum.
\end{itemize}

\begin{figure}[ht]
\centering
\includegraphics[width=0.7\textwidth]{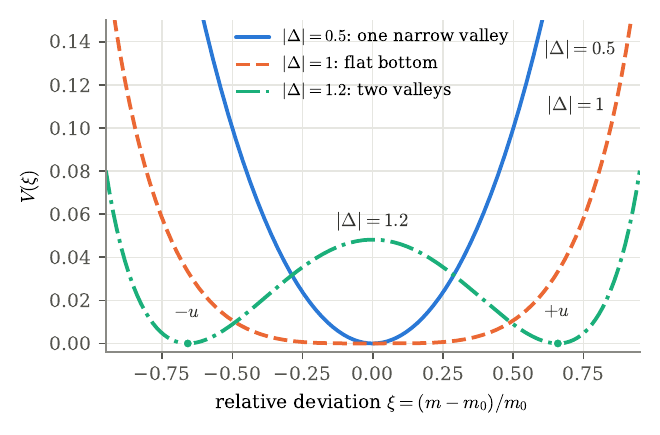}
\caption{
\revision{ Leading large-$N$ probability potential landscape $V(\xi)$ of Eq.~\eqref{eq:logistic_potential}, shifted to zero at its minimum. The exact finite-$N$ distribution is Eq.~\eqref{eq:logistic_law}. For $N=4000$, the distribution is concentrated near the minima of $V$. For $|\Delta|<1$, one narrow valley gives Gaussian fluctuations, described by the corrected equation. For $|\Delta|=1$, the valley has a flat bottom, and the fluctuations are wide and non-Gaussian. For $|\Delta|>1$, two valleys appear at $\xi=\pm u$, Eq.~\eqref{eq:two_state_amplitude}. The landscape is the same for $\Delta$ and $-\Delta$.}}
\label{fig:sigmoid_landscape}
\end{figure}

\begin{table}[ht]
\centering
\small
\begin{tabular}{@{}l c cc cc@{}}
\toprule
& master equation & \multicolumn{2}{c}{corrected} & \multicolumn{2}{c}{published} \\
\cmidrule(lr){2-2}\cmidrule(lr){3-4}\cmidrule(l){5-6}
$\Delta$ & exact & $c$ & ratio & $c$ & ratio \\
\midrule
$+1.20$ (one well) & $2.642$ & $2.635$ & $0.997$ & $2.213$ & $0.838$ \\
$+1.00$ & $131.692$ & $\infty$ & -- & $\infty$ & -- \\
$+0.99$ & $77.090$ & $125.628$ & $1.630$ & $125.000$ & $1.621$ \\
$+0.90$ & $13.028$ & $13.158$ & $1.010$ & $12.500$ & $0.959$ \\
$+0.50$ & $3.333$ & $3.333$ & $1.000$ & $2.500$ & $0.750$ \\
\midrule
$-0.50$ & $3.333$ & $3.333$ & $1.000$ & $0.833$ & $0.250$ \\
$-0.90$ & $13.028$ & $13.158$ & $1.010$ & $0.658$ & $0.050$ \\
$-0.99$ & $77.090$ & $125.628$ & $1.630$ & $0.628$ & $8.1\times10^{-3}$ \\
$-1.00$ & $131.692$ & $\infty$ & -- & $0.625$ & $4.7\times10^{-3}$ \\
$-1.20$ & $4333.946$ & $\infty$ & -- & $0.568$ & $1.3\times10^{-4}$ \\
$-2.00$ & $9168.197$ & $\infty$ & -- & $0.417$ & $4.5\times10^{-5}$ \\
\bottomrule
\end{tabular}
\caption{\revision{
Stationary variance (Hz$^2$) of one population with the sigmoid~\eqref{eq:sigmoid_F} ($N=4000$, $T=5\,\mathrm{ms}$, $m_0=1/(2T)=100\,\mathrm{Hz}$, $A^{-1}=2.5\,\mathrm{Hz}^2$). Except in the row marked "one well", the exact column is the variance of the full stationary law~\eqref{eq:logistic_law}, whose mean is $m_0$. The $+1.20$ row instead conditions that law on $m>m_0$ and compares it with local moment solutions near the upper stable state; symmetry gives the same within-well variance below $m_0$. The unconditioned variance at $+1.20$ is $4333.946\,\mathrm{Hz}^2$, identical to that at $-1.20$. Ratios use the exact quantity specified in their own row. The symbol $\infty$ denotes failure of the indicated local closure to give a finite stationary variance at $m_0$, not an infinite variance or absence of a stationary distribution in the bounded master equation.}}
\label{tab:covariance_closure_sigmoid}
\end{table}
 
For $|\Delta|>1$, the stationary distribution is the same for $\Delta$ and $-\Delta$, but what a single network does, their dynamics, is very different.
\begin{itemize}
    \item \emph{Strong inhibitory feedback, $\Delta<-1$.} The decreasing transfer function sends the activity to the other side of $m_0$ at each transition: the activity alternates between $m_0+a$ and $m_0-a$, a period-two cycle, with random holding times. One network visits both states, so its time-averaged statistics are those of the two-peaked distribution, with a variance close to $a^2$.
    \item \emph{Strong excitatory feedback, $\Delta>1$.} The two states $m_0\pm a$ are both stable fixed points of $\mathcal{F}$, with a local slope $\mathcal{F}'(m_0\pm a)=\Delta(1-u^2)=0.68$ at $\Delta=1.2$. A network stays in the state where it started: for $N=4000$, the probability of crossing $m_0$ in one transition is very small. Its time-averaged statistics are those of the distribution~\eqref{eq:logistic_law} restricted to one valley\footnote{Because the chain satisfies detailed balance, the restricted distribution is the exact stationary distribution of the dynamics in which transitions across $m_0$ are suppressed.}, which Table~\ref{tab:covariance_closure_sigmoid} reports. Noise or adaptation is what turns this regime into Up-Down alternation.
\end{itemize}
The corrected local covariance equation has no finite stationary solution at $m_0$ when $|\Delta|>1$. For $\Delta<-1$, the published equation still gives a finite value there, but it does not describe the variance of the two-peaked stationary law. For $\Delta>1$, both closures can instead be evaluated locally near either stable state.

The stationary probability landscape bifurcates symmetrically. The underlying deterministic redraw map has a pitchfork for positive gain and a period-doubling transition for negative gain; the continuous-time process has random holding times. The picture also explains the behaviour of the closures. The corrected equation describes a single Gaussian valley. It is accurate when the valley is narrow, it degrades when the bottom of the valley flattens ($|\Delta|\to1$), and it has no stationary solution when the valley turns into a hill ($|\Delta|>1$). This highlights a clear limitation of the second-order closure. The two valleys, representing two activity levels ($34$ Hz and $166$ Hz for $\Delta=1.2$), cannot be captured by the closure that tracks only a mean and a variance around one point (which amounts to one Gaussian bump). It works locally, in one valley, but the closure cannot follow where the activity goes. Putative solutions for enhancing the model would be to go beyond second-order near criticality, or to evaluate the nonlinear averages $\langle \mathcal{F} \rangle$, $\text{Var}(\mathcal{F})$ and $\langle A^{-1} \rangle$ under an assumed distribution (e.g., a Beta distribution that can capture a U-shape) instead of Taylor-expanding them. Because $\mathcal{F}$ saturates, such closures keep the variance bounded.

\paragraph{Why the corrected equation becomes unstable, and why a stable published equation is not better:}
Tables~\ref{tab:covariance_closure_affine} and~\ref{tab:covariance_closure_sigmoid} show that the corrected equation loses its stationary solution as soon as $|\Delta|\ge1$, whereas the published one keeps a finite, stable solution for every $\Delta<1$, including $\Delta\le-1$. This is not a defect of the corrected equation, and it is not an advantage of the published one.

Table~\ref{tab:covariance_closure_sigmoid} shows that the corrected equation is more accurate. For $|\Delta|\le0.9$, it agrees with the exact variance within $0.1\%$ for the affine function and within $1\%$ for the sigmoid, for both signs of $\Delta$. The published equation underestimates the variance by the factor $(1+\Delta)/2$, whatever the shape of $\mathcal{F}$: by $25\%$ at $\Delta=0.5$, and by factors of $4$ and $20$ at $\Delta=-0.5$ and $-0.9$. The exact and corrected variances are symmetric in $\Delta$; the published one is not. The published equation therefore gets the effect of the sign of the feedback wrong: it predicts that inhibitory feedback suppresses fluctuations, whereas in the master equation an inhibitory feedback amplifies them exactly as much as an excitatory feedback of the same strength.
\emph{Near criticality, both closures fail.} For the sigmoid at $|\Delta|=0.99$, condition~\eqref{eq:sigmoid_validity} is violated: the fluctuations feel the saturation of $\mathcal{F}$, and the corrected equation overestimates the variance by $63\%$. For $\Delta=+0.99$, the published equation is equally wrong; for $\Delta=-0.99$ it is about $120$ times too small. Around a new stable state, the corrected equation works again. At $\Delta=1.2$, where $m_0$ is unstable, the corrected equation agrees with the exact variance in each stable state within $0.3\%$, whereas the published one is $16\%$ too small, close to the factor $(1+0.68)/2=0.84$ of the linear analysis.

Linearized around a fixed point, the mean equation, $T\partial_t\langle\delta\mathbf{m}\rangle=(J-I)\langle\delta\mathbf{m}\rangle$, is stable when $\operatorname{Re}\zeta_a<1$ for every eigenvalue $\zeta_a$ of $J$. The fluctuations of the master equation, however, evolve by the discrete map $\delta\mathbf{m}^{(k+1)}=J\,\delta\mathbf{m}^{(k)}+\boldsymbol{\xi}^{(k)}$ at each transition, which is stable only when $|\zeta_a|<1$. The two criteria differ when an eigenvalue lies outside the unit circle while $\operatorname{Re}\zeta_a<1$, e.g.\ $\zeta=\Delta<-1$. There, each realization moves away from the fixed point, overshooting it with alternating signs, while the ensemble mean still relaxes towards it, because realizations jump at random times and averaging over these times smooths out the alternation. The instability is invisible in the mean. It lives in the fluctuations, and only the covariance equation can detect it.
\emph{The corrected equation detects it.} The linear part of Eq.~\eqref{eq:covariance_consistent_closure}, $C\mapsto JCJ^{\mathsf T}-C$, has eigenvalues $\zeta_a\bar\zeta_b-1$, which become positive exactly when $|\zeta_a|>1$. Its divergence is the correct prediction that the fixed point is not a stationary state of the master equation. The exact process leaves it, towards a period-two oscillation ($\Delta<-1$) or one of two new stable states ($\Delta>1$). No single-peaked second-order description can represent the oscillation, but around the new stable states the corrected equation is accurate again. \emph{The published equation cannot detect it.} The linear part of Eq.~\eqref{eq:covariance_published_closure}, $C\mapsto(J-I)C+C(J-I)^{\mathsf T}$, has eigenvalues $(\zeta_a-1)+(\bar\zeta_b-1)$, which are negative whenever $\operatorname{Re}\zeta_a<1$. This is the covariance equation of the Ornstein--Uhlenbeck process with the same mean dynamics: it inherits the stability of the mean, not that of the fluctuations. Its finite solutions for $\Delta\le-1$ are spurious. At $\Delta=-1.2$ it predicts $0.57\,\mathrm{Hz}^2$, whereas the network oscillates with a variance of $4334\,\mathrm{Hz}^2$, i.e.\ $7600$ times larger. At the critical point $|\Delta|=1$, the divergence of the corrected equation has a different origin. The linear restoring force vanishes, and only the saturation of $\mathcal{F}$ bounds the fluctuations, which a second-order truncation cannot describe. The exact variance is finite but large, and it scales anomalously, Eq.~\eqref{eq:critical_variance}. A stable published equation is therefore not a sign of a correct model: it can hide an instability of the master equation.

\quad The mean-value closure (published version \cite{el2009master}) omission can become small and legitimate under additional conditions, either further approximation of the transfer function or on the stochastic process. For instance, if the transfer map is close to the identity, then a transition changes the fluctuations only slightly, the state dependence of the second jump moment becomes negligible and the linear-noise approximation becomes accurate. It can also be legitimate for small jumps. The activity resolution $1/(N_\gamma T)$, which Supplementary~\ref{subsec:transfer_funtion_ME} and El Boustani and Destexhe~\cite{el2009master} required to be small, is the change of activity caused by one additional spike. It describes a population activity that changes by small increments. The transition operator~\eqref{eq:transition_operator_continuous_definition}, however, replaces the whole state at once. To separate the two, consider a family of transitions indexed by $0<\kappa\le1$, occurring at rate $1/(\kappa T)$:
\begin{equation}
m_\mu\;\longrightarrow\;m'_\mu=(1-\kappa)\,m_\mu+\frac{K'_\mu}{N_\mu T},
\qquad
K'_\mu\sim\mathcal{B}\big(N_\mu,\;\kappa T\mathcal{F}_\mu(\mathbf{m})\big).
\label{eq:kappa_family}
\end{equation}
For $\kappa=1$, this is exactly the master equation~\eqref{eq:master_equation}. For $\kappa\to0$, the update interval $\kappa T$ vanishes, and $m_\mu$ becomes the population spike train filtered by an exponential kernel of time constant $T$:
\begin{equation}
T\,dm_\mu=-m_\mu\,dt+\frac{dS_\mu}{N_\mu},
\label{eq:small_jump_limit}
\end{equation}
where $S_\mu$ counts the spikes of population $\mu$, emitted as a Poisson process of rate $N_\mu\mathcal{F}_\mu(\mathbf{m})$. Each spike then moves $m_\mu$ by exactly $1/(N_\mu T)$. In every member of the family, $T$ remains the relaxation time of the mean activity; only the update interval changes. The jump moments follow from the binomial moments, as in Eqs.~\eqref{eq:final_drift_moments}-\eqref{eq:final_diffusion_moments}:
\begin{equation*}
\begin{aligned}
a_\mu
&=\frac{1}{\kappa T}\,\mathbb{E}\big[m'_\mu-m_\mu\mid\mathbf{m}\big]
=\frac{1}{\kappa T}\left[-\kappa m_\mu+\frac{N_\mu\kappa T\mathcal{F}_\mu}{N_\mu T}\right]
=\frac{\mathcal{F}_\mu-m_\mu}{T},\\
a_{\mu\nu}
&=\frac{1}{\kappa T}\,\mathbb{E}\big[(m'_\mu-m_\mu)(m'_\nu-m_\nu)\mid\mathbf{m}\big]
=\frac{\kappa}{T}\,(\mathcal{F}_\mu-m_\mu)(\mathcal{F}_\nu-m_\nu)
+\delta_{\mu\nu}\,\frac{\mathcal{F}_\mu(1-\kappa T\mathcal{F}_\mu)}{N_\mu T^2}. \,\ \footnotemark
\end{aligned}
\end{equation*}
\footnotetext{The conditional mean of the jump is $\kappa(\mathcal{F}_\mu-m_\mu)$ and its conditional variance is $N_\mu\kappa T\mathcal{F}_\mu(1-\kappa T\mathcal{F}_\mu)/(N_\mu T)^2=\kappa\mathcal{F}_\mu(1-\kappa T\mathcal{F}_\mu)/(N_\mu T)$. The second moment is the variance plus the squared mean, and both are divided by the update interval $\kappa T$.}%
The drift does not depend on $\kappa$, so the mean equation is the same for all members of the family. The displacement product in $a_{\mu\nu}$, which is the origin of terms (ii) and (iii), is multiplied by $\kappa$. The following $\mathcal O(N^{-2})$ remainder estimate presupposes a regular system-size expansion with $C=\mathcal O(N^{-1})$ and appropriately small higher central moments. It is not uniform over strong transients or critical regimes. Repeating the derivation above therefore gives, for any $\kappa$,
\begin{equation*}
\begin{aligned}
T\partial_t c_{\mu\nu}
={}&\delta_{\mu\nu}\frac{\mathcal{F}_\mu(1-\kappa T\mathcal{F}_\mu)}{N_\mu T}
+(\partial_\lambda\mathcal{F}_\mu-\delta_{\mu\lambda})c_{\nu\lambda}
+(\partial_\lambda\mathcal{F}_\nu-\delta_{\nu\lambda})c_{\mu\lambda}\\
&+\kappa\Big[(\partial_\lambda\mathcal{F}_\mu-\delta_{\mu\lambda})\,c_{\lambda\eta}\,(\partial_\eta\mathcal{F}_\nu-\delta_{\nu\eta})
+T^2\,\partial_t\langle m_\mu\rangle\,\partial_t\langle m_\nu\rangle\Big]
+\mathcal{O}(N^{-2}).
\end{aligned}
\end{equation*}
For $\kappa=1$ this is Eq.~\eqref{eq:covariance_consistent_closure} (the drift terms and term (ii) combine as above). For $\kappa\to0$, term (ii) and the source term both disappear, and the second-order truncation becomes the standard system-size (linear-noise) expansion of van Kampen (see, e.g.,~\cite{bressloff2010stochastic,buice2010systematic}), whose errors are $\mathcal{O}(N^{-2})$:
\begin{empheq}[box=\fbox]{equation}
\begin{aligned}
    T\partial_t c_{\mu\nu}
=\delta_{\mu\nu}\frac{\mathcal{F}_\mu}{N_\mu T}
+(\partial_\lambda\mathcal{F}_\mu-\delta_{\mu\lambda})\,c_{\nu\lambda}
+(\partial_\lambda\mathcal{F}_\nu-\delta_{\nu\lambda})\,c_{\mu\lambda}.
\end{aligned}
\label{eq:covariance_small_jump}
\end{empheq}
This is the structure of the mean-value closure~\eqref{eq:covariance_published_closure}, now obtained consistently. The mean-value closure is thus the small-jump equation~\eqref{eq:covariance_small_jump} plus two remnants of the $\kappa=1$ construction: the source term $(\mathcal{F}_\mu-\langle m_\mu\rangle)(\mathcal{F}_\nu-\langle m_\nu\rangle)$, which is absent for small jumps, and the binomial factor $1-T\mathcal{F}_\mu$ in the noise source, which turns $\mathcal{F}_\mu/(N_\mu T)$ into $A^{-1}_{\mu\mu}$. For one population with the affine transfer function~\eqref{eq:affine_F}, for instance, the moment equations of the small-jump model close exactly\footnote{The spike rate $N\mathcal{F}(m)$ is affine in $m$, so the equations for the first two moments of Eq.~\eqref{eq:small_jump_limit} involve no higher moment.}, its stationary variance is $c=m_0/[2NT(1-\Delta)]$, and the published values of Table~\ref{tab:covariance_closure_affine} are exactly $1-Tm_0=0.5$ times these values, for every slope.

\quad These results suggest several ways to improve the second-order description, depending on what the model is meant to represent.
Replacing Eq.~\eqref{eq:covariance_published_closure} by Eq.~\eqref{eq:covariance_consistent_closure} costs nothing, and it removes the slope-dependent error. When some eigenvalue of $J$ satisfies $|\zeta_a|\ge1$, the divergence of the covariance should be read as a diagnostic, not as a numerical failure: the fixed point is not a stationary state of the master equation. When condition~\eqref{eq:sigmoid_validity} fails, both closures fail. Possible remedies are cumulant closures at third or fourth order, or closures that evaluate the nonlinear averages $\langle\mathcal{F}\rangle$, $\operatorname{Var}(\mathcal{F})$ and $\langle A^{-1}\rangle$ under an assumed distribution instead of Taylor-expanding them.

\paragraph{What this implies:}

The complete expansion corrects an omitted leading covariance contribution of the stated redraw master equation. With the transition operator~\eqref{eq:transition_operator_continuous_definition}, the covariance equation is Eq.~\eqref{eq:covariance_consistent_closure}, not Eq.~\eqref{eq:covariance_published_closure}. Two changes are needed. The gain terms $\partial_\lambda\mathcal{F}_\mu c_{\nu\lambda}+\partial_\lambda\mathcal{F}_\nu c_{\mu\lambda}-2c_{\mu\nu}$ become $\partial_\lambda\mathcal{F}_\mu\partial_\eta\mathcal{F}_\nu c_{\lambda\eta}-c_{\mu\nu}$ (term (ii)), and the source term $(\mathcal{F}_\mu-\langle m_\mu\rangle)(\mathcal{F}_\nu-\langle m_\nu\rangle)$ becomes the product of the complete right-hand sides of the mean equations, Eq.~\eqref{eq:consistent_source_term} (term (iii)). The stationary fluctuations are larger, by a factor of about $2/(1+\Delta)$ for one population, and their stability requires $|\zeta_a|<1$ instead of $\operatorname{Re}\zeta_a<1$. Its corrected form is useful when the additional omissions and Taylor remainders are small.

The first-moment equation retains its algebraic form, \emph{but its solution can change through covariance feedback}. The first-order equations obtained by omitting that feedback retain their right-hand sides. The linearized lag-covariance equation~\eqref{eq:diff_equation_correlation_populations} also retains its form, although its initial covariance changes. These algebraic statements do not establish that every published simulation used a first-order configuration or that its reported results are unchanged.

The filtered-spike construction is a separate model. Its covariance equation has noise source $\mathcal F_\mu/(N_\mu T)$ and no squared-displacement source. Only with further restrictions, such as stationary low-occupancy conditions, can it approximate the published expression. Consequences for empirical or parent-network comparisons require matching the implemented model, parameters and activity observable; they cannot be directly inferred by relabeling the original covariance.

}

\subsubsection{Correlation Matrix}

Let's derive the differential equation of the correlation between two populations $\mu$ and $\nu$ with time delay $\tau$, with the network in a stationary state. $\partial_{\tau}Corr_{\mu\nu}(\tau)$ is the rate at which the correlation between population $\mu$ at time $t$ and population $\nu$ at time $t+\tau$ changes as the time lag $\tau$ increases.

\begin{equation*}
\begin{aligned}
\mathrm{Corr}_{\mu\nu}(\tau)
&=\mathbb E[\delta m_\mu(t)\delta m_\nu(t+\tau)],
\qquad \tau\geq0.\\
\partial_{\tau} \mathrm{Corr}_{\mu\nu}(\tau)
&= \partial_{\tau} \left\langle \left(m_\mu(t) - \langle m_\mu(t) \rangle\right)\left(m_\nu(t+\tau) - \langle m_\nu(t+\tau) \rangle\right) \right\rangle \\
&= \partial_{\tau} \left( \langle m_\mu(t) m_\nu(t+\tau) \rangle - \langle m_\mu(t) \rangle \langle m_\nu(t+\tau) \rangle \right).
\end{aligned}
\end{equation*}

This is an unnormalized lagged covariance; $\mathrm{Corr}_{\mu\nu}(0)=c_{\mu\nu}$. Let's start with the first term:
\begin{equation*}
\begin{aligned}
\partial_{\tau}\langle m_{\mu}(t)m_{\nu}(t+\tau) \rangle 
&= \partial_{\tau} \prod_{\alpha = 1}^{P} \int^{\frac{1}{T}}_{0} dm_\alpha  \prod_{\beta = 1}^{P} \int^{\frac{1}{T}}_{0} dm'_\beta m_\mu m'_\nu \,\ P(\{m'_\gamma \}, t+\tau \mid \{m_\gamma \}, t)P_t(\{m_\gamma \}). \,\ \footnotemark \\
&= \prod_{\alpha = 1}^{P} \int^{\frac{1}{T}}_{0} dm_\alpha m_\mu  \left( \prod_{\beta = 1}^{P} \int^{\frac{1}{T}}_{0} dm'_\beta  m'_\nu \,\ \partial_{\tau} P(\{m'_\gamma \}, t+\tau \mid \{m_\gamma \}, t) \right) P_t(\{m_\gamma \}). \,\ \\
&= \prod_{\alpha = 1}^{P} \int^{\frac{1}{T}}_{0} dm_\alpha m_\mu \Big( a_{\nu}(\{m_{\gamma}(t+\tau)\}) \Big) P_t(\{m_\gamma\}). \,\ \footnotemark \\
&= \langle m_{\mu}(t)a_{\nu}(\{m_{\gamma}(t+\tau)\}) \rangle
\end{aligned}
\end{equation*}
\footnotetext{The expected value of the product of activities at two different times is the probability of the network being in state $\{m_{\gamma}\}$ at time $t$ multiplied by the conditional probability that it transitions to state $\{m_{\gamma}'\}$ at time $t+\tau$}
\footnotetext{$P(\{m_{\gamma}^{\prime}\},t+\tau|\{m_{\gamma}\},t)$ is solution of the network's master equation with respect to the forward time variable $\tau$, so we apply Eq.~\ref{eq:mean_activity_drift} trick to turn it into step moment function}

The second term is straightforward using previous Eq.~\ref{eq:mean_activity_drift}:
\begin{equation*}
\begin{aligned}
\partial_{\tau}\langle m_{\mu}(t)\rangle\langle m_{\nu}(t+\tau)\rangle 
&= \langle m_{\mu}(t)\rangle \partial_{\tau}\langle m_{\nu}(t+\tau)\rangle  \\
&= \langle m_{\mu}(t)\rangle\langle a_{\nu}(\{m_{\gamma}(t+\tau)\})\rangle
\end{aligned}
\end{equation*}

Combining the two, we get:
$$\partial_{\tau}Corr_{\mu\nu}(\tau) = \langle (m_{\mu}(t) - \langle m_{\mu}(t) \rangle)a_{\nu}(\{m_{\gamma}(t+\tau)\}) \rangle$$
This means that the change in correlation over time $\tau$ depends entirely on how the initial fluctuation of population $\mu$ away from its mean interacts with the expected direction of change ($a_{\nu}$) of population $\nu$ at the later time.

\revision{We now expand $a_\nu$ around the stationary mean activities. The original derivation~\cite{el2009master} (their Appendix~B) motivates this step by considering time scales beyond the decay time of the network activity correlations. However, although a large lag can reduce lagged correlations, it does not make the fluctuations smaller: in a stationary state, $\{m_\gamma(t+\tau)\}$ has the stationary distribution at every lag, with covariance $c_{\mu\nu}$. The expansion relies on the same narrow-distribution assumption as the equal-time moments (Supplementary~\ref{sub:generic_second_order_master_eq_moments}). Keeping the quadratic term, with all derivatives at the stationary mean,
$$a_{\nu}(\{m_{\gamma}(t+\tau)\}) = a_{\nu}(\{\langle m_{\gamma} \rangle\}) + \partial_{\lambda}a_{\nu}\,\delta m_{\lambda}(t+\tau) + \frac{1}{2}\partial_{\lambda}\partial_{\eta}a_{\nu}\,\delta m_{\lambda}(t+\tau)\,\delta m_{\eta}(t+\tau) + \mathcal{O}(\delta m^3).$$
Multiplied by $\delta m_\mu(t)$ and averaged, the first term vanishes, the second gives $\partial_\lambda a_\nu\,Corr_{\mu\lambda}(\tau)$, and the third gives the three-point correlation $\frac{1}{2}\partial_\lambda\partial_\eta a_\nu\,\langle\delta m_\mu(t)\,\delta m_\lambda(t+\tau)\,\delta m_\eta(t+\tau)\rangle$. For finite-size fluctuations, this three-point term is of order $N^{-2}$ at every lag, against $N^{-1}$ for the retained term, and it vanishes for Gaussian fluctuations; for an affine transfer function, $a_\nu$ is linear and the resulting equation is exact. The linear equation below is therefore the leading order at every lag $\tau\geq0$. The time-scale condition concerns the model itself: the coarse-grained Markov description, hence Eq.~\eqref{eq:diff_equation_correlation_populations}, can only be compared with a spiking network at lags of order $T$ or longer.

Finally, reinjecting everything, we get:

\begin{equation}
\begin{aligned}
\partial_{\tau}Corr_{\mu\nu}(\tau) 
&= \left\langle (m_{\mu}(t) - \langle m_{\mu}(t) \rangle)\, a_{\nu}(\{m_{\gamma}(t+\tau)\}) \right\rangle \\
&\simeq \partial_{\lambda}a_{\nu}(\{\langle m_{\gamma}(t+\tau)\rangle\})\,
Corr_{\mu\lambda}(\tau).
\end{aligned}
\label{eq:diff_equation_correlation_populations}
\end{equation}

}

\subsection{Master Equation formalism applied to different neuron models}
\label{sub:application_to_specific_SNN}

\subsubsection{Linear model}
\label{subsec:master_equation_linear_model}

This section assumes a linear transfer function for a two-population excitatory/inhibitory network with homogeneous neurons: $\mathcal{F}_\mu(\{ m_\gamma \})= \mathcal{F}(\{ m_\gamma \}) = \nu_0 + \sum_\lambda w_{\mu\lambda} m_\lambda = \nu_0 + k_{E}m_{exc} + k_{I}m_{inh}$.

This is a simplified homogeneous network in which all neurons share the same linear transfer function and respond to the same population-level inputs. While this is not biologically realistic, the primary purpose of the linear toy model is to mathematically isolate and study the stability of the network and the behavior of the covariances. Because the transfer function is purely linear, its second derivative is strictly zero ($\partial^2 \mathcal{F}_\mu = 0$). Therefore, the mean activity becomes completely decoupled from the covariance, allowing for a clean separation of the first-order and second-order structures.

Thanks to this decoupling, the exact fixed points of the network can be analytically calculated. This enables the study of the system's behavior near the critical point of instability. The mean equation becomes $T\partial_t \langle m_\mu \rangle = \nu_0 + k_{E}\langle m_{exc}\rangle + k_{I} \langle m_{inh}\rangle - \langle m_\mu \rangle$, and the stationary solution of the system is given by:
\begin{equation*}
\begin{aligned}
0
&= \nu_0 + k_{E}\langle m_{exc}\rangle + k_{I} \langle m_{inh}\rangle - \langle m_{exc} \rangle \\
0
&= \nu_0 + k_{E}\langle m_{exc}\rangle + k_{I} \langle m_{inh}\rangle - \langle m_{inh} \rangle \\
\text{so,} \
m_0 
&= \langle m_{exc}^{FP} \rangle =\langle m_{inh}^{FP} \rangle = \frac{\nu_0}{1-\Delta}, \ \Delta = k_{E} + k_{I}
\end{aligned}
\end{equation*}

The stability of the system is dictated by the sign of the eigenvalues of the system's Jacobian matrix:

$$T \frac{d}{dt} \begin{pmatrix} \langle m_E \rangle \\ \langle m_I \rangle \end{pmatrix} = {\begin{pmatrix} k_E - 1 & k_I \\ k_E & k_I - 1 \end{pmatrix}} \begin{pmatrix} \langle m_E \rangle \\ \langle m_I \rangle \end{pmatrix} + \begin{pmatrix} \nu_0 \\ \nu_0 \end{pmatrix}$$

The eigenvalues are found by solving the characteristic equation:
$$\det \begin{pmatrix} k_E - 1 - \lambda & k_I \\ k_E & k_I - 1 - \lambda \end{pmatrix} = - k_E - k_I + 1 - \lambda(k_E + k_I - 2) + \lambda^2 = 0$$
The roots of this equation are $\lambda_1 = -1$ and $\lambda_2 = \Delta - 1$.

Thus, the stability of the \revision{mean dynamics} requires $\Delta < 1$. If the total synaptic gain $\Delta$ equals or exceeds $1$, this critical eigenvalue becomes positive, and the macroscopic firing rate of the network explodes to infinity. Such linear behavior should not be extrapolated beyond the range where the transfer function defines admissible firing probabilities.

\revision{Regarding the second-order moments, we use the corrected covariance equation~\eqref{eq:covariance_consistent_closure}:
$$T\partial_t c_{\mu\nu} = \delta_{\mu\nu} \frac{\mathcal{F}_\mu(1/T - \mathcal{F}_\mu)}{N_\mu} + \partial_\lambda\mathcal{F}_\mu\,\partial_\eta\mathcal{F}_\nu\,c_{\lambda\eta} - c_{\mu\nu} + T^2\,\partial_t\langle m_\mu\rangle\,\partial_t\langle m_\nu\rangle .$$
Near the fixed point, $\partial_t\langle m_\mu\rangle=0$ and the source term vanishes. Because both populations share the same transfer function, the two rows of the gain matrix are identical, $\partial_\lambda\mathcal{F}_\mu=(k_E,k_I)$ for $\mu=E,I$. The gain term is then the same for every pair $(\mu,\nu)$, $\partial_\lambda\mathcal{F}_\mu\,\partial_\eta\mathcal{F}_\nu\,c_{\lambda\eta}=s$ with $s=k_E^2c_{EE}+k_I^2c_{II}+2k_Ek_I\,c_{EI}$, which leaves the system
$$
T \frac{d}{dt} \begin{pmatrix} c_{EE} \\ c_{II} \\ c_{EI} \end{pmatrix}
=
\underbrace{\begin{pmatrix} k_E^2-1 & k_I^2 & 2k_Ek_I \\ k_E^2 & k_I^2-1 & 2k_Ek_I \\ k_E^2 & k_I^2 & 2k_Ek_I-1 \end{pmatrix}}_{M}
\begin{pmatrix} c_{EE} \\ c_{II} \\ c_{EI} \end{pmatrix}
+
\underbrace{\begin{pmatrix} A^{-1}_{E} \\ A^{-1}_{I} \\ 0 \end{pmatrix}}_{D},
\qquad
A^{-1}_{\mu}=\frac{m_0(1/T - m_0)}{N_\mu}.
$$
The matrix $M$ is a rank-one matrix minus the identity, so its eigenvalues are $-1$ (twice) and $\Delta^2-1$. Stationary covariances therefore exist only if $|\Delta|<1$, and are given by
\begin{equation}
c_{EE}=A^{-1}_{E}+s,\qquad
c_{II}=A^{-1}_{I}+s,\qquad
c_{EI}=s,\qquad
s=\frac{k_E^2A^{-1}_{E}+k_I^2A^{-1}_{I}}{1-\Delta^2}.
\label{eq:linear_model_stationary_covariances}
\end{equation}
With the published equation~\eqref{eq:covariance_published_closure}, the covariance Jacobian has the eigenvalues $-2$, $2(\Delta-1)$ and $\Delta-2$, and the stationary covariances, which exist for every $\Delta<1$, solve $c_{\mu\nu}=\frac{1}{2}\big(\delta_{\mu\nu}A^{-1}_{\mu}+r_\mu+r_\nu\big)$ with $r_\mu=k_E\,c_{E\mu}+k_I\,c_{I\mu}$.

The covariance equations show that macroscopic fluctuations are a finite-size phenomenon, since all covariance terms vanish as $N \to \infty$. Near $|\Delta|=1$, however, they grow as $(1-\Delta^2)^{-1}$, indicating that the network becomes increasingly sensitive to demographic noise, a behavior out of the scope of first-order models. Unlike the published equation, the corrected one introduces a new stability threshold: for $\Delta<-1$, the mean dynamics are stable but the fluctuations are not. Each transition then overshoots the fixed point by more than the previous deviation, alternately above and below it, and the fixed point is no longer a stationary state of the master equation (Supplementary~\ref{subsubsec:covariance_closure_benchmarks}). The two equations also differ quantitatively in inhibition-dominated networks. For $k_E=0.8$, $k_I=-1.3$ ($\Delta=-0.5$), $\nu_0=150\,\mathrm{Hz}$ (so that $m_0=100\,\mathrm{Hz}$), $T=5\,\mathrm{ms}$, $N_E=4000$ and $N_I=1000$, Eq.~\eqref{eq:linear_model_stationary_covariances} gives $(c_{EE},c_{II},c_{EI})=(27.2,\,34.7,\,24.7)\,\mathrm{Hz}^2$, within $0.3\%$ of the values $(27.1,\,34.6,\,24.6)\,\mathrm{Hz}^2$ of the complete equation~\eqref{eq:covariance_complete}, which is exact for a linear transfer function as long as the kernel remains admissible. The published equation gives $(4.5,\,2.3,\,0.27)\,\mathrm{Hz}^2$, i.e.\ $6$, $15$ and $92$ times too small. This affine function is only a local approximation: at some
corners of $[0,1/T]^2$ it gives $T\mathcal F<0$ or
$T\mathcal F>1$. The values $(27.1,34.6,24.6)\,\mathrm{Hz}^2$
from the complete affine moment equations are therefore formal
algebraic values, not an exact stochastic benchmark for an
admissible global kernel. A bounded continuation would define a
different model requiring a separate comparison.

The correlation equation can also be explicitly calculated. Its eigenvalues match those governing the mean dynamics, $-1$ and $\Delta-1$: correlations decay on the timescale of the mean dynamics, and correlation times diverge only at $\Delta=1$. Their value at zero lag is the equal-time covariance of Eq.~\eqref{eq:linear_model_stationary_covariances}.}

\subsubsection{Spiking Neural Network}
\label{subsec:master_equation_SNN_case}

After extracting a transfer function $\mathcal{F}_\mu$ from a single microscopic spiking neuron, will the mesoscopic ODEs accurately predict the macroscopic behavior of a massive network of those neurons?

El Boustani and Destexhe \cite{el2009master} tested this on two standard paradigms of spiking neural networks. They considered sparsely connected Excitatory-Inhibitory (E/I) networks with random connectivity, though they note that perfect randomness is not essential and some spatial locality can be tolerated as long as correlations do not become too strong. Each neuron in population $\mu$ receives, on average, $C_{\alpha\mu}$ random synaptic inputs from population $\alpha$, where the connection probability is $\frac{C_{\alpha\mu}}{N_\alpha} = p_{\alpha\mu}^{conn} \ll 1$ (typically $< 10\%$).

\textbf{Current-Based (CUBA) Integrate-and-Fire (IF) Models}
\label{subsec:master_equation_CUBA}

In a CUBA network, incoming spikes inject a fixed amount of current into the neuron, independent of the neuron's current membrane voltage.

$$
\tau_\mu^{mem}\frac{d}{dt}V_\mu^i(t) = -(V_\mu^i(t)-V_\mu^{rest})+R_\mu I_\mu^i(t), \,\ \mu \in \llbracket 1,P\rrbracket, \,\ \cdot_\mu^i \in \llbracket 1,N_\mu\rrbracket
$$
This linear equation is supplemented with standard threshold, reset, and refractory period mechanisms. The total current consists of recurrent input from other neurons in the network, $I_\mu^{i,int}(t)$, and external drive, $I_\mu^{i,ext}(t)$. Because the populations are homogeneous, we can drop the index $i$. The internal current $I_\mu^{int}(t)$ is modeled as a sum of synaptic events, treated as Poisson point processes $N_{\alpha \mu}(ds)$ describing the spike train\footnote{The Poisson distribution assumption is made because the synaptic bombardment consists of a massive number of independent events arriving at random times; therefore, the arrival of these events at the receiver approximates a Poisson process.}, convolved with postsynaptic-potential kernel $PSP_{\alpha \mu}(t)$ from population $\alpha$ to population $\mu$. The external current $I_\mu^{ext}(t)$ can also be modeled as Poisson spike trains with rate $m_\alpha^{ext}$.

Assuming incoming spikes are highly frequent and numerous, we can use the Central Limit Theorem (specifically Donsker's Theorem\footnote{Donsker's theorem states that as a random walk gets faster (more steps) and the steps get smaller (tiny PSPs), the entire trajectory of the random walk converges into a continuous Brownian motion.}) to approximate the synaptic bombardment as continuous Brownian motion (a random walk). This allows one to replace the discrete synaptic inputs with a continuous background current characterized by two properties: (1) the drift $\langle V_\mu \rangle$, which is the average trend of the voltage, and (2) the diffusion $\sigma^2(V_\mu)$, which is the amplitude of the continuous fluctuations around the mean. This diffusion approximation allows the membrane voltage $V(t)$ to be treated as a continuous Ornstein--Uhlenbeck process\footnote{Ornstein--Uhlenbeck was originally created to model the velocity of a massive Brownian particle under the influence of friction.} and to define a probability density function $P(V,t)$ that yields the probability of finding the neuron at state $V$ at time $t$. The time evolution of $P(V, t)$ is governed by a partial differential equation known as the Fokker--Planck Equation (FPE) (see Supplementary~\ref{subsec:fokker_planck_reminder} for details).

To simplify the system, they calculated the \textit{free} membrane potential (ignoring the threshold and spike mechanisms), such that the steady-state solution of the FPE ($P_\mu(V)$) is approximated as a Gaussian distribution \cite{amit1997model}. To estimate the firing rate, they use a heuristic: the probability that a neuron is in a firing state is roughly equal to the fraction of this Gaussian probability distribution that lies above $V_{thresh}$, divided by the membrane time constant $\tau^{mem}_\mu$ (see Supplementary~\ref{subsec:fokker_planck_microscopic_level} for more details):

\begin{equation}
\begin{aligned}
\mathcal{F}_\mu 
&= \frac{1}{\tau_\mu^{mem}} \int_{V_{thresh_\mu}}^{\infty} P_\mu(V) dV \\
&= \frac{1}{\tau_\mu^{mem}} \int_{V_{thresh_\mu}}^{\infty} \frac{1}{\sqrt{2\pi}\sigma(V)}\exp{\left( -\frac{(V-\langle V \rangle)^2}{2\sigma(V)^2}\right)} dV \\
&= \frac{1}{\tau_\mu^{mem} \sqrt{\pi}} \int_{\frac{V_{thresh} - \langle V_\mu \rangle}{\sqrt{2}\sigma(V_\mu)}}^{\infty} \exp{(-z^2)} dz \\
&= \frac{1}{2\tau_\mu^{mem}} \left[ 1 + \text{erf}\left( \frac{\langle V_\mu \rangle - V_{thresh}}{\sqrt{2}\sigma(V_\mu)} \right) \right]. \,\ \footnotemark
\end{aligned}
\end{equation}
\footnotetext{$$\int_{x}^{\infty} \text{Gaussian} = \frac{1}{2} \left[ 1 - \text{erf}\left( \frac{x - \mu}{\sqrt{2}\sigma} \right) \right] = \frac{1}{2} \left[\text{erfc}\left( \frac{x - \mu}{\sqrt{2}\sigma} \right) \right]$$}

It means that the more the fluctuating membrane voltage sits above threshold, the more the neuron fires. Hence, a diffusion approximation, via Fokker--Planck equations, provides a simple analytical approximation of the transfer function $\mathcal{F}_\mu$. This approximation relies on the fact that the membrane time constant leads the dynamics, which corresponds to AI regime. More exact expressions exist only for more restricted classes of neuron and synapse models.

To fully determine the framework, we must establish the relationship between the incoming spike train statistics and the membrane potential probability distribution; this means calculating the membrane-potential mean $\langle V_\mu \rangle$ and standard deviation $\sigma(V_\mu)$ resulting from Poisson synaptic bombardment. This is achieved using Campbell’s theorem, yielding:\footnote{Campbell's theorem states that for a Poisson point process $N$ with intensity $\Lambda$ and a measurable function $f:\mathbb{R}^d \to \mathbb{R}$, the mean of the random sum $S=\sum_{x \in N}f(x)$ is $E(S) = \int_{\mathbb{R}^d}f(x)\Lambda(dx)$ and the variance of $S$ is $Var(S)= \int_{\mathbb{R}^d}f(x)^2\Lambda(dx)$.}

\begin{equation}
\begin{aligned}
\langle V_\mu \rangle
&= V_\mu^{\mathrm{rest}} + \sum_{\alpha=1,\ldots,P}
C_{\alpha\mu}\left(m_\alpha + m_\alpha^{\mathrm{ext}}\right)
\int_{\mathbb{R}} PSP_{\alpha\mu}(t) \, dt
\\
\sigma^2(V_\mu)
&=\sum_{\alpha=1,\ldots,P}
C_{\alpha\mu}\left(m_\alpha + m_\alpha^{\mathrm{ext}}\right)
\int_{\mathbb{R}} PSP_{\alpha\mu}^2(t) \, dt
\end{aligned}
\end{equation}

Where the intensity of the Poisson point process is $\Lambda=C_{\alpha\mu}\left(m_\alpha + m_\alpha^{\mathrm{ext}}\right)$, representing the total aggregate spike train. This also assumes that without inputs, the variance drops to zero.

They also explicitly computed the membrane potential probability distribution for specific synapse families (e.g., \textit{Dirac synapses, exponential synapses, alpha synapses}), though the full derivations are outside the scope of this study. Nevertheless, this step is conceptually important because it demonstrates that $\langle V_\mu \rangle$ and $\sigma(V_\mu)$ depend directly on the chosen synapse type. Thus, the macroscopic global framework incorporates microscopic ingredients of both individual neuron and synapse models.

Therefore, while the CUBA neuron models are not the most realistic, one can derive a relatively tractable analytical transfer function to perform parameter space exploration of the framework.
By plugging this transfer function into the mesoscopic ODEs, they obtain good agreement with direct numerical simulations over a broad region of the AI regime, especially for current-based networks of sufficient size. \revision{These comparisons used the published covariance equation~\eqref{eq:covariance_published_closure}. The mean activities are affected by the correction only through the $\mathcal{O}(N^{-1})$ curvature term of the mean equation, whereas the predicted fluctuations change (Supplementary~\ref{subsubsec:covariance_closure_benchmarks}); the agreement of the fluctuations should therefore be re-examined with the corrected equation~\eqref{eq:covariance_consistent_closure}, using the observable that matches the reading of $T$ (binned or filtered activity).}

\textbf{Conductance-Based (COBA) Integrate-and-Fire Models}
\label{subsec:master_equation_COBA}

In a COBA network, incoming spikes open ion channels, changing the membrane's conductance. This means the effect of a synapse changes depending on the current membrane voltage, and the effective membrane time constant fluctuates dynamically with network activity.

\begin{equation}
\begin{aligned}
C_\mu^{mem}\frac{d}{dt}V_\mu(t) = G_\mu^L(V_\mu^{rest}-V_\mu(t))+\sum_\alpha G_{\alpha\mu}(t)(E_\alpha-V_\mu(t)), \,\ \mu \in \llbracket 1,P\rrbracket, \,\ \cdot_\mu^i \in \llbracket 1,N_\mu\rrbracket
\end{aligned}
\label{eq:COBA_membrane_potential_equation}
\end{equation}

where $C_\mu^{mem}$ and $G_\mu^L$ are the membrane capacitance and the leak conductance, such that $\frac{C_\mu^{mem}}{G_\mu^L}=\tau_\mu^{mem}$. So the effective membrane time constant depends on the ongoing synaptic conductances. $G_{\alpha\mu}$ is the total conductance of the synaptic set $\alpha$ and $E_\alpha$ the corresponding reversal potential. Then, similarly to CUBA model, the synaptic input can be modeled by Poisson processes, and the total conductance of the synapses $\alpha$ is calculated with the convolution of $g_{\alpha\mu}$ (the conductance time course elicited by an incoming spike from population $\alpha$), with the spike train $N_{\alpha\mu}$.

COBA networks are significantly more biologically realistic, but they are mathematically intractable, compared to CUBA model, no exact analytical transfer function exists for Eq.~\eqref{eq:COBA_membrane_potential_equation}. (see Supplementary~\ref{subsec:fokker_planck_microscopic_level} for more details)

Instead, they introduced an approximation inspired by Kuhn et al. (2004) \cite{kuhn2004neuronal}, which replaces the fluctuating conductances by effective averages and uses an effective current-based description. Without going into the mathematical details, they showed that even though the biophysics can get harder, the outer mesoscopic structure remains the same; only the formulas feeding $\mathcal{F}_\mu$, $\partial_\lambda \mathcal{F}_\mu$, and $\partial_\lambda \partial_\eta \mathcal{F}_\mu$ become more complicated. This emphasizes that the framework can deal with complex neuron models, as long as one has a workable effective transfer function

However, they found that the approximation is only qualitatively good, and then introduced an optimized phenomenological transfer function, with 2 fitted parameters ($\tau$: the time constant, $\Delta h$: a corrective term), plugged back into the macroscopic ODEs, in order to improve the quantitative agreement between the mesoscopic model and network simulations (see Section~\ref{sec:semi-analytical_TF}). With the optimized effective transfer function, the mesoscopic ODEs reproduce the mean activity and fluctuation structure much more accurately in the conductance-based AI regime\revision{; as in the current-based case, the fluctuation comparison used the published covariance equation}.

This makes this section both as an extension and a warning that the mesoscopic formalism is general, but its success depends heavily on the quality of the transfer function.

Therefore, by succeeding in both the CUBA and COBA paradigms, El Boustani and Destexhe \cite{el2009master} showed that the formalism is flexible and can be instantiated for different neuron and synapse models, provided a sufficiently accurate transfer function is available and the assumptions of the mesoscopic framework remain valid.

\subsection{From Binomial to Gaussian distribution}
\label{subsec:from_Binomial_to_Gaussian_distrbution}

The Binomial distribution is:
$$P(k) = \frac{N!}{k!(N-k)!} p^k q^{N-k}$$

Taking the logarithm, we get:
$$\ln P(k) = \ln(N!) - \ln(k!) - \ln((N-k)!) + k \ln p + (N-k) \ln q$$

Stirling's approximation states that for large $n$:
$\ln(n!) \approx n \ln n - n + \frac{1}{2}\ln(2\pi n)$

Substituting this approximation into the log-distribution gives:
\begin{equation*}
\begin{aligned}
\ln P(k) 
&\approx \left( N \ln N - N + \frac{1}{2}\ln(2\pi N) \right) - \left( k \ln k - k + \frac{1}{2}\ln(2\pi k) \right) \\
&\quad - \left( (N-k) \ln(N-k) - (N-k) + \frac{1}{2}\ln(2\pi (N-k)) \right) + k \ln p + (N-k) \ln q \\
\\
&\approx N\ln N + \frac{1}{2}\ln\left(\frac{N}{2\pi k(N-k)}\right)  - k \ln\left(\frac{k}{p}\right) 
      - (N-k) \ln\left(\frac{N-k}{q}\right) \\
\\
&\approx \frac{1}{2}\ln\left(\frac{N}{2\pi k(N-k)}\right) - k \ln\left(\frac{k}{Np}\right) 
      - (N-k) \ln\left(\frac{N-k}{Nq}\right).
\end{aligned}
\end{equation*}

Let’s define the small fluctuation $\delta$ as the number of spikes $k$ that fluctuate closely around its expected mean $k = Np + \delta$. Similarly, the number of failures is $N-k = Nq - \delta$.

So we get:
$$k \ln\left(\frac{k}{Np}\right) = (Np + \delta) \ln\left(1 + \frac{\delta}{Np}\right) = (Np + \delta) \left( \frac{\delta}{Np} - \frac{\delta^2}{2N^2p^2} + o(\delta^2) \right) \approx \delta + \frac{\delta^2}{2Np}$$

$$(N-k) \ln\left(\frac{N-k}{Nq}\right) = (Nq - \delta) \ln\left(1 - \frac{\delta}{Nq}\right) = (Nq - \delta) \left( -\frac{\delta}{Nq} - \frac{\delta^2}{2N^2q^2} + o(\delta^2) \right) \approx -\delta + \frac{\delta^2}{2Nq}$$

Reinjecting these results back into the log-distribution gives:
\begin{equation*}
\begin{aligned}
\ln P(k) 
&\approx \frac{1}{2}\ln\left(\frac{N}{2\pi k(N-k)}\right) - \frac{\delta^2}{2N} \left( \frac{1}{p} + \frac{1}{q} \right)\\
&\approx \ln\left(\frac{1}{\sqrt{2\pi Npq}}\right) - \frac{\delta^2}{2Npq}. \,\ \footnotemark
\end{aligned}
\end{equation*}
\footnotetext{Since $\frac{1}{p} + \frac{1}{q} = \frac{1}{pq}$, and assuming $k \approx Np$ and $(N-k) \approx Nq$.}

Finally, after exponentiating:
$$P(k) \approx \frac{1}{\sqrt{2\pi Npq}} \exp\left( - \frac{\delta^2}{2Npq} \right)$$

\subsection{Calculation of the subthreshold membrane-potential fluctuations}
\label{subsec:calculs_subthreshold_membrane_statistics}

To evaluate the transfer function, one must compute the subthreshold fluctuation variables\\ $(\langle V \rangle,   \sigma(V), \tau_V^N, \langle g \rangle)$ from presynaptic firing rates $(\nu_e,\nu_i)$.

The synaptic input is modeled as the superposition of two Poisson shot-noise processes, one excitatory and one inhibitory, each convolved with an exponential waveform to generate synaptic conductance time courses. If a presynaptic spike arrives at a synapse of type \(s\in\{e,i\}\), the resulting conductance transient is assumed to rise instantaneously and decay exponentially:
\[
g_s(t)=Q_s e^{-t/\tau_s}\,\mathcal H(t),
\]
where \(Q_s\) is the quantal conductance (peak conductance increment produced by one synaptic event, in siemens), \(\tau_s\) is the synaptic decay time constant, and \(\mathcal H(t)\) is the Heaviside function. Such events arrive at rate \(K_s \nu_s\), where $(K_s = pN_s)$ is the number of incoming synapses of type \(s\). Thus, the total synaptic bombardment is a Poisson shot-noise process.

Using Campbell's theorem\footnote{Here again, the synaptic bombardment is seen as a sum of many random synaptic pulses. This can be approximated by a Poisson shot-noise process, which enables the use of Campbell's theorem. It states that the mean is the event rate times the area of one pulse, while the variance is the event rate times the integral of the squared pulse. Recall: mean $\sim \lambda \int_{0}^{\infty} g_s(t)dt = \lambda Q_s \tau_s$, variance $\sim \lambda \int_{0}^{\infty} g_s(t)^2dt = \lambda Q_s^2 \frac{\tau_s}{2}$.}, we can compute the mean and standard deviation of population conductance:

\begin{equation}
\begin{aligned}
\langle G_e \rangle 
&=\nu_e K_e Q_e \tau_e, \,\ 
\sigma(G_e) = Q_e \sqrt{\frac{\nu_e K_e \tau_e}{2}}, \,\ \footnotemark \\
\langle G_i \rangle 
&= \nu_i K_i Q_i \tau_i, \,\ 
\sigma(G_i) =  Q_i \sqrt{\frac{\nu_i K_i \tau_i}{2}}.
\end{aligned}
\label{eq:mean_variance_conductances}
\end{equation}
\footnotetext{This is the correct formula derived by Zerlaut et al. \cite{zerlaut2016heterogeneous}. A dimensional inconsistency appears in the printed expression for the conductance standard deviation in Sacha et al. \cite{sacha2025computational}: the expression \(\sigma_{G_s}=\sqrt{\nu_s K_s \tau_s^2 Q_s}\) is not dimensionally homogeneous.}

The total mean conductance is then $\langle g \rangle = \langle G_e\rangle + \langle G_i \rangle + g_L$ which defines the effective membrane time constant  $\tau_m^{eff} = \tau_m(\nu_e,\nu_i) = \frac{C_m}{\langle g \rangle}$.

Following Kuhn et al. (2004) \cite{kuhn2004neuronal}, the mean membrane potential $\langle V \rangle$ is obtained by taking the stationary solution of the passive membrane equation under the mean synaptic bombardment:
$$ C_m \frac{dV}{dt} = -g_L(V-E_L)-G_e(V-E_e) - G_i(V-E_i)$$

Under the mean-conductance approximation, stationarity gives the following voltage proxy. This step neglects $\operatorname{Cov}(G_e,V)+\operatorname{Cov}(G_i,V)$, as explained in Supplementary~\ref{subsec:adaptation_master_equation_formalism}:
$0 = -g_L(\langle V \rangle -E_L)- \langle G_e \rangle(\langle V \rangle -E_e) - \langle G_i \rangle(\langle V \rangle -E_i)$, 
so that the mean membrane potential $\langle V \rangle$ is a conductance-weighted average of reversal potentials:
$$\langle V \rangle = \frac{g_L E_L + \langle G_e \rangle E_e + \langle G_i \rangle E_i}{g_L + \langle G_e \rangle + \langle G_i \rangle}$$
Therefore, the mean voltage is entirely determined by the balance between leak, excitation, and inhibition.

To compute the standard deviation \(\sigma(V)\) and the global autocorrelation time \(\tau_V\) of the membrane fluctuations, Zerlaut et al.~\cite{zerlaut2016heterogeneous} used the power spectral density (PSD) of the membrane response. Around the mean voltage \(\langle V \rangle\), the membrane is approximated as a passive RC filter with effective time constant \(\tau_m^{\mathrm{eff}}\). A postsynaptic potential event of type \(s\) then obeys:
$$
\tau_m \frac{d PSP_s}{dt} + PSP_s = U_s\mathcal{H}(t)e^{-t/\tau_s}, \,\ \footnotemark
$$
\footnotetext{Using Ohm’s law, the effective PSP amplitude is: $U_s = \frac{Q_s}{\langle g \rangle} (E_s - \langle V \rangle)$ where $(E_s - \langle V \rangle)$ is the driving force.}

Solving the linear differential equation gives:
$$\text{PSP}_s(t) = U_s \frac{\tau_s}{\tau_m - \tau_s} \left( e^{-t/\tau_m} - e^{-t/\tau_s} \right) \mathcal{H}(t)$$

It's easier to compute the power spectrum of the noise in the frequency domain (converted with Fourier Transform), such that the PSD $P_V(f)$ of the membrane is: 

$$P_V(f) = \sum_{s \in \{e,i\}} K_s \nu_s \Arrowvert \widehat{\text{PSP}}_s(f) \Arrowvert^2, \,\ \footnotemark$$
\footnotetext{It means that the PSD of a system bombarded by independent Poisson events at rate $\lambda$ is the arrival rate multiplied by the square modulus of the Fourier transform of a single event.}

Using Parseval's theorem\footnote{Parseval's theorem states that the total power of a signal in the time domain is equal to the total power in the frequency domain, meaning $\int_{-\infty}^{\infty} |\widehat{f}(\xi)|^2d\xi= \int_{-\infty}^{\infty} |f(x)|^2dx$}, we get:
$$\sigma^2(V) = \int_{\mathbb{R}} P_V(f) df = \int_{\mathbb{R}} \sum_{s \in \{e,i\}} K_s \nu_s \left| \widehat{\text{PSP}}_s(f) \right|^2 df$$

Also, 
\begin{equation*}
\begin{aligned}
\int_{\mathbb{R}} \left| \widehat{\text{PSP}}_s(f) \right|^2 df 
&= \frac{1}{2\pi}\int_{\mathbb{R}} (U_s \frac{\tau_s}{\tau_m - \tau_s})^2 \left( \frac{\tau_m^2}{1 + \omega^2\tau_m^2} + \frac{\tau_s^2}{1 + \omega^2\tau_s^2} - \frac{2\tau_m\tau_s(1 + \omega^2\tau_m\tau_s)}{(1 + \omega^2\tau_m^2)(1 + \omega^2\tau_s^2)} \right) d\omega \\ &= \cdots = \frac{(U_s \tau_s)^2}{2(\tau_m + \tau_s)}
\end{aligned}
\end{equation*}

which gives the final closed expression:
$$ \sigma(V(\nu_e,\nu_i)) = \sqrt{\sum_s K_s\nu_s \frac{(U_s\tau_s)^2}{2(\tau_m+\tau_s)}}$$
Therefore, the fluctuation amplitude depends on both the synaptic rates and the strength of the effective voltage deflections produced by individual synaptic events.

Finally, following Zerlaut et al. (2016) \cite{zerlaut2016heterogeneous}, the global autocorrelation time of the membrane fluctuations is defined by

$$ \tau_V(\nu_e,\nu_i) = \frac{1}{2} \left ( \frac{\int_{\mathbb{R}} P_V(f) df}{P_V(0)}  \right)^{-1} = \frac{\sum_s K_s\nu_s (U_s\tau_s)^2}{\sum_s K_s\nu_s \frac{(U_s\tau_s)^2}{(\tau_m+\tau_s)}}$$

\subsection{Adding an adaptation term to the Master Equation formalism}
\label{subsec:adaptation_master_equation_formalism}

\subsubsection{Average membrane potential with adaptation}
\label{subsub:average_membrane_potential_with_adaptation}

Following Kuhn et al. (2004) \cite{kuhn2004neuronal}, the mean membrane potential $\langle V \rangle$ is obtained from the stationary subthreshold dynamics. For an AdEx neuron, the membrane equation can be written as
\begin{equation}
\begin{aligned}
C_m \frac{dV}{dt} 
&= -g_L(V-E_L) + I_{syn} + g_L \Delta_T e^{\frac{V-V_{thresh}}{\Delta_T}} - w, \\
I_{syn} 
&= -G_e(V-E_e) - G_i(V-E_i), \\
\frac{dw}{dt} 
&= \frac{a(V - E_L) - w}{\tau_w} + b \sum_{t_s \in \{ t_{spike} \}}  \delta(t - t_{s})
\end{aligned}
\label{eq:AdEx_model_2equations}
\end{equation}

\revision{Taking stationary averages, neglecting the exponential spike-initiation term in the subthreshold regime, and factorizing the averages of products of conductance and voltage, $\langle G_sV\rangle\simeq\langle G_s\rangle\langle V\rangle$, yields:}
\begin{equation*}
\begin{aligned}
0
&= -g_L \langle V \rangle + g_L E_L - \langle G_e \rangle \langle V \rangle + \langle G_e \rangle E_e - \langle G_i \rangle \langle V \rangle + \langle G_i \rangle E_i - \langle w \rangle, \\
0
&= \frac{a(\langle V \rangle - E_L) - \langle w \rangle}{\tau_w} + b\nu_{out}, \, \, \text{  so that, } \langle w \rangle = a(\langle V \rangle - E_L) + \nu_{out}\tau_w b
\end{aligned}
\end{equation*}
We stress that here $w$ is the stationary mean adaptation used inside the transfer function calculation, not yet the mesoscopic dynamical equation for $W(t)$.

Substituting this expression into the stationary voltage balance gives:
\begin{equation}
\begin{aligned}
\langle V(\nu_e,\nu_i,w) \rangle = \frac{g_L E_L + \langle G_e \rangle E_e + \langle G_i \rangle E_i - w}{g_L + \langle G_e \rangle + \langle G_i \rangle} = \frac{g_L E_L + \langle G_e \rangle E_e + \langle G_i \rangle E_i - \nu_{out}\tau_wb +aE_L }{\langle g \rangle + a},
\end{aligned}
\label{eq:average_membrane_equation_DiVolo_adaptation}
\end{equation}

\revision{\paragraph{Neglected conductance-voltage covariance:}
The factorization $\langle G_sV\rangle\simeq\langle G_s\rangle\langle V\rangle$, $s\in\{e,i\}$, neglects the covariances $\operatorname{Cov}(G_s,V)=\langle G_sV\rangle-\langle G_s\rangle\langle V\rangle$. We estimate their effect in four steps.

\textit{Exact mean:} Averaging the subthreshold membrane equation at stationarity without factorizing gives, at given $w$,
\begin{equation}
\langle V\rangle=\langle V\rangle_0-\frac{1}{\langle g\rangle}\sum_{s\in\{e,i\}}\operatorname{Cov}(G_s,V),
\label{eq:conductance_voltage_covariance_shift}
\end{equation}
where $\langle V\rangle_0$ is the factorized value, given by the first expression of Eq.~\eqref{eq:average_membrane_equation_DiVolo_adaptation}, and $\langle g\rangle=g_L+\langle G_e\rangle+\langle G_i\rangle$. Once $w$ is eliminated, $\langle g\rangle$ is replaced by $\langle g\rangle+a$.

\textit{Linearized fluctuations:} Writing $V=\langle V\rangle+\delta V$ and $G_s=\langle G_s\rangle+\delta G_s$, and subtracting the averaged equation, the voltage fluctuation obeys
\begin{equation*}
C_m\frac{d\,\delta V}{dt}=-\langle g\rangle\,\delta V+\sum_{s}(E_s-\langle V\rangle)\,\delta G_s-\sum_{s}\big[\delta G_s\,\delta V-\operatorname{Cov}(G_s,V)\big].
\end{equation*}
The last term is a product of two fluctuations. Neglecting it is the linearization already used in Supplementary~\ref{subsec:calculs_subthreshold_membrane_statistics}, where each synaptic event acts with the driving force at the mean voltage, $U_s=Q_s(E_s-\langle V\rangle)/\langle g\rangle$. The membrane then low-pass filters the conductance fluctuations with the effective time constant $\tau_m^{\mathrm{eff}}=C_m/\langle g\rangle$:
\begin{equation*}
\delta V(t)=\frac{1}{C_m}\int_0^{\infty}e^{-u/\tau_m^{\mathrm{eff}}}\sum_{s}(E_s-\langle V\rangle)\,\delta G_s(t-u)\,du .
\end{equation*}

\textit{Covariance:} Multiplying by $\delta G_s(t)$ and averaging, only the synapses of type $s$ contribute, because excitatory and inhibitory inputs are independent. For Poisson arrivals and exponential conductance transients, $\langle\delta G_s(t)\,\delta G_s(t-u)\rangle=\sigma^2(G_s)\,e^{-|u|/\tau_s}$, with $\sigma^2(G_s)=K_s\nu_sQ_s^2\tau_s/2$ (Campbell's theorem, Eq.~\eqref{eq:mean_variance_conductances}). Hence
\begin{equation*}
\operatorname{Cov}(G_s,V)=\frac{E_s-\langle V\rangle}{C_m}\,\sigma^2(G_s)\int_0^{\infty}e^{-u/\tau_m^{\mathrm{eff}}}\,e^{-u/\tau_s}\,du
=\frac{(E_s-\langle V\rangle)\,\sigma^2(G_s)\,\tau_s}{\langle g\rangle\,(\tau_m^{\mathrm{eff}}+\tau_s)} .
\end{equation*}
Its sign is that of the driving force: an excitatory conductance fluctuation depolarizes the membrane, an inhibitory one hyperpolarizes it. The factor $\tau_s/(\tau_m^{\mathrm{eff}}+\tau_s)$ is the fraction of the conductance fluctuation that the membrane follows.

\textit{Mean shift:} Inserting this covariance into Eq.~\eqref{eq:conductance_voltage_covariance_shift}, and using $U_s^2=Q_s^2(E_s-\langle V\rangle)^2/\langle g\rangle^2$, gives
\begin{equation}
\langle V\rangle-\langle V\rangle_0=-\sum_{s\in\{e,i\}}\frac{\sigma_s^2(V)}{E_s-\langle V\rangle},
\qquad
\sigma_s^2(V)=\frac{K_s\nu_s\,(U_s\tau_s)^2}{2(\tau_m^{\mathrm{eff}}+\tau_s)},
\label{eq:conductance_voltage_shift_identity}
\end{equation}
where $\sigma_s^2(V)$ is the contribution of the synapses of type $s$ to the voltage variance $\sigma^2(V)$ of Supplementary~\ref{subsec:calculs_subthreshold_membrane_statistics}. When $G_s$ fluctuates upwards, the voltage moves towards $E_s$ and the driving force decreases, so that the mean synaptic current is weaker than its factorized estimate: excitation lowers the mean voltage, and inhibition raises it (relative to the factorized estimate).

\textit{Size and consequences:} Divided by its driving force $(E_s-\langle V\rangle)$, each contribution equals $-[\sigma_s(V)/(E_s-\langle V\rangle)]^2$, the square of a small ratio: the shift is of second order in the voltage fluctuations. This is consistent with the result that the effective-time-constant approximation is the leading-order solution of the conductance-based membrane equation, the corrections due to conductance fluctuations being of the same order as those due to the shot-noise nature of the input, which the calculation of Supplementary~\ref{subsec:calculs_subthreshold_membrane_statistics} also neglects~\cite{richardson2005synaptic}. For the default parameters of the TVB implementation of this model, at its asynchronous operating point~\cite{sanz2013virtual,goldman2023comprehensive}, the shift is about $+0.2\,\mathrm{mV}$ (about $-0.1\,\mathrm{mV}$ from excitation and $+0.35\,\mathrm{mV}$ from inhibition), against $\sigma(V)\simeq3.9\,\mathrm{mV}$. At a fixed threshold, such a shift would not be negligible for the firing rate: the operating point lies about two standard deviations below the effective threshold, and the $\mathrm{erfc}$ formula would give a rate about $14\%$ higher. The transfer function can nevertheless remain unaffected, because its effective threshold is fitted to single-neuron simulations with $\langle V\rangle$ computed in the same factorized way (Section~\ref{sec:semi-analytical_TF}). The shift varies smoothly with the inputs, and the quadratic threshold polynomial reproduces it to within $0.01$\,mV over a broad range of inputs. The transfer function can therefore absorb part of the discrepancy, and neglecting the covariance will mostly only change the fitted coefficients. The same estimate applies to the mean voltage of Supplementary~\ref{subsec:calculs_subthreshold_membrane_statistics}.}

\subsubsection{Extension of the Markovian state space}
\label{subsubsec:extension_markovian_state_space}

The state of the network is now defined by $\{m_\gamma, W_\gamma\}$. The variable $W_\gamma(t) = (1/N_\gamma)\sum_i^{N_\gamma} w_{\gamma,i}(t)$ where $w_{\gamma,i}$ is the adaptation of the $i$-th neuron in population $\gamma$. To maintain mathematical tractability, Di Volo et al. \cite{di2019biologically} utilized an adiabatic approximation such that adaptation dynamics are much slower than the mean-field time scale $T$, $(\tau_w \gg T)$.

This allows the adaptation to be treated as stationary within the Markovian transition step $T$ while still evolving over the full duration of the stimulation. The resulting coupled system for a population $\mu$ still consists of the firing rate $\nu_\mu$ and demographic covariance $c_{\mu\nu}$, plus the new macroscopic adaptation term $W_\mu$.
The network behavior is characterized by the transition probability $P_T(\{m_\gamma, W_\gamma\} \mid \{m'_\gamma, W'_\gamma\})$.
By making the same assumption that the population-conditional probabilities are independent beyond the time scale of $T$, we get:
$$
P_T(\{m_\gamma, W_\gamma\} \mid \{m'_\gamma, W'_\gamma\}) = \prod_{\alpha=1}^{P} P_T(m_\alpha, W_\alpha \mid \{m'_\gamma, W'_\gamma\}), \,\ \footnotemark
$$
\footnotetext{\revision{Conditional independence of the populations given the previous state is the same assumption as without adaptation. The condition $\tau_w\gg T$ plays a different role: it allows $W$ to be treated as constant within one bin, so that the transfer function, computed at fixed $W$, applies during the bin. Adaptation and firing remain coupled through the transfer function, and $W$ still evolves from bin to bin.}}

Applying the chain rule, we get:
\begin{equation}
\begin{aligned}
P_T(\{m_\gamma, W_\gamma\} | \{m'_\gamma, W'_\gamma\}) = P_T(\{m_\gamma \}|\{ W_\gamma, m'_\gamma, W'_\gamma\}) \cdot P_T(\{W_\gamma \}| \{m'_\gamma, W'_\gamma\})
\end{aligned}
\label{eq:chain_rule_transition_probability_adaptation}
\end{equation}

\revision{During one bin, $W_\gamma$ changes by a relative amount of order $T/\tau_w$, and this change is set by the current state: the spikes emitted during the bin (through $b$) and the current voltage (through $a$). We therefore describe the transition of $W_\gamma$ by a deterministic Euler step, with $f(W_\gamma,\mathbf m)\equiv-W_\gamma+b\tau_w m_\gamma+a(\bar V_\gamma-E_L)$ the right-hand side of the adaptation equation~\eqref{eq:final_adaptation_ODE}. The transition density of $W_\gamma$ is then a Dirac delta centred on this Euler update.}
The transition probability is therefore written as
\begin{equation*}
\begin{aligned}
P_T(\{W_\gamma \}| \{m'_\gamma, W'_\gamma\}) = \delta\left( W_\gamma - \left[ W'_\gamma + \frac{T}{\tau_w} f(W'_\gamma, m'_\gamma) \right] \right) ,\,\ \footnotemark
\end{aligned}
\end{equation*}
\footnotetext{This means that the probability is 100\% when the new state $W_\gamma$ is equal to the Euler update. As a recall, Euler integration is such that $y(t+T) = y(t) + T \cdot \frac{dy}{dt}$ and $\delta(x-y)$ is nonzero only when $x=y$.}
which states that the activity variables remain stochastic, whereas the slow adaptation variable follows a deterministic Euler update over one time bin. This does not mean that firing and adaptation are uncoupled; rather, it means that adaptation fluctuations within one bin are neglected. \revision{The update is deterministic given the current state, but $W_\gamma$ is not deterministic: it integrates the stochastic activity and inherits its fluctuations. These inherited fluctuations are neglected only later, when the adaptation equation is closed at first order, Eq.~\eqref{eq:final_adaptation_ODE}, which amounts to setting to zero the variance of $W_\gamma$ and its covariances with the activities.}

Thus, Eq.~\eqref{eq:chain_rule_transition_probability_adaptation} becomes:

\begin{equation}
\begin{aligned}
P_T(\{m_\gamma, W_\gamma\} | \{m'_\gamma, W'_\gamma\}) = P_T(\{m_\gamma \}|\{m'_\gamma, W'_\gamma\}) \cdot \delta\!\left(W_\gamma - \left[W'_\gamma + \frac{T}{\tau_w} f(W'_\gamma, m'_\gamma)\right]\right)
\end{aligned}
\end{equation}

\revision{\paragraph{Adaptation currents and time-scale separation:}
The AdEx variable $w$ is a phenomenological adaptation current. Its subthreshold coupling $a$ and its spike-triggered increment $b$ lump together the slow potassium currents responsible for spike-frequency adaptation, and $\tau_w$ is their effective decay time. In cortical neurons, such currents act on time scales from hundreds of milliseconds to tens of seconds. In slices of ferret visual cortex, prolonged firing produces spike-frequency adaptation with time constants of $1$--$10\,\mathrm{s}$ and a slow afterhyperpolarization lasting $12$--$75\,\mathrm{s}$, to which Na$^+$- and Ca$^{2+}$-activated K$^+$ currents contribute~\cite{sanchezvives2000longlasting}. In neocortical neurons, spike-triggered adaptation lasts more than $20\,\mathrm{s}$ and decays over multiple time scales as a power law~\cite{pozzorini2013temporal}. The value $\tau_w=500\,\mathrm{ms}$ used by Di Volo et al.~\cite{di2019biologically} lies in this range. They evaluated changes in that parameter against their parent network. Their reported deterioration for faster adaptation supports a regime-dependent approximation, not a universal separation theorem.

Retaining only the mean adaptation is a separate closure:
the model does not evolve $\operatorname{Var}(W)$ or mixed
rate-adaptation covariances. Its adequacy requires their feedback
on the transfer functions to be small. Slow adaptation and
population averaging can help, but do not guarantee this in
correlated or slowly switching regimes. Near instabilities or in slowly switching regimes, including Up-Down dynamics, activity fluctuations can persist on time scales comparable with adaptation. The condition $\tau_w\gg T$ then does not by itself justify omitting adaptation variance and rate-adaptation covariances. Their feedback must be assessed for the regime of interest; retaining only mean adaptation remains a heuristic closure rather than a guaranteed approximation. Consistently, Di Volo et al. reproduced Up-Down transitions in the mean field by adding an external Ornstein--Uhlenbeck noise to the drive~\cite{di2019biologically} (Section~\ref{sec:DiVolo_Adaptation_MFT}). A consistent extension would add the covariances $c_{\mu W}$ and $c_{WW}$ to the second-order closure; finite-size population theories that keep spike-history effects such as adaptation stochastic offer an alternative~\cite{schwalger2017towards}.

}

\subsubsection{Macroscopic adaptation dynamics}

\revision{
We define the instantaneous population spike train and mean voltage by
\[
\rho_\mu^N(t)=\frac1{N_\mu}\sum_{i=1}^{N_\mu}
\sum_s\delta(t-t_{s,i}),\qquad
\bar V_\mu(t)=\frac1{N_\mu}\sum_i V_{\mu i}(t).
\]
}

At the microscopic level, each excitatory AdEx neuron $k$ obeys:
$$\tau_w\frac{dw_k}{dt} = -w_k + b \tau_w \sum_{t_{sp}(k)} \delta(t - t_{sp}(k)) + a(V_k - E_L)$$

To find the macroscopic dynamics of $W_\mu$, we take the derivative with respect to time and average the right side of the single-neuron equation across the population:

\begin{equation*}
\begin{aligned}
\tau_w\partial_t W_\mu 
&= \frac{1}{N_\mu} \sum_{k=1}^{N_\mu} \left( -w_k + b\tau_w\sum_{t_s \in t_{spike}(k)}\delta(t-t_{s}(k)) + a(V_k - E_L) \right), \\
&= -W_\mu + b\tau_w \rho_\mu^N + a(\bar{V}_\mu - E_L)
\end{aligned}
\end{equation*}

\revision{The population rate $\rho_\mu^N(t)$ is a sum of Dirac pulses. It is not defined at the resolution $T$ of the master equation, whose activity variable counts the spikes in a bin. We therefore integrate over the bin $(t-T,t]$ and divide by $T$, such that:
\begin{equation}
\tau_w\,\frac{W_\mu(t)-W_\mu(t-T)}{T}
= -W_{\mu,T}(t) + b\tau_w\, m_\mu(t) + a\big(\bar{V}_{\mu,T}(t) - E_L\big),
\label{eq:adaptation_bin_average}
\end{equation}
where $m_\mu(t)=\frac{1}{T}\int_{t-T}^{t}\rho^N_\mu(s)\,ds$ is the number of spikes emitted by population $\mu$ in the bin divided by $N_\mu T$, i.e.\ the activity variable of the master equation, and $W_{\mu,T}$ and $\bar V_{\mu,T}$ are the bin averages of $W_\mu$ and of the population-mean voltage $\bar V_\mu$. The derivative approximation replaces the backward difference on the left-hand side by $\partial_tW_\mu(t)$, and the bin averages $W_{\mu,T}$ and $\bar V_{\mu,T}$ by their values at time $t$. Since $W_\mu(t-T)=W_\mu(t)-T\,\partial_tW_\mu(t)+\frac{T^2}{2}\partial_t^2W_\mu(t)+\dots$, these replacements are accurate to first order in $T/\tau_{\mathrm{var}}$, where $\tau_{\mathrm{var}}$ is the time scale over which $W_\mu$ varies. During one bin, the spikes change $W_\mu$ by about $b\,m_\mu T$, whereas $W_\mu\simeq b\tau_w m_\mu$: unless the activity departs strongly from its average, the relative change of $W_\mu$ per bin is of order $T/\tau_w$, i.e.\ $1$--$4\%$ for $T=5$--$20\,\mathrm{ms}$ and $\tau_w=500\,\mathrm{ms}$. This error is of the same order as that of the Euler update of $W$ in the transition kernel (Supplementary~\ref{subsubsec:extension_markovian_state_space}). In this sense, $\partial_tW_\mu$ in Eq.~\eqref{eq:final_adaptation_ODE} is the rate of change of $W_\mu$ at the resolution $T$, driven by the binned activity rather than by the instantaneous spike train:
\begin{equation*}
\tau_w\partial_t W_\mu \simeq -W_{\mu} + b\tau_w m_\mu(t) + a(\bar{V}_{\mu} - E_L).
\end{equation*}}

When $a=0$, the first-moment adaptation equation closes directly, because its right-hand side is affine in $W_\mu$ and $m_\mu$. \revision{Since the Euler update of the transition kernel changes $W_\mu$ by $(T/\tau_w)f$ at rate $1/T$, the drift of $W_\mu$ is $f/\tau_w$, and} $\partial_t\langle W_\mu \rangle = \tau_w^{-1}\langle f(W_\mu, \mathbf{m}) \rangle = -\frac{ \langle W_\mu \rangle}{\tau_w} + b \langle m_\mu \rangle$.
The complete network is still nonlinear because $\nu_\mu$ depends nonlinearly on the network state.

If $a\neq0$, the system cannot be closed directly, because the population-mean voltage is not a state variable of the Markovian system. \revision{It is closed by identifying $\bar V_\mu$ with the stationary mean free membrane potential of Eq.~\eqref{eq:average_membrane_equation_DiVolo_adaptation}, evaluated at the current mean rates and adaptation, $\langle V(\nu_e,\nu_i,W_\mu)\rangle$. This is a heuristic, quasi-static closure rather than a controlled expansion: it assumes that the mean membrane potential relaxes within one bin ($\tau_m^{\mathrm{eff}}$ of a few milliseconds, not longer than $T$), it ignores the contribution of spikes and resets to the population-mean voltage, and it evaluates the voltage at the mean rates, i.e.\ it neglects the curvature correction $\frac{1}{2}\partial_\lambda\partial_\eta\langle V\rangle\,c_{\lambda\eta}$, which is of order $N^{-1}$.}
Finally, we get:
\begin{equation}
\begin{aligned}
\tau_w\partial_t \langle W_\mu\rangle 
&= -\langle W_\mu \rangle + b\tau_w \langle m_\mu\rangle+ \langle a( \bar{V}_{\mu}(\nu_e, \nu_i, W_\mu) - E_L)\rangle \\
&= -\langle W_\mu \rangle + b\tau_w \nu_\mu+  a(\langle V(\nu_e, \nu_i, W_\mu)\rangle - E_L)
\end{aligned}
\label{eq:final_adaptation_ODE}
\end{equation}

One may notice that this is not the final form found in Di Volo et al. 2019 \cite{di2019biologically}. They explicitly used first-order (mean-only) for the equation of $W$, assuming that its dynamics is not strongly affected by fluctuations. Because it is both spatially averaged (across $N$) and temporally smoothed (across $\tau_w$), the actual variance of the macroscopic variable $W_\mu$ around its expected value $\langle W_\mu \rangle$ is considered negligible, so we get $W_\mu \simeq \langle W_\mu \rangle$. \revision{We stress that evaluating transfer functions at that mean, without evolving adaptation variance or mixed covariances, is not an implication, but an additional mean-only adaptation closure used by Di Volo et al.~\cite{di2019biologically}.}

\subsubsection{Final ODE system}

Using the same master equation derivation as in Section~\ref{sec:MFT_NN_formalism}, but with transfer functions now depending on $W$, the final closed system reads as below.

The first jump moment becomes:
\begin{equation}
\begin{aligned}
a_\mu({ \{ \langle m_\gamma \rangle, \langle W_\gamma \rangle \} }) 
&= \prod_{\alpha = 1}^{P} \int dW'_\alpha \int (m'_\mu - \langle m_\mu \rangle) \mathcal{W}(\{m'_\gamma, W'_\gamma\} \mid \{\langle m_\gamma \rangle, \langle W_\gamma \rangle\}) dm'_\alpha\\
&= \frac{1}{T} \prod_{\alpha = 1}^{P} \int dW'_\alpha \int dm'_\alpha \, (m'_\mu - \langle m_\mu \rangle) P_T( m'_\alpha |\{ \langle m_\gamma \rangle, \langle W_\gamma \rangle \}) \\
&\quad \cdot \delta \left(W'_\alpha - \left[\langle W_\alpha \rangle + \frac{T}{\tau_w} f(\langle W_\alpha \rangle, \langle m_\alpha \rangle )\right]\right) \\
&= \frac{1}{T} \int dm_\mu' (m'_\mu - \langle m_\mu \rangle) P_T(m'_\mu \mid \{\langle m_\gamma \rangle, \langle W_\gamma \rangle\}) \\
&\quad \times \left[ \prod_{\alpha \neq \mu} \int dm'_\alpha P_T(m'_\alpha \mid \{\langle m_\gamma \rangle, \langle W_\gamma \rangle\}) \right] \times \left[ \prod_{\alpha = 1}^{P} \int dW'_\alpha \delta\!\left(W'_\alpha - \left[\dots\right]\right) \right], \,\ \footnotemark \\
&= \frac{1}{T} \left[ \int m'_\mu P_T(m'_\mu \mid \{\langle m_\gamma \rangle, \langle W_\gamma \rangle\}) dm'_\mu - \langle m_\mu \rangle \int  P_T(m'_\mu \mid \{\langle m_\gamma \rangle, \langle W_\gamma \rangle\}) dm'_\mu \right]\\
&= \frac{1}{T} (\mathcal{F}_\mu - \langle m_\mu \rangle). \\
\end{aligned}
\end{equation}
\footnotetext{When ($\alpha \neq \mu$), the integrals are exactly 1. This is also the case for all $\alpha$ with the $\delta$ function.}

The second jump moment becomes:
\begin{equation}
\begin{aligned}
a_{\mu\nu}(\{ \langle m_\gamma \rangle, \langle W_\gamma \rangle \})
&= \prod_{\alpha = 1}^{P} \int dW'_\alpha \int dm'_\alpha \, (m'_\mu - \langle m_\mu \rangle) (m'_\nu - \langle m_\nu \rangle) \mathcal{W}(\{m'_\gamma, W'_\gamma\} \mid \{\langle m_\gamma \rangle, \langle W_\gamma \rangle\}) \\
&= \frac{1}{T} \prod_{\alpha = 1}^{P} \int dm'_\alpha (m'_\mu - \langle m_\mu \rangle)(m'_\nu - \langle m_\nu \rangle) P_T(m'_\alpha \mid \{\langle m_\gamma \rangle, \langle W_\gamma \rangle\}) \\
&\quad \times \left[ \prod_{\alpha = 1}^{P} \int dW'_\alpha \delta(W'_\alpha - [\dots]) \right] \\
&= \frac{1}{T} \int^{\infty}_{-\infty} dm'_\mu \int^{\infty}_{-\infty} \left. dm'_\nu \right|_{\nu \neq \mu}  (m'_\mu-\langle m_\mu \rangle)(m'_\nu-\langle m_\nu \rangle) P_T(m'_\mu \mid \cdots) P_T(m'_\nu \mid \cdots)\\
&= \frac{\delta_{\mu\nu}}{T} \int_{-\infty}^{\infty} dm'_\mu (m'^2_\mu + \mathcal{F}^2_\mu - \mathcal{F}^2_\mu - 2m'_\mu \langle m_\mu \rangle + \langle m_\mu \rangle^2) P_T(m'_\mu \mid \{\langle m_\gamma \rangle, \langle W_\gamma \rangle\}) \\
&\quad + \frac{(1 - \delta_{\mu\nu})}{T} (\mathcal{F}_\mu - \langle m_\mu \rangle)(\mathcal{F}_\nu - \langle m_\nu \rangle) \\
&= \frac{1}{T} \left[ \delta_{\mu\nu} \frac{\mathcal{F}_\mu(1/T - \mathcal{F}_\mu)}{N_\mu} + (\mathcal{F}_\mu - \langle m_\mu \rangle)(\mathcal{F}_\nu - \langle m_\nu \rangle) \right],
\qquad\mathcal F_\mu=\mathcal F_\mu(\boldsymbol\nu,\mathbf W). \\
\end{aligned}
\label{eq:final_diffusion_moments_with_adaptation}
\end{equation}

\revision{Equation~\eqref{eq:final_diffusion_moments_with_adaptation} gives the second jump moment at the mean state. As without adaptation, its average over the activity distribution differs from this value by a curvature term of the same order as the retained terms. The same diagonal/off-diagonal distinction as in
Eq.~\eqref{eq:step_drift_diffusion_moment_functions} applies.
Under the additional mean-only adaptation closure, the rate
covariance derivation applies with derivatives taken at fixed
retained $W_\mu$, subject to the approximation conditions in
Supplementary~\ref{subsubsec:covariance_matrix}.
This restriction follows from omitting adaptation moments,
not merely from freezing adaptation during a single bin.} Therefore, using the same derivation as in Section~\ref{sec:MFT_NN_formalism} and with Eq.~\eqref{eq:final_adaptation_ODE}, we get the final set of differential equations governing the Markovian system:
\begin{empheq}[box=\doubleboxeq]{equation}
\begin{aligned}
T\partial_t \nu_\mu
&= ( \mathcal{F}_\mu - \nu_\mu) + \frac{1}{2}\partial_\lambda\partial_\eta \mathcal{F}_\mu c_{\lambda\eta} \\
T\partial_t c_{\mu\nu}
&= \delta_{\mu\nu} A_{\mu\mu}^{-1} \revision{+ \partial_\lambda \mathcal{F}_\mu\,\partial_\eta \mathcal{F}_\nu\,c_{\lambda\eta} - c_{\mu\nu} + T^2\,\partial_t\nu_\mu\,\partial_t\nu_\nu} \\
\tau_w\partial_t W_\mu
&= -W_\mu + b\tau_w\nu_\mu + a(\langle V(\nu_e, \nu_i, W_\mu)\rangle - E_L)
\end{aligned}
\label{eq:final_ODEs_adaptation}
\end{empheq}
\revision{with $\mathcal{F}_\mu=\mathcal{F}_\mu(\nu_e,\nu_i,W_\mu)$. Di Volo et al.~\cite{di2019biologically} used the published covariance equation~\eqref{eq:covariance_published_closure} in this system.}
Adaptation is no longer hidden inside a stationary transfer function only: it becomes an explicit mesoscopic dynamical variable coupled to the population rates and covariances. Thus, the system becomes a second-order closure coupled to a first-order deterministic adaptation equation.

\revision{

\subsubsection{Adaptation variance and rate-adaptation covariances}
\label{subsubsec:adaptation_second_order_extension}

System~\eqref{eq:final_ODEs_adaptation} treats the activities at second order (means and covariances), but the adaptation at first order (mean only). As noted in Supplementary~\ref{subsubsec:extension_markovian_state_space}, $W$ integrates the spikes of the population and therefore fluctuates; the first-order treatment sets its variance and its covariances with the activities to zero. This subsection sketches what the system becomes when these moments are kept. It is a putative analytical extension: it uses the same master equation and the same approximations as the covariances of the activities, but it has not been tested against simulations. $\operatorname{Var}(W)$ is the variance, across realizations of the network, of the population-averaged adaptation $W=(1/N_e)\sum_i w_i$; it does not describe how the adaptation currents $w_i$ differ from neuron to neuron. We stress again here that "Second order" refers to the moments that are kept: the equations remain first-order differential equations.
}

\revision{
\paragraph{State and moments:}
As in the RS-FS network, only excitatory neurons adapt ($a_i=b_i=0$), and we write $W\equiv W_e$, $a\equiv a_e$ and $b\equiv b_e$. The state of the network is $\mathbf{x}=(m_e,m_i,W)$. Its mean is $(\nu_e,\nu_i,\langle W\rangle)$, and its covariance matrix contains the three covariances of the activities and three new moments:
\begin{equation*}
\begin{pmatrix}
c_{ee} & c_{ei} & c_{eW}\\
c_{ei} & c_{ii} & c_{iW}\\
c_{eW} & c_{iW} & c_{WW}
\end{pmatrix},
\qquad
c_{\mu W}=\operatorname{Cov}(m_\mu,W),\qquad c_{WW}=\operatorname{Var}(W).
\end{equation*}
The indices $\lambda,\eta,\mu,\nu\in\{e,i\}$ refer to the activities, as before, and the indices $A,B,C,D\in\{e,i,W\}$ to all three variables; repeated indices are summed. The first-order treatment of $W$ considered the setting $c_{eW}=c_{iW}=c_{WW}=0$.
}

\revision{

\paragraph{Transition rule:}

We denote the states before and after a transition by $\mathbf{x}_k$ and $\mathbf{x}_{k+1}$. Here $k$ indexes transitions, not fixed observation times: the continuous-time master equation below applies this bin-inspired update at rate $1/T$. The number $K_{\mu,k+1}$ of neurons of population $\mu$ that fire during the bin is binomial, and
\begin{equation}
K_{\mu,k+1}\sim\mathcal{B}\big(N_\mu,\,T\mathcal{F}_\mu(\mathbf{x}_k)\big),
\qquad
m_{\mu,k+1}=\frac{K_{\mu,k+1}}{N_\mu T}.
\label{eq:adaptation_ext_activity_update}
\end{equation}
The only difference is that $\mathcal{F}_e$ now also depends on $W_k$, which here is retained as a fluctuating state variable. We rewrite Eq.~\eqref{eq:adaptation_bin_average} as an update. Multiplied by $T/\tau_w$, it reads 

\begin{equation}
\begin{aligned}
W(t)
&=W(t-T)+\frac{T}{\tau_w}\Big[-W_T(t)+a\big(\bar V_T(t)-E_L\big)\Big]+bT\,m_e(t),\\
W_{k+1}
&=\underbrace{W_k+\frac{T}{\tau_w}\Big[-W_k+a\big(\mathcal{V}(\mathbf{x}_k)-E_L\big)\Big]}
\;+\; \underbrace{\frac{b}{N_e}\,K_{e,k+1}} .
\end{aligned}
\label{eq:adaptation_ext_W_update}
\end{equation}
where $W_T$ and $\bar V_T$ are averages over the bin $(t-T,t]$. First, we approximate $W$ changes by a relative amount of order $T/\tau_w$ within one bin, so that $W_T$ can be replaced by $W_k$ at the accuracy of Eq.~\eqref{eq:adaptation_bin_average}. Second, during the bin the neurons receive the inputs set by $\mathbf{x}_k$, and the quasi-stationary approximation used for the transfer function gives the mean voltage $\langle V\rangle$ of Eq.~\eqref{eq:average_membrane_equation_DiVolo_adaptation} evaluated at $\mathbf{x}_k$. We write this function of the state as $\mathcal{V}(\mathbf{x})=\langle V(m_e,m_i,W)\rangle$.

The first part of Eq.~\eqref{eq:adaptation_ext_W_update} is the Euler step of the leak and of the subthreshold coupling. The second part adds the spike-triggered increments of the bin: each of the $K_{e,k+1}$ spikes increases the adaptation of its neuron by $b$, hence the population average $W$ by $b/N_e$. We stress that even when $\mathbf{x}_k$ is known, $W_{k+1}$ is defined by the new activity $m_{e,k+1}$ and therefore $W_{k+1}$ is random and is correlated with $m_{e,k+1}$.

The Euler kernel of Supplementary~\ref{subsubsec:extension_markovian_state_space} computes the increment with the activity of the previous bin, $bT\,m_{e,k}$, so that $W_{k+1}$ is fixed once $\mathbf{x}_k$ is known (the Dirac delta of that kernel). Both versions give the same mean adaptation equation~\eqref{eq:final_adaptation_ODE} at the accuracy of Eq.~\eqref{eq:adaptation_bin_average}, but not the same fluctuations. We use Eq.~\eqref{eq:adaptation_ext_W_update} because it follows Eq.~\eqref{eq:adaptation_bin_average} and the network: the spikes that increment $W$ are those counted in the activity of the same bin, and the adaptation reacts to them in the next bin, like the synaptic coupling.\footnote{With the previous-bin increment, the adaptation reacts one bin later than the synaptic coupling. For one population with $a=0$ and an affine transfer function of slope $\Delta$, this extra delay changes $c_{mW}$ at leading order (by about $bT\,c_{mm}$), and it makes the fluctuations unstable when $g_w>\Delta+(1-\Delta)\tau_w/T$ ($g_w$ is defined below), although the mean dynamics remains stable.}

\paragraph{Mean and Covariance of one transition:}
We now compute the mean and the covariance of the new state $\mathbf{x}_{k+1}$ for a given current state $\mathbf{x}_k=\mathbf{x}$.

For the activities, equation~\eqref{eq:supp_conditional_moments} gives the conditional mean $\mathcal{F}_\mu(\mathbf{x})$ of $m_{\mu,k+1}$ and its conditional variance $A^{-1}_{\mu\mu}(\mathbf{x})=\mathcal{F}_\mu(1/T-\mathcal{F}_\mu)/N_\mu$. The two populations are drawn independently, so $m_{e,k+1}$ and $m_{i,k+1}$ have no conditional covariance.

For the adaptation: in Eq.~\eqref{eq:adaptation_ext_W_update}, the first part is a fixed number once $\mathbf{x}$ is known, and the only random term is $bT\,m_{e,k+1}$. Taking the conditional mean therefore simply replaces the new activity $m_{e,k+1}$ by its expected value $\mathcal{F}_e(\mathbf{x})$, while the current state $\mathbf{x}$ stays as it is. We call $\mathcal{G}(\mathbf{x})$ the expected new state:
\begin{equation}
\mathcal{G}_\mu(\mathbf{x})=\mathcal{F}_\mu(\mathbf{x}),
\qquad
\mathcal{G}_W(\mathbf{x})=W+\frac{T}{\tau_w}\Big[-W+b\tau_w\,\mathcal{F}_e(\mathbf{x})+a\big(\mathcal{V}(\mathbf{x})-E_L\big)\Big].
\label{eq:adaptation_ext_G}
\end{equation}
$\mathcal{G}_W$ is thus the Euler step of Eq.~\eqref{eq:final_adaptation_ODE}, with the activity of the coming bin replaced by its expected value. Since $W_{k+1}$ fluctuates only through $bT\,m_{e,k+1}$, its conditional variance is $(bT)^2$ times that of $m_{e,k+1}$, its conditional covariance with $m_{e,k+1}$ is $bT$ times the variance of $m_{e,k+1}$, and its conditional covariance with $m_{i,k+1}$ is zero. Let $\Sigma$ collect these conditional covariances with rows and columns ordered $(e,i,W)$:
\begin{equation}
\Sigma(\mathbf{x})=
\begin{pmatrix}
A^{-1}_{ee} & 0 & bT\,A^{-1}_{ee}\\
0 & A^{-1}_{ii} & 0\\
bT\,A^{-1}_{ee} & 0 & b^2T^2A^{-1}_{ee}
\end{pmatrix}.
\label{eq:adaptation_ext_Sigma}
\end{equation}
$\Sigma$ is the noise created by one transition when the current state is known; it plays the role of $\delta_{\mu\nu}A^{-1}_{\mu\mu}$ in Eq.~\eqref{eq:covariance_consistent_closure}. This is why it contains $A^{-1}_{ee}$ and $A^{-1}_{ii}$, but not $c_{ee}$, $c_{ei}$ or $c_{ii}$: the latter describe the fluctuations of the current state, and they enter through $\mathcal{G}$ below. It is also why only the factor $bT$ appears for $W$: the leak and subthreshold terms of Eq.~\eqref{eq:adaptation_ext_W_update} carry no transition noise.

}

\revision{
\paragraph{Exact moment equations:}
In the master equation, transitions occur at rate $1/T$. The mean of any function $\varphi$ of the state therefore changes at rate $1/T$ times the average change produced by one transition:
\begin{equation*}
T\partial_t\langle\varphi(\mathbf{x})\rangle=\big\langle\mathbb{E}[\varphi(\mathbf{x}_{k+1})\mid\mathbf{x}_k=\mathbf{x}]\big\rangle-\langle\varphi(\mathbf{x})\rangle ,
\end{equation*}
which is the identity used to derive Eq.~\eqref{eq:covariance_exact_redraw}. With $\varphi=x_A$, it gives the mean equations
\begin{equation}
T\partial_t\langle x_A\rangle=\langle\mathcal{G}_A(\mathbf{x})\rangle-\langle x_A\rangle .
\label{eq:adaptation_ext_exact_mean}
\end{equation}
With $\varphi=x_Ax_B$, and since the conditional mean of a product is the conditional covariance plus the product of the conditional means, $\mathbb{E}[x_{A,k+1}x_{B,k+1}\mid\mathbf{x}]=\Sigma_{AB}(\mathbf{x})+\mathcal{G}_A(\mathbf{x})\mathcal{G}_B(\mathbf{x})$, it gives
\begin{equation*}
T\partial_t\langle x_Ax_B\rangle=\langle\Sigma_{AB}(\mathbf{x})\rangle+\langle\mathcal{G}_A(\mathbf{x})\,\mathcal{G}_B(\mathbf{x})\rangle-\langle x_Ax_B\rangle .
\end{equation*}
We then use $c_{AB}=\langle x_Ax_B\rangle-\langle x_A\rangle\langle x_B\rangle$, write $\langle\mathcal{G}_A\mathcal{G}_B\rangle=\operatorname{Cov}(\mathcal{G}_A,\mathcal{G}_B)+\langle\mathcal{G}_A\rangle\langle\mathcal{G}_B\rangle$, and replace $\langle\mathcal{G}_A\rangle$ by $\langle x_A\rangle+T\partial_t\langle x_A\rangle$ (Eq.~\eqref{eq:adaptation_ext_exact_mean}). All products of means then cancel, except one, as for Eq.~\eqref{eq:covariance_exact_redraw}:
\begin{equation}
T\partial_t c_{AB}=\langle\Sigma_{AB}(\mathbf{x})\rangle+\operatorname{Cov}\big(\mathcal{G}_A(\mathbf{x}),\mathcal{G}_B(\mathbf{x})\big)-c_{AB}+T^2\,\partial_t\langle x_A\rangle\,\partial_t\langle x_B\rangle .
\label{eq:adaptation_ext_exact_cov}
\end{equation}
Both equations are exact consequences of the transition~\eqref{eq:adaptation_ext_activity_update}-\eqref{eq:adaptation_ext_W_update}. The four terms of Eq.~\eqref{eq:adaptation_ext_exact_cov} have the same meaning as for the activities: noise created by the transition; fluctuations of the current state, transmitted to the next state; erasure of the current state; and spreading of the realizations while the mean moves. For $A=W$, Eq.~\eqref{eq:adaptation_ext_exact_mean} reads $\tau_w\partial_t\langle W\rangle=-\langle W\rangle+b\tau_w\langle\mathcal{F}_e\rangle+a\big(\langle\mathcal{V}\rangle-E_L\big) \, \footnotemark$. The same equation for $A=e$ gives $\langle\mathcal{F}_e\rangle=\nu_e+T\partial_t\nu_e$; the difference with $\nu_e$ only shifts the spike input of $W$ by one bin, and we neglect it at the accuracy of Eq.~\eqref{eq:adaptation_bin_average}. For $a=0$, this is Eq.~\eqref{eq:final_adaptation_ODE}, which contains no covariance: the adaptation moments do not change the mean adaptation equation directly, but they change the activities.
\footnotetext{For the sake of simplicity, we neglect the transient correction term $+a\big(\tfrac{1}{2}\partial_\lambda\partial_\eta\mathcal{V} \,c_{\lambda\eta} +\partial_\lambda\partial_W\mathcal{V}\,c_{\lambda W}\big)$ that requires an additional approximation concerning the rate
variation over $T$.}

}

\revision{

\paragraph{Second-order closure:}
As for Eq.~\eqref{eq:covariance_consistent_closure}, we expand the averages around the mean state $(\nu_e,\nu_i,\langle W\rangle)$ and keep terms up to the covariances. All functions and derivatives below are evaluated at this mean state; $\partial_\lambda$ denotes the derivative with respect to the activity $\lambda$ at fixed $W$, and $\partial_W$ the derivative with respect to $W$ at fixed activities. Three approximations are made, as for the activities.
\begin{itemize}
\item \emph{Means:} With $\delta x_C=x_C-\langle x_C\rangle$, the second-order Taylor expansion $\mathcal{F}_\mu(\mathbf{x})\simeq\mathcal{F}_\mu+\partial_C\mathcal{F}_\mu\,\delta x_C+\frac12\partial_C\partial_D\mathcal{F}_\mu\,\delta x_C\delta x_D$, averaged with $\langle\delta x_C\rangle=0$ and $\langle\delta x_C\delta x_D\rangle=c_{CD}$, gives
\begin{equation*}
\langle\mathcal{F}_\mu\rangle\simeq\mathcal{F}_\mu+\tfrac12\partial_\lambda\partial_\eta\mathcal{F}_\mu\,c_{\lambda\eta}
+\partial_\lambda\partial_W\mathcal{F}_\mu\,c_{\lambda W}+\tfrac12\partial_W^2\mathcal{F}_\mu\,c_{WW}.
\end{equation*}
The first correction is the curvature correction already present for the activities. The two others are new: when $m_e$ and $W$ fluctuate together, the curved transfer function does not average to its value at the mean. They vanish for $\mu=i$, since $\mathcal{F}_i$ does not depend on $W$. The same expansion applies to $\langle\mathcal{V}\rangle$, with $\partial_W^2\mathcal{V}=0$ since $\mathcal{V}$ is affine in $W$.
\item \emph{Transition noise:} $\langle\Sigma_{AB}(\mathbf{x})\rangle\simeq\Sigma_{AB}$, evaluated at the mean state, as $\langle A^{-1}_{\mu\mu}\rangle\simeq A^{-1}_{\mu\mu}$ for the activities.
\item \emph{Transmitted fluctuations:} Linearizing $\mathcal{G}$ around the mean state, $\mathcal{G}_A(\mathbf{x})\simeq\mathcal{G}_A+J_{AC}\,\delta x_C$, with the gain matrix $J_{AC}=\partial\mathcal{G}_A/\partial x_C$, gives $\operatorname{Cov}(\mathcal{G}_A,\mathcal{G}_B)\simeq J_{AC}J_{BD}\,c_{CD}$, exactly as $\operatorname{Cov}(\mathcal{F}_\mu,\mathcal{F}_\nu)\simeq\partial_\lambda\mathcal{F}_\mu\,\partial_\eta\mathcal{F}_\nu\,c_{\lambda\eta}$ for the activities.
\end{itemize}
The covariance equation therefore keeps the form of Eq.~\eqref{eq:covariance_consistent_closure}, now with indices running over $\{e,i,W\}$:
\begin{equation}
T\partial_t c_{AB}=\Sigma_{AB}+J_{AC}J_{BD}\,c_{CD}-c_{AB}+T^2\,\partial_t\langle x_A\rangle\,\partial_t\langle x_B\rangle .
\label{eq:adaptation_ext_closure}
\end{equation}
The entries of the gain matrix follow from Eq.~\eqref{eq:adaptation_ext_G}:
\begin{itemize}
\item $J_{\mu\lambda}=\partial_\lambda\mathcal{F}_\mu$ is the gain matrix of the activities used so far.
\item $J_{eW}=\partial_W\mathcal{F}_e<0$: a larger adaptation lowers the excitatory rate. Since $\mathcal{F}_i$ does not depend of $W$, we have $J_{iW}=0$. Since $W$ enters $\mathcal{F}_e$ only through $\langle V\rangle$ (Section~\ref{sec:DiVolo_Adaptation_MFT}), we have $\partial_W\mathcal{F}_e=\frac{-1}{\langle g\rangle}\, \frac{d\mathcal{F}_e}{d\langle V\rangle}$, and the new derivatives require no new fit.
\item $J_{W\lambda}=\partial_\lambda\mathcal{G}_W=\frac{T}{\tau_w}\big(b\tau_w\,\partial_\lambda\mathcal{F}_e+a\,\partial_\lambda\mathcal{V}\big)=bT\,\partial_\lambda\mathcal{F}_e+\frac{aT}{\tau_w}\,\partial_\lambda\mathcal{V}$: a fluctuation of the current activities changes the expected number of spikes in the coming bin, and the subthreshold drive of $W$.
\item $J_{WW}=\partial_W\mathcal{G}_W=1+\frac{T}{\tau_w}\big(-1+b\tau_w\,\partial_W\mathcal{F}_e+a\,\partial_W\mathcal{V}\big)$. With $\partial_W\mathcal{V}=-1/\langle g\rangle$ (Eq.~\eqref{eq:average_membrane_equation_DiVolo_adaptation}) and the dimensionless loop gain $g_w\equiv-b\tau_w\,\partial_W\mathcal{F}_e\geq0$,
\begin{equation*}
J_{WW}=1-\frac{T}{\tau_w^{\mathrm{eff}}},
\qquad
\tau_w^{\mathrm{eff}}=\frac{\tau_w}{1+g_w+a/\langle g\rangle}.
\end{equation*}
The quantity $\tau_w^{\mathrm{eff}}$ describes the direct response
of the expected adaptation update to $W$, with the current
activities held fixed. It shows that a fluctuation of $W$ is largely kept from one bin to the next. It decays through the leak of $W$, but also because a larger $W$ lowers the firing rate, hence the spikes that feed $W$ ($g_w$), and lowers the voltage ($a/\langle g\rangle$); its time constant is therefore $\tau_w^{\mathrm{eff}}<\tau_w$. The gain $g_w$ measures the strength of spike-triggered adaptation: a sustained increase $\delta\nu_e$ builds up $\delta W=b\tau_w\,\delta\nu_e$, which lowers $\mathcal{F}_e$ by $g_w\,\delta\nu_e$.
\end{itemize}

\paragraph{Final system:}
Writing Eq.~\eqref{eq:adaptation_ext_closure} for the six covariances, and inserting the expanded means in Eq.~\eqref{eq:adaptation_ext_exact_mean}, gives the putative second-order system for one RS-FS node ($\mu,\nu\in\{e,i\}$; $\lambda,\eta\in\{e,i\}$ summed):
\begin{empheq}[box=\fbox]{equation}
\begin{aligned}
T\partial_t\nu_\mu
&=\mathcal{F}_\mu-\nu_\mu+\tfrac{1}{2}\partial_\lambda\partial_\eta\mathcal{F}_\mu\,c_{\lambda\eta}+\partial_\lambda\partial_W\mathcal{F}_\mu\,c_{\lambda W}+\tfrac{1}{2}\partial_W^2\mathcal{F}_\mu\,c_{WW},\\[3pt]
\tau_w\partial_t\langle W\rangle
&=-\langle W\rangle+b\tau_w\nu_e+a\big(\mathcal{V}-E_L\big)+a\big(\tfrac{1}{2}\partial_\lambda\partial_\eta\mathcal{V}\,c_{\lambda\eta}+\partial_\lambda\partial_W\mathcal{V}\,c_{\lambda W}\big),\\[3pt]
T\partial_t c_{\mu\nu}
&=\delta_{\mu\nu}A^{-1}_{\mu\mu}+J_{\mu\lambda}J_{\nu\eta}\,c_{\lambda\eta}-c_{\mu\nu}+T^2\partial_t\nu_\mu\,\partial_t\nu_\nu\\
&\quad+\big(J_{\mu\lambda}J_{\nu W}+J_{\mu W}J_{\nu\lambda}\big)c_{\lambda W}+J_{\mu W}J_{\nu W}\,c_{WW},\\[3pt]
T\partial_t c_{\mu W}
&=\delta_{\mu e}\,bT A^{-1}_{ee}+J_{\mu\lambda}\big(J_{W\eta}\,c_{\lambda\eta}+J_{WW}\,c_{\lambda W}\big)+J_{\mu W}\big(J_{W\eta}\,c_{\eta W}+J_{WW}\,c_{WW}\big)\\
&\quad-c_{\mu W}+T^2\partial_t\nu_\mu\,\partial_t\langle W\rangle,\\[3pt]
T\partial_t c_{WW}
&=b^2T^2A^{-1}_{ee}+J_{W\lambda}J_{W\eta}\,c_{\lambda\eta}+2J_{WW}J_{W\lambda}\,c_{\lambda W}-\big(1-J_{WW}^2\big)c_{WW}+T^2\big(\partial_t\langle W\rangle\big)^2 .
\end{aligned}
\label{eq:adaptation_ext_system}
\end{empheq}
These are nine equations instead of six. In the first three, we recover the original terms of Eq.~\eqref{eq:final_ODEs_adaptation}, with the additional terms on their right-hand side. The last two equations are new. Read from top to bottom:
\begin{itemize}
\item In the mean activities (1), the new moments add a curvature correction to $\mathcal{F}_e$ only.
\item In the adaptation equation (2), the additional terms are the curvature of the voltage proxy, which Eq.~\eqref{eq:final_ODEs_adaptation} neglects; they vanish for $a=0$.
\item In the covariances of the activities (3), the additional terms transmit the fluctuations of $W$, and its correlations with the activities, to the activities through $J_{eW}$.
\item In $c_{\mu W}$ (4), the source $bT A^{-1}_{ee}$ expresses that the spikes of a bin increase both $m_e$ and $W$; the other terms transmit the current fluctuations to the next state.
\item In $c_{WW}$ (5), the sources are the spike noise of each transition, $b^2T^2A^{-1}_{ee}$, and the fluctuations of the activities, which change the expected number of spikes, $J_{W\lambda}J_{W\eta}c_{\lambda\eta}$. Since $1-J_{WW}^2\simeq2T/\tau_w^{\mathrm{eff}}$, the variance of $W$ is forgotten only over about $\tau_w^{\mathrm{eff}}/2$, i.e.\ over many bins: $W$ accumulates the small noise of successive transitions.
\end{itemize}
With $c_{eW}=c_{iW}=c_{WW}=0$, the system reduces to Eq.~\eqref{eq:final_ODEs_adaptation} (up to the voltage curvature terms). As for the activities, stationary solutions exist only if all eigenvalues of the gain matrix $J_{AB}$ lie inside the unit circle.

\paragraph{Expected size of the new terms:}
This is only indicative; further analysis and numerical simulations are required for an accurate estimate of the weight of each term. $W$ sums the increments of the spikes emitted by $N_e$ neurons over a time of order $\tau_w$, i.e.\ of about $N_e\nu_e\tau_w$ spikes. If these spikes were independent Poisson events, Campbell's theorem, as for the conductances in Eq.~\eqref{eq:mean_variance_conductances}, would give a relative fluctuation $\sqrt{c_{WW}}/\langle W\rangle=(2N_e\nu_e\tau_w)^{-1/2}$, i.e.\ $0.8\%$ for $N_e=8000$, $\nu_e=2\,\mathrm{Hz}$ and $\tau_w=500\,\mathrm{ms}$.\footnote{With recurrent coupling, for one population with $a=0$ and an affine transfer function of slope $\Delta$, Eq.~\eqref{eq:adaptation_ext_system} gives at a stationary state, to leading order in $T/[(1-\Delta)\tau_w]$, $c_{WW}/\langle W\rangle^2\simeq[2N\nu\tau_w(1-\Delta)(1-\Delta+g_w)]^{-1}$: recurrent excitation increases the fluctuations of $W$, and the adaptation feedback decreases them.} This is a reference estimate, not an
exact result for the proposed binomial transition. The new terms of the mean equations should then be smaller than the existing curvature correction by a factor of order $T/\tau_w$, and their effect on the covariances of the activities should be of order $\tau_c/\tau_w$, where $\tau_c$ is the correlation time of the activity fluctuations ($\tau_c=T/(1-\Delta)$ for one population). In asynchronous regimes, $\tau_c$ is a few milliseconds, and we expect changes of a few percent at most. The extension may matter close to a slow instability of the activities, such as the onset of Up-Down dynamics, where $\tau_c$ approaches $\tau_w$. This is, however, also where the second-order closure is least reliable (Supplementary~\ref{subsubsec:covariance_closure_benchmarks}), and where bimodal distributions escape any single-peaked closure.

Whether the extension is worth its cost (three more equations per node, and the derivatives $\partial_W\mathcal{F}_e$, $\partial_\lambda\partial_W\mathcal{F}_e$ and $\partial_W^2\mathcal{F}_e$) is a numerical question. The six- and nine-equation systems could be compared with Monte Carlo simulations of the Markov chain defined by Eqs.~\eqref{eq:adaptation_ext_activity_update}-\eqref{eq:adaptation_ext_W_update}, which isolate the error of the closure. It should also be compared with the parent AdEx network, with activities counted in bins of length $T$ and $W$ averaged over neurons, which measures the error of the whole reduction. Natural observables are the moments themselves, the mean rates, transient responses and Up-Down statistics, while varying the drive, $b$ and $\tau_w$. The practical benefit would be an improvement exceeding sampling uncertainty at an acceptable additional computational cost; no such improvement has yet been established.

}

\section{Complementary neural-modeling frameworks: large-scale spiking networks and population-level reductions}
\label{sec:complementary_neural_modeling_appendix}

\subsection{Large-scale spiking network models}
\label{subsec:appendix_background_snn_capabilities}

It would be misleading to present detailed neural-network models as merely local toy systems. Recent years have seen impressive progress in large-scale mechanistic simulation. 

In \textit{C. elegans}, Zhao et al. \cite{zhao2024integrative} introduced BAAIWorm, a recent OpenWorm-inspired integrative model \cite{sarma2018openworm} that couples a biophysically detailed multicompartment neural model to a 3D body-environment simulation. In contrast to earlier OpenWorm-style integrations, which remained largely open-loop, BAAIWorm implements closed-loop brain--body--environment interactions: sensory neurons receive input from the simulated environment, while motor neurons drive muscle activation and body deformation. This architecture produces worm-like zigzag locomotion toward an attractor. However, despite being an impressive organism-scale simulation, its current brain component models a restricted sensory-locomotor circuit of 136 neurons rather than the complete set of 302 neurons of the adult hermaphrodite nervous system. It should therefore be understood as an integrative brain--body--environment model of a specific behavioral regime, rather than as a complete whole-nervous-system digital twin of \textit{C. elegans}.

For \textit{Drosophila}, the scale is even more striking. Shiu et al. \cite{shiu2024drosophila} built a leaky integrate-and-fire (LIF) model of the adult fly brain using the FlyWire connectome, comprising 127,400 proofread neurons and more than 50 million synaptic connections. Implemented in Brian2, this model was used to generate experimentally testable predictions about feeding and grooming circuits. The underlying FlyWire adult brain connectome \cite{dorkenwald2024neuronal} was produced through large-scale electron-microscopy reconstruction, AI-assisted segmentation, and extensive community proofreading and annotation. This therefore represents a genuine whole-brain spiking model, although of a compact nervous system and with highly simplified neuronal dynamics. This modeling strategy contrasts with BAAIWorm, where it aims to build a biophysically detailed, embodied, organism-scale model of \textit{C. elegans}, whereas Shiu et al. instead built a simple but brainwide connectome-based spiking model of the \textit{Drosophila} brain. Despite omitting many biological details, such as neuronal morphology, receptor dynamics, gap junctions, neuromodulation and internal state, the model made experimentally valid predictions about sensorimotor circuits: across 164 empirically tested predictions, 91\% were consistent with experimental results.

At the scale of mammalian cortical microcircuits, detailed circuit models have also become increasingly ambitious. Laquitaine et al. \cite{laquitaine2024spike} used a biophysically detailed model of rat primary somatosensory cortex, based on the Blue Brain neocortical microcircuit framework \cite{reimann2026modeling, isbister2026modeling}, to generate synthetic Neuropixels-like extracellular recordings with known ground-truth spike times. The simulated subvolume contained 30,190 multi-compartment neurons distributed across all six cortical layers and connected by 36.7 million synaptic connections, with 60 morphological classes, 11 electrical firing classes, and realistic thalamocortical inputs. Extracellular signals were generated from the transmembrane currents of the detailed morphologies using NEURON/Neurodamus and the BlueRecording pipeline, then sorted with several modern spike sorters. This illustrates how large-scale biophysically detailed simulations can now serve as benchmarking environments for experimental methods. However, 10 minutes of extracellular recording typically took 6 days on large clusters containing more than 120 CPUs, which clearly limits applications.

In parallel, Pronold et al. \cite{pronold2024multi} developed a multi-scale spiking network model of one human cortical hemisphere, illustrating that human-scale spiking simulations are becoming possible in highly structured, reduced settings. The model covers the 34 areas of the Desikan--Killiany parcellation, with each area represented by a 1 mm² layer-resolved cortical column. Layers 2/3, 4, 5, and 6 are modeled with excitatory and inhibitory LIF populations, yielding 3.47 million neurons and 42.8 billion synapses. Local connectivity relies on a rescaled cortical microcircuit blueprint, whereas long-range cortico-cortical connectivity is derived from human diffusion-imaging data and augmented with predictive connectomics to determine laminar source and target patterns. The resulting framework links multiple scales, from single-neuron spiking statistics to area-level functional connectivity. The authors found good agreement with human medial-frontal spiking data and resting-state fMRI functional connectivity, though only after significantly strengthening inter-areal synapses, suggesting that robust long-range coupling is critical for reproducing large-scale resting-state structures. While such simulations demonstrate that hemisphere-wide spiking models are now computationally tractable, they do not replace neural-mass or mean-field whole-brain models. Simulations performed in NEST on a 768-core supercomputer required approximately 200 core-hours to generate just 10 seconds of biological activity. Therefore, the model's high dimensionality, extensive foundational assumptions, and massive computational cost make systematic parameter exploration, subject-specific fitting, and mechanistic interpretation substantially more difficult than in reduced whole-brain models.

These examples show the scientific value of detailed spiking simulations. They can generate synthetic data, test mechanistic hypotheses, validate analysis pipelines, and reveal how structure shapes dynamics. Their main limitation, however, is often not conceptual but practical: even when feasible, they remain computationally heavy, difficult to invert, often empirical data-hungry, cumbersome to scan over large parameter spaces, and poorly suited to repeated whole-brain fitting. They are therefore very powerful for local circuit investigation and for selected large-scale demonstrations, but they are not yet the most convenient framework for receptor-to-whole-brain inference in humans. The challenge is therefore to find reduced descriptions that remain biologically meaningful without retaining the full cost of explicit spiking simulation.

\subsection{Wilson--Cowan model (1972)}
\label{subsec:wilson_cowan_model}

The Wilson--Cowan model \cite{wilson1972excitatory,wilson1973mathematical} is one of the foundational population models in theoretical neuroscience. It is best understood as a phenomenological neural-mass model, rather than as a strict microscopic derivation from spiking-neuron dynamics. Its purpose is to describe how interacting excitatory and inhibitory populations evolve over time, and in its spatial extension, over cortical tissue.

In a notation consistent with the present review, one may write the model as coupled differential equations:
\begin{equation}
\begin{aligned}
\tau_e \frac{d\nu_e(x,t)}{dt} 
&= -\nu_e(x,t) + [1-r_e \nu_e(x,t)]\mathcal{F}_e(w_{ee}*\nu_e - w_{ei}*\nu_i + I_e(x,t)) \\
\tau_i \frac{d\nu_i(x,t)}{dt} 
&= -\nu_i(x,t) + [1-r_i \nu_i(x,t)]\mathcal{F}_i(w_{ie}*\nu_e - w_{ii}*\nu_i + I_i(x,t)) \\
w_{jk}*\nu_k 
&= \int_{\Omega} w_{jk}(x-y)\nu_k(y,t)dy
\end{aligned}
\label{eq:wilson_cowan_ODEs}
\end{equation}

Here, \(\nu_e\) and \(\nu_i\) denote the activities of the excitatory and inhibitory populations, \(\tau_e\) and \(\tau_i\) are characteristic relaxation times, \(r_e\) and \(r_i\) represent refractory effects, \(I_e\) and \(I_i\) are external inputs, and \(w_{jk}\) are coupling kernels. Strictly speaking, the original Wilson--Cowan variables correspond more closely to fractions of active neurons than to exact firing rates; however, the firing-rate interpretation is now standard in modern neural-mass formulations. Propagation delays are neglected because the average time lags delays is supposed negligible next to the membrane time constant, therefore the conduction velocity $v$ is considered as infinite (though delay-extended formulations are also common).

The nonlinear functions \(\mathcal{F}_e\) and \(\mathcal{F}_i\) play the role of population response functions. They transform the net synaptic input into an output activity level. In practice, they are usually chosen as sigmoidal functions, because such shapes capture three basic qualitative features of neuronal population responses: a threshold for weak inputs, a high-gain intermediate regime, and saturation at high input levels. The refractory prefactor \([1-r_\mu \nu_\mu]\) further limits the maximal activity, preventing the prediction of impossible high activity by reducing the fraction of neurons available to respond. In many later simplified formulations, this factor is omitted or absorbed into an effective transfer function \cite{pinto1996quantitative}.

The convolution term in Eq.~\eqref{eq:wilson_cowan_ODEs} expresses spatial coupling. Instead of assuming that activity at position \(x\) depends only on a local scalar weight, Wilson and Cowan introduced a spatial kernel \(w_{jk}(x-y)\), which measures how activity at location \(y\) contributes to the drive at location \(x\). In practice, these kernels are often taken to decay with distance, for instance in a Gaussian-like fashion, reflecting the fact that nearby populations tend to interact more strongly than distant ones.

From a dynamical point of view, each Wilson--Cowan equation has the structure of a first-order relaxation system:
$$
\tau \frac{d\nu}{dt} = -\nu + \text{nonlinear drive}.
$$
The first term, \(-\nu\), represents the tendency of the population activity to relax back toward baseline in the absence of input, whereas the second term represents the effective synaptic drive produced by recurrent excitation, inhibition, and external stimulation. This structure makes the model mathematically analogous to a low-pass filter: fast input fluctuations are attenuated, while slower components can be tracked by the population dynamics.

The main strength of the Wilson--Cowan formalism is therefore its simplicity. It captures excitation-inhibition competition, thresholded nonlinear responses, and spatial propagation with a very compact set of equations. Its limitation is that the transfer functions and coupling parameters are not derived directly from explicit single-neuron biophysics. In that sense, it provides a powerful phenomenological description of mesoscopic population dynamics, but not the explicit microscopic-to-macroscopic bridge sought in more biophysically grounded mean-field approaches such as the master-equation lineage studied in the present review.

\subsection{Wong--Wang model}
\label{subsec:Wong_Wang_model}

The Wong--Wang model \cite{wong2006recurrent, deco2013resting} is a biophysically motivated reduced population model that occupies an intermediate position between detailed spiking-network simulations and purely phenomenological neural-mass models.

Wong and Wang demonstrated how to reduce a highly nonlinear, 2,000-neuron spiking network (originally developed by Wang \cite{wang2002probabilistic}) down to a two-variable system while preserving its core biophysics. Unlike standard Wilson--Cowan-type descriptions, its state variables are not directly the firing rates themselves, but rather slow synaptic gating variables, specifically associated with NMDA-mediated excitation. The central idea is that fast spiking and membrane fluctuations can be adiabatically reduced, while the slow synaptic dynamics dominate the collective behavior on the timescales relevant for decision-making, working memory, and resting-state whole-brain activity.

In its generic reduced form, the local dynamics can be written as:
\begin{equation}
\begin{aligned}
\frac{dS_i}{dt}
&= -\frac{S_i}{\tau_S} + (1-S_i)\,\gamma\,H(x_i), \\
x_i
&= J_{ii}S_i-\sum_{j\neq i}J_{ij}S_j + I_0 + I_i + I_{noise,i},
\end{aligned}
\label{eq:Wong_Wang_model}
\end{equation}
where \(S_i\) denotes the fraction of open slow excitatory synaptic gates in population \(i\), \(\tau_S\) is the NMDA synaptic decay time constant, \(\gamma\) is a kinetic factor linking the firing rate to NMDA channel opening, and \(x_i\) is the total effective input current. The term \(I_{noise,i}\) is an \textit{Ornstein--Uhlenbeck} (OU) process representing the stochastic background chatter of the brain. Because of the inclusion of this noise term, the Wong--Wang model constitutes a system of stochastic differential equations.

The derivation bridges mean-field theory and Wilson--Cowan-type population dynamics. It treats the discrete synaptic inputs received by a single neuron as a continuous Gaussian random process. This assumption allows the steady-state firing rate of a LIF neuron to be calculated using the first-passage time formula derived from the Fokker--Planck equation. They assume a constant driving force, which effectively reduces conductance-based synapses to current-based ones, so that the first-passage-time formula used for current-based LIF networks (e.g., Brunel 2000~\cite{brunel2000dynamics}) applies. Because the exact Fokker--Planck solution involves a computationally heavy integration, the transfer function is approximated by the closed-form analytical expression \(H(x_i)\).

A widely used choice for this transfer function is the Abbott and Chance formulation:
\begin{equation}
H(x)=\frac{ax-b}{1-\exp[-d(ax-b)]},
\label{eq:Wong_Wang_transfer}
\end{equation}
where \(a\), \(b\), and \(d\) are phenomenological parameters fitted to the analytical integral formula to reproduce the stationary firing-rate response of the underlying spiking network. This function plays a role analogous to the gain function in Wilson--Cowan models, but here it is interpreted as an effective summary of fast population activity conditioned on the slow synaptic input.

The mathematical reduction of the model from four neural populations (two selective, one non-selective, one inhibitory) to just two variables relies on three main approximations. First, because the firing rate of the non-selective excitatory population remains empirically stable, it is treated as a constant. Second, the input-output response of the inhibitory interneurons is linearized; this allows the inhibitory firing rate to be solved explicitly and mathematically absorbed into the effective mutual inhibition between the two selective populations. Finally, the model relies on a separation of timescales. AMPA and GABA synapses, membrane dynamics, and individual spikes are assumed to relax rapidly compared with NMDA-mediated excitation. As a result, the fast variables are not tracked explicitly; instead, they are assumed to be in their steady states (\(S_{AMPA,\ GABA} \approx \tau_{AMPA,\ GABA} H(x_i)\)) and their net effect is absorbed into the nonlinear function \(H(x)\). The reduced system therefore evolves on the slow manifold defined by the synaptic gating variables \(S_i\). In that sense, the Wong--Wang formalism is not an exact reduction in the mathematical sense of Montbrió--Pazó--Roxin, but rather a biologically grounded mean-field approximation built around slow synaptic currents.

This structure makes the Wong--Wang framework especially useful for cognitive and whole-brain applications. Because the dynamics are governed by slow recurrent excitation, the model naturally accounts for persistent activity, attractor competition, and the low-frequency fluctuations relevant for fMRI-scale dynamics. In whole-brain implementations, each node is typically described by one or several Wong--Wang populations and coupled to other regions through long-range excitatory interactions derived from a structural connectome. Deco and colleagues \cite{deco2013resting} later scaled the Wong--Wang model up to the whole-brain level, connecting 66 local regions via a structural connectome to investigate resting-state functional connectivity. Instead of simulating computationally expensive, long noisy trajectories of \(S\) to extract the simulated fMRI covariance matrix, they employed a dynamic mean-field approach. They utilized the Fokker--Planck equation to describe the temporal evolution of the probability density \(P(S,t)\). Because solving the Fokker--Planck equation for the highly nonlinear Wong--Wang system is mathematically intractable, they assumed that, at rest, the brain operates near a spontaneous stable, low-firing fixed point. By linearizing the deterministic dynamics around this spontaneous state using a first-order Taylor expansion, they were able to apply the moments method (solving the Lyapunov equation) to directly and analytically compute the stationary covariance matrix (Functional Connectivity). This provided a highly tractable mathematical framework to study how the resting brain's functional networks are dynamically shaped by its anatomical structure.

The main strength of the Wong--Wang framework is therefore that it preserves a clear biological interpretation while remaining computationally lightweight. Its main limitation is that the reduction is tied to specific assumptions: slow synaptic gating must dominate the collective timescale, and the transfer function \(H(x)\) is an effective fitted object rather than a fully explicit microscopic derivation. In addition, standard versions rely on LIF neurons and current-based interactions. The framework is therefore more specific about NMDA kinetics than the lineage studied in the present review, which pools AMPA and NMDA into one excitatory conductance, but less flexible regarding the neuron model and conductance-based effects.

\subsection{Montbrió--Pazó--Roxin model}
\label{subsec:MPR_model}

The Montbrió--Pazó--Roxin (MPR) model \cite{montbrio2015macroscopic} is an elegant example of an exact macroscopic reduction of a spiking-neuron network. Starting from a population of all-to-all coupled Quadratic Integrate-and-Fire (QIF) neurons, they demonstrated that, in the thermodynamic limit, the collective dynamics can be expressed exactly in terms of only two macroscopic variables: the population firing rate $r$ and the mean membrane potential $v$.

The microscopic QIF neuron is a one-dimensional differential equation system governed by:
\begin{equation}
\tau_m \frac{dV_j}{dt} = V_j^2 + \eta_j + Js(t) + I(t), \,\ \footnotemark
\label{eq:QIF_neuron}
\end{equation}
\footnotetext{One may notice that this ODE is dimensionally inhomogeneous, mixing several units. The model uses mathematical notation with dimensionless variables for almost everything except time and firing rate.}
where \(V_j\) is the membrane potential of neuron \(j\), \(\eta_j\) is its intrinsic excitability, \(J\) is the synaptic coupling strength, $s(t)$ is the recurrent mean synaptic activations, and \(I(t)\) is an external driving current. In the standard MPR formulation, synaptic interactions are instantaneous ($\tau \to 0$) and current-based, so that in the large-population limit the recurrent input is proportional to the firing rate $Js(t) = J\tau_m r(t)$. Whenever \(V_j\) diverges to \(+\infty\), a spike is emitted and the membrane potential is instantly reset to \(-\infty\). A key assumption of the model is that the excitability parameters \(\eta_j\) are heterogeneous and distributed according to a Lorentzian (Cauchy) distribution $g(\eta)$ with center \(\bar{\eta}\) and half-width \(\Delta\).

Under these assumptions, the network admits the following exact macroscopic description:
\begin{equation}
\begin{aligned}
\tau_m \frac{dr}{dt}
&= \frac{\Delta}{\pi\tau_m} + 2rv, \\
\tau_m \frac{dv}{dt}
&= v^2 + \bar{\eta} + I(t) + J\tau_m r - (\pi\tau_m r)^2,
\end{aligned}
\label{eq:MPR_model}
\end{equation} 
These equations are exact in the infinite-size limit: no moment closure, diffusion approximation, or phenomenological transfer-function fitting is required. The reduction is possible because the QIF model is mathematically equivalent, via the change of variables $V = \tan(\theta/2)$, to a population of $\theta$-neurons. As the population approaches the thermodynamic limit, the corresponding population density $\rho(V \mid \eta, t)$ admits a Lorentzian ansatz, which is closely related to the Ott--Antonsen reduction for coupled phase oscillators. By substituting the Lorentzian ansatz into the continuity equation, the partial differential equation cleanly splits into two exact, coupled ordinary differential equations for any given subpopulation. A final complex contour integration over the heterogeneous background currents \(g(\eta)\) yields the macroscopic equations for the entire network. 

Therefore, the global state variables \(r(t)\) and \(v(t)\), representing the population firing rate and the mean membrane potential, mathematically correspond to the average of the width and the center of the subpopulation Lorentzian membrane voltages over the excitability distribution $g(\eta)$. Thus, in the standard MPR derivation, both the voltage variable $V$ and the excitability parameter $\eta$ are assumed to follow Lorentzian distributions. This is crucial for the derivation: the former emerges naturally from the dynamics of coupled phase oscillators in the thermodynamic limit, while the latter is an analytic assumption chosen for convenience. Because a Lorentzian distribution possesses exactly one complex pole in the upper half-plane, the contour-integral calculation becomes analytically tractable, allowing the infinite-dimensional population density dynamics to collapse onto a two-dimensional invariant manifold. In contrast, assuming a Gaussian distribution, while arguably more biologically realistic, requires additional approximations and series expansions to achieve a closed form. 

The exactness of the reduction gives the MPR model major theoretical advantages. First, it preserves a rigorous correspondence between the microscopic spiking dynamics and the macroscopic variables. Second, because the reduced system is strictly two-dimensional, one can perform phase-plane, fixed-point, and bifurcation analyses efficiently. Third, unlike purely phenomenological firing-rate models (such as the Wilson--Cowan equations), the formalism retains both the firing rate $r$ and a voltage-like collective variable $v$, making the macroscopic description richer and more tightly connected to the underlying population dynamics. In contrast to the Wong--Wang framework, the MPR model does not rely on a separation of timescales or the assumption of slow NMDA-receptor dynamics. In fact, due to the exact derivation, the MPR model naturally captures fast transient dynamics, spike synchronization, and ringing oscillations. This makes it well suited to studying fast electrophysiological signals, which operate on much shorter timescales than fMRI BOLD signals, although prediction of EEG or MEG still requires an explicit observation model.

At the same time, this exactness comes at the price of strong structural assumptions. The standard MPR model requires QIF/\(\theta\)-neuron dynamics, all-to-all coupling, instantaneous current-based interactions, the thermodynamic limit, and Lorentzian heterogeneity in the excitabilities.\footnote{Moreover, the Ott--Antonsen/Lorentzian ansatz inherently assumes that the network's phase distribution remains unimodal. If neurons start forming distinct cluster states (e.g., synchronized sub-groups firing out of phase), the ansatz breaks down, and the population must be split to be accurately modeled.} These conditions make the model mathematically elegant but arguably less flexible than semi-analytical conductance-based mean-field approaches when researchers wish to incorporate realistic synaptic kinetics, spike-frequency adaptation, receptor-specific effects, or more complex cellular biophysics. Thus, while the MPR framework provides one of the clearest available examples of an exact microscopic-to-macroscopic bridge, it is not immediately suited to the receptor-aware, conductance-based whole-brain programs discussed elsewhere in the present review. 

Nevertheless, while the Montbrió--Pazó--Roxin model is not the only exact mean-field derivation of a spiking network, it remains the paradigmatic exact low-dimensional neural-mass reduction for heterogeneous QIF/\(\theta\)-neuron networks. Subsequent next-generation neural mass models have successfully extended this same mathematical framework to include synaptic filtering, transmission delays, multi-population architectures, adaptation, gap junctions, and certain conductance-based interactions \cite{coombes2023next}. However, these exact low-dimensional reductions remain largely tied to QIF/\(\theta\)/Riccati-type dynamics and to Lorentzian/Cauchy analyticity assumptions. For LIF, AdEx, Hodgkin--Huxley, and more general conductance-based spiking networks, exact descriptions typically exist only at the level of population-density, master-equation, or Fokker--Planck formalisms, whereas their closed low-dimensional neural masses still require approximations.

\subsection{Fokker--Planck formalism: from single neurons to population densities}
\label{subsec:fokker_planck_reminder}

\subsubsection{General principles of the Fokker--Planck formalism}

The Fokker--Planck Equation (FPE) is a partial differential equation widely used in statistical physics to describe the time evolution of the probability density function under the influence of both deterministic forces and random fluctuations. Rather than tracking a single noisy trajectory of a state variable $x(t)$ via a stochastic equation, the FPE tracks the probability $P(x,t)$ of finding the system in state $x$ at time $t$, thus it compresses individual stochastic realizations into a probability density that evolves according to deterministic linear evolution equation \cite{longtin2010stochastic}.

For a general stochastic process defined by the Langevin equation $dx = f(x)dt + g(x)dW_t$, the corresponding one dimensional FPE is given by:
\begin{equation}
\frac{\partial P(x,t)}{\partial t} = -\frac{\partial}{\partial x} \left[ D_1(x) P(x,t) \right] + \frac{1}{2} \frac{\partial^2}{\partial x^2} \left[ D_2(x) P(x,t) \right]
\end{equation}
The equation is governed by the Drift coefficient ($D_1(x)=f(x)$), representing the deterministic dynamic of the system, pulling the state variable toward specific attractors; and the Diffusion coefficient ($D_2(x)=g(x)^2$), representing the variance introduced by stochastic noise, causing the probability distribution to spread over time.

FPE can be very powerful in fields like computational neuroscience because it is a mathematical framework reusable at different levels of description. The state variable $x$ can represent either a microscopic quantity like a neuron's membrane potential or a macroscopic quantity like average firing rate of an entire network. However, these advantages come at some costs: one typically moves from ODEs to PDEs with boundary conditions at threshold and reset, which are mathematically and numerically heavier. For this reason, Fokker--Planck methods are often used either as an intermediate analytical tool or as a reference description from which lower-dimensional approximations are later derived \cite{gerstner2000population, gerstner2014neuronal}.

\subsubsection{Fokker--Planck at microscopic (single-neuron) level}
\label{subsec:fokker_planck_microscopic_level}

The classical single-neuron use of the FP formalism starts from a LIF neuron receiving many weak, approximately independent synaptic inputs. Under the diffusion approximation, the total synaptic bombardment can be replaced by a continuous noisy drive, so that the membrane potential behaves as a stochastic diffusion process rather than as a deterministic trajectory with discrete jumps. In this case, one can write the FPE for the voltage density $P(V,t)$~\cite{amit1997model}-\cite{brunel2000dynamics}.

Below, we show the derivation steps for CUBA and COBA LIF neuron models.

\paragraph{For CUBA LIF neuron,} a standard differential equation governing the model is:
$$\tau_m \frac{dV}{dt} = -(V - V_{rest}) + R_m I_{syn}(t)$$

Under the diffusion approximation, the bombardment of discrete synaptic events is approximated as a continuous Gaussian process. Therefore, we can write the synaptic current $R_m I_{syn}(t) \approx \mu_{syn} + \tau_m \ c\ \eta(t)$.

Thus, the CUBA ODE becomes a stochastic Langevin equation:
\begin{equation*}
\begin{aligned}
dV 
&= - \frac{(V - \mu)}{\tau_m}dt + c\ dW_t, \\
&= - \frac{(V - \mu)}{\tau_m}dt + \sqrt{\frac{2\sigma^2}{\tau_m}}\ dW_t, \footnotemark \\
\frac{dV}{dt} 
&= - \frac{(V - \mu)}{\tau_m} +  \sqrt{\frac{2\sigma^2}{\tau_m}}\eta(t).
\end{aligned}
\end{equation*}
\footnotetext{$\mu = V_{rest} + \mu_{syn}$. Using Itô calculus, the steady-state variance of the OU process is $2\sigma^2 = c^2 \cdot \tau_m$. $\eta(t)$ is a white noise, assumed to be the derivative of the Wiener process.}

Instead of tracking a single fluctuating neuron over time, FPE tracks the time evolution of the probability density function $P(V, t)$. Substituting the drift and diffusion coefficients of the Langevin equation into the FPE yields:
$$
\frac{\partial P}{\partial t} = - \frac{\partial}{\partial V} \left[ \left( -\frac{V-\mu}{\tau_m} \right) P \right]
+ \frac{1}{2} \frac{\partial^2}{\partial V^2} \left[ \left( \frac{2 \sigma^2}{\tau_m} \right) P \right]
$$
Under steady-state assumptions, the long-term equilibrium is obtained for $\frac{\partial P}{\partial t} = 0$. Using the free-membrane potential assumption, because firing threshold and reset mechanisms are ignored, the voltage has no boundaries, but the probability density must drop to 0 as the voltage goes to $\pm \infty$. This implies that the probability current is zero everywhere:
\begin{equation*}
\begin{aligned}
0 
&= \frac{\partial}{\partial V} \left[ \left(-\frac{V - \mu}{\tau_m}\right) P  - \frac{1}{2} \frac{\partial}{\partial V} \left[ \left(\frac{2\sigma^2}{\tau_m}\right) P \right] \right] \\
0
&= \left(-\frac{V - \mu}{\tau_m}\right) P - \frac{1}{2} \frac{\partial}{\partial V} \left[ \left(\frac{2\sigma^2}{\tau_m}\right) P \right]
\end{aligned}
\end{equation*}

We now solve the differential equation for P(V):
\begin{equation*}
\begin{aligned}
\frac{1}{P} dP 
&= -\frac{V - \mu}{\sigma^2} dV \\
\ln P(V) 
&= -\frac{(V - \mu)^2}{2\sigma^2} + C \\
P(V) 
&= \exp(C) \cdot \exp\left( -\frac{(V - \mu)^2}{2\sigma^2} \right) \\
\end{aligned}
\end{equation*}
Because $P(V)$ is a probability distribution, the total area under the curve over full range of $V$ must equal 1, yielding the normalization constant $\exp(C)=(\sqrt{2\pi}\sigma)^{-1}$. We end up with the probability density of the membrane potential describing a Gaussian distribution.
\begin{equation}
\begin{aligned}
P(V) 
&= \frac{1}{\sqrt{2\pi}\sigma} \exp\left( -\frac{(V - \mu)^2}{2\sigma^2} \right) \\
\end{aligned}
\end{equation}
Therefore, the Fokker--Planck formalism mathematically proves that if a CUBA LIF neuron is subjected to continuous random noise without firing boundary, its membrane potential will follow a Gaussian distribution. Then, the firing rate can be approximated by taking the probability mass above the threshold and dividing by the membrane time constant.

\paragraph{For COBA LIF neuron,} a standard differential equation governing the model is:
$$C_m \frac{dV}{dt} = -g_L(V - E_L) - g_{syn}(t)(V - E_{syn})$$

In contrast to CUBA model, the diffusion approximation is applied to conductance: $g_{syn}(t) \approx \mu_g + \sigma_g \ \eta(t)$. Substituting this into the ODE yields:
\begin{equation}
\begin{aligned}
\frac{dV}{dt} = - \frac{g_L(V-E_L) + \mu_g(V-E_{syn})}{C_m} - \frac{\sigma_g}{C_m}(V-E_{syn})\eta(t)
\end{aligned}
\end{equation}

Because the noise amplitude depends on the state variable $V$, this constitutes a \textit{multiplicative noise} instead of \textit{additive noise} for CUBA.
Substituting the drift $D_1$ and diffusion $D_2$ coefficients of the Langevin equation into the steady-state, free-membrane FPE yields:
\begin{equation*}
\begin{aligned}
0
&= D_1(V) P - \frac{1}{2} \frac{d}{d V} \left[ D_2(V) P \right], \\
dD_2(V)P 
&= 2D_1(V) P \frac{D_2(V)}{D_2(V)}dV, \\
D_2(V)P 
&= \exp{(C)}\exp{\left( \int \frac{2 D_1(V)}{D_2(V)}dV \right)}, \\
P(V) 
&= \frac{\exp{C}}{D_2(V)}\exp{\left( \int \frac{2 D_1(V)}{D_2(V)}dV \right)}
\end{aligned}
\end{equation*}

The term $\frac{D_1(V)}{D_2(V)}$ takes the structural form of $\frac{Au+B}{u^2}$ where $u=V-E_{syn}$. Integration yields:
$$\int \frac{2 D_1(V)}{D_2(V)}dV = a \ln{|V-E_{syn}|} + \frac{b}{V-E_{syn}} $$

Substituting this back into the probability equation, we obtain
\begin{equation*}
\begin{aligned}
P(V)
\propto \frac{1}{(V-E_{syn})^2}| V-E_{syn}|^{a}\exp{ \left( \frac{b}{V-E_{syn}} \right)}
\end{aligned}
\end{equation*}

Therefore, the Fokker--Planck solution demonstrates that when a COBA LIF neuron is subjected to continuous random conductance noise without firing boundary, its membrane potential follows a highly skewed distribution. To find the analytical transfer function $\mathcal{F_\mu}$, one would have to integrate this distribution from $V_{thresh}$ to infinity, which is generally analytically intractable or does not yield the same simple closed-form approximation.

To recover the Gaussian distribution and a solvable transfer function, one convenient Gaussian approximation freezes the driving force by replacing $V$ with $\langle V \rangle$ in the noise term, which becomes $\frac{\sigma_g}{C_m} (\langle V \rangle - E_{syn}) \eta(t)$, effectively converting the multiplicative noise back into additive noise.

This microscopic Fokker–Planck description is one of the foundations of first-order mean-field theories: it converts the stochastic dynamics of a representative neuron into a transfer function, which can then be closed self-consistently to obtain stationary rates and phase diagrams, as in Brunel (2000) \cite{brunel2000dynamics}. However, this leading-order theory assumes a thermodynamic/asynchronous limit in which finite-size fluctuations of the population activity are neglected, such that inputs to different neurons are effectively uncorrelated.

\subsubsection{Fokker--Planck and population-density approaches at the population level}
\label{subsec:fokker_planck_population_level}

At the population level, the role of the FP formalism becomes broader than in the single-neuron case. In the classical sparse-network work of Brunel (2000) \cite{brunel2000dynamics}, the FP equation is used within a self-consistent mean-field description of large excitatory-inhibitory networks. The central object is the population density of membrane potentials, from which one can compute stationary firing rates and analyze the stability of asynchronous or oscillatory regimes by linear perturbation. This is therefore already a population-level theory, but it remains primarily a deterministic infinite-size description rather than a compact mesoscopic model of finite-size fluctuations. Several closely related approaches were developed, which also treated the density of neurons over membrane potential as the main dynamical object \cite{nykamp2000population, omurtag2000simulation}. The strength of this family of methods is its close connection to microscopic spiking dynamics; its drawback is that the model remains a partial differential equation, which is often costly to solve and difficult to reduce when additional biological variables are included.

This difficulty becomes particularly evident when one introduces more realistic single-neuron mechanisms such as spike-frequency adaptation. In that case, the FP description must be extended to a higher-dimensional state space, which greatly complicates both analysis and simulation. To overcome this limitation, Augustin et al. (2017) \cite{augustin2017low} derived low-dimensional spike-rate models for AdEx populations directly from the FP framework. Using separately spectral decomposition and linear-nonlinear reduction methods, they obtained first- and second-order ODE approximations that preserve the dependence on underlying neuronal parameters while being much cheaper to simulate than the full population-density PDE.

A second important limitation of classical FP formalisms is that they are naturally suited to the thermodynamic limit, where population activity becomes deterministic. Finite-size fluctuations are therefore not straightforward to include and usually require additional approximations or stochastic closures. In a different, though related, population-density approach, Schwalger et al. (2017) \cite{schwalger2017towards} introduced a mesoscopic theory based on refractory density, that is, the distribution of neurons with respect to their last spike time or age. By tracking the survival time of homogeneous populations of finite-size generalized integrate-and-fire (GIF) neurons, they derived stochastic mesoscopic integral-equations that preserve normalization and reproduce collective effects such as finite-size-induced switching in bistable networks.

Thus, Fokker--Planck and population-density approaches form a major alternative lineage of mechanistic population modeling. They are especially powerful when one wishes to retain a close link with microscopic spiking dynamics, but they often remain mathematically heavier than reduced ODE formalisms.

\subsubsection{Synthesis: Master Equation and Fokker--Planck approaches}

Given that the Kramers--Moyal expansion shows how a master equation (ME) can, under suitable assumptions, be approximated by a Fokker--Planck equation, one may ask why El Boustani and Destexhe (2009) \cite{el2009master} used a ME formalism rather than directly constructing a macroscopic FP theory.

In this construction, the relevant distinction is not between continuous and discrete time, but between the state variables represented and the closure assumptions employed. Continuous-time point-process descriptions and master equations for finite spike counts can be mathematically well defined. El Boustani and Destexhe \cite{el2009master} instead introduced a mesoscopic coarse-graining at a finite time resolution $T$, chosen so that population activity could be treated as approximately Markovian, and used a master equation to describe transitions between discrete population-activity states. A second-order moment closure then yields differential equations for the mean activity and covariance matrix, together with stationary lagged correlations. \revision{Truncating the Kramers--Moyal expansion of this master equation at second order would give a Fokker--Planck equation with the same first- and second-moment identities, and hence the same complete covariance equation. The published covariance equation instead uses a mean-value closure of the second jump moment. Its structure resembles a linear-noise closure, but the consistent small-jump limit has a different noise source and no squared-displacement term (Supplementary~\ref{subsubsec:covariance_matrix}).} Fokker--Planck equations, by contrast, evolve continuous-state probability densities and can themselves be formulated at the population level, as discussed above. In the master-equation lineage considered here, however, Fokker--Planck theory enters mainly upstream, as one possible route to approximate the single-neuron transfer function by describing the voltage density of a representative noisy neuron. The resulting transfer function $\mathcal{F}_\mu$ determines, through the binwise firing probability $\mathcal{F}_\mu T$, the transition kernel of the population-level master equation. Once $\mathcal{F}_\mu$ is specified, the master-equation formalism propagates this single-neuron input-output relation into explicit mesoscopic equations for finite-size activity fluctuations and covariances.

This modularity is precisely the conceptual advantage of the El Boustani--Destexhe approach. Classical macroscopic population-density PDEs can be very powerful, but they are naturally oriented toward deterministic, infinite-size limits and remain mathematically heavy when attempting to inject finite-size noise, unless additional approximations and tricks. The ME formalism, by contrast, takes the single-neuron transfer function as a plug-and-play input and directly builds a compact mesoscopic description of finite-size population variance. In that sense, the ME formalism does not replace the FP formalism; it uses a transfer function that may be informed by FP calculations within a distinct mesoscopic closure designed to retain finite-size population fluctuations.

We stress that while this ME approach offers a highly intuitive bottom-up derivation, it is not inherently superior to modern population-density theories. As previously discussed, exact macroscopic reductions (e.g., Montbrió et al., 2015 \cite{montbrio2015macroscopic}) and stochastic refractory-density equations (e.g., Schwalger et al., 2017 \cite{schwalger2017towards}) have emerged as highly efficient continuous competitors. At this stage, systematic benchmarking of these diverse mesoscopic reduced models against large-scale spiking simulations remains the next logical step (see Section~\ref{sec:benchmarks_surrogates}).

\section{Algorithmic cost analysis and benchmarking of spiking, mean-field, and learned surrogate models}
\label{sec:algorithmic_cost_analysis_neural_simulation_appendix}

\subsection{Algorithmic cost analysis for neural simulation}

We define the computational cost required to simulate one biological second as
\begin{equation}
W_{1\mathrm{s}} = \sum_{\mathrm{steps}} W_{\Delta t} =
\sum_{\mathrm{steps}} \left[ n_{\mathrm{RHS}} W_{\mathrm{RHS}}
+ W_{\mathrm{solv\_upd}} \right].
\end{equation}
Here, \(W_{\mathrm{RHS}}\) is the cost of one full evaluation of the right-hand side (RHS) vector field, \(n_{\mathrm{RHS}}\) is the number of RHS evaluations per numerical step, and \(W_{\mathrm{solv\_upd}}\) is the additional algebraic cost of the numerical solver, such as vector additions and scalar multiplications.

To compare ODE-based models, we assume that the dynamical system
$\dot{x}=f(x)$ is integrated with a fixed-step first-order Euler method, $x(t+\Delta t)=x(t)+\Delta t f(x(t))$, as in the operation-count comparison of Izhikevich~\cite{izhikevich2004model}. For a \(d\)-dimensional system, the Euler update requires one multiplication and one addition per state variable. Therefore, $W_{\Delta t} = W_f + 2d$. This common Euler step defines a standardized operation-count workload; it does not imply equal numerical error or an optimal solver for each model. Runtime or efficiency comparisons require either model-specific convergence tests or a common error tolerance.

We assume that time-constant subexpressions, such as \(1/\tau_w\), and fixed normalization factors, are precomputed. To remain consistent with Izhikevich's type of operation-count convention, exponentials are assigned a cost $C_{\exp}=10$. We use the following FLOP-equivalent accounting convention:
\[
C_{+}=C_{-}=C_{\times}=1,\qquad
C_{\div}=4,\qquad
C_{\sqrt{\cdot}}=4,\qquad
C_{\exp}=10,\qquad
C_{\mathrm{erfc}}=2C_{\exp}=20.
\]
These values are not intended to represent exact hardware runtimes. They are an explicit accounting convention used to avoid treating special functions as single elementary arithmetic operations.

We define a source-level array-access proxy over one biological second, for fixed-step Euler integration, as
\begin{equation}
M_{1\mathrm{s}} = \sum_{\mathrm{steps}}M_{\Delta t} =
\frac{T_{\mathrm{bio}}}{\Delta t}M_{\Delta t}.
\end{equation}
The proxy assumes that each dynamical state is loaded and stored once per step, that the explicitly identified intermediate quantities are written once, and that each nonzero connectome weight and required delayed value is streamed once per coupling evaluation. Unless stated otherwise, it excludes solver temporaries, derivative arrays, connectivity metadata, random-number generation, inter-process communication, recording, and post-processing. Multiplication by \(B=8\) bytes corresponds to double-precision floating-point state and weight values; integer indices need not have the same size.

Accordingly, \(M\) is an explicit array-access scenario rather than measured traffic to main memory. Cache reuse can reduce DRAM traffic, whereas cache-line effects, temporary arrays, delay interpolation, queues, and metadata can increase it. The ratio \(I=W/M\) is therefore a source-level work-to-access ratio, not a hardware Roofline operational intensity~\cite{williams2009roofline}. It cannot by itself establish whether an implementation is compute- or memory-bound.

\subsubsection{First-order adaptive master equation formalism}

\paragraph{Node-local cost:} 

The first-order adaptive system is:
\begin{equation}
\begin{aligned}
T\partial_t \nu_\mu 
&= F_\mu - \nu_\mu\\
\tau_w\partial_t W_\mu 
&= -W_\mu + b\tau_w\nu_\mu + a(\langle V(\nu_e, \nu_i, W_\mu)\rangle - E_L)
\end{aligned}
\end{equation}
Here, \(\mu=1,\ldots,P\) indexes the populations of a local node. Let \(P_w\leq P\) denote the number of populations carrying an explicit adaptation variable. The rate, transfer-function, and state-dependent calculations are performed for all \(P\) output populations, whereas the adaptation equation contributes only for the \(P_w\) adaptive populations (adaptation is usually present only in the excitatory population). We use here the convention \(P_w=P\) as a conservative upper scenario independent of the type of populations.

The cost per node is decomposed as: $W_{node}^{1st} = P\left( W_\nu+W_{\mathcal F_\mu}+W_{\mathrm{states}} \right) +
P_w W_W$ where $W_\nu$ is the cost of the rate equation, $W_W$ is the cost of the adaptation equation, $W_{\mathcal F_\mu}$ is the final transfer-function evaluation cost, and $W_{\mathrm{states}}$ is the cost of computing the state-dependent quantities entering the transfer function.

The state-dependent cost includes the subthreshold moments $(\langle V\rangle,\sigma(V),\tau_V)$. We estimate their arithmetic cost as $W_{\mathrm{moments}} \approx 15P+10$.

It also includes the effective threshold \(V_{\mathrm{thresh}}^{\mathrm{eff}}\), represented by a second-order polynomial expression and is estimated to cost $W_{V_{\mathrm{thresh}}^{\mathrm{eff}}} \approx 36$. Therefore, $W_{\mathrm{states}} =
W_{\mathrm{moments}} + W_{V_{\mathrm{thresh}}^{\mathrm{eff}}} \approx 15P+46$.


The transfer function costs $W_{\mathcal{F}_\mu} = W_{erfc(z)} + W_z + 6  \approx 32$. The rate and adaptation equations are estimated to cost approximately 2 and 6 respectively.

Therefore $W_{node}^{1st} \approx 15P^2+80P+6P_w$. For the standard E-I node with $P=2$, $W_{node}^{1st} \approx 232$ FLOP-equivalent operations per node and per RHS evaluation. This estimate does not depend on the number of microscopic neurons represented by the node. It depends only on the number of populations and on the cost of evaluating the semi-analytical transfer functions. The quadratic term is a conservative upper-bound contribution: it assumes that each of the \(P\) populations computes distinct synaptic and membrane moments from every source population. If some moment calculations are shared across populations, the local cost is reduced. With \(P_w=1\) as in the model described in Section~\ref{sec:Sacha_Molecular_MFT}, it would give approximately \(226\) FLOP-equivalents per RHS evaluation.

\paragraph{Long-range coupling cost:}

For a whole-brain network with $K$ nodes, each region receives delayed excitatory input from other regions. Let $\rho$ be the density of the directed connectome. The number of nonzero long-range edges is approximately $\rho K^2$.

The delayed coupling term has a generic form $G \sum_j C_{ij} \nu_e(j,t-\tau_{ij})$. If delayed values are already available and no interpolation is required, each edge requires only one addition and multiplication. Therefore, $c_{edge} \approx 2$. One may argue that for $P$ populations, $c_{edge} \approx 2P$ because each population communicates its states to other nodes, but traditionally populations are excitatory/inhibitory and inhibitory stays local. We keep that convention for simplicity for all models.

The first-order RHS cost is then:$$W_{RHS}^{1st} \approx K W_{node}^{1st} + c_{edge} \rho K^2$$

For a dense connectome, $\rho \simeq 1$, and the coupling term scales as $K^2$. For fixed-step explicit Euler integration, the cost per biological second is: $W_{1s} \approx \frac{T_{bio}}{\Delta t}( W_{RHS} + 2d) $ where Euler update contributes one multiplication and one addition per state variable. With $P$ rate variables and $P_w$ adaptation variables per node, $d=(P+P_w)K$ ($=2PK$ for $P_w=P$).

For $\Delta t =0.1$ ms, the number of steps is $\frac{T_{bio}}{\Delta t }= 10^4$.

The algorithmic cost over one biological second becomes
\begin{equation}
\boxed{ W_{1\mathrm{s}}^{\mathrm{1st}} \approx 10^4\left[ \left(15P^2+82P+8P_w\right)K +2\rho K^2 \right]. }
\end{equation}

\paragraph{Streaming-access estimation:}

We now estimate the streaming-access traffic per integration step. We count the number of scalar values read from or written to memory and multiply by \(B=8\) bytes for double precision. We denote by \(a_{\mathrm{states}}\) the number of scalar memory accesses per node associated with local state variables and stored local quantities, by \(a_{\mathrm{edge}}\) the number of scalar memory accesses per long-range edge, and by \(a_{\mathrm{hist}}\) the number of scalar writes per node into the delay buffer.

The memory traffic per Euler step can be approximated as
\begin{equation}
M_{\Delta t}^{\mathrm{1st}} \approx 8\left[ a_{\mathrm{states}}K
+ a_{\mathrm{edge}}\rho K^2 + a_{\mathrm{hist}}K \right].
\end{equation}
The three terms correspond respectively to local node state, connectome coupling, and delay-buffer writing.

\emph{State-variable traffic:} Each node with $P$ populations is considered to read and write the dynamical variables rate and adaptation $(\nu_\mu, W_\mu)$, and to write the moments $(\langle V \rangle, \sigma(V), \tau_V)$. This makes $a_{\mathrm{states}} \approx 7P$. Additional local parameters may also be read, but because they are few, repeatedly reused, we consider them cache-resident. A more conservative estimate would add a term $a_{\mathrm{param}}PK$ as the number of parameter scalars read per population and per step.

\emph{Connectome and delayed-state traffic:} For each nonzero long-range edge, the coupling term requires reading at least the structural connectivity weight and the delayed source activity: $C_{ij},\  \nu_e(j,t-\tau_{ij})$. If delayed values are read directly from a delay buffer without interpolation, this gives $a_{\mathrm{edge}}=2$ scalar reads per edge. This is a streaming-memory estimate: every connectome read is counted as memory traffic. In practice, for a standard whole-brain parcellation, the dense connectome can be small enough to remain cache-resident. For example, a dense \(68\times 68\) double-precision connectome contains \(68^2\) entries, corresponding to approximately \(37\,\mathrm{kB}\). In such a case, the effective DRAM traffic may be lower than the streaming estimate. We retain this deliberately pessimistic streaming-access convention for scaling illustrations. It is not an implementation-independent bound; actual DRAM traffic may be either lower through reuse or higher through metadata, queues, cache-line transfers and temporary arrays.

At each time step, the current excitatory rate of each node must be written into the delay buffer so that it can be used by future delayed coupling evaluations. If only the excitatory rate is used for long-range coupling, this requires one scalar write per node: $a_{\mathrm{hist}}=1$.

Combining the state, edge, and delay-buffer terms, with  \(\Delta t=0.1\,\mathrm{ms}\), gives
\begin{equation}
\boxed{M_{1\mathrm{s}}^{\mathrm{1st}} \approx 8\times 10^4
\left[ (5P+2P_w+1)K + 2\rho K^2 \right]\ \mathrm{bytes}}.
\end{equation}

Under the streaming-memory convention, connectome traffic dominates local state traffic when $a_{\mathrm{edge}}\rho K^2
\gtrsim (7P+1)K$. For a dense connectome \(\rho=1\), \(P=2\), this threshold is $K \gtrsim 7.5$. Thus, for dense whole-brain coupling, memory traffic is rapidly dominated by reading connectivity weights and delayed source states. In reality,  this threshold would be significantly higher, depending on the available cache size.

\subsubsection{Second-order adaptive master-equation formalism}

\paragraph{Node-local cost:}

The second-order formalism augments the population rates with their covariances. For a node of $P$ populations indexed by $\mu,\nu,\lambda,\eta=1,\ldots,P$, the local system is
\begin{equation}
\begin{aligned}
T\partial_t \nu_\mu
&=(F_\mu-\nu_\mu) + \tfrac{1}{2}\,\partial_\lambda\partial_\eta F_\mu\,c_{\lambda\eta}, \\
T\partial_t c_{\mu\nu}
&=\delta_{\mu\nu} A_{\mu\mu}^{-1}\revision{+\partial_\lambda F_\mu\,\partial_\eta F_\nu\,c_{\lambda\eta}-c_{\mu\nu}
+T^2\,\partial_t\nu_\mu\,\partial_t\nu_\nu},\\
\tau_w\partial_t W_\mu
&=-W_\mu+b\tau_w\nu_\mu+a\left(\langle V(\nu_e,\nu_i,W_\mu)\rangle-E_L\right),
\end{aligned}
\label{eq:second_order_local}
\end{equation}
where $\delta_{\mu\nu}A_{\mu\mu}^{-1}=\delta_{\mu\nu}F_\mu(1/T-F_\mu)/N_\mu$ is the diagonal finite-size noise term\revision{, and the covariance equation is the corrected one, Eq.~\eqref{eq:covariance_consistent_closure}, whose last term reuses the right-hand sides of the mean equations}. The covariance matrix is symmetric, so a node carries $R=P(P+1)/2$ unique covariance variables. Relative to the first-order node, the extra cost has two sources: evaluating the transfer-function derivatives $\partial_\lambda F_\mu$ and $\partial_\lambda\partial_\eta F_\mu$, and propagating the covariances.

\emph{Transfer-function derivatives (analytic):} Writing $F_\mu=\tfrac{1}{2\tau_V}\mathrm{erfc}(z_\mu)$, the $\mathrm{erfc}(z_\mu)$ already computed for $F_\mu$ is reused, and since $\partial_z\mathrm{erfc}\propto e^{-z^2}$ and $\partial_z^2\mathrm{erfc}\propto z\,e^{-z^2}$, the entire derivative tower needs only one new Gaussian factor $g_\mu=e^{-z_\mu^2}$ per population, with $W_g\approx 13$. The gradient ($P$ entries) and Hessian ($R$ unique entries) then reduce to chain-rule algebra through the moment derivatives $\partial\langle V\rangle,\partial\sigma_V,\partial\tau_V$ (and their second derivatives) and $\partial V_{\mathrm{thresh}}^{\mathrm{eff}}$. Estimating this at ${\approx}14$ FLOP-equivalents per gradient direction and ${\approx}18$ per unique Hessian entry,
\begin{equation*}
W_{\mathrm{der},\mu}\approx W_g+14P+18R\approx 13+14P+18R,
\end{equation*}
i.e. $W_{\mathrm{der},\mu}\approx 95$ for $P=2,\,R=3$. Finite-difference derivatives are substantially costlier: the Hessian alone needs $\mathcal{O}(P^2)$ re-evaluations\footnotemark of the full moment-and-transfer pipeline, giving $W_{\mathrm{node}}^{\mathrm{2nd}}\sim(1+2P^2)\,W_{\mathrm{node}}^{\mathrm{1st}}\approx 2.1\times 10^3$ at $P=2$. \revision{The figures assume this analytic-derivative accounting scenario. The coefficients above are approximate cost allowances, not an audited count for an implemented derivative kernel.}
\footnotetext{For all $\lambda, \eta$, $\frac{\partial^2 F_\mu}{\partial \nu_\lambda \partial \nu_\eta} \approx \frac{F_\mu(\nu_\lambda + h, \nu_\eta + h) - F_\mu(\nu_\lambda + h, \nu_\eta) - F_\mu(\nu_\lambda, \nu_\eta + h) + F_\mu(\nu_\lambda, \nu_\eta)}{h^2}$}

\emph{Covariance terms:} The curvature correction $\tfrac12\partial_\lambda\partial_\eta F_\mu\,c_{\lambda\eta}$ in each mean equation is a symmetric contraction over the $R$ unique entries, ${\approx}2R+1$ operations; over $P$ populations, $W_{\mathrm{mean,cov}}\approx P(2R+1)$. \revision{The gain term of the covariance equation, $\partial_\lambda F_\mu\,\partial_\eta F_\nu\,c_{\lambda\eta}=(JCJ^{\mathsf T})_{\mu\nu}$, is a triple matrix product: forming $B=CJ^{\mathsf T}$ costs $P^2(2P-1)$ operations, and the $R$ unique entries of $JB$ cost $R(2P-1)$. In each of the $R$ covariance equations, the source term $T^2\partial_t\nu_\mu\,\partial_t\nu_\nu$ reuses the right-hand sides of the mean equations (one multiply), and the $-c_{\mu\nu}$ term plus the final sums add ${\approx}2$; the diagonal finite-size term contributes ${\approx}3P$ over the $P$ diagonal entries (one subtraction and two multiplies each, with $1/T,1/N_\mu$ precomputed). Hence, $W_c\approx P^2(2P-1)+R(2P+2)+3P$.}

Adding these to the first-order node cost,$\ W_{\mathrm{node}}^{\mathrm{2nd,local}} \approx W_{\mathrm{node}}^{\mathrm{1st}}+P\,W_{\mathrm{der},\mu}+P(2R+1)+W_c$.

The Hessian term ($P\cdot R\sim P^3/2$ entries) \revision{and the triple product of $W_c$ control} the asymptotic dependence on P, so the node cost grows as $\mathcal{O}(P^3)$, against $\mathcal{O}(P^2)$ for first order. For the E/I node ($P=2,\,R=3$), $\,W_{\mathrm{node}}^{\mathrm{2nd,local}}\approx 232+2(95)+2(7)+\revision{36\approx 4.7\times 10^2}\ $
FLOP-equivalent operations per node and per RHS evaluation: roughly twice the first-order node and, like it, independent of the number of microscopic neurons represented.

\paragraph{Long-range coupling (local covariance).}

If covariances are kept local to each node, the network structure is unchanged from the first-order case: with $K$ nodes, connectome density $\rho$, and ${\approx}\rho K^2$ delayed edges of cost $c_{\mathrm{edge}}\approx 2$, $ \ W_{\mathrm{RHS}}^{\mathrm{2nd,local}}\approx K\,W_{\mathrm{node}}^{\mathrm{2nd,local}}+2\rho K^2 $.

The per-node state now contains \(P+P_w+R\) dynamical variables: \(P\) rates, \(P_w\) adaptation variables, and \(R\) covariances, so the Euler update contributes $2d_{\mathrm{local}}=2K(2P+R)=K(4P+2R)$. With $\Delta t=0.1\,\mathrm{ms}$,
\begin{equation}
\boxed{\,W_{1\mathrm{s}}^{\mathrm{2nd,local}}\approx 10^4\!\left[K\,W_{\mathrm{node}}^{\mathrm{2nd,local}}+2\rho K^2+2K(P+P_w+R)\right]\,}
\end{equation}
which for E/I is $\approx 10^4\!\left[486K+2\rho K^2\right]$. The network scaling $\mathcal{O}(K)+\mathcal{O}(\rho K^2)$ is identical to first order; only the node prefactor grows.

\paragraph{Streaming-access estimation (local covariance).}

Each node reads and writes its $2P+R$ dynamical variables ($4P+2R$ accesses) and writes the $3P$ stored moments $(\langle V\rangle,\sigma_V,\tau_V)$, so $a_{\mathrm{states}}^{\mathrm{2nd,local}}=7P+2R$. With direct delayed reads $a_{\mathrm{edge}}=2$ and one delay-buffer write $a_{\mathrm{hist}}=1$,
\begin{equation}
\boxed{\,M_{1\mathrm{s}}^{\mathrm{2nd,local}}\approx 8\times 10^4\!\left[(5P+2P_w+2R+1)K+2\rho K^2\right]\ \mathrm{bytes}\,}
\end{equation}
i.e. $8\times 10^4\!\left[21K+2\rho K^2\right]$ for $P=2,\,R=3$. As in the first-order case, connectome streaming dominates the local-state term beyond a handful of regions.

\paragraph{Global covariance: instantaneous, sparse, and delayed cases.}

Let \(Q=PK\) denote the number of population-rate variables across the \(K\) regions. Under the adaptation closure used in Eq.~\eqref{eq:second_order_local}, adaptation is treated as a deterministic slow variable, so the covariance matrix considered here contains rate--rate covariances only. It therefore has
\begin{equation*}
R_{\mathrm{global}}
=
\frac{Q(Q+1)}{2}
=
\frac{PK(PK+1)}{2}
=
\mathcal{O}(P^2K^2)
\end{equation*}
unique entries. Including covariances involving adaptation would define a different closure and would increase the covariance dimension. If we first neglect the delayed propagation and consider an instantaneous, Markovian global mean map such that
\[
\mathbf F:\mathbb R^Q\rightarrow\mathbb R^Q,
\qquad
J_F=\frac{\partial\mathbf F}{\partial\boldsymbol{\nu}}.
\]
then the covariance-propagation terms in Eq.~\eqref{eq:second_order_local} can be written as \revision{$J_FCJ_F^{\mathsf T}-C$}.

Because \(C=C^{\mathsf T}\), the cost depends on how the Jacobian structure is represented. The corrected covariance equation also contains \revision{the outer product $\mathbf{v}\mathbf{v}^{\mathsf T}$ of the mean velocities $\mathbf{v}=T\partial_t\boldsymbol{\nu}$} and the diagonal finite-size noise term. These contribute
\(\Theta(Q^2)\) and \(\Theta(Q)\) work, respectively. They do not alter the leading dense or sparse scaling and are included in the \(\Theta(Q^2)\) elementwise term below.

\emph{Dense-algebra case.}
If \(J_F\) is stored and multiplied as a dense \(Q\times Q\) matrix, the covariance propagation requires
\begin{equation}
W_{\mathrm{prop}}^{\mathrm{dense}} = \Theta(Q^3) = \Theta(P^3K^3).
\end{equation}
\revision{The product \(J_FCJ_F^{\mathsf T}\) requires two dense products, \(B=CJ_F^{\mathsf T}\) and \(J_FB\), i.e.\ approximately \(4Q^3\) classical multiply-add operations, or approximately \(3Q^3\) if only the symmetric half of \(J_FB\) is formed.} These constants are implementation choices.

\emph{Block-sparse Jacobian case.}
The Jacobian of the whole-brain mean dynamics is not generically dense. Its local contribution contains at most \(K\) dense \(P\times P\) blocks. Let \(E\simeq\rho K^2\) be the number of nonzero directed regional edges, and let \(\ell\) be the number of population-level Jacobian entries generated by one regional edge. Then
\begin{equation*}
\operatorname{nnz}(J_F) = \mathcal{O}\!\left(KP^2+\ell E\right) = \mathcal{O}\!\left(KP^2+\ell\rho K^2\right).
\end{equation*}

\revision{Each of the two products by this sparse Jacobian, \(CJ_F^{\mathsf T}\) and then \(J_F(CJ_F^{\mathsf T})\), costs \(\Theta(\operatorname{nnz}(J_F)Q)\), so that}
\begin{equation}
W_{\mathrm{prop}}^{\mathrm{sparse}} =
\Theta\!\left(\operatorname{nnz}(J_F)Q+Q^2\right) =
\mathcal{O}\!\left( P^3K^2 +\ell\rho P K^3 +P^2K^2 \right).
\end{equation}
For the long-range convention used in the manuscript, one excitatory source variable is transmitted by each regional edge. If that source affects all \(P\) target rate equations, then \(\ell=P\).

The curvature correction in the global mean equation also requires separate treatment. The local derivative coefficient \(W_{\mathrm{der},\mu}\), derived for \(P\) local input directions, does not include this global contraction. Its cost depends on the Hessian structure. A nonlinear transfer function applied after summing long-range inputs can generate mixed derivatives between distinct incoming regions, although low-rank or outer-product structure may permit a more efficient contraction than explicitly forming a dense Hessian. Consequently, the global RHS work is more appropriately decomposed as
\begin{equation}
W_{\mathrm{RHS}}^{\mathrm{2nd,global}}
=
W_{\mathrm{mean+adapt}}
+
W_J
+
W_{\mathrm{curv}}
+
W_{\mathrm{prop}},
\end{equation}
where \(W_{\mathrm{prop}}\) is chosen from the dense or sparse cases above, and \(W_J\) and \(W_{\mathrm{curv}}\) must be derived from the implemented analytic derivatives.

Thus, for any specified RHS implementation,
\begin{equation}
W_{1\mathrm{s}}^{\mathrm{2nd,global}}
\approx
\frac{T_{\mathrm{bio}}}{\Delta t}
\left[
W_{\mathrm{RHS}}^{\mathrm{2nd,global}}
+
2\left(
Q+P_wK+R_{\mathrm{global}}
\right)
\right].
\end{equation}
A lower bound on streaming-access estimation per biological second, ignoring the extra streaming of the dense $J$ and $C$ in the matrix products, is
\begin{equation}
M_{1\mathrm{s}}^{\mathrm{2nd,global}}\gtrsim 8\times 10^4\!\left[(5P+2P_w+1)K+2R_{\mathrm{global}}+2\rho K^2\right]\ \mathrm{bytes}.
\end{equation}
This is not the total traffic of the covariance propagation: accesses generated by \revision{\(J_FCJ_F^{\mathsf T}\)}, Hessian contractions, temporary arrays, and cache blocking depend on the implementation and memory hierarchy.

A delayed global second-order implementation must either propagate the required lagged covariance structure, introduce a finite-dimensional Markovian approximation to the delays, or explicitly neglect delay-induced covariance propagation. Its work and storage then depend on the number of distinct delays, history discretization, and chosen approximation.

In summary, the robust global-covariance result is the
\(\mathcal{O}(P^2K^2)\) rate-covariance footprint. The
\(\mathcal{O}(P^3K^3)\) work shown in the benchmark figures is a dense-algebra scenario for an instantaneous or Markovized closure. A block-sparse Jacobian gives \(\mathcal{O}(\operatorname{nnz}(J_F)PK)\) covariance-propagation work, while the delayed model requires an additional, explicitly specified lag-covariance or augmented-state treatment.

\paragraph{\revision{Scope of the plotted global-work proxy.}}

\revision{The figures use $P_w=P$, $\rho=1$, $Q=PK$, $R=P(P+1)/2$ and $10^4$ Euler steps per biological second. Its global-covariance work curve is generated by}
\revision{\[
\widehat W_{1\mathrm{s}}^{\mathrm{global}}
=10^4\!\Bigl[
  K W_{\mathrm{node}}^{\mathrm{1st}}
  +QW_{\mathrm{der},\mu}
  +2Q^2+4Q^3+2K^2
  +2\!\left(2Q+\frac{Q(Q+1)}{2}\right)
\Bigr].
\]}

\revision{Here $W_{\mathrm{node}}^{\mathrm{1st}}=15P^2+86P$ and $W_{\mathrm{der},\mu}=13+14P+18R$ reuse the local accounting above. The $4Q^3$ term represents the leading work of the two dense matrix products \revision{needed for $J_FCJ_F^{\mathsf T}$}. The quadratic allowance is schematic, and reusing the local derivative coefficient does not account explicitly for the global Jacobian and Hessian contractions. Accordingly, this curve illustrates a dense-algebra scaling scenario; its ordinate is not a complete operation count for a specified global closure. Such a count requires explicit $W_J$, $W_{\mathrm{curv}}$ and delay treatment. The plotted global array-access curve likewise includes the state/covariance-access baseline only, excluding the additional accesses required by these matrix and derivative operations.}

\subsubsection{Wong--Wang mean-field}

\paragraph{Node-local cost:}

The Wong--Wang (WW) model, Eqs.~\eqref{eq:Wong_Wang_model}-\eqref{eq:Wong_Wang_transfer}, is a two-variable reduction of a biophysical spiking network of decision-making~\cite{wong2006recurrent}. The firing rate enters only algebraically, through the input-output function \(H(x_i)\).

For a local node with \(P\) interacting populations indexed by \(\mu=1,\ldots,P\), Eq.~\eqref{eq:Wong_Wang_model} reads
\begin{equation}
\begin{aligned}
\frac{dS_\mu}{dt}
&= -\frac{S_\mu}{\tau_S} + (1-S_\mu)\,\gamma\,H(x_\mu), \\
x_\mu
&= \sum_{\nu=1}^{P} \tilde{J}_{\mu\nu} S_\nu + I_0 + I_\mu + I_{noise,\mu},
\end{aligned}
\label{eq:WW_general}
\end{equation}
where the effective local weights absorb the sign convention of Eq.~\eqref{eq:Wong_Wang_model}: \(\tilde{J}_{ii}=J_{ii}>0\) and \(\tilde{J}_{ij}=-J_{ij}<0\) for \(j\neq i\).

We assume that time-constant subexpressions are precomputed, consistent with the convention of the previous section, and we ignore the stochastic-input generation. The cost per node decomposes as $W_{node}^{WW} = W_{coupling} + P\left(W_x^{\mathrm{loc}} + W_{H} + W_{S}\right)$, where \(W_{coupling}\) is the within-node synaptic mixing, \(W_x^{\mathrm{loc}}\) is the per-population assembly of the remaining inputs into \(x_i\), \(W_{H}\) is the transfer-function evaluation, and \(W_{S}\) is the cost of the gating equation.

The transfer function \(H\) gives $W_{H} = 4\,(C_{+}/C_{-}/C_{\times}) + C_{\exp} + C_{\div} \approx 18$,
dominated by the single exponential and the division. The gating equation \(dS_i/dt = -S_i/\tau_S + (1-S_i)\gamma H\) is \(W_{S}\approx 5\). The recurrent sum \(\sum_j \tilde{J}_{ij}S_j\) requires \(P\) multiplications and \(P-1\) additions per population, so \(W_{coupling}\approx 2P^2\) across the node; assembling the remaining inputs \(I_0+I_i+I_{noise,i}\) costs \(W_x^{\mathrm{loc}}\approx 3\) per population. Therefore
\[
W_{node}^{WW} \approx 2P^2 + 26P.
\]
For the standard two-population node with \(P=2\), \(W_{node}^{WW} \approx 60\) FLOP-equivalent operations per node and per RHS evaluation. As in the master-equation estimate, this does not depend on the number of microscopic neurons represented by the node. Compared with the first-order adaptive node (\(W_{node}^{1st}\approx 15P^2+86P\approx 232\) at \(P=2\)), the WW node is roughly \(4\times\) cheaper, the difference being the absence of the \(\mathrm{erfc}\)-based transfer function, of the subthreshold-moment block, and of the effective threshold. Compared with the Montbrió--Pazó--Roxin node (\(\approx 30\) at \(P=2\)), it is about \(2\times\) more expensive, the difference being concentrated in the single exponential of \(H\).

\paragraph{Long-range coupling cost:}

The long-range structure is identical to the master-equation case. For a whole-brain network of \(K\) nodes with directed connectome density \(\rho\), the number of nonzero edges is \(\approx \rho K^2\), and each region receives an additional input of the generic form \(G\sum_j C_{ij} S_e(j,t-\tau_{ij})\), entering the synaptic drive \(x_i\), where \(S_e\) is the excitatory gating variable used for long-range projection. If delayed values are already buffered and require no interpolation, each edge costs \(c_{edge}\approx 2\). The RHS cost is then $W_{RHS}^{WW} \approx K W_{node}^{WW} + c_{edge}\,\rho K^2
\approx (2P^2+26P)K + 2\rho K^2$.
For fixed-step explicit Euler integration, the update contributes one multiplication and one addition per state variable, i.e. \(2d\) with \(d = PK\) (a single gating variable per population), adding \(2P\) per node. With \(\Delta t = 0.1\,\mathrm{ms}\) (\(T_{bio}/\Delta t = 10^4\) steps) and a dense connectome (\(\rho\simeq 1\)), the cost over one biological second is
\begin{equation}
\boxed{
W_{1s}^{WW} \approx 10^4\left[(2P^2+28P)K + 2\rho K^2\right]}.
\end{equation}
For \(P=2\) this gives \(W_{1s}^{WW} \approx 10^4\left[64K + 2K^2\right]\), versus \(\approx 10^4\left[240K + 2K^2\right]\) for the first-order adaptive system and \(\approx 10^4\left[38K + 2K^2\right]\) for the Montbrió--Pazó--Roxin model. The three expressions share the same \(O(K^2)\) coupling term: the WW node-local term is intermediate, so for small-to-moderate or sparse networks the WW model has substantially fewer FLOP-equivalent operations under this fixed-step convention than the adaptive formalism, whereas for large dense connectomes all three become coupling-dominated and the per-node differences become marginal relative to the shared \(K^2\) edge cost. If long-range coupling is taken to be instantaneous, as is common in whole-brain WW implementations, the delay-buffer bookkeeping below is removed but the arithmetic edge cost is unchanged.

\paragraph{Streaming-access estimation:}

We again count scalar transfers per Euler step and multiply by \(B=8\) bytes for double precision. With \(a_{\mathrm{states}}\) the per-node state accesses, \(a_{\mathrm{edge}}\) the per-edge accesses, and \(a_{\mathrm{hist}}\) the per-node delay-buffer writes,
\begin{equation}
M_{\Delta t}^{WW} \approx 8\left[ a_{\mathrm{states}}K
+ a_{\mathrm{edge}}\rho K^2 + a_{\mathrm{hist}}K \right].
\end{equation}

Each population reads and writes the single dynamical variable \(S_\mu\). If the firing rate \(H(x_\mu)\) is stored explicitly, this adds one write per population. This makes \(a_{\mathrm{states}} \approx 3P\). As before, the few local parameters are taken to be cache-resident. For each nonzero long-range edge, the coupling reads the structural weight and the delayed source gating, \(C_{ij}\) and \(S_e(j,t-\tau_{ij})\), giving \(a_{\mathrm{edge}}=2\) under the streaming-memory convention. At each step, the excitatory gating variable of every node is written into the delay buffer, so \(a_{\mathrm{hist}}=1\).

Combining the state, edge, and delay-buffer terms with \(\Delta t = 0.1\,\mathrm{ms}\),
\begin{equation}
\boxed{M_{1s}^{WW} \approx 8\times 10^4
\left[ (3P+1)K + 2\rho K^2 \right]\ \mathrm{bytes}}.
\end{equation}
Connectome traffic dominates local state traffic when \(a_{\mathrm{edge}}\rho K^2 \gtrsim (3P+1)K\), i.e. for \(\rho=1\), \(P=2\), when \(K \gtrsim 3.5\).

\subsubsection{Montbrió--Pazó--Roxin mean-field}

The Montbrió--Pazó--Roxin (MPR) equations provide an exact macroscopic reduction, in the thermodynamic limit, for the class of heterogeneous all-to-all coupled QIF networks and Lorentzian parameter distributions considered in the original derivation~\cite{montbrio2015macroscopic}. The \(P\)-population system below is used here as a computational-cost template. Its interpretation as an exact reduction requires the corresponding assumptions for each population and coupling term; arbitrary finite-size, sparse, or delayed extensions are not automatically covered by the original exactness result. In contrast to the master-equation formalism, the macroscopic dynamics are closed in only two state variables per population, and the RHS is a low-order polynomial. It contains no semi-analytic transfer function. This is the structural origin of its low node-local cost.

For a local node with \(P\) interacting populations indexed by \(\mu=1,\ldots,P\), Eq.~\eqref{eq:MPR_model} generalizes to
\begin{equation}
\begin{aligned}
\tau_m \frac{dr_\mu}{dt}
&= \frac{\Delta_\mu}{\pi\tau_m} + 2 r_\mu v_\mu, \\
\tau_m \frac{dv_\mu}{dt}
&= v_\mu^2 + \bar{\eta}_\mu + I_\mu(t)
+ \tau_m\sum_{\nu=1}^{P} J_{\mu\nu} r_\nu - (\pi\tau_m r_\mu)^2,
\end{aligned}
\label{eq:MPR_EI}
\end{equation}
where \(J_{\mu\nu}\) is the local synaptic weight from population \(\nu\) onto \(\mu\).

We assume that time-constant subexpressions are precomputed, consistent with the convention of the previous section. The cost per node decomposes as $W_{node}^{MPR} = P\left(W_r + W_v\right) + W_{coupling}$, where \(W_r\) and \(W_v\) are the per-population costs of the rate and mean-potential equations excluding the recurrent coupling, and \(W_{coupling}\) is the within-node synaptic mixing.

The rate equation costs \(W_r \approx 4\). 
The intrinsic part of the mean-potential equation costs \(W_v \approx 7\). 
The recurrent term \(\sum_\nu J_{\mu\nu} r_\nu\) requires \(P\) multiplications and \(P\) additions per population, giving \(W_{coupling} \approx 2P^2\) across the node. Therefore
\[
W_{node}^{MPR} \approx 2P^2 + 11P.
\]
For the standard E-I node with \(P=2\), \(W_{node}^{MPR} \approx 30\) FLOP-equivalent operations per node and per RHS evaluation. This does not depend on the number of microscopic neurons represented by the node. The quadratic term \(2P^2\) reflects the all-to-all synaptic mixing between the \(P\) local populations; it is a genuine multiply-add count and carries no special-function overhead. Compared with the first-order adaptive node (\(W_{node}^{1st} \approx 15P^2+86P \approx 232\) at \(P=2\)), the MPR node is roughly \(8\times\) cheaper, the difference being dominated by the absence of the \(\mathrm{erfc}\)-based transfer function and of the subthreshold-moment block.

\paragraph{Long-range coupling cost:}

The long-range structure is identical to the master-equation case. For a whole-brain network of \(K\) nodes with directed connectome density \(\rho\), the number of nonzero edges is \(\approx \rho K^2\), and each region receives a delayed term of the generic form \(G \sum_j C_{ij} r_e(j,t-\tau_{ij})\), where \(r_e\) is the excitatory firing rate used for long-range projection. If delayed values are already buffered and require no interpolation, each edge costs \(c_{edge} \approx 2\). The RHS cost is then
\[
W_{RHS}^{MPR} \approx K W_{node}^{MPR} + c_{edge}\,\rho K^2
\approx (2P^2+11P)K + 2\rho K^2.
\]
For fixed-step explicit Euler integration, the update contributes one multiplication and one addition per state variable, i.e. \(2d\) with \(d = 2PK\), adding \(4P\) per node. With \(\Delta t = 0.1\,\mathrm{ms}\) (\(T_{bio}/\Delta t = 10^4\) steps) and a dense connectome (\(\rho \simeq 1\)), the cost over one biological second is
\begin{equation}
\boxed{
W_{1s}^{MPR} \approx 10^4\left[(2P^2+15P)K + 2\rho K^2\right]}.
\end{equation}
For \(P=2\) this gives \(W_{1s}^{MPR} \approx 10^4\left[38K + 2K^2\right]\), versus \(\approx 10^4\left[240K + 2K^2\right]\) for the first-order adaptive system. The two expressions share the same \(O(K^2)\) coupling term: the MPR advantage is concentrated entirely in the node-local \(O(K)\) term, where it is about \(6\times\) cheaper. Under the present fixed-step FLOP-equivalent convention, this model has a smaller node-local operation count. This ratio is not a wall-clock speedup: actual performance additionally depends on numerical accuracy, vectorization, special-function implementation, cache behavior, parallelization, and communication. Nevertheless, for very large dense connectomes, all models become coupling-dominated and the per-node savings become marginal relative to the shared  \(K^2\) edge cost (assuming same coupling convention).

\paragraph{Streaming-access estimation:}
We again count scalar transfers per Euler step and multiply by \(B=8\) bytes for double precision. With \(a_{\mathrm{states}}\) the per-node state accesses, \(a_{\mathrm{edge}}\) the per-edge accesses, and \(a_{\mathrm{hist}}\) the per-node delay-buffer writes,
\begin{equation}
M_{\Delta t}^{MPR} \approx 8\left[ a_{\mathrm{states}}K
+ a_{\mathrm{edge}}\rho K^2 + a_{\mathrm{hist}}K \right].
\end{equation}

Each population reads and writes only the two dynamical variables \((r_\mu, v_\mu)\), with no intermediate moments to store. This makes \(a_{\mathrm{states}} \approx 4P\), compared with \(7P\) for the master-equation node. As before, the few local parameters (\(\bar{\eta}_\mu, \Delta_\mu\)) are taken to be cache-resident.
For each nonzero long-range edge, the coupling reads the structural weight and the delayed source activity, \(C_{ij}\) and \(r_e(j,t-\tau_{ij})\), giving \(a_{\mathrm{edge}}=2\) under the streaming-memory convention. At each step, the excitatory rate of every node is written into the delay buffer, so \(a_{\mathrm{hist}}=1\).

Combining the state, edge, and delay-buffer terms with \(\Delta t = 0.1\,\mathrm{ms}\),
\begin{equation}
\boxed{M_{1s}^{MPR} \approx 8\times 10^4
\left[ (4P+1)K + 2\rho K^2 \right]\ \mathrm{bytes}}.
\end{equation}
Connectome traffic dominates local state traffic when \(a_{\mathrm{edge}}\rho K^2 \gtrsim (4P+1)K\), i.e. for \(\rho=1\), \(P=2\), when \(K \gtrsim 4.5\). As in the master-equation case, dense whole-brain coupling is therefore rapidly dominated by reading connectivity weights and delayed source states, although in practice a small parcellation can keep the connectome cache-resident and the effective DRAM traffic correspondingly lower.

Because the MPR right-hand side is arithmetically lean, its source-level work-to-access ratio is lower than that of the adaptive master-equation node. In the node-local regime, $\frac{W_{1\mathrm{s}}}{M_{1\mathrm{s}}} \approx \frac{2P^2+15P}{8(4P+1)} \approx 0.5$ FLOP-equivalent per counted byte at \(P=2\), compared with approximately \(2.0\) for the first-order adaptive node. In the coupling-dominated regime, both proxies approach \(1/8\) FLOP-equivalent per counted byte. At this stage, these ratios describe the adopted source-level accounting and do not determine a hardware bottleneck without cache-filtered traffic and a machine-specific Roofline analysis~\cite{williams2009roofline}, but they reveal a substantial difference of workload between models.

\subsubsection{SNN computational cost analysis}

In contrast to the macroscopic models above, a spiking neural network (SNN) is the microscopic system that those mean-field equations approximate. Its cost therefore no longer reduces to a fixed per-population figure but scales explicitly with the number of neurons. We consider a single recurrent network of \(N\) neurons: there is no nodal or regional decomposition here, the whole system is one flat spiking network. The connectivity is sparse, with connection probability \(p\approx 0.1\), so the network contains \(\approx pN^2\) synapses. This synaptic matrix is very large (millions of nonzeros) even though it is sparse, which has direct consequences for memory traffic.

Let \(\bar{k}\) be the mean number of outgoing synapses per neuron. The expected work of one integration step is
\begin{equation}
W_{\Delta t}^{\mathrm{SNN}}
=
\underbrace{c_{\mathrm{neur}}N}_{\text{clock-driven}}
+
\underbrace{
c_{\mathrm{syna}}\langle\nu\rangle\Delta t\,N\bar{k}
}_{\text{event-driven}},
\label{eq:SNN_step}
\end{equation}
where the event term counts the expected number of emitted spikes, \(\langle\nu\rangle\Delta t\,N\), multiplied by their mean fan-out \(\bar{k}\). For a fixed connection probability \(p\), \(\bar{k}=pN\) and the event term scales as \(pN^2\). For fixed mean out-degree, it instead scales linearly in \(N\). The coefficient \(c_{\mathrm{syna}}\) describes only the arithmetic assigned to one static synaptic delivery; delay queues, receptor-specific updates, plasticity, and recording require additional terms. \(\langle\nu\rangle\) is the mean firing rate (assumed \(\approx 5\,\mathrm{Hz}\)) and \(c_{\mathrm{syna}}\approx 2\) FLOP-equivalents is the cost of delivering one spike across one synapse. The first term is paid every step for every neuron (state-variable integration); the second is paid only when spikes are emitted (synaptic transmission). This hybrid accounting follows the clock-driven/event-driven decomposition of network simulators~\cite{brette2007simulation}: per step the network emits \(\langle\nu\rangle\Delta t\,N\) spikes, each delivered to its \(\approx pN\) postsynaptic targets, giving \(\langle\nu\rangle\Delta t\,pN^2\) synaptic events.

\paragraph{Neuron-local (clock-driven) cost:}

For a fixed-step Euler scheme the per-neuron cost follows the same convention as the macroscopic models, \(c_{\mathrm{neur}} = n_{\mathrm{RHS}}W_{\mathrm{RHS}} + 2 d_{\mathrm{neur}}\), with \(n_{\mathrm{RHS}}=1\). Here \(d_{\mathrm{neur}}\) is the number of intrinsic state variables of the neuron, namely the membrane potential plus any intrinsic gating or adaptation variables. Synaptic inputs are delivered as event-driven jumps and are accounted for separately, so \(d_{\mathrm{neur}}\) does not include synaptic variables and \(W_{\mathrm{RHS}}\) is the intrinsic single-neuron vector field. Using the FLOP-equivalent convention of this Supplementary Note, the dominant cost is the number of special-function evaluations and state variables.

\paragraph{Choice of the integration step:}

A subtlety is that the appropriate integration step \(\Delta t\) is not the same for every neuron model, and the number of steps per biological second, \(T_{\mathrm{bio}}/\Delta t\), is itself part of the cost. The step is set by the fastest timescale that must be resolved, under three constraints: stability of the explicit scheme (a stiff vector field forces a small \(\Delta t\)); truncation accuracy, which for Euler is \(O(\Delta t)\) and is largest precisely during the spike upstroke; and the precision of spike-emission times, which a coarse grid smears by \(\sim\Delta t\). 

For the illustrative accounting below, we adopt \(\Delta t=1\,\mathrm{ms}\) for LIF and Izhikevich neurons and
\(\Delta t=0.1\,\mathrm{ms}\) for AdEx and Hodgkin--Huxley neurons. These resolutions reproduce the historical operation-count scenario \cite{izhikevich2004model, brette2005adaptive}; they are not universal accuracy or stability limits.


\begin{center}
\begin{tabular}{lcccc}
\hline
Neuron model & \(d_{\mathrm{neur}}\) & \(c_{\mathrm{neur}}\) (FLOP-eq/step) & typical \(\Delta t\) & clock cost \(/\mathrm{s}\) (per neuron) \\
\hline
LIF        & 1 & \(\approx 5\)   & \(1\,\mathrm{ms}\)   & \(\approx 5\times10^{3}\) \\
Izhikevich & 2 & \(\approx 13\)  & \(1\,\mathrm{ms}\)   & \(\approx 1.3\times10^{4}\) \\
AdEx       & 2 & \(\approx 28\)  & \(0.1\,\mathrm{ms}\) & \(\approx 2.8\times10^{5}\) \\
Hodgkin-Huxley & 4 & \(\approx 120\) & \(0.1\,\mathrm{ms}\) & \(\approx 1.2\times10^{6}\) \\
\hline
\end{tabular}
\end{center}
The per-step figures are for the intrinsic neuron with a minimal current-based synaptic input. The LIF neuron updates a single linear membrane variable, hence \(c_{\mathrm{neur}}\approx 5\). The Izhikevich neuron adds a recovery variable and a quadratic term but no special function, giving \(c_{\mathrm{neur}}\approx 13\). The AdEx neuron contains a single exponential whose cost dominates and gives \(c_{\mathrm{neur}}\approx 28\). The Hodgkin--Huxley neuron evaluates six voltage-dependent rate functions, each containing an exponential, so the exponentials alone account for the bulk of \(c_{\mathrm{neur}}\approx 120\); this per-step figure corresponds to the per-millisecond count (\(\sim 1200\)) of Izhikevich~\cite{izhikevich2004model} evaluated at \(\Delta t = 0.1\,\mathrm{ms}\). The final column anticipates the cost per biological second derived below: it combines the per-step cost with the model's own step count \(T_{\mathrm{bio}}/\Delta t\), and shows that the two effects compound. Reading the per-step column alone, the LIF-to-Hodgkin--Huxley ratio is \(\approx 24\); once the smaller step required by the detailed models is included, the illustrative ratio under the adopted Euler resolutions widens to \(\approx 240\). 

The integration resolutions in the table are illustrative fixed-step Euler choices, selected to reproduce the historical operation-count comparison and common simulation practice. They are not intrinsic constants of the neuron models. Stability and accuracy depend on model parameters, synaptic kinetics, spike-time tolerance, and the numerical solver. Linear LIF subthreshold dynamics may, for example, be integrated exactly, whereas adaptive or implicit solvers can alter the step requirements of AdEx and conductance-based models. The table should therefore be interpreted as one explicit numerical scenario rather than a common-error benchmark.

Under a simple fixed-step Euler implementation with one first-order conductance variable per receptor type, two excitatory/inhibitory conductance states add approximately eight FLOP-equivalents per neuron and per step under the present convention. This is an illustrative coefficient rather than a model-independent constant: the increment depends on receptor kinetics, the number of rise and decay variables, exact versus numerical decay updates, and the numerical solver.

\paragraph{Synaptic (event-driven) cost:}

The second term of Eq.~\eqref{eq:SNN_step} carries an explicit factor \(\langle\nu\rangle\Delta t\), so the relative weight of the two regimes depends on the network size and on the step. The clock-driven and event-driven costs are equal, over one biological second, at a crossover size $N^{\star} = c_{\mathrm{neur}}/(c_{\mathrm{syna}}\,\langle\nu\rangle\,\Delta t\,p)$.
For \(c_{\mathrm{syna}}=2\), \(\langle\nu\rangle=5\,\mathrm{Hz}\), and \(p=0.1\), the denominator is \(c_{\mathrm{syna}}\langle\nu\rangle p \,\Delta t = \Delta t\) in SI units, so \(N^{\star}=c_{\mathrm{neur}}/\Delta t\). At each model's own step this gives \(N^{\star}\approx 5\times10^{3}\) (LIF, \(1\,\mathrm{ms}\)), \(\approx 1.3\times10^{4}\) (Izhikevich, \(1\,\mathrm{ms}\)), \(\approx 2.8s\times10^{5}\) (AdEx, \(0.1\,\mathrm{ms}\)), and \(\approx 1.2\times10^{6}\) (Hodgkin--Huxley, \(0.1\,\mathrm{ms}\)); numerically equal to the clock-cost-per-neuron column above, since the event prefactor is unity. Below these thresholds single-neuron integration dominates; only for very large or networks with simpler neuron models does synaptic delivery become the bottleneck. Note the direction of the step dependence: a coarser \(\Delta t\) lowers \(N^{\star}\), because it reduces the clock-driven work while leaving the event-driven work unchanged. A LIF network run at \(1\,\mathrm{ms}\) therefore becomes synapse-dominated at a smaller size (\(N^{\star}\approx 5\times10^{3}\)) than the same network forced to \(0.1\,\mathrm{ms}\) (\(N^{\star}\approx 5\times10^{4}\)).

\paragraph{Cost over one biological second:}

Accumulating over \(T_{\mathrm{bio}}/\Delta t\) steps,
\begin{equation}
W_{1\mathrm{s}}^{\mathrm{SNN}}
\approx \frac{T_{\mathrm{bio}}}{\Delta t}\,c_{\mathrm{neur}}\,N
+ T_{\mathrm{bio}}\,c_{\mathrm{syna}}\,\langle\nu\rangle\,p\,N^2.
\end{equation}
The factor \(\Delta t\) cancels in the event-driven term: over a fixed biological duration the synaptic cost is set by the total number of spikes (\(\propto \langle\nu\rangle T_{\mathrm{bio}}\)), not by the number of steps, and is therefore independent of the integration step. Only the clock-driven term scales as \(1/\Delta t\), and this is where the model-dependent step enters. With \(T_{\mathrm{bio}}=1\,\mathrm{s}\) and \(c_{\mathrm{syna}}\langle\nu\rangle p = 1\),
\begin{equation}
\boxed{
W_{1\mathrm{s}}^{\mathrm{SNN}} \approx \frac{c_{\mathrm{neur}}}{\Delta t}\,N + N^2\ \text{FLOP-equivalents}},
\end{equation}
so that \(c_{\mathrm{neur}}/\Delta t\) is the per-neuron clock coefficient tabulated above. Using each model's appropriate \(\Delta t\) gives the cost under the adopted illustrative resolutions per biological second: the step count is then part of the cost, and the detailed models pay twice, through both a higher per-step cost and a finer step.

\paragraph{Streaming-access estimation:}

Let \(b_{\mathrm{state}}\) be the bytes moved per neuron and per clock-driven step, and \(b_{\mathrm{event}}\) the bytes moved per delivered synaptic event. The latter includes the connectivity representation, weight, and postsynaptic update and need not be an integer multiple of the floating-point state size because target indices may use a different precision. The array-access proxy over one biological second is
\begin{equation}
\boxed{
M_{1\mathrm{s}}^{\mathrm{SNN}}
\approx
\frac{T_{\mathrm{bio}}}{\Delta t}
b_{\mathrm{state}}N
+
T_{\mathrm{bio}}
b_{\mathrm{event}}
\langle\nu\rangle N\bar{k}.
}
\end{equation}
For the illustrative choices \(T_{\mathrm{bio}}=1\,\mathrm{s}\), \(\bar{k}=pN\), \(p=0.1\), \(\langle\nu\rangle=5\,\mathrm{Hz}\), \(b_{\mathrm{state}}=8c_{\mathrm{states}}\), and \(b_{\mathrm{event}}=32\) bytes, this becomes
\[
M_{1\mathrm{s}}^{\mathrm{SNN}}
\approx
\frac{8c_{\mathrm{states}}N}{\Delta t}
+16N^2\ \mathrm{bytes},
\]
with \(\Delta t\) expressed in seconds. Here \(c_{\mathrm{states}}\) denotes the number of double-precision scalar accesses per neuron and per clock-driven step, typically about twice the number of clock-driven state variables when each is read and written once.

The resulting work-to-access ratio is
\[
I(N,\Delta t)
=
\frac{
c_{\mathrm{neur}}N/\Delta t+
c_{\mathrm{syna}}\langle\nu\rangle N\bar{k}
}{
b_{\mathrm{state}}N/\Delta t+
b_{\mathrm{event}}\langle\nu\rangle N\bar{k}
}.
\]
It generally depends on \(\Delta t\) in the crossover regime. In the clock-driven limit it approaches \(c_{\mathrm{neur}}/b_{\mathrm{state}}\), while in the event-driven limit it approaches \(c_{\mathrm{syna}}/b_{\mathrm{event}}\); these two limiting plateaus are independent of \(\Delta t\). Neither plateau establishes a hardware bottleneck without cache-filtered traffic and a machine-specific Roofline analysis \cite{williams2009roofline}.

\subsubsection{TRIBE v2 cost analysis}

TRIBE v2~\cite{d2026foundation} differs fundamentally from the preceding mechanistic models. Rather than integrating a dynamical system, it predicts BOLD responses from multimodal stimulus representations. Its computational pipeline comprises two distinct workloads: feature extraction from raw text, audio, and video, followed by fMRI prediction from the resulting cached features. For comparison with the preceding models, the quantities below are expressed as approximate FLOP-equivalents per second of input stimulus.

\paragraph{Raw stimuli to cached features.}
Frozen modality-specific encoders are applied at different native resolutions. Llama-3.2-3B embeds each timed word using up to 1,024 preceding words; Wav2Vec-BERT-2.0 processes $60$ s audio chunks; and Video-JEPA-2-Giant processes 64 frames spanning 4s at $f_{\mathrm{stim}}=2\,\mathrm{Hz}$. A simple per-second cost proxy is therefore
\begin{equation}
W_{\mathrm{raw}\rightarrow\mathrm{cache}}
\approx
r_{\mathrm{word}}C_{\mathrm{text}}^{(1024)}
+\frac{C_{\mathrm{audio}}^{(60\,\mathrm{s})}}{60}
+f_{\mathrm{stim}}C_{\mathrm{video}}^{(64,\mathrm{frames})},
\label{eq:TRIBE_feature_cost}
\end{equation}
where $r_{\mathrm{word}}$ is the number of timed words per second and each $C$ denotes the arithmetic cost of one corresponding encoder evaluation. Terms associated with absent modalities are omitted. The audio term assumes non-overlapping chunks, whereas the video term accounts for one overlapping $4\,\mathrm{s}$ evaluation every $0.5\,\mathrm{s}$.

This expression is intentionally symbolic. Encoder parameter counts alone do not provide a reliable FLOP estimate because the workloads depend on sequence length, tokenization, overlapping contexts, native sampling rates, batching, and implementation. The authors report that feature extraction for the complete dataset required approximately $24\,\mathrm{h}$ on $128$ V100 GPUs with $32\,\mathrm{GB}$ of memory each. The extracted representations were stored as memory-mapped arrays, so this cost is incurred once for a given stimulus set and can be amortized when the same features are reused across subjects, models, or analyses.

\paragraph{Cached features to fMRI.}

For each modality, selected intermediate encoder layers are first averaged within $l$ layer groups. \revision{We assumed $l=3$ for the plotted accounting scenario, as implemented in the repository}. The three group-level representations are then concatenated and mapped by a trainable projector to $384$ dimensions. The projected text, audio, and video representations are subsequently concatenated into a $D_{\mathrm{model}}=3\times384=1152$ dimensional vector at each point of the $f_{\mathrm{stim}}=2\,\mathrm{Hz}$ temporal grid. A context window of duration $T=100\,\mathrm{s}$ therefore contains $n_{\mathrm{ctx}} = f_{\mathrm{stim}}T = 200$ temporal tokens. These tokens are processed jointly by an eight-layer, eight-head temporal transformer. In the architecture described in the article, adaptive average pooling subsequently reduces the temporal sampling rate from $f_{\mathrm{stim}}=2\,\mathrm{Hz}$ to $f_{\mathrm{fMRI}}=1\,\mathrm{Hz}$, after which a subject-conditioned linear readout predicts either $N_{\mathrm{targets}}=20{,}484$ cortical vertices or $N_{\mathrm{targets}}=8{,}802$ subcortical voxels (these correspond to separate output configurations).

Let $N_{\mathrm{proj}}$ and $N_{\mathrm{tr}}$ denote the numbers of weights in the modality projectors and temporal transformer, respectively. Assuming non-overlapping $100\,\mathrm{s}$ windows and counting one multiplication and one addition as two FLOP-equivalents~\cite{kaplan2020scaling}, the approximate paper-level inference cost per second of input stimulus is

\begin{equation}
\boxed{
W_{\mathrm{cache}\rightarrow\mathrm{fMRI}}
\approx
2f_{\mathrm{stim}}
\left(
N_{\mathrm{proj}}+N_{\mathrm{tr}}
\right)
+
4n_{\mathrm{layer}}
\frac{n_{\mathrm{ctx}}^{2}}{T}
D_{\mathrm{model}}
+
2f_{\mathrm{fMRI}}
D_{\mathrm{model}}
N_{\mathrm{targets}}.
}
\label{eq:TRIBE_cached_cost}
\end{equation}

The first term accounts for the dense operations in the modality projectors and transformer layers. The second accounts for the two sequence-length-dependent matrix products in self-attention, $QK^{\mathsf{T}}$ and $\mathrm{softmax}(QK^{\mathsf{T}})V$. The third represents the linear fMRI readout. Biases, normalization, nonlinear activations, softmax, positional embeddings, and adaptive pooling are neglected.

Using the feature dimensions reported in the article gives $N_{\mathrm{proj}} \simeq
3\times384 \left(2048+1024+1280 \right) \simeq 5.0\times10^{6}$, where the leading factor of $3$ corresponds to the three retained layer groups. For a standard transformer with a feed-forward expansion factor of $4$,$N_{\mathrm{tr}}\simeq 12n_{\mathrm{layer}}D_{\mathrm{model}}^{2} \simeq 1.27\times10^{8}.$

The target-independent projector, transformer, and attention terms therefore contribute approximately $5.44\times10^{8}$ FLOP-equivalents per second of input stimulus. The spatial readout contributes approximately $4.72\times10^{7}$ FLOP-equivalents for cortex or $2.03\times10^{7}$ for subcortex, yielding $W_{\mathrm{cache}\rightarrow\mathrm{fMRI}}
\simeq 5.92\times10^{8} \text{ for cortical prediction},\, 5.65\times10^{8},\text{for subcortical prediction}.$

At the published spatial resolutions, the arithmetic cost is therefore dominated by the modality projectors and temporal transformer, although the readout cost remains linear in $N_{\mathrm{targets}}$.

Equation~\eqref{eq:TRIBE_cached_cost} is a transparent proxy based on the architecture described in the article rather than an exact profile of the released software. The public implementation includes an additional projection to $D_{\mathrm{head}}=2048$ dimensions and applies the target readout before temporal pooling, not accounted here. These implementation choices change the numerical readout cost but not its linear dependence on $N_{\mathrm{targets}}$.

Within the same paper-level approximation, the inference-weight footprint for one subject is $M_{\mathrm{weights}}
\approx B\left( N_{\mathrm{proj}} + N_{\mathrm{tr}} + D_{\mathrm{model}}N_{\mathrm{targets}} \right),$
where $B$ is the number of bytes used to store each parameter. This quantity represents stored weights, not memory traffic, and excludes activations and temporary workspace. For half-precision weights, $B=2$ bytes, giving approximately $312\,\mathrm{MB}$ for the cortical predictor or $285\,\mathrm{MB}$ for the subcortical predictor. During multi-subject training, the subject-specific block instead scales as $\mathcal{O}(S D_{\mathrm{model}}N_{\mathrm{targets}})$, where $S$ is the number of training subjects.

For prediction from a novel raw stimulus, the two inference workloads add:

\begin{equation}
W_{\mathrm{TRIBE}}
=
W_{\mathrm{raw}\rightarrow\mathrm{cache}}
+
W_{\mathrm{cache}\rightarrow\mathrm{fMRI}}.
\end{equation}

For previously cached stimuli, only $W_{\mathrm{cache}\rightarrow\mathrm{fMRI}}$ is incurred. Training is excluded from both estimates. These expressions provide a deliberately naive arithmetic comparison with mechanistic models: the models do not perform equivalent scientific tasks, and the low fMRI sampling rate should not be interpreted as implying a proportionally low computational workload. Empirical profiling remains necessary for comparisons of wall-clock time, memory traffic, and energy consumption.

\subsection{Regional SNNs versus matched mean-field models: an open benchmark}
\label{subsec:regional_snn_meanfield_benchmark}

The preceding SNN analysis considers a single flat recurrent network. It therefore cannot be applied directly to a whole-brain system by simply replacing $N$ with $Kn$, where $K$ is the number of regions and $n$ the number of neurons per region: this would describe a brain-wide random graph rather than $K$ local microcircuits coupled through a regional connectome. A meaningful intermediate architecture would instead place an SNN of $n$ neurons in each region and compare it with a mean-field network built from the same regional circuit. Thus, "$K$ nodes versus $K$ nodes" is not by itself a matched comparison: a mean-field node contains a few population variables, whereas an SNN region explicitly instantiates neurons and synapses.

For local connection probability $p$, mean firing rate $\langle\nu\rangle$, and biological duration $T_{\mathrm{bio}}$, such a regional SNN contains
\begin{equation}
N_{\mathrm{tot}}=Kn,\qquad
N_{\mathrm{syn,loc}}\simeq Kpn^{2},\qquad
N_{\mathrm{event,loc}}\simeq Kpn^{2}\langle\nu\rangle T_{\mathrm{bio}},
\end{equation}
and its approximate work is
\begin{equation}
W_{\mathrm{SNN}}^{\mathrm{reg}}
\simeq K\left[\frac{T_{\mathrm{bio}}}{\Delta t}c_{\mathrm{neur}}n+
T_{\mathrm{bio}}c_{\mathrm{syna}}\langle\nu\rangle pn^{2}
\right]+W_{\mathrm{LR}}.
\end{equation}
The long-range term $W_{\mathrm{LR}}$ depends critically on the implementation. Regional rate-mediated coupling can retain an approximately $\mathcal{O}(\rho K^{2})$ cost, whereas explicit microscopic projections scale with their number of long-range synapses and spike deliveries. Fixed long-range out-degree and fixed pairwise connection probability consequently define very different workloads.

Setting \(W_{\mathrm{LR}}=0\) to isolate local-microcircuit work, and taking \(K=50\), \(p=0.1\), and \(\langle\nu\rangle=5\,\mathrm{Hz}\), increasing \(n\) from \(10^3\) to \(10^4\) gives \(5\times10^4\) to \(5\times10^5\) neurons, \(5\times10^6\) to \(5\times10^8\) local synapses, and \(2.5\times10^7\) to \(2.5\times10^9\) local synaptic deliveries per biological second. This assumes a rate-mediated regional coupling such that the omission of $W_{\mathrm{LR}}$ remains small, but for a fixed microscopic connection probability across nonzero
regional blocks, $W_{\mathrm{LR}}$ may dominate and can recover a quadratic \(\mathcal{O}((Kn)^2)\) workload comparable in scaling to the flat-network scenario. At $8\text{-}16$ bytes per stored synapse, local connectivity alone would occupy approximately $40\text{-}80\mathrm{MB}$ to $4\text{-}8\mathrm{GB}$. Under the FLOP-equivalent conventions used above, the selected 50-node mean-field systems require approximately $0.07\text{-}\revision{0.29}\times10^{9}$ FLOP-equivalents per biological second. The corresponding regional SNN estimate ranges from $0.30$ to $7.5\times10^{9}$ for LIF neurons and from $14.1$ to $145\times10^{9}$ for AdEx neurons as $n$ increases from $10^{3}$ to $10^{4}$. These estimates show that a small LIF microcircuit per region can overlap with an elaborate local mean-field closure in abstract work count, whereas larger circuits or more detailed neuron models can add orders of magnitude. They do not predict wall-clock time or establish real-time feasibility.

These values compare local arithmetic scenarios only. They exclude \(W_{\mathrm{LR}}\), communication, initialization, random-number generation, recording, and post-processing, and they do not yet constitute a matched parent-SNN/mean-field benchmark. Their purpose is to show that the workload depends jointly on regional circuit size, neuron and synapse models, connectivity scaling, and long-range implementation.

\paragraph{Biological informativeness versus detail.}
An SNN is more microscopically detailed, but it is not necessarily more informative, identifiable, or reliable. It can explicitly represent spike timing, finite-size fluctuations, trial variability, cellular heterogeneity, sparse assemblies, microscopic correlations, reset and refractory effects, and spike-dependent plasticity. Mean-field models can nevertheless already represent firing rates, membrane-potential statistics, receptor and adaptation dynamics, oscillations, Up/Down states, delayed propagation, perturbation responses, and, with suitable closures, approximate population covariances. For observables dominated by slow population dynamics, such as BOLD activity or coarse functional connectivity, microscopic variables may average out without improving predictions. Additional parameters can also introduce degeneracy and weaken identifiability.

A defensible comparison should therefore use a regional SNN and its derived or calibrated mean-field surrogate, with matched regional scale, connectome, delays, inputs, operating point, perturbations, biological duration, observation model, and validation targets. Numerical convergence should be established independently rather than imposing an identical time step. Construction, warm-up, integration, communication, recording, and post-processing should be timed separately, and wall time, real-time factor, peak memory, energy, and input/output should be reported for no-recording, regional-rate, and full-spike workloads. Scientific agreement should be assessed using population rates and spectra, variability, synchrony, correlations, state transitions, propagation, and responses to identical interventions, while the total ensemble cost required for fitting or sensitivity analysis should be distinguished from the cost of one forward trajectory.

This comparison raises several further questions: what is the smallest $n$ for which the target macroscopic observables converge; under which regimes do spike timing, heterogeneity, or finite-size correlations alter whole-brain predictions; how should long-range spikes be mapped between regional microcircuits; and can hybrid architectures restrict explicit spiking dynamics to regions in which the mean-field closure fails? The present analysis identifies a feasible and informative benchmark program, but does not establish that regional whole-brain SNNs are either scientifically superior or sufficiently tractable for inference.

\begin{figure}[hbt!]
    \centering
    \includegraphics[width=0.7\textwidth]{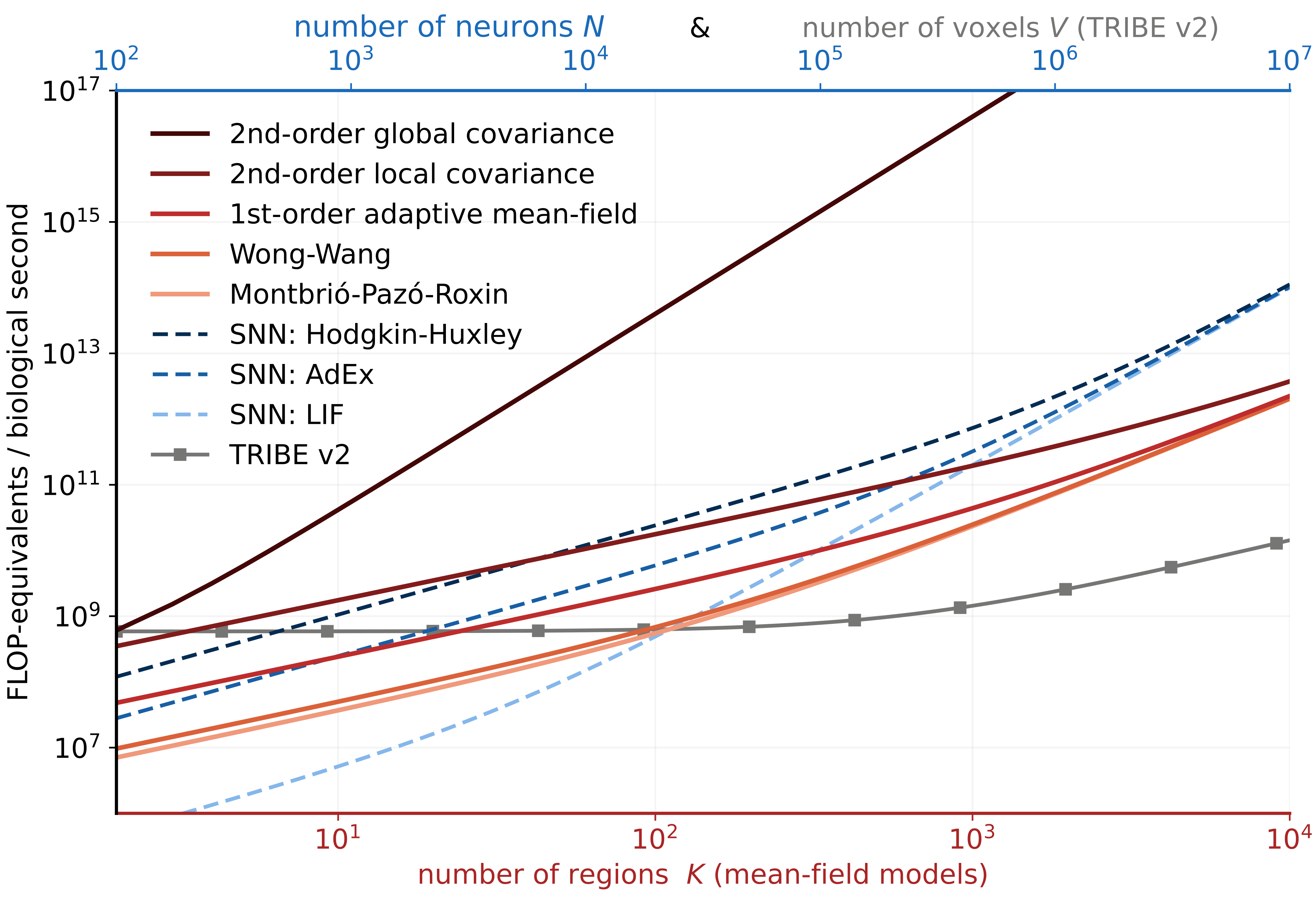}
    \caption{\revision{Illustrative work scenarios with $P=P_w=10$ populations per regional node and $\rho=1$. Red curves use region count $K$; blue dashed SNN curves use neuron count $N$; gray squares use the number of fMRI prediction targets $V$ (cortical vertices or subcortical voxels, abbreviated as "voxels" on the axis). The ordinate is normalized per simulated biological second for dynamical models and per second of input stimulus for cached-feature prediction; raw-feature extraction and training are excluded. Regional models use $0.1\,\mathrm{ms}$ Euler steps, while SNN resolutions are specified in Supplementary~\ref{sec:algorithmic_cost_analysis_neural_simulation_appendix}. Increasing $P$ changes the regional workload only; the SNN and prediction scenarios are unchanged. Local second-order work assumes analytic derivatives, and the global curve is a schematic dense-algebra proxy with leading $\mathcal{O}(P^3K^3)$ work. Its continuation exceeds the displayed range. Separate size axes do not define cross-family equivalence, break-even points or scientific-efficiency rankings.} Horizontal positions should be compared only within the same color family; the purpose of the figure is to show the scaling regimes per family and orders of magnitude rather than a one-to-one equivalence between \(K\), \(N\), and \(V\).}
    \label{fig:appendix_cost_compute_P10}
\end{figure}

\begin{figure}[hbt!]
    \centering
    \includegraphics[width=0.8\textwidth]{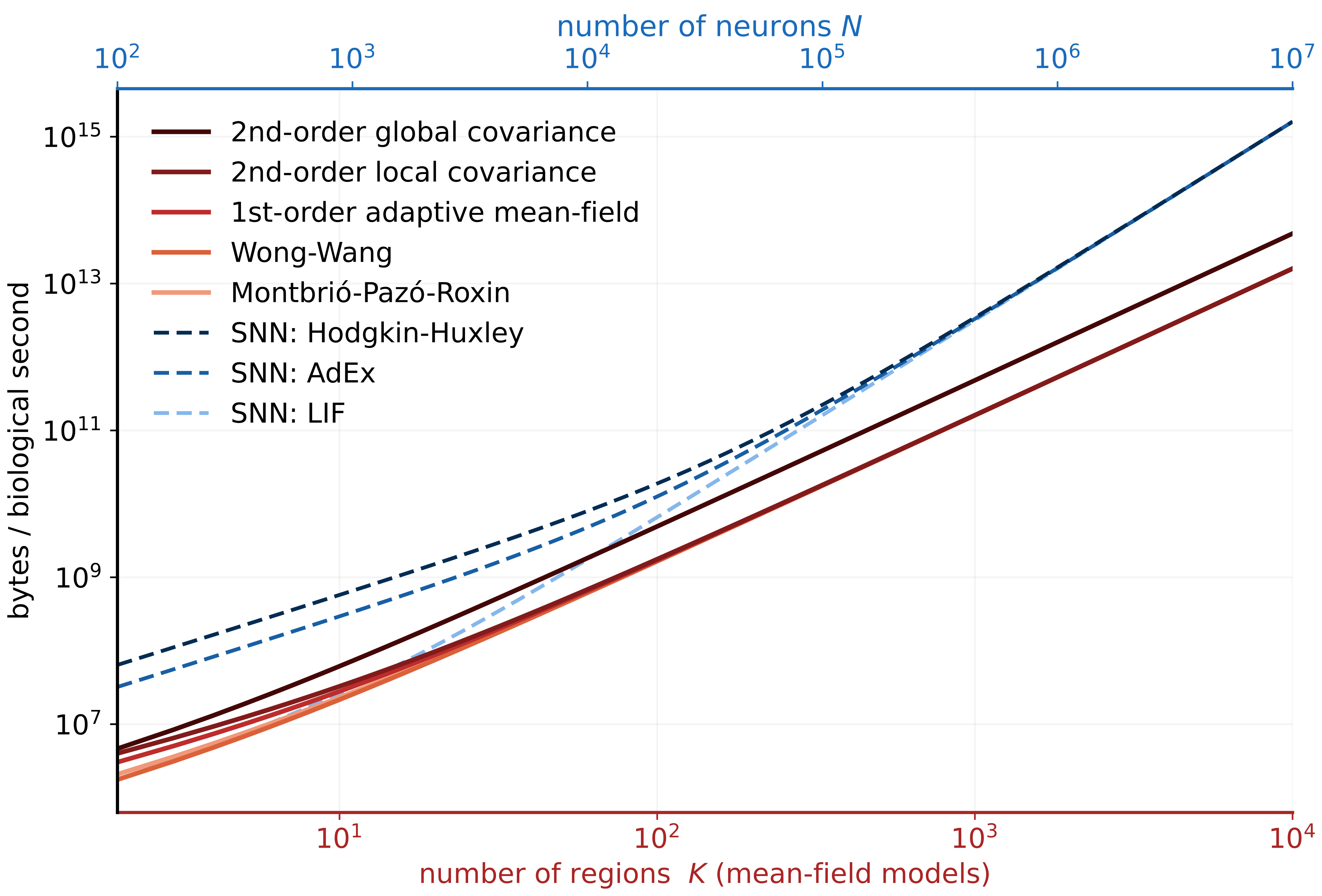}
    \caption{\revision{Estimated source-level array accesses in bytes per simulated biological second, with $P=P_w=2$ and $\rho=1$ for regional models. Red curves use region count $K$; blue dashed curves use neuron count $N$ for flat SNNs. The accounting counts specified state/intermediate accesses, connectivity and delayed-state reads, and synaptic deliveries under Supplementary~\ref{sec:algorithmic_cost_analysis_neural_simulation_appendix}; it does not measure DRAM traffic or memory capacity. The global-covariance curve includes only the state/covariance-access baseline and excludes additional matrix-product and derivative-array accesses. Other material exclusions include solver temporaries, communication, recording and post-processing. At fixed connectivity probability, the large-size quadratic terms arise from regional edge accesses or SNN synaptic deliveries under their respective conventions. The shared display summarizes separate workloads; intersections do not establish equivalent model sizes, runtime break-even points or scientific-efficiency rankings.}  Horizontal positions should be compared only within the same color family; the purpose of the figure is to show the scaling regimes and orders of magnitude rather than a one-to-one equivalence between \(K\) and \(N\).}
    \label{fig:appendix_memory_traffic_P2}
\end{figure}

\begin{figure}[hbt!]
    \centering
    \includegraphics[width=0.8\textwidth]{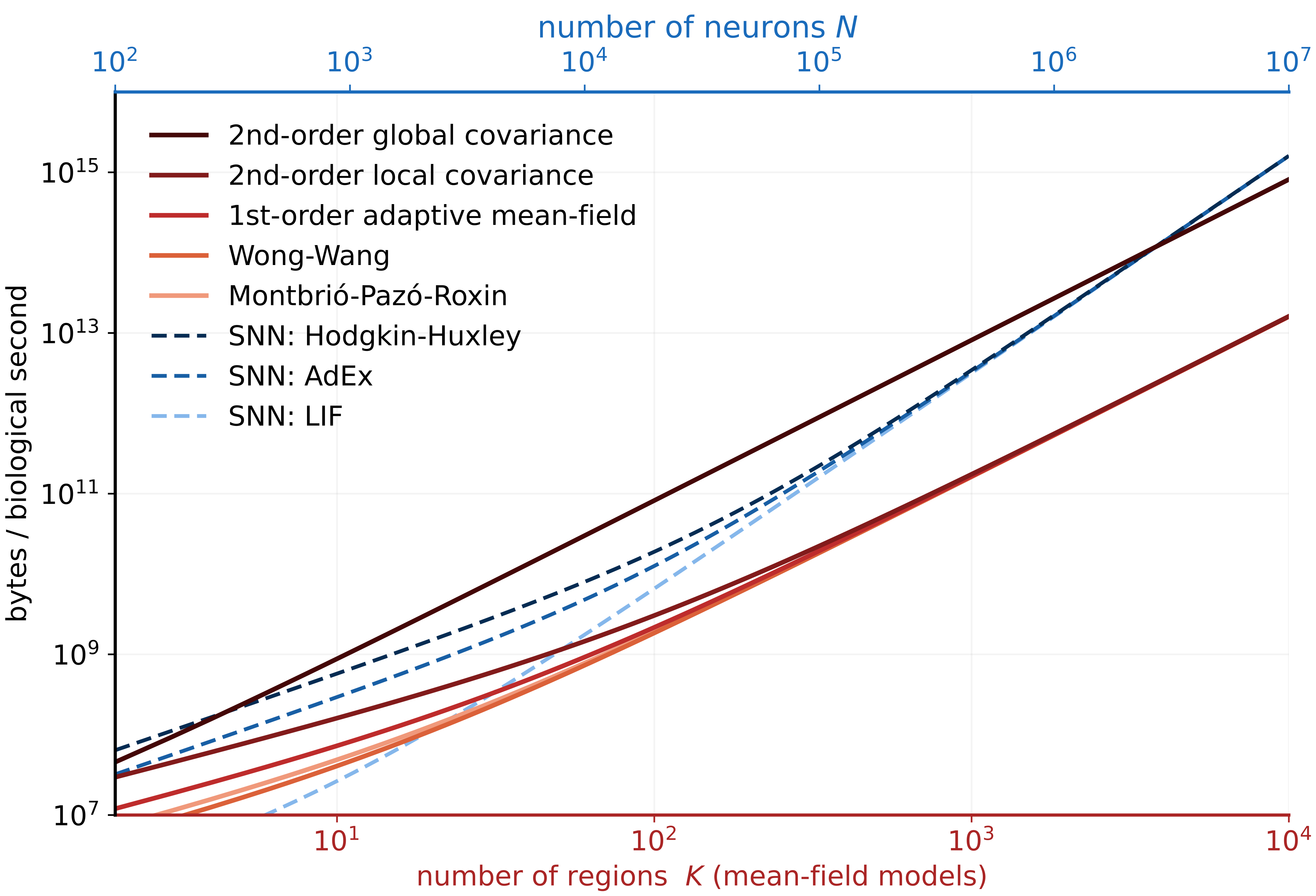}
    \caption{\revision{Estimated source-level array accesses in bytes per simulated biological second, with $P=P_w=10$ and $\rho=1$ for regional models. Red curves are indexed by regions $K$ and blue dashed SNN curves by neurons $N$, under Supplementary~\ref{sec:algorithmic_cost_analysis_neural_simulation_appendix}. Increasing $P$ enlarges regional local-state accesses; the global rate-covariance baseline additionally contains $\mathcal{O}(P^2K^2)$ entries read and written per step. That curve omits the additional matrix-product and derivative-array accesses needed for global covariance propagation. These accounting scenarios are neither measured DRAM traffic nor memory-capacity estimates; solver temporaries, communication, recording and post-processing are also excluded. The SNN scenarios are unchanged from $P=2$. Each family must be read against its own size axis: horizontal alignment and curve intersections do not establish biological equivalence or runtime advantage.} Horizontal positions should be compared only within the same color family; the purpose of the figure is to show the scaling regimes and orders of magnitude rather than a one-to-one equivalence between \(K\) and \(N\).}
    \label{fig:appendix_memory_traffic_P10}
\end{figure}

\newpage

\end{document}